\documentclass[rmp,aps,reprint,twocolumn,floatfix]{revtex4}
\usepackage{longtable}
\usepackage{dcolumn}
\usepackage{epsf}
\usepackage{bm}
\usepackage{amsmath}
\usepackage{graphicx}
\usepackage{xr}
\usepackage{afterpage}
\begin{document}
\newcommand{\greeksym}[1]{{\usefont{U}{psy}{m}{n}#1}}
\newcommand{\rmssmu}{\mbox{\scriptsize{\greeksym{m}}}}
\newcommand{\rmsstau}{\mbox{\scriptsize{\greeksym{t}}}}
\newcommand{\rmgamma}{\mbox{\greeksym{g}}}
\newcommand{\rmssgamma}{\mbox{\scriptsize{\greeksym{g}}}}
\newcommand{\rmmu}{\mbox{\greeksym{m}}}
\newcommand{\rmtau}{\mbox{\greeksym{t}}}
\newcommand{\rmalpha}{\mbox{\greeksym{a}}}
\newcommand{\rmssalpha}{\mbox{\scriptsize{\greeksym{a}}}}
\newcommand{\rmpi}{\mbox{\greeksym{p}}}
\newcommand{\rmrho}{\mbox{\greeksym{r}}}
\newcommand{\rmsigma}{\mbox{\greeksym{s}}}
\newcommand{\rmsspi}{\mbox{\scriptsize{\greeksym{p}}}}
\newcommand{\rmphi}{\mbox{\greeksym{f}}}
\newcommand\T{\rule{0pt}{2.4ex}}
\newcommand\B{\rule[-0.8ex]{0pt}{0pt}}
\newcommand{\s}[1]{\hbox to #1 pt {}}
\def\lbar{\lambda\hskip-4.5pt\vrule height4.6pt depth-4.3pt width4pt}
\def\moles{$^{\rm o}\hskip-4.85pt\vrule height4.98pt depth-4.89pt
  width4.4pt\hskip4.85pt$}
\def\fr#1#2{{\textstyle{#1\over#2}}}
\def\nothing#1{\phantom{#1}}
\title{
       CODATA Recommended Values of the Fundamental Physical
       Constants: 2010\footnote{
  This report was prepared by the authors under the auspices of
  the CODATA Task Group on Fundamental Constants.
  The members of the task group are: \\
  F. Cabiati, Istituto Nazionale di Ricerca Metrologica, Italy \\
  J. Fischer, Physikalisch-Technische Bundesanstalt, Germany \\
  J. Flowers, National Physical Laboratory, United Kingdom \\
  K. Fujii, National Metrology Institute of Japan, Japan \\
  S. G. Karshenboim, Pulkovo Observatory, Russian Federation \\
  P. J. Mohr, National Institute of Standards and Technology, United States of
  America \\
  D. B. Newell, National Institute of Standards and Technology, United States of
  America \\
  F. Nez, Laboratoire Kastler-Brossel, France \\
  K. Pachucki, University of Warsaw, Poland \\
  T. J. Quinn, Bureau international des poids et mesures \\
  B. N. Taylor, National Institute of Standards and Technology, United States of
  America \\
  B. M. Wood, National Research Council, Canada \\
  Z. Zhang, National Institute of Metrology, China (People's Republic of) \\
  }
  }
\author{
Peter J. Mohr\footnote{Electronic address: mohr@nist.gov},
Barry N. Taylor\footnote{Electronic address: barry.taylor@nist.gov}, and
David B. Newell\footnote{Electronic address: dnewell@nist.gov},}
\affiliation{
    National Institute of Standards and Technology,
    Gaithersburg, Maryland 20899-8420, USA
   }

\date{\today}

\begin{abstract}
This paper gives the 2010 self-consistent set of values of the basic
constants and conversion factors of physics and chemistry recommended by
the Committee on Data for Science and Technology (CODATA) for
international use.  The 2010 adjustment takes into account the data
considered in the 2006 adjustment as well as the data that became
available from 1 January 2007, after the closing date of that
adjustment, until 31 December 2010, the closing date of the new
adjustment.  Further, it describes in detail the adjustment of the
values of the constants, including the selection of the final set of
input data based on the results of least-squares analyses.  The 2010 set
replaces the previously recommended 2006 CODATA set and may also be
found on the World Wide Web at physics.nist.gov/constants.
\end{abstract}
\maketitle

\tableofcontents
\vbox to 3cm {}

%
\section{Introduction}
\label{sec:intro}

\subsection{Background}
\label{ssec:bk}

This article reports work carried out under the auspices of the
Committee on Data for Science and Technology (CODATA) Task Group on
Fundamental Constants.\footnote{CODATA was established in 1966 as an
interdisciplinary committee of the International Council of Science.
The Task Group was founded 3 years later.}  It describes in detail the
CODATA 2010 least-squares adjustment of the values of the constants, for
which the closing date for new data was 31 December 2010.  Equally
important, it gives the 2010 self-consistent set of over 300 CODATA
recommended values of the fundamental physical constants based on the
2010 adjustment.  The 2010 set, which replaces its immediate predecessor
resulting from the CODATA 2006 adjustment \cite{2008145}, first became
available on 2 June 2011 at physics.nist.gov/constants, a Web
site of the NIST Fundamental Constants Data Center (FCDC).

The World Wide Web has engendered a sea change in expectations regarding
the availability of timely information.  Further, in recent years new
data that influence our knowledge of the values of the constants seem to
appear almost continuously.  As a consequence, the Task Group decided at
the time of the 1998 CODATA adjustment to take advantage of the
extensive computerization that had been incorporated in that effort to
issue a new set of recommended values every 4 years; in the era of the
Web, the 12-13 years between the first CODATA set of 1973 \cite{1973003}
and the second CODATA set of 1986 \cite{1987004}, and between this
second set and the third set of 1998 \cite{2000035}, could no longer be
tolerated.  Thus, if the 1998 set is counted as the first of the new
4-year cycle, the 2010 set is the 4th of that cycle.

Throughout this article we refer to the detailed reports describing the
1998, 2002, and 2006 adjustments as CODATA-98, CODATA-02, and CODATA-06,
respectively \cite{2000035,2005191,2008145}.  To keep the paper to a
reasonable length, our data review focuses on the new results that
became available between the 31 December 2006 and 31 December 2010
closing dates of the 2006 and 2010 adjustments; the reader should
consult these past reports for detailed discussions of the older data.
These past reports should also be consulted for discussions of
motivation, philosophy, the treatment of numerical calculations and
uncertainties, etc.  A rather complete list of acronyms and symbols may
be found in the nomenclature section near the end of the paper.

To further achieve a reduction in the length of this report compared to
the lengths of its three most recent predecessors, it has been decided
to omit extensive descriptions of new experiments and calculations and
to comment only on their most pertinent features; the original
references should be consulted for details.  For the same reason,
sometimes the older data used in the 2010 adjustment are not given in
the portion of the paper that discusses the data by category, but are
given in the portion of the paper devoted to data analysis.  For
example, the actual values of the 16 older items of input data recalled
in Sec.~\ref{sec:elmeas} are given only in Sec.~\ref{sec:ad}, rather
than in both sections as done in previous adjustment reports.

As in all previous CODATA adjustments, as a working principle, the
validity of the physical theory underlying the 2010 adjustment is
assumed.  This includes special relativity, quantum mechanics, quantum
electrodynamics (QED), the standard model of particle physics, including
$CPT$ invariance, and the exactness (for all practical
purposes--see Sec.~\ref{sec:elmeas}) of the relationships between the
Josephson and von Klitzing constants $K_{\rm J}$ and $R_{\rm K}$ and the
elementary charge $e$ and Planck constant $h$, namely, $K_{\rm
J} = 2e/h$ and $R_{\rm K} = h/e^2$.

Although the possible time variation of the constants continues to be an
active field of both experimental and theoretical research, there is no
observed variation relevant to the data on which the 2010 recommended
values are based; see, for example, the recent reviews by 
\citet{2011109} and \citet{2011229}.  Other
references may be found in the FCDC bibliographic database at
physics.nist.gov/constantsbib using, for example, the keywords
``time variation'' or ``constants.''

With regard to the 31 December closing date for new data, a datum was
considered to have met this date if the Task Group received a preprint
describing the work by that date and the preprint had already been, or
shortly would be, submitted for publication.  Although results are
identified by the year in which they were published in an archival
journal, it can be safely assumed that any input datum labeled with an
``11'' or ``12'' identifier was in fact available by the closing date.
However, the 31 December 2010 closing date does not apply to clarifying
information requested from authors; indeed, such information was
received up to shortly before 2 June 2011, the date the new values were
posted on the FCDC Web site.  This is the reason that some “private
communications” have 2011 dates.

\subsection{Brief overview of CODATA 2010 adjustment}

The 2010 set of recommended values is the result of applying the same
procedures as in previous adjustments and is based on a least-squares
adjustment with, in this case, $N = 160$ items of input data, $M = 83$
variables called {\it adjusted constants}, and $\nu = N-M = 77$ degrees
of freedom.  The statistic ``chi-squared'' is $\chi^2 = 59.1$ with
probability $p(\chi^2|\nu) = 0.94$ and Birge ratio $R_{\rm B} = 0.88$.

A significant number of new results became available for consideration,
both experimental and theoretical, from 1 January 2007, after the
closing date of the 2006 adjustment, to 31 December 2010, the closing
date of the current adjustment.  Data that affect the determination of
the fine-structure constant $\alpha$, Planck constant $h$, molar gas
constant $R$, Newtonian constant of gravitation $G$, Rydberg constant
$R_\infty$, and rms proton charge radius $r_{\rm p}$ are the focus of
this brief overview, because of their inherent importance and, in the
case of $\alpha$, $h$, and $R$, their impact on the determination of the
values of many other constants.  (Constants that are not among the
directly adjusted constants are calculated from appropriate combinations
of those that are directly adjusted.)

\subsubsection{Fine-structure constant $\alpha$}

An improved measurement of the electron magnetic moment anomaly $a_{\rm
e}$, the discovery and correction of an error in its theoretical
expression, and an improved measurement of the quotient $h/m(^{87}{\rm
Rb})$ have led to a 2010 value of $\alpha$ with a relative standard
uncertainty of $3.2 \times 10^{-10}$ compared to $6.8 \times 10^{-10}$
for the 2006 value.  Of more significance, because of the correction of
the error in the theory, the 2010 value of $\alpha$ shifted
significantly and now is larger than the 2006 value by $6.5$ times the
uncertainty of that value.  This change has rather profound
consequences, because many constants depend on $\alpha$, for example,
the molar Planck constant $N_{\rm A}h$.

\subsubsection{Planck constant $h$}

A new value of the Avogadro constant $N_{\rm A}$ with a relative
uncertainty of $3.0 \times 10^{-8}$ obtained from highly enriched
silicon with amount of substance fraction $x(^{28}{\rm
Si}) \approx 0.999\,96$ replaces the 2006 value based on natural silicon
and provides an inferred value of $h$ with essentially the same
uncertainty.  This uncertainty is somewhat smaller than
$3.6 \times 10^{-8}$, the uncertainty of the most accurate directly
measured watt-balance value of $h$.  Because the two values disagree,
the uncertainties used for them in the adjustment were increased by a
factor of two to reduce the inconsistency to an acceptable level; hence
the relative uncertainties of the recommended values of $h$ and $N_{\rm
A}$ are $4.4 \times 10^{-8}$, only slightly smaller than the
uncertainties of the corresponding 2006 values.  The 2010 value of $h$
is larger than the 2006 value by the fractional amount
$9.2 \times 10^{-8}$ while the 2010 value of $N_{\rm A}$ is smaller than
the 2006 value by the fractional amount $8.3 \times 10^{-8}$.  A number
of other constants depend on $h$, for example, the first radiation
constant $c_1$, and consequently the 2010 recommended values of these
constants reflect the change in $h$.

\subsubsection{Molar gas constant $R$}

Four consistent new values of the molar gas constant together with the
two previous consistent values, with which the new values also agree,
have led to a new 2010 recommended value of $R$ with an uncertainty of
$9.1 \times 10^{-7}$ compared to $1.7 \times 10^{-6}$ for the 2006
value.  The 2010 value is smaller than the 2006 value by the fractional
amount $1.2 \times 10^{-6}$ and the relative uncertainty of the 2010
value is a little over half that of the 2006 value.  This shift and
uncertainty reduction is reflected in a number of constants that depend
on $R$, for example, the Boltzmann constant $k$ and the Stefan-Boltzmann
constant $\sigma$.

\subsubsection{Newtonian constant of gravitation $G$}

Two new values of $G$ resulting from two new experiments each with
comparatively small uncertainties but in disagreement with each other
and with earlier measurements with comparable uncertainties led to an
even larger expansion of the a priori assigned uncertainties of the data
for $G$ than was necessary in 2006.  In both cases the expansion was
necessary to reduce the inconsistencies to an acceptable level.  This
increase has resulted in a 20\,\% increase in uncertainty of the 2010
recommended value compared to that of the 2006 value: $12$ parts in
$10^{5}$ vs. $10$ parts in $10^{5}$.  Furthermore, the 2010 recommended
value of $G$ is smaller than the 2006 value by the fractional amount
$6.6 \times 10^{-5}$.

\subsubsection{Rydberg constant $R_\infty$ and proton radius $r_{\rm p}$}

New experimental and theoretical results that have become available in
the past 4 years have led to the reduction in the relative uncertainty
of the recommended value of the Rydberg constant from
$6.6 \times 10^{-12}$ to $5.0 \times 10^{-12}$, and the reduction in
uncertainty from $0.0069$ fm to $0.0051$ fm of the proton rms charge
radius based on spectroscopic and scattering data but not muonic
hydrogen data.  Data from muonic hydrogen, with the assumption that the
muon and electron interact with the proton at short distances in exactly
the same way, are so inconsistent with the
other data that they have not been included in the determination of
$r_{\rm p}$ and thus do not have an influence on $R_\infty$.  The 2010
value of $R_\infty$ exceeds the 2006 value by the fractional amount
$1.1 \times 10^{-12}$ and the 2010 value of $r_{\rm p}$ exceeds the 2006
value by $0.0007$ fm.

\subsection{Outline of the paper}

Section~\ref{sec:squ} briefly recalls some constants that have exact
values in the International System of Units (SI) \cite{bipmsi}, the unit
system used in all CODATA adjustments.
Sections~\ref{sec:ram}-\ref{sec:xeq} discuss the input data with a
strong focus on those results that became available between the 31
December 2006 and 31 December 2010 closing dates of the 2006 and 2010
adjustments.  It should be recalled (see especially Appendix E of
CODATA-98) that in a least-squares analysis of the constants, both the
experimental and theoretical numerical data, also called observational
data or input data, are expressed as functions of a set of independent
variables called directly adjusted constants (or sometimes simply
adjusted constants).  The functions themselves are called observational
equations, and the least-squares procedure provides best estimates, in
the least-squares sense, of the adjusted constants.  In essence, the
procedure determines the best estimate of a particular adjusted constant
by automatically taking into account all possible ways of determining
its value from the input data.  As already noted, the recommended values
of those constants not directly adjusted are calculated from the
adjusted constants.

Section~\ref{sec:ad} describes the analysis of the data.  The analysis
includes comparison of measured values of the same quantity, measured
values of different quantities through inferred values of another
quantity such as $\alpha$ or $h$, and by the method of least-squares.
The final input data used to determine the adjusted constants, and hence
the entire 2010 CODATA set of recommended values, are based on these
investigations.

Section~\ref{sec:2010crv} provides, in several tables, the set of over
300 recommended values of the basic constants and conversion factors of
physics and chemistry, including the covariance matrix of a selected
group of constants.  Section~\ref{sec:c} concludes the report with a
comparison of a small representative subset of 2010 recommended values
with their 2006 counterparts, comments on some of the more important
implications of the 2010 adjustment for metrology and physics, and
suggestions for future experimental and theoretical work that will
improve our knowledge of the values of the constants.  Also touched upon
is the potential importance of this work and that of the next CODATA
constants adjustment (expected 31 December 2014 closing date) for the
redefinition of the kilogram, ampere, kelvin, and mole currently under
discussion internationally \cite{2011191}.

\section{Special quantities and units}
\label{sec:squ}

As a consequence of the SI definitions of the meter, the ampere, and the
mole, $c$, $\mu_0$ and $\epsilon_0$, and $M(^{12}$C) and $M_{\rm u}$,
have exact values; see Table~\ref{tab:exact}.  Since the relative atomic
mass $A_{\rm r}(X)$ of an entity $X$ is defined by $A_{\rm r}(X) =
m(X)/m_{\rm u}$, where $m(X)$ is the mass of $X$, and the (unified)
atomic mass constant $m_u$ is defined according to $m_{\rm u}$ =
$m(^{12}{\rm C})/12$, $A_{\rm r}(^{12}{\rm C}) = 12$ exactly, as shown
in the table.  Since the number of specified entities in one mole is
equal to the numerical value of the Avogadro constant $N_{\rm A} \approx
6.022 \times 10^{23}$/mol, it follows that the molar mass of an entity
$X$, $M(X)$, is given by $M(X) = N_{\rm A}m(X) = A_{\rm r}(X)M_{\rm u}$
and $M_{\rm u} = N_{\rm A}m_{\rm u}$.  The (unified) atomic mass unit u
(also called the dalton, Da), is defined as 1 u $= m_{\rm u} \approx
1.66 \times 10^{-27}$ kg.  The last two entries in
Table~\ref{tab:exact}, $K_{\rm J-90}$ and $R_{\rm K-90}$, are the
conventional values of the Josephson and von Klitzing constants
introduced on 1 January 1990 by the International Committee for Weights
and Measures (CIPM) to foster worldwide uniformity in the measurement of
electrical quantities.  In this paper, those electrical quantities
measured in terms of the Josephson and quantum Hall effects with the
assumption that $K_{\rm J}$ and $R_{\rm K}$ have these conventional
values are labeled with a subscript 90.

Measurements of the quantity $K_{\rm J}^2R_{\rm K} = 4/h$ using a moving
coil watt balance (see Sec.~\ref{sec:elmeas}) require the determination
of the local acceleration of free fall $g$ at the site of the balance
with a relative uncertainty of a few parts in $10^9$.  That currently
available absolute gravimeters can achieve such an uncertainty if
properly used has been demonstrated by comparing different instruments
at essentially the same location.  An important example is the periodic
international comparison of absolute gravimeters (ICAG) carried out at
the International Bureau of Weights and Measures (BIPM), S\`evres,
France \cite{2011112}.  The good agreement obtained between a commercial
optical interferometer-based gravimeter that is in wide use and a cold
atom, atomic interferometer-based instrument  also provides evidence
that the claimed uncertainties of determinations of $g$ are realistic
\cite{2010132}.  However, not all gravimeter comparisons have obtained
such satisfactory results \cite{2011113}.  Additional work in this area
may be needed when the relative uncertainties of watt-balance
experiments reach the level of 1 part in $10^8$.

\def\vsp#1{\noalign{\vbox to #1 pt {}}}
\def\hsp{\hbox to 44pt{}}
\begin{table*}
\caption{Some exact quantities relevant to the 2010 adjustment.}
\label{tab:exact}
\begin{tabular}{l@{\hsp}l@{\hsp}l}
\toprule
\vbox to 10 pt {}
Quantity & Symbol
& Value \\
\colrule
\vsp{3}
speed of light in vacuum
&$c$, $c_0$
&$299\,792\,458 \ {\rm m \  s^{-1}}$ \\
magnetic constant
&$\mu_0$
& $4\rmpi\times10^{-7} \  {\rm N \  A^{-2}}$
$= 12.566\,370\,614... \times 10 ^{-7} \ {\rm N \  A^{-2}}$ \\
electric constant
&$\epsilon_0$
&$(\mu_0c^2)^{-1} $
$=8.854\,187\,817... \ \times10^{-12} \ {\rm F \  m^{-1}}$ \\
molar mass of $^{12}$C
&$M(^{12}{\rm C})$
& $12\times10^{-3} \ {\rm kg \  mol^{-1}}$ \\
molar mass constant
&$M_{\rm u}$
& $10^{-3} \ {\rm kg \  mol^{-1}}$ \\
relative atomic mass of $^{12}$C
&$A_{\rm r}(^{12}{\rm C})$
& $12$ \\
conventional value of Josephson constant
&$K_{{\rm J}-90}$
& $483\,597.9 \ {\rm GHz \  V^{-1}}$ \\
conventional value of von Klitzing constant
&$R_{{\rm K}-90}$
& $25\,812.807 \ {\rm \Omega}$ \\
\botrule
\end{tabular}
\end{table*}

\section{Relative atomic masses}
\label{sec:ram}

The directly adjusted constants include the relative atomic masses
$A_{\rm r}(X)$ of a number of particles, atoms, and ions.  Further,
values of $A_{\rm r}(X)$ of various atoms enter the calculations of
several potential input data.  The following sections and Tables
~\ref{tab:rmass03} to ~\ref{tab:rmcovvd} summarize the relevant
information.

\begin{table}[t]
\caption{Values of the relative atomic masses of
the neutron and
various atoms as given in the
2003 atomic mass evaluation
together with the defined value for $^{12}$C.}
\label{tab:rmass03}
\begin{tabular}{cD{.}{.}{8.20}l}
\toprule
\vbox to 10pt{}
Atom     &  \multicolumn{1}{c}{Relative atomic}
& Relative standard \\
& \multicolumn{1}{c}{\text{mass $A_{\rm r}({\rm X})$}}
& \hskip 8pt uncertainty $u_{\rm r}$  \\
\colrule
n \vbox to 10pt{} &   1.008\,664\,915\,74(56) & \hskip 16pt $ 5.6\times 10^{-10}
$ \\
$^{1}$H              &   1.007\,825\,032\,07(10) & \hskip 16pt $ 1.0\times
10^{-10}$ \\
$^{2}$H              &    2.014\,101\,777\,85(36)    & \hskip 16pt $ 1.8\times
10^{-10}$    \\
$^{3}$H           &    3.016\,049\,2777(25)   & \hskip 16pt $ 8.2\times 10^{-10}
$    \\
$^{3}$He            &    3.016\,029\,3191(26)   & \hskip 16pt $ 8.6\times
10^{-10}$   \\
$^{4}$He            &    4.002\,603\,254\,153(63)   & \hskip 16pt $ 1.6\times
10^{-11}$   \\
$^{12}$C            &   12             & \hskip 16pt    (exact)          \\
$^{16}$O            &    15.994\,914\,619\,56(16)   & \hskip 16pt $ 1.0\times
10^{-11}$   \\
$^{28}$Si           &    27.976\,926\,5325(19)  & \hskip 16pt $ 6.9\times
10^{-11}$  \\
$^{29}$Si           &    28.976\,494\,700(22)  & \hskip 16pt $ 7.6\times
10^{-10}$  \\
$^{30}$Si           &    29.973\,770\,171(32)  & \hskip 16pt $ 1.1\times 10^{-9}
$  \\
$^{36}$Ar           &    35.967\,545\,105(28)  & \hskip 16pt $ 7.8\times
10^{-10}$  \\
$^{38}$Ar           &    37.962\,732\,39(36)  & \hskip 16pt $ 9.5\times 10^{-9}$
\\
$^{40}$Ar           &    39.962\,383\,1225(29)  & \hskip 16pt $ 7.2\times
10^{-11}$  \\
$^{87}$Rb           &    86.909\,180\,526(12)  & \hskip 16pt $ 1.4\times
10^{-10}$  \\
$^{107}$Ag        &    106.905\,0968(46) & \hskip 16pt $ 4.3\times 10^{-8}$ \\
$^{109}$Ag        &    108.904\,7523(31) & \hskip 16pt $ 2.9\times 10^{-8}$ \\
$^{133}$Cs        &    132.905\,451\,932(24) & \hskip 16pt $ 1.8\times 10^{-10}$
\\
\botrule
\end{tabular}
\end{table}

\begin{table}[t]
\caption{Values of the relative atomic masses of
various atoms that have become
available since the 2003 atomic mass evaluation.}
\label{tab:rmass10}
\begin{tabular}{cD{.}{.}{8.20}l}
\toprule
\vbox to 10pt{}
Atom     &   \multicolumn{1}{c}{Relative atomic}
& Relative standard \\
& \multicolumn{1}{c}{mass $A_{\rm r}(X)$}  & \hskip 8pt uncertainty $u_{\rm r}$
\\
\colrule
$^{2}$H         &    2.014\,101\,778\,040(80)    & \hskip 16pt $ 4.0\times
10^{-11}$  \vbox to 10pt{} \\
$^{4}$He         &    4.002\,603\,254\,131(62)    & \hskip 16pt $ 1.5\times
10^{-11}$   \\
$^{16}$O         &    15.994\,914\,619\,57(18)    & \hskip 16pt $ 1.1\times
10^{-11}$   \\
$^{28}$Si         &    27.976\,926\,534\,96(62)   & \hskip 16pt $ 2.2\times
10^{-11}$   \\
$^{29}$Si         &    28.976\,494\,6625(20)   & \hskip 16pt $ 6.9\times
10^{-11}$   \\
$^{87}$Rb         &    86.909\,180\,535(10)   & \hskip 16pt $ 1.2\times 10^{-10}
$   \\
$^{133}$Cs         &    132.905\,451\,963(13)   & \hskip 16pt $ 9.8\times
10^{-11}$   \\
\botrule
\end{tabular}
\end{table}

\subsection{Relative atomic masses of atoms}
\label{ssec:rama}

Table~\ref{tab:rmass03}, which is identical to Table II in CODATA-06, gives
values of $A_{\rm r}(X)$ taken from the 2003 atomic mass evaluation
(AME2003) carried out by the Atomic Mass Data Center (AMDC), Centre de
Spectrom\'etrie Nucl\'eaire et de Spectrom\'etrie de Masse (CSNMS),
Orsay, France \cite{2003252,2003253,amdc06}. However, not all of
these values are actually used in the adjustment; some are given for
comparison purposes only. Although these values are correlated to a
certain extent, the only correlation  that needs to be taken into
account in the current adjustment is that between $A_{\rm r}(^{1}{\rm
H})$ and $A_{\rm r}(^{2}{\rm H})$; their correlation coefficient is
0.0735 \cite{awcovs03}.

Table~\ref{tab:rmass10} lists seven values of $A_{\rm r}(X)$ relevant to
the 2010 adjustment obtained since the publication of ASME2003. It is
the updated version of Table IV discussed in CODATA-06. The changes made
are the deletion of the $^{3}{\rm H}$ and $^{3}{\rm He}$ values obtained
by the SMILETRAP group at Stockholm University (StockU), Sweden; and the
inclusion of values for $^{28}{\rm Si}$, $^{87}{\rm Rb}$, and
$^{133}{\rm Cs}$ obtained by the group at Florida State University
(FSU), Tallahassee, FL, USA \cite{2008038,2010182}. This group uses
the method initially developed at the Massachusetts Institute of
Technology, Cambridge, MA, USA \cite{2005195}.  In the MIT approach,
which eliminates or reduces a number of systematic effects and their
associated uncertainties, mass ratios are determined by directly
comparing the cyclotron frequencies of two different ions simultaneously
confined in a Penning trap. (The  value of $A_{\rm r}(^{29}{\rm Si})$ in
Table ~\ref{tab:rmass10} is given in the supplementary information of
the last cited reference. The MIT atomic mass work was transferred to
FSU a number of years ago.)

The deleted SMILETRAP results are not discarded but are included in the
adjustment in a more fundamental way, as described in
Sec.~\ref{ssec:smtr}. The values of $A_{\rm r}(^{2}{\rm H})$, $A_{\rm
r}(^{4}{\rm He})$, and $A_{\rm r}(^{16}{\rm O})$ in Table
~\ref{tab:rmass10} were obtained by the University of Washington (UWash)
group, Seattle, WA, USA and were used in the 2006 adjustment.  The three
values are correlated and their variances, covariances, and correlation
coefficients are given in Table ~\ref{tab:rmcovvd}, which is identical
to Table IV in CODATA-06

\begin{table}
\caption{The variances, covariances, and correlation coefficients of the
University of Washington values of the relative atomic masses of
deuterium, helium 4, and oxygen 16.  The numbers in bold above the main
diagonal are $10^{20}$ times the numerical values of the covariances;
the numbers in bold on the main diagonal are $10^{20}$ times the
numerical values of the variances; and the numbers in italics below the
main diagonal are the correlation coefficients.}
\label{tab:rmcovvd}
\def\hsp{\hbox to 26pt{}}
\begin{tabular} {l@{\hsp}|@{\hsp}r@{\hsp}r@{\hsp}r}
\toprule
\vbox to 10 pt {}
& $A_{\rm r}(^2{\rm H})$ \hbox to 0.5pt{}
& $A_{\rm r}(^4{\rm He})$ \hbox to 0.5pt{}
& $A_{\rm r}(^{16}{\rm O})$ \hbox to 0.5pt{}  \\
\colrule
\vbox to 10 pt {}
$A_{\rm r}(^2{\rm H})$ & ${\bf  0.6400}$ & ${\bf  0.0631}$ & ${\bf  0.1276}$ \\
$A_{\rm r}(^4{\rm He})$  & ${\it  0.1271}$ & ${\bf  0.3844}$ & ${\bf  0.2023}$
\\
$A_{\rm r}(^{16}{\rm O})$  & ${\it  0.0886}$ & ${\it  0.1813}$ & ${\bf  3.2400}$
\\
\botrule
\end{tabular}
\end{table}

The values of $A_{\rm r}(X)$ from Table ~\ref{tab:rmass03} initially used
as input data for the 2010 adjustment are  $A_{\rm r}(^{1}{\rm H})$,
$A_{\rm r}(^{2}{\rm H})$, $A_{\rm r}(^{87}{\rm Rb})$, and $A_{\rm
r}(^{133}{\rm Cs})$; and from Table ~\ref{tab:rmass10}, $A_{\rm
r}(^{2}{\rm H})$, $A_{\rm r}(^{4}{\rm He})$, $A_{\rm r}(^{16}{\rm O})$,
$A_{\rm r}(^{87}{\rm Rb})$, and $A_{\rm r}(^{133}{\rm Cs})$.  These
values are items $B1$, $B2.1$, $B2.2$, and $B7$ to $B10.2$ in Table
~\ref{tab:pdata}, Sec.~\ref{sec:ad}.  As in the 2006 adjustment, the
ASME2003 values for $A_{\rm r}(^{3}{\rm H})$, and $A_{\rm r}(^{3}{\rm
He})$ in Table  ~\ref{tab:rmass03} are not used because they were
influenced by an earlier $^{3}{\rm He}$ result of the UWash group that
disagrees with their newer, more accurate result \cite{pcvd2010}.
Although not yet published, it can be said that it agrees well with the
value from the SMILETRAP group; see Sec.~\ref{ssec:smtr}.

Also as in the 2006 adjustment, the UWash group's values for $A_{\rm
r}(^{4}{\rm He})$ and $A_{\rm r}(^{16}{\rm O})$ in  Table
~\ref{tab:rmass10} are used in place of the corresponding ASME2003
values in Table ~\ref{tab:rmass03} because the latter are based on a
preliminary analysis of the data while those in Table ~\ref{tab:rmass10}
are based on a thorough reanalysis of the data  \cite{2006036}.

Finally, we note that the $A_{\rm r}(^{2}{\rm H})$ value of the UWash
group in  Table ~\ref{tab:rmass10} is the same as used in the 2006
adjustment. As discussed in CODATA-06, it is a near-final result with a
conservatively assigned uncertainty based on the analysis of 10 runs
taken over a 4-year period privately communicated to the Task Group in
2006 by R. S. Van Dyck. A final result completely consistent with it
based on the analysis of 11 runs but with an uncertainty of about half
that given in the table should be published in due course together with
the final result for $A_{\rm r}(^{3}{\rm He})$ \cite{pcvd2010}.

\subsection{Relative atomic masses of ions and nuclei}
\label{ssec:ramnuc}

For a neutral atom $X$, $A_{\rm r}(X)$ can be expressed in terms of
$A_{\rm r}$ of an ion of the atom formed by the removal of $n$ electrons
according to
\begin{eqnarray}
A_{\rm \rm r} (X)
&=&A_{\rm r}(X^{n+}) + n A_{\rm r} ({\rm e})
\nonumber\\ &&
- \frac{E_{\rm b}(X)
- E_{\rm b}(X^{n+})}{m_{\rm u} c^2} \ .
\label{eq:araxn}
\end{eqnarray}
In this expression, $E_{\rm b}(X)/m_{\rm u}c^{2}$ is the
relative-atomic-mass equivalent of the total binding energy of the $Z$
electrons of the atom and $Z$ is the atom's atomic number (proton
number). Similarly, $E_{\rm b}(X^{n+})/m_{\rm u}c^{2}$ is the
relative-atomic-mass equivalent of the binding energy of the $Z-n$
electrons of the $X^{n+}$ ion. For an ion that is fully stripped $n = Z$
and $X^{Z+}$ is simply $N$, the nucleus of the atom. In this case
$E_{\rm b}(X^{Z+})/m_{\rm u}c^{2} = 0$ and Eq.~(\ref{eq:araxn}) becomes
of the form of the first two equations of Table~\ref{tab:pobseqsb1},
Sec.~\ref{sec:ad}.

The binding energies $E_{\rm b}$ employed in the 2010 adjustment are the
same as those used in that of 2002 and 2006; see Table IV of CODATA-02.
As noted in CODATA-06, the binding energy for tritium, $^{3}{\rm H}$, is
not included in that table.  We employ the value used in the 2006
adjustment, $1.097\,185\,439\times10^{7}~{\rm m^{-1}}$,  due to
\citet{pcsk2006}.  For our purposes here, the uncertainties of the
binding energies are negligible.

\subsection{Relative atomic masses of the proton, triton, and helion}
\label{ssec:smtr}

The focus of this section is the cyclotron frequency ratio measurements
of the SMILETRAP group that lead to values of $A_{\rm r}({\rm p})$,
$A_{\rm r}({\rm t})$, and $A_{\rm r}({\rm h})$, where the triton t and
helion h are the nuclei of $^{3}{\rm H}$ and $^{3}{\rm He}$. As noted in
Sec.~\ref{ssec:rama} above, the reported values of \citet{2006032} for
$A_{\rm r}(^{3}{\rm H})$ and $A_{\rm r}(^{3}{\rm He})$ were used as
input data in the 2006 adjustment but are not used in this adjustment.
Instead, the actual cyclotron frequency ratio results underlying those
values are used as input data.  This more fundamental way of handling
the SMILETRAP group's results is motivated by the similar but more
recent work of the group related to the proton, which we discuss before
considering the earlier work.

\citet{2008161} used the Penning-trap mass spectrometer SMILETRAP,
described in detail by  \citet{2002157}, to measure the ratio of the
cyclotron frequency $f_{\rm c}$ of the ${\rm H_2}^{+*}$ molecular ion to
that of the deuteron d, the nucleus of the $^{2}{\rm H}$ atom. (The
cyclotron frequency of an ion of charge $q$ and mass $m$ in a magnetic
flux density $B$ is given by $f_{\rm c}=qB/2\rmpi{m}$.) Here the
asterisk indicates that the singly ionized ${\rm H_2}$ molecules are in
excited vibrational states as a result of the 3.4 keV electrons used to
bombard neutral ${\rm H_2}$ molecules in their vibrational ground state
in order to ionize them.  The reported result is
\begin{eqnarray}
\frac{f_{\rm c}({\rm H}_2^{+*})}{f_{\rm c}({\rm d})} &=&
0.999\,231\,659\,33(17)
\qquad [ 1.7\times 10^{-10}] \, .  \qquad
\label{eq:rSMTp08}
\end{eqnarray}

This value was obtained using a two-pulse Ramsey technique to excite the
cyclotron frequencies, thereby enabling a more precise determination of
the cyclotron resonance frequency line-center than was possible with the
one-pulse excitation used in earlier work \cite{2007189,2007339}. The
uncertainty is essentially all statistical; components of uncertainty
from systematic effects such as ``$q/A$ asymmetry'' (difference of
charge-to-mass ratio of the two ions), time variation of the 4.7 T
applied magnetic flux density, relativistic mass increase, and ion-ion
interactions were deemed negligible by comparison.

The frequency ratio $f_{\rm c}({\rm H_2}^{+*})/{f_{\rm c}(\rm d)}$ can
be expressed in terms of adjusted constants and ionization and binding
energies that have negligible uncertainties in this context. Based on
Sec.~\ref{ssec:ramnuc} we can write
\begin{eqnarray}
A_{\rm r}({\rm H}_2) &=& 2A_{\rm r}({\rm H})
- E_{\rm B}({\rm H}_2)/m_{\rm u} c^2 \, ,
\\[6 pt]
A_{\rm r}({\rm H}) &=& A_{\rm r}({\rm p}) + A_{\rm r}({\rm e})
- E_{\rm I}({\rm H})/m_{\rm u} c^2 \, ,
\\[6 pt]
A_{\rm r}({\rm H}_2) &=& A_{\rm r}({\rm H}_2^+) + A_{\rm r}({\rm e})
- E_{\rm I}({\rm H}_2)/m_{\rm u} c^2 \, ,
\\[6 pt]
A_{\rm r}({\rm H}_2^{+*}) &=& A_{\rm r}({\rm H}_2^+)
+ E_{\rm av}/m_{\rm u} c^2 \, ,
\end{eqnarray}
which yields
\begin{eqnarray}
A_{\rm r}({\rm H}_2^{+*}) &=& 2A_{\rm r}({\rm p}) + A_{\rm r}({\rm e}) -
E_{\rm B}({\rm H}_2^{+*})/m_{\rm u} c^2
\label{eq:arhrelat} \, ,
\end{eqnarray}
where
\begin{eqnarray}
E_{\rm B}({\rm H}_2^{+*}) &=& 2E_{\rm I}({\rm H}) + E_{\rm B}({\rm H}_2)
- E_{\rm I}({\rm H}_2) - E_{\rm av}
\end{eqnarray}
is the binding energy of the ${\rm H}_2^{+*}$ excited molecule.  Here
$E_{\rm I}({\rm H})$is the ionization energy of hydrogen, $E_{\rm
B}({\rm H_{2}})$ is the disassociation energy of the ${\rm H_{2}}$
molecule, $E_{\rm I}({\rm H}_{2})$ is the single electron ionization
energy of ${\rm H}_{2}$, and $E_{\rm av}$ is the average vibrational
excitation energy of an ${\rm H}_{2}^{+}$ molecule as a result of the
ionization of ${\rm H}_{2}$ by 3.4 keV electron impact.

The observational equation for the frequency ratio is thus
\begin{eqnarray}
\frac{f_{\rm c}({\rm H}_2^{+*})}{f_{\rm c}({\rm d})}
&=&
\frac{A_{\rm r}({\rm d})}{2A_{\rm r}({\rm p}) + A_{\rm r}({\rm e}) -
E_{\rm B}({\rm H}_2^{+*}) /m_{\rm u} c^2} \, .
\label{eq:obsh2psd}
\end{eqnarray}
We treat $E_{\rm av}$ as an adjusted constant in addition to
$A_{\rm r}({\rm e})$, $A_{\rm r}({\rm p})$, and $A_{\rm r}({\rm d})$ in
order to take its uncertainty into account in a consistent way,
especially since it enters into the observational equations for the
frequency ratios to be discussed below.

The required ionization and binding energies as well as $E_{\rm av}$
that we use are as given by \citet{2008161} and except for $E_{\rm av}$,
have negligible uncertainties:
\begin{eqnarray}
E_{\rm I}({\rm H}) &=&  13.5984 \mbox{ eV} =  14.5985
\times 10^{-9} \,m_{\rm u} c^2 \, ,
\\[6 pt]
E_{\rm B}({\rm H}_2) &=&  4.4781 \mbox{ eV} =  4.8074
\times 10^{-9} \,m_{\rm u} c^2 \, ,
\\[6 pt]
E_{\rm I}({\rm H}_2) &=&  15.4258 \mbox{ eV} =  16.5602
\times 10^{-9} \,m_{\rm u} c^2 \, ,
\\[6 pt]
E_{\rm av} &=&  0.740(74) \mbox{ eV} =  0.794(79)
\times 10^{-9} \,m_{\rm u} c^2 \, . \qquad
\label{eq:bengh2ps}
\end{eqnarray}

We now consider the SMILETRAP results of \citet{2006032} for the
ratio of the cyclotron frequency of the triton t and of the $^{3}{\rm
He}^{+}$ ion to that of the ${\rm H_2}^{+*}$ molecular ion.  These
authors report for the triton
\begin{eqnarray}
\frac{f_{\rm c}({\rm t})}{f_{\rm c}({\rm H}_2^{+*})} &=&
0.668\,247\,726\,86(55)
\quad [ 8.2\times 10^{-10}]\qquad
\label{eq:rSMTt06}
\end{eqnarray}
and for the $^{3}{\rm He}^{+}$ ion
\begin{eqnarray}
\frac{f_{\rm c}(^3{\rm He}^+)}{f_{\rm c}({\rm H}_2^{+*})} &=&
0.668\,252\,146\,82(55)
\quad [ 8.2\times 10^{-10}]\, .\qquad
\label{eq:rSMTh06}
\end{eqnarray}
The relative uncertainty of the triton ratio consists of the following
uncertainty components in parts in 10$^{9}$: 0.22 statistical, and 0.1,
0.1, 0.77, and 0.1 due to relativistic mass shift, ion number
dependence, $q/A$ asymmetry, and contaminant ions, respectively. The
components for the $^{3}{\rm He}^{+}$ ion ratio are the same except the
statistical uncertainty is 0.24.  All of these components are
independent except the 0.77$\times10^{-9}$ component due to $q/A$
asymmetry; it leads to a correlation coefficient between the two
frequency ratios of 0.876.

Observational equations for these frequency ratios are
\begin{eqnarray}
\frac{f_{\rm c}({\rm t})}{f_{\rm c}({\rm H}_2^{+*})} &=& \frac{2A_{\rm r}({\rm
p}) +
A_{\rm r}({\rm e}) - E_{\rm B}({\rm H}_2^{+*})/m_{\rm u} c^2}{A_{\rm r}({\rm t}
)} \qquad
\end{eqnarray}
and
\begin{eqnarray}
\frac{f_{\rm c}(^3{\rm He}^+)}{f_{\rm c}({\rm H}_2^{+*})} &=& \frac{2A_{\rm r}
({\rm p}) +
A_{\rm r}({\rm e}) -
E_{\rm B}({\rm H}_2^{+*})/m_{\rm u} c^2}{A_{\rm r}({\rm h})
+A_{\rm r}({\rm e})-E_{\rm I}(^3{\rm He}^+)/m_{\rm u} c^2} \, , \qquad
\end{eqnarray}
where
\begin{eqnarray}
A_{\rm r}(^{3}{\rm He}^{+}) = A_{\rm r}({\rm
h}) + A_{\rm r}({\rm e}) - E_{\rm I}(^{3}{\rm He}^{+})/m_{\rm u}c^{2}
\end{eqnarray}
and
\begin{eqnarray} E_{\rm I}(^3{\rm He}^+) = 51.4153 \mbox{ eV} =
58.4173 \times 10^{-9} \,m_{\rm u} c^2
\end{eqnarray}
is the ionization energy of the $^3{\rm He}^+$ ion, based on Table IV of
CODATA-02.

The energy $E_{\rm av}$ and the three frequency ratios given in
Eqs.~(\ref{eq:rSMTp08}), (\ref{eq:rSMTt06}), and (\ref{eq:rSMTh06})
are items $B3$ to $B6$ in Table~\ref{tab:pdata}.

\subsection{Cyclotron resonance measurement of the electron relative atomic
mass}
\label{ssec:ptmare}

As in the 2002 and 2006 CODATA adjustments, we take as an input datum
the Penning-trap result for the electron relative atomic mass $A_{\rm
r}({\rm e})$ obtained by the University of Washington group
\cite{1995160}:
\begin{eqnarray}
A_{\rm r}({\rm e})=  0.000\,548\,579\,9111(12)
\quad [ 2.1\times 10^{-9}]\, .
\label{eq:arexp}
\end{eqnarray}
This is item $B11$ of Table~\ref{tab:pdata}.

\section{Atomic transition frequencies}
\label{sec:tf}

Measurements and theory of transition frequencies in hydrogen,
deuterium, anti-protonic helium, and muonic hydrogen provide information
on the Rydberg constant, the proton and deuteron charge radii, and the
relative atomic mass of the electron.  These topics as well as hyperfine
and fine-structure splittings are considered in this section.

\subsection{Hydrogen and deuterium transition frequencies,
the Rydberg constant $\bm{R_\infty}$, and the proton and deuteron
charge radii $\bm{r_{\rm p}, r_{\rm d}}$}
\label{ssec:ryd}

Transition frequencies between states $a$ and $b$ in hydrogen and
deuterium are given by
\begin{eqnarray}
\nu_{ab} &=& \frac{E_b - E_a}{h}\ ,
\end{eqnarray}
where $E_a$ and $E_b$ are the energy levels of the states.  The energy
levels divided by $h$ are given by
\begin{eqnarray}
\frac{E_a}{h} &=& -\frac{\alpha^2m_{\rm e}c^2}{2n_a^2h}
\left(1 + \delta_a\right)
= -\frac{R_\infty c}{n_a^2} \left(1 + \delta_a\right)  ,
\end{eqnarray}
where $R_\infty c$ is the Rydberg constant in frequency units, $n_a$ is
the principle quantum number of state $a$, and $\delta_a$ is a small
correction factor (\,$|\delta_a|\ll 1$\,) that contains the details of
the theory of the energy level, including the effect of the finite size
of the nucleus as a function of the rms charge radius $r_{\rm p}$ for
hydrogen or $r_{\rm d}$ for deuterium.  In the following summary,
corrections are given in terms of the contribution to the energy level,
but in the numerical evaluation for the least-squares adjustment,
$R_\infty$ is factored out of the expressions and is an adjusted
constant.

\subsubsection{Theory of hydrogen and deuterium energy levels}
\label{sssec:hdel}

Here we provide the information necessary to determine theoretical
values of the relevant energy levels, with the emphasis of the
discussion on results that have become available since the 2006
adjustment.  For brevity, most references to earlier work, which can be
found in \citet{2001057,2007353}, for example, are not included here.

Theoretical values of the energy levels of different states are highly
correlated.  In particular, uncalculated terms for S states are
primarily of the form of an unknown common constant divided by $n^3$.
We take this fact into account by calculating covariances between energy
levels in addition to the uncertainties of the individual levels (see
Sec.~\ref{par:teu}).  The correlated uncertainties are denoted by $u_0$,
while the uncorrelated uncertainties are denoted by $u_n$.

\paragraph{Dirac eigenvalue}
\label{par:dev}

The Dirac eigenvalue for an electron in a Coulomb field is
\begin{eqnarray}
E_{\rm D} = f(n,j)\, m_{\rm e}c^2 \ ,
\label{eq:diracen}
\end{eqnarray}
where
\begin{eqnarray}
f(n,j) = \left[ 1+{(Z\alpha)^2\over (n-\delta)^2} \right]^{-1/2} \ ,
\label{eq:diracev}
\end{eqnarray}
$n$ and $j$ are the principal quantum number and total
angular momentum of the state, respectively, and
\begin{eqnarray}
\delta = j+\fr{1}{2}-\left[(j+\fr{1}{2})^2-(Z\alpha)^2\right]^{1/2} \ .
\label{eq:delta}
\end{eqnarray}
In Eqs.~(\ref{eq:diracev}) and (\ref{eq:delta}), $Z$ is the charge
number of the nucleus, which for hydrogen and deuterium is 1.  However,
we shall retain $Z$ as a parameter to classify the various
contributions.

Equation~(\ref{eq:diracen}) is only valid for an infinitely heavy
nucleus.   For a nucleus with a finite mass $m_{\rm N}$ that expression
is replaced by
\cite{1955001,1990020}:
\begin{eqnarray}
E_M(H) &=& Mc^2 +[f(n,j)-1]m_{\rm r}c^2 -[f(n,j)-1]^2{m_{\rm r}^2c^2\over
2M} \nonumber\\
&&+\, {1-\delta_{\ell 0}\over \kappa(2\ell+1)} {(Z\alpha)^4m_{\rm r}^3c^2\over
2 n^3 m_{\rm N}^2} +\cdots
\label{eq:relred}
\end{eqnarray}
for hydrogen or by \cite{1995073}
\begin{eqnarray}
E_M(D) &=& Mc^2 +[f(n,j)-1]m_{\rm r}c^2 -[f(n,j)-1]^2{m_{\rm r}^2c^2\over
2M} \nonumber\\
&&+\, {1\over \kappa(2\ell+1)} {(Z\alpha)^4m_{\rm r}^3c^2\over
2 n^3 m_{\rm N}^2} +\cdots
\label{eq:relredd}
\end{eqnarray}
for deuterium.  In Eqs.~(\ref{eq:relred}) and (\ref{eq:relredd}) $\ell$
is the nonrelativistic orbital angular momentum quantum number, $\kappa
= (-1)^{j-\ell+1/2}(j+\fr{1}{2})$ is the angular-momentum-parity quantum
number, $M = m_{\rm e} + m_{\rm N}$, and $m_{\rm r} = m_{\rm e}m_{\rm
N}/(m_{\rm e}+m_{\rm N})$ is the reduced mass.

Equations~(\ref{eq:relred}) and (\ref{eq:relredd}) differ in that the
Darwin-Foldy term proportional to $\delta_{\ell 0}$ is absent in
Eq.~(\ref{eq:relredd}), because it does not occur for a spin-one nucleus
such as the deuteron \cite{1995073}.  In the three previous adjustments,
Eq.~(\ref{eq:relred}) was used for both hydrogen and deuterium and the
absence of the Darwin-Foldy term in the case of deuterium was accounted
for by defining an effective deuteron radius given by Eq.~(A56) of
CODATA-98 and using it to calculate the finite nuclear size correction
given by Eq.~(A43) and the related equations.  The extra term in the
size correction canceled the Darwin-Foldy term in
Eq.~(\ref{eq:relred}).  See also Sec.~\ref{par:nucsize}.

\paragraph{Relativistic recoil}

The leading relativistic-recoil correction, to lowest order in $Z\alpha$
and all orders in $m_{\rm e}/m_{\rm N}$, is \cite{1977002,1990020}
\begin{eqnarray}
E_{\rm S} &=& {m_{\rm r}^3\over m_{\rm e}^2 m_{\rm N}}{(Z\alpha)^5\over \rmpi
n^3}
m_{\rm e} c^2 \nonumber\\
&&\times \bigg\{\fr{1}{3}\delta_{\ell 0}\ln (Z\alpha)^{-2} -\fr{8}{3}\ln
k_0(n,\ell) -\fr{1}{9}\delta_{\ell 0}-\fr{7}{3}a_n \nonumber\\
&&- \, {2\over m_{\rm N}^2-m_{\rm e}^2}\delta_{\ell 0} \left[m_{\rm N}^2\ln
\Big({m_{\rm e}\over m_{\rm r}}\Big) - m_{\rm e}^2\ln\Big({m_{\rm N}\over m_{\rm
r}}
\Big)\right]\bigg\}, \nonumber\\
\label{eq:salp}
\end{eqnarray}
where
\begin{eqnarray}
a_n &=& -2\left[\ln\Big({2\over n}\Big) + \sum_{i=1}^n{1\over i} +
1 - \frac{1}{2n}\right]\delta_{\ell0} \nonumber\\ &&
+ \, {1-\delta_{\ell0}\over \ell(\ell+1)(2\ell+1)} \,.
\end{eqnarray}

To lowest order in the mass ratio, the next two orders in $Z\alpha$ are
\begin{eqnarray}
E_{\rm R} &=& {m_{\rm e}\over m_{\rm N}}
{(Z\alpha)^6\over n^3}m_{\rm e}c^2
\nonumber\\ && \times
\left[D_{60}  + D_{72}Z\alpha \ln^2{(Z\alpha)^{-2}} + \cdots \right]
\ ,
\label{eq:err60etc}
\end{eqnarray}
where for $n{\rm S}_{1/2}$ states
\cite{1995071,1997084,1999164,1999182}
\begin{eqnarray}
D_{60} &=&
4\ln 2 -{7\over 2}  \ ,
\label{eq:errd60s}
\\[0 pt]
D_{72} &=& -{11\over 60 \rmpi} \ ,
\label{eq:err72}
\end{eqnarray}
and for states with $\ell \ge 1$ \cite{1995070,1996171,1996109}
\begin{eqnarray}
D_{60} &=&
\left[3-{\ell(\ell+1)\over n^2}\right]
{2\over(4\ell^2-1)(2\ell+3)}   \ .
\label{eq:errlg1}
\end{eqnarray}

Based on the general pattern of the magnitudes of higher-order 
coefficients, the uncertainty for S states is taken to be 10\,\% of
Eq.~(\ref{eq:err60etc}), and for states with $\ell\ge 1$, it is taken to
be 1\,\%.  Numerical values for Eq.~(\ref{eq:err60etc}) to all orders in
$Z\alpha$ have been obtained by \citet{1998031}, and although they
disagree somewhat with the analytic result, they are consistent within
the uncertainty assigned here.  We employ the analytic equations in the
adjustment.  The covariances of the theoretical values are calculated by
assuming that the uncertainties are predominately due to uncalculated
terms proportional to $(m_{\rm e}/m_{\rm N})/n^3$.

\paragraph{Nuclear polarizability}
\label{par:nucpol}

For hydrogen, we use the result \cite{2000054}
\begin{eqnarray}
E_{\rm P}({\rm H}) &=&  -0.070(13) h {\delta_{l0}\over n^3} \ {\rm kHz}
\ .
\label{eq:hnucpol}
\end{eqnarray}
More recent results are a model calculation by \citet{2008081} and a
slightly different result than Eq.~(\ref{eq:hnucpol}) calculated by
\citet{2006186}.

For deuterium, the sum of the proton polarizability, the neutron
polarizibility \cite{1998110}, and the dominant nuclear structure
polarizibility \cite{1997130}, gives
\begin{eqnarray}
E_{\rm P}({\rm D}) &=&  -21.37(8) h{\delta_{l0}\over n^3} \ {\rm kHz} \ .
\label{eq:dnucpol}
\end{eqnarray}

Presumably the polarization effect is negligible for states of higher
$\ell$ in either hydrogen or deuterium.

\paragraph{Self energy}
\label{par:selfen}

The one-photon self energy of the bound electron is
\begin{eqnarray}
E_{\rm SE}^{(2)} = {\alpha\over\rmpi}{(Z\alpha)^4\over n^3} F(Z\alpha) \, m_{\rm
e} c^2 \ ,
\label{eq:selfen}
\end{eqnarray}
where
\begin{eqnarray}
F(Z\alpha) &=& A_{41}\ln(Z\alpha)^{-2}+A_{40}+A_{50}\,(Z\alpha)
\nonumber\\
&&+A_{62}\,(Z\alpha)^2\ln^2(Z\alpha)^{-2}
+A_{61}\,(Z\alpha)^2\ln(Z\alpha)^{-2}
\nonumber\\
&&+G_{\rm SE}(Z\alpha)\,(Z\alpha)^2 \ .
\label{eq:sepow}
\end{eqnarray}
From \citet{1965012} and earlier papers cited therein,
\begin{eqnarray}
A_{41} &=& \fr{4}{3} \, \delta_{\ell0} \, ,
\nonumber\\
A_{40} &=&
-\fr{4}{3}\ln k_0(n,\ell)
+\fr{10}{9} \, \delta_{\ell0}-{1\over 2\kappa(2\ell+1)}(1-\delta_{\ell0})
\, ,
\nonumber\\
A_{50} &=& \left(\fr{139}{32} - 2 \ln 2 \right) \rmpi \, \delta_{\ell0}
\, ,
\label{eq:sepows}
\\
A_{62} &=& - \delta_{\ell0} \, ,
\nonumber\\
A_{61} &=& \left[4\left(1 + \frac{1}{2} + \cdots + \frac{1}{n}\right)
+{28\over3}\ln{2}-4\ln{n}\right.
\nonumber\\
&-&\left.{601\over180} - {77\over 45n^2}\right]\delta_{\ell0}
+\left(1-{1\over n^2}\right)
\left({2\over15}+{1\over3}\delta_{j{1\over2}}\right)\delta_{\ell1}
\, ,
\nonumber\\
&+&{\left[96n^2-32\ell(\ell+1)\right] \left(1-\delta_{\ell0}\right)
\over 3 n^2(2\ell-1)(2\ell)(2\ell+1)(2\ell+2)(2\ell+3)})\ .
\nonumber
\end{eqnarray}
The Bethe logarithms $\ln k_0(n,\ell)$ in Eq.~(\ref{eq:sepows})
are given in Table~\ref{tab:bethe} \cite{1990002}.

\begin{table}
\def\sb{\hbox to 5mm{}}
\caption{Relevant values of the Bethe logarithms $\ln k_0(n,l)$.}
\label{tab:bethe}
\begin{tabular}{r@{\sb}c@{\sb}c@{\sb}c}
\toprule
\vbox to 10 pt {}
$n$ & S & P & D \\
\colrule
1 & $ 2.984\,128\,556$ & \vbox to 10pt{} & \\
2 & $ 2.811\,769\,893$ & $ -0.030\,016\,709$ & \\
3 & $ 2.767\,663\,612$ &                  &                  \\
4 & $ 2.749\,811\,840$ & $ -0.041\,954\,895$ & $ -0.006\,740\,939$ \\
6 & $ 2.735\,664\,207$ &                  & $ -0.008\,147\,204$ \\
8 & $ 2.730\,267\,261$ &                  & $ -0.008\,785\,043$ \\
12&                  &                  & $ -0.009\,342\,954$ \\
\botrule
\end{tabular}
\end{table}

For S and P states with $n \le 4$, the values we use here for $G_{\rm
SE}(Z\alpha)$ in Eq.~(\ref{eq:sepow}) are listed in Table~\ref{tab:gse}
and are based on direct numerical evaluations by
\citet{1999001,2001072,2004090,2005127}.  The values of $G_{\rm
SE}(\alpha)$ for the 6S and 8S states are based on the low-$Z$ limit
$G_{\rm SE}(0)=A_{60}$ \cite{2005212} together with extrapolations of
the results of complete numerical calculations of $F(Z\alpha)$ in
Eq.~(\ref{eq:selfen}) at higher $Z$ \cite{pc06skpm}.  A calculation of
the constant $A_{60}$ for various D states, including 12D states, has
been done by \citet{2008045}.  In CODATA-06 this constant was obtained
by extrapolation from lower-$n$ states.  The more recent calculated
values are
\begin{eqnarray}
A_{60}(12\mbox{D}_{3/2}) &=& 0.008\,909\,60(5) \, , \\
A_{60}(12\mbox{D}_{5/2}) &=& 0.034\,896\,67(5) \, .
\end{eqnarray}
To estimate the corresponding value of $G_{\rm SE}(\alpha)$, we use the
data from \citet{hdelse2005} given in Table~\ref{tab:hdel}.  It is
evident from the table that
\begin{eqnarray}
G_{\rm SE}(\alpha) - A_{60} \approx 0.000\,22
\end{eqnarray}
for the $n\mbox{D}_{3/2}$ and $n\mbox{D}_{5/2}$ states for
$n=4,5,6,7,8$, so we make the approximation
\begin{eqnarray}
G_{\rm SE}(\alpha) = A_{60} + 0.000\,22 \, ,
\end{eqnarray}
with an uncertainty given by $0.000\,09$ and $0.000\,22$ for the
$12\mbox{D}_{3/2}$ and $12\mbox{D}_{5/2}$ states, respectively.
This yields
\begin{eqnarray}
G_{\rm SE}(\alpha) &=& 0.009\,13(9)\qquad\ \,
\mbox{ for }12\mbox{D}_{3/2} \, , \\
G_{\rm SE}(\alpha) &=& 0.035\,12(22)\qquad
\mbox{ for }12\mbox{D}_{5/2} \, .
\end{eqnarray}
All values for $G_{\rm SE}(\alpha)$ that we use here are listed in
Table~\ref{tab:gse}.  The uncertainty of the self energy contribution to
a given level arises entirely from the uncertainty of $G_{\rm
SE}(\alpha)$ listed in that table and is taken to be type $u_n$.

\begin{table*}
\def\sb{\hbox to 6mm{}}
\caption{Values of the function $G_{\rm SE}(\alpha)$.}
\label{tab:gse}
\begin{tabular}{r @{\sb} l @{\sb} l @{\sb} l @{\sb} l @{\sb} l}
\toprule
\vbox to 10 pt {}
$n$& \ \ \ \ S$_{1/2}$& \ \ \ \ P$_{1/2}$& \ \ \ \ P$_{3/2}$ & \ \ \ \ D$_{3/2}
$& \ \ \ \ D$_{5/2}$\\
\colrule
\vbox to 10 pt {}
$ 1  $&$  -30.290\,240(20) $&&&&\\
$ 2  $&$  -31.185\,150(90) $&$  -0.973\,50(20) $&$  -0.486\,50(20) $&&\\
$ 3  $&$  -31.047\,70(90) $&$                $&$                 $&$
$&$                 $ \\
$ 4  $&$  -30.9120(40) $&$  -1.1640(20) $&$  -0.6090(20) $&$                $&$
0.031\,63(22) $ \\
$ 6  $&$  -30.711(47) $&$                $&$                 $&$
$&$  0.034\,17(26) $ \\
$ 8  $&$  -30.606(47) $&$                $&$                 $&$
0.007\,940(90) $&$  0.034\,84(22) $ \\
$ 12 $&$                 $&$                $&$                 $&$
0.009\,130(90) $&$  0.035\,12(22) $ \\
\botrule
\end{tabular}
\end{table*}

\begin{table*}
\caption{Data from \citet{hdelse2005} and the deduced values of $G_{\rm
SE}(\alpha)$ for $n=12$.}
\label{tab:hdel}
\begin{tabular}{r@{\qquad}l@{\qquad}l@{\qquad}l@{\qquad}l@{\qquad}l@{\qquad}l}
\toprule
\vbox to 10 pt {}
& \multicolumn{2}{c}{$A_{60}\ \ $}
& \multicolumn{2}{c}{$G_{\rm SE}(\alpha)\ \ \ $}
& \multicolumn{2}{c}{$G_{\rm SE}(\alpha)-A_{60}$} \\
$n$
& $\qquad\rm D_{3/2}$ & $\qquad\rm D_{5/2}$
& $\quad\rm D_{3/2}$ & $\quad\rm D_{5/2}$
& $\quad\rm D_{3/2}$ & $\quad\rm D_{5/2}$ \\
\colrule
\vbox to 10 pt {}
$ 3$ & $0.005\,551\,575(1)$ & $0.027\,609\,989(1)$ & $
0.005\,73(15)$ & $0.027\,79(18)$ & $0.000\,18(15)$ & $0.000\,18(18)$ \\
$ 4$ & $0.005\,585\,985(1)$ & $0.031\,411\,862(1)$ & $
0.005\,80(9)$ & $0.031\,63(22)$ & $0.000\,21(9)$ & $0.000\,22(22)$ \\
$ 5$ & $0.006\,152\,175(1)$ & $0.033\,077\,571(1)$ & $
0.006\,37(9)$ & $0.033\,32(25)$ & $0.000\,22(9)$ & $0.000\,24(25)$ \\
$ 6$ & $0.006\,749\,745(1)$ & $0.033\,908\,493(1)$ & $
0.006\,97(9)$ & $0.034\,17(26)$ & $0.000\,22(9)$ & $0.000\,26(26)$ \\
$ 7$ & $0.007\,277\,403(1)$ & $0.034\,355\,926(1)$ & $
0.007\,50(9)$ & $0.034\,57(22)$ & $0.000\,22(9)$ & $0.000\,21(22)$ \\
$ 8$ & $0.007\,723\,850(1)$ & $0.034\,607\,492(1)$ & $
0.007\,94(9)$ & $0.034\,84(22)$ & $0.000\,22(9)$ & $0.000\,23(22)$ \\
$12$ & $0.008\,909\,60(5)$ & $0.034\,896\,67(5)$ & $
0.009\,13(9)$ & $0.035\,12(22)$ & $0.000\,22(9)$ & $0.000\,22(22)$ \\
\botrule
\end{tabular}
\end{table*}

The dominant effect of the finite mass of the nucleus on the self energy
correction is taken into account by multiplying each term of
$F(Z\alpha)$ by the reduced-mass factor $(m_{\rm r}/m_{\rm e})^3$,
except that the magnetic moment term $-1/[2\kappa(2\ell+1)]$ in $A_{40}$ is
instead multiplied by the factor $(m_{\rm r}/m_{\rm e})^2$.  In
addition, the argument $(Z\alpha)^{-2}$ of the logarithms is replaced by
$(m_{\rm e}/m_{\rm r})(Z\alpha)^{-2}$ \cite{1990020}.

\paragraph{Vacuum polarization}
\label{par:vacpol}

The second-order vacuum-polarization level shift is
\begin{eqnarray}
E_{\rm VP}^{(2)} = {\alpha\over\rmpi}{(Z\alpha)^4\over n^3} H\!(Z\alpha) \,
m_{\rm e} c^2 \ ,
\label{eq:vacpol}
\end{eqnarray}
where the function $H\!(Z\alpha)$ consists of the Uehling potential
contribution $H^{(1)}\!(Z\alpha)$ and a higher-order remainder
$H^{({\rm R})}\!(Z\alpha)$:
\begin{eqnarray}
H^{(1)}\!(Z\alpha) &=& V_{40}+V_{50}\,(Z\alpha) +V_{61}\,(Z\alpha)^2\ln(Z
\alpha)^{-2}
\nonumber\\[0 pt]
&&+ \, G_{\rm VP}^{(1)}(Z\alpha)\,(Z\alpha)^2  \, ,
\label{eq:uehpow}
\\[10 pt]
H^{({\rm R})}\!(Z\alpha) &=& G_{\rm VP}^{({\rm R})}(Z\alpha)\,(Z\alpha)^2 \ ,
\label{eq:wkvp}
\end{eqnarray}
with
\begin{eqnarray}
V_{40} &=& -\frac{4}{15} \, \delta_{\ell0} \, , \nonumber\\
V_{50} &=& \frac{5}{48}\rmpi \, \delta_{\ell0} \, , \\
V_{61} &=& -\frac{2}{15} \, \delta_{\ell0}\,. \nonumber
\end{eqnarray}

Values of $G_{\rm VP}^{(1)}(Z\alpha)$ are given in Table~\ref{tab:gvp1}
\cite{1982004,2002148}.  The Wichmann-Kroll contribution $G_{\rm
VP}^{({\rm R})} (Z\alpha)$ has the leading powers in $Z\alpha$ given by
\cite{1956001,1975028,1983005}
\begin{eqnarray}
G_{\rm VP}^{\rm (R)}(Z\alpha) &=& \left(\frac{19}{45} - \frac{\rmpi^2}{27}
\right)
\delta_{\ell0}\nonumber\\
&&+ \left(\frac{1}{16} - \frac{31\rmpi^2}{2880}\right)\rmpi(Z\alpha)
\delta_{\ell0} + \cdots \ . \qquad
\label{eq:hovp}
\end{eqnarray}
Higher-order terms are negligible.

\begin{table*}
\def\sb{\hbox to 5mm{}}
\caption{Values of the function $G_{\rm VP}^{(1)}(\alpha)$.  (The minus
signs on the zeros in the last two columns indicate that the values are
nonzero negative numbers smaller than the digits shown.)}
\label{tab:gvp1}
\begin{tabular}{r @{\sb} l @{\sb} l @{\sb} l @{\sb} l @{\sb} l}
\toprule
\vbox to 10 pt {}
$n$& \ \ \ \ S$_{1/2}$& \ \ \ \ P$_{1/2}$& \ \ \ \ P$_{3/2}$ & \ \ \ \ D$_{3/2}
$& \ \ \ \ D$_{5/2}$\\
\colrule
\vbox to 10 pt {}
$ 1  $&$  -0.618\,724 $&&&&\\
$ 2  $&$  -0.808\,872 $&$  -0.064\,006 $&$  -0.014\,132 $&&\\
$ 3  $&$  -0.814\,530 $&$                $&$                 $&$
$&$                 $ \\
$ 4  $&$  -0.806\,579 $&$  -0.080\,007 $&$  -0.017\,666 $&$                $&$
-0.000\,000 $ \\
$ 6  $&$  -0.791\,450 $&$                $&$                 $&$
$&$  -0.000\,000 $ \\
$ 8  $&$  -0.781\,197 $&$                $&$                 $&$  -0.000\,000
$&$  -0.000\,000 $ \\
$ 12 $&$                 $&$                $&$                 $&$
-0.000\,000 $&$  -0.000\,000 $ \\
\botrule
\end{tabular}
\end{table*}

The finite mass of the nucleus is taken into account by multiplying
Eq.~(\ref{eq:vacpol}) by $(m_{\rm r}/m_{\rm e})^3$ and including a
factor of $(m_{\rm e}/m_{\rm r})$ in the argument of the logarithm in
Eq.~(\ref{eq:uehpow}).

Vacuum polarization from ${\rmmu}^+{\rmmu}^-$ pairs is
\cite{1995102,1995005}
\begin{eqnarray}
E_{{\rmssmu}{\rm VP}}^{(2)} = {\alpha\over\rmpi}{(Z\alpha)^4\over n^3}
\left(-\frac{4}{15}\,\delta_{\ell0}\right)
\left({m_{\rm e}\over m_{\rmssmu}}\right)^2
\left({m_{\rm r}\over m_{\rm e}}\right)^3
m_{\rm e} c^2 \ ,
\nonumber\\
\label{eq:vacpolmu}
\end{eqnarray}
and the effect of ${\rmtau}^+{\rmtau}^-$ pairs is negligible.

Hadronic vacuum polarization gives \cite{1999042}
\begin{eqnarray}
E_{\rm had \, VP}^{(2)} =  0.671(15) E_{{\rmssmu}{\rm VP}}^{(2)} \ ,
\end{eqnarray}
where the uncertainty is of type $u_0$.

The muonic and hadronic vacuum polarization contributions are negligible
for higher-$\ell$ states.

\paragraph{Two-photon corrections}
\label{par:tpc}

The two-photon correction, in powers of $Z\alpha$, is
\begin{eqnarray}
E^{(4)} &=& \left({\alpha\over\rmpi}\right)^2 {(Z\alpha)^4\over n^3} m_{\rm e}
c^2
F^{(4)}(Z\alpha) \ ,
\label{eq:total4}
\end{eqnarray}
where
\begin{eqnarray}
F^{(4)}(Z\alpha) &=&
B_{40} + B_{50}\,(Z\alpha)
+ B_{63}\,(Z\alpha)^2\ln^3(Z\alpha)^{-2}
\nonumber\\
&& + B_{62}\,(Z\alpha)^2\ln^2(Z\alpha)^{-2}
\nonumber\\
&& + B_{61}\,(Z\alpha)^2\ln(Z\alpha)^{-2}
+ B_{60}\,(Z\alpha)^2
\nonumber\\ &&+\cdots \ .
\end{eqnarray}

The leading term $B_{40}$ is
\begin{eqnarray}
B_{40} &=& \left[\frac{3\rmpi^2}{2}\ln 2-\frac{10\rmpi^2}{27}-\frac{2179}{648}
-\frac{9}{4}\zeta(3)
\right] \delta_{\ell0} \nonumber\\
&&+ \left[\frac{\rmpi^2\ln 2}{2}-\frac{\rmpi^2}{12}-\frac{197}{144}
-\frac{3\zeta(3)}{4}
\right] {1 - \delta_{\ell0} \over \kappa(2\ell+1)} \ ,
\nonumber\\
\label{eq:b40}
\end{eqnarray}
where $\zeta$ is the Riemann zeta function \cite{2010075},
and the next term is
\cite{1993025,1997027,1995102,1994092,2010037}
\begin{eqnarray}
B_{50} &=& -21.554\,47(13)\delta_{\ell0} \ .
\label{eq:2phb50}
\end{eqnarray}

The leading sixth-order coefficient is
\cite{1993019,2000101,2000094,2001059}
\begin{eqnarray}
B_{63} = -{8\over 27}\delta_{\ell0} \ .
\label{eq:b63}
\end{eqnarray}
For S states $B_{62}$ is \cite{1996012,2001059}
\begin{eqnarray}
B_{62} &=& {16\over 9} \left[{71\over60}-\ln{2} + \rmgamma
+ \psi(n) - \ln n - {1\over n} + {1\over4n^2}
\right] \ ,
\nonumber\\
\label{eq:b62s}
\end{eqnarray}
where $\rmgamma = 0.577...$ is Euler's constant
and $\psi$ is the psi function \cite{2010075}.
For P states \cite{1996012,2002155}
\begin{eqnarray}
B_{62} = {4\over 27} {n^2 - 1 \over n^2} \ ,
\label{eq:b62p}
\end{eqnarray}
and $B_{62}=0$ for $\ell\ge2$.

For S states $B_{61}$ is \cite{2001059,2005212}
\begin{eqnarray}
B_{61} &=& {413\,581\over 64\,800} +
{4N(n{\rm S})\over3} + {2027\rmpi^2\over864} - {616\,\ln{2}\over 135}
\nonumber\\&&
- {2\rmpi^2\ln{2}\over 3}
+ {40\ln^2{2}\over 9}
+ \zeta(3)
+\left(\frac{304}{135}-\frac{32\,\ln{2}}{9}\right)
\nonumber\\ && \times
\left[{3\over4} + \rmgamma
+ \psi(n) - \ln n - {1\over n} + {1\over4n^2}
\right] \ .
\label{eq:b61s}
\end{eqnarray}
For P states \cite{2005212,2003118}
\begin{eqnarray}
B_{61}(n{\rm P}_{1/2}) &=& \frac{4}{3}\,N(n{\rm P})
+ \frac{n^2-1}{n^2}\left(\frac{166}{405}-\frac{8}{27}\,\ln{2}\right)
,
\qquad
\\
B_{61}(n{\rm P}_{3/2}) &=& \frac{4}{3}\,N(n{\rm P})
+ \frac{n^2-1}{n^2}\left(\frac{31}{405}-\frac{8}{27}\,\ln{2}\right)
,
\qquad
\end{eqnarray}
and $B_{61}=0$ for $\ell\ge2$.  Values for $B_{61}$ used in the
adjustment are listed in Table~\ref{tab:b61}

\begin{table*}
\newpage
\caption{Values of $B_{61}$ used in the 2010 adjustment.}
\label{tab:b61}
\def\spa{\hbox to 18 pt{}}
\begin{tabular}{c@{\spa}c@{\spa}c@{\spa}c@{\spa}c@{\spa}c@{\spa}c}
\toprule
$n$  & $B_{61}(n$S$_{1/2})$  &  ${B}_{61}(n$P$_{1/2})$ &
${B}_{61}(n$P$_{3/2})$ & ${B}_{61}(n$D$_{3/2})$ &
${B}_{61}(n$D$_{5/2})$ \\
\colrule
\vbox to 10 pt {}
1&$ 48.958\,590\,24(1)$&&&&\\
2&$ 41.062\,164\,31(1)$&$ 0.004\,400\,847(1)$&$ 0.004\,400\,847(1)$&&\\
3&$ 38.904\,222(1)$&&&&\\
4&$ 37.909\,514(1)$&$ -0.000\,525\,776(1)$&$ -0.000\,525\,776(1)$&&$ 0.0(0)$\\
6&$ 36.963\,391(1)$&&&&$ 0.0(0)$\\
8&$ 36.504\,940(1)$&&&$ 0.0(0)$&$ 0.0(0)$\\
12&&&&$ 0.0(0)$&$ 0.0(0)$\\
\botrule
\end{tabular}
\end{table*}

For the 1S state, the result of a perturbation theory estimate for the
term $B_{60}$ is
\cite{2001059,2003160}
\begin{eqnarray}
B_{60}(1{\rm S}) = -61.6(9.2) \, .
\label{eq:oldb60}
\end{eqnarray}
All-order numerical calculations of the two-photon correction have also
been carried out.  The diagrams with closed electron loops have been
evaluated by \citet{2008148}.  They obtained results for the 1S,
2S, and 2P states at $Z=1$ and higher $Z$, and obtained a value for the
contribution of the terms of order $(Z\alpha)^6$ and higher.  The
remaining contributions to $B_{60}$ are from the self-energy diagrams.
These have been evaluated by
\citet{2005346,2005205,2003167,2007142} for the 1S state for
$Z=10$ and higher $Z$, and more recently, \citet{2010164} has done
an all-order calculation of the 1S-state no-electron-loop two-loop self
energy correction for $Z\ge10$.  His extrapolation of the higher-$Z$
values to obtain a value for $Z=1$ yields a contribution to $B_{60}$,
including higher-order terms, given by $-86(15)$.  This result combined
with the result for the electron-loop two-photon diagrams, reported by
\citet{2008148}, gives a total of $B_{60} + \dots = -101(15)$, where the
dots represent the contribution of the higher-order terms.  This may be
compared to the earlier evaluation which gave $-127(39)$
\cite{2005346,2005205,2003167,2007142}.  The new value also differs
somewhat from the result in Eq.~(\ref{eq:oldb60}).  In view of this
difference between the two calculations, to estimate $B_{60}$ for the
2010 adjustment, we use the average of the analytic value of $B_{60}$
and the numerical result for $B_{60}$ with higher-order terms included,
with an uncertainty that is half the difference.  The higher-order
contribution is small compared to the difference between the results of
the two methods of calculation.  The average result is
\begin{eqnarray}
B_{60}(1{\rm S}) =  -81.3( 0.3)( 19.7) \ .
\label{eq:b601s}
\end{eqnarray}
In Eq.~(\ref{eq:b601s}), the first number in parentheses is the
state-dependent uncertainty $u_n(B_{60})$ associated with the two-loop
Bethe logarithm, and the second number in parentheses is the
state-independent uncertainty $u_0(B_{60})$ that is common to all
S-state values of $B_{60}$.  Two-loop Bethe logarithms needed to
evaluate $B_{60}(n{\rm S})$ have been given for $n=1$ to 6
\cite{2003160,2004148}, and a value at $n=8$ may be obtained by a simple
extrapolation from the calculated values [see Eq.~(43) of CODATA-06].
The complete state dependence of $B_{60}(nS)$ in terms of the two-loop
Bethe logarithms has been calculated by \citet{2005093,2005212}.  Values
of $B_{60}$ for all relevant S-states are given in Table~\ref{tab:b60}.

\begin{table*}
\newpage
\caption{Values of $B_{60}$, $\overline{B}_{60}$, or $\Delta B_{71}$ used in the
2010 adjustment}
\label{tab:b60}
\def\spa{\hbox to 18 pt{}}
\begin{tabular}{c@{\spa}c@{\spa}c@{\spa}c@{\spa}c@{\spa}c@{\spa}c@{\spa}c}
\toprule
$n$  & $B_{60}(n$S$_{1/2})$  &  $\overline{B}_{60}(n$P$_{1/2})$ &
$\overline{B}_{60}(n$P$_{3/2})$ & $\overline{B}_{60}(n$D$_{3/2})$ &
$\overline{B}_{60}(n$D$_{5/2})$ & $\Delta B_{71}(n$S$_{1/2})$\vbox to
10 pt {}\\
\colrule
1&$ -81.3( 0.3)( 19.7)$&&&&& \vbox to 10 pt {}\\
2&$ -66.2( 0.3)( 19.7)$&$ -1.6(3)$&$ -1.7(3)$&&&$ 16(8)$\\
3&$ -63.0( 0.6)( 19.7)$&&&&&$ 22(11)$\\
4&$ -61.3( 0.8)( 19.7)$&$ -2.1(3)$&$ -2.2(3)$&&$ -0.005(2)$&$ 25(12)$\\
6&$ -59.3( 0.8)( 19.7)$&&&&$ -0.008(4)$&$ 28(14)$\\
8&$ -58.3( 2.0)( 19.7)$&&&$ 0.015(5)$&$ -0.009(5)$&$ 29(15)$\\
12&&&&$ 0.014(7)$&$ -0.010(7)$\\
\botrule
\end{tabular}
\end{table*}

For higher-$\ell$ states, an additional consideration is necessary.  The
radiative level shift includes contributions associated with decay to
lower levels.  At the one-loop level, this is the imaginary part of the
level shift corresponding to the resonance scattering width of the
level.  At the two-loop level there is an imaginary contribution
corresponding to two-photon decays and radiative corrections to the
one-photon decays, but in addition there is a real contribution from the
square of one-photon decay width.  This can be thought of as the
second-order term that arises in the expansion of the resonance
denominator for scattering of photons from the atom in its ground state
in powers of the level width \cite{2002218}.  As such, this term should
not be included in the calculation of the resonant line center shift of
the scattering cross section, which is the quantity of interest for the
least-squares adjustment.  The leading contribution of the square of the
one-photon width is of order $\alpha(Z\alpha)^6m_{\rm e} c^2/\hbar$.
This correction vanishes for the 1S and 2S states, because the 1S level
has no width and the 2S level can only decay with transition rates that
are higher order in $\alpha$ and/or $Z\alpha$.  The higher-$n$ S states
have a contribution from the square of the one-photon width from decays
to lower P states, but for the 3S and 4S states for which it has been
separately identified, this correction is negligible compared to the
uncertainty in $B_{60}$ \cite{2004148,2006316}.  We assume the
correction for higher S states is also negligible compared to the
numerical uncertainty in $B_{60}$.  However, the correction is taken
into account in the 2010 adjustment for P and D states for which it is
relatively larger \cite{2002218,2006316}.

Calculations of $B_{60}$ for higher-$\ell$ states have been made by
\citet{2006316}.  The results can be expressed as
\begin{eqnarray}
B_{60}(nL_j) &=& a(nL_j) + b_{\rm L}(nL) \, ,
\end{eqnarray}
where $a(nL_j)$ is a precisely calculated term that depends on $j$, and
the  two-loop Bethe logarithm $b_{\rm L}(nL)$ has a larger numerical
uncertainty but does not depend on $j$.  \citet{2006316} gives
semianalytic formulas for $a(nL_j)$ that include numerically calculated
terms.  The information needed for the 2010 adjustment is in Eqs.~(22a),
(22b), (23a), (23b), Tables~VII, VIII, XI, and X of \citet{2006316}
and Eq.~(17) of \citet{2003118}.  Two corrections to Eq.~(22b) are
\begin{eqnarray}
&&-\frac{73321}{103680} + \frac{185}{1152n} + \frac{8111}{25920n^2}
\nonumber\\[4 pt]&&\qquad\qquad
\rightarrow
-\frac{14405}{20736} + \frac{185}{1152n} + \frac{1579}{5184n^2}
\qquad\qquad
\end{eqnarray}
on the first line and
\begin{eqnarray}
- \frac{3187}{3600n^2} \rightarrow + \frac{3187}{3600n^2}
\end{eqnarray}
on the fourth line \cite{pc11udj}.

Values of the two-photon Bethe logarithm $b_{\rm L}(nL)$ may be divided
into a contribution of the ``squared level width'' term $\delta^2
B_{60}$ and the rest $\overline{b}_{\rm L}(nL)$, so that
\begin{eqnarray}
b_{\rm L}(nL) &=& \delta^2 B_{60} + \overline{b}_{\rm L}(nL) \, .
\end{eqnarray}
The corresponding value $\overline{B}_{60}$ that represents the shift of
the level center is given by
\begin{eqnarray}
\overline{B}_{60}(nL_j) &=& a(nL_j) + \overline{b}_{\rm L}(nL) \, .
\end{eqnarray}
Here we give the numerical values for $\overline{B}(nL_j)$ in
Table~\ref{tab:b60} and refer the reader to \citet{2006316} for the
separate values for $a(nL_j)$ and $\overline{b}_{\rm L}(nL)$.  The
D-state values for $n=6,8$ are extrapolated from the corresponding
values at $n=5,6$ with a function of the form $a+b/n$.  The values in
Table~\ref{tab:b60} for S states may be regarded as being either
$B_{60}$ or $\overline{B}_{60}$, since the difference is expected to be
smaller than the uncertainty.  The uncertainties listed for the P- and
D-state values of $\overline{B}(nL_j)$ in that table are predominately
from the two-photon Bethe logarithm which depends on $n$ and $L$, but
not on $j$ for a given $n,L$.  Therefore there is a large covariance
between the corresponding two values of $\overline{B}(nL_j)$.  However,
we do not take this into consideration when calculating the uncertainty
in the fine structure splitting, because the uncertainty of higher-order
coefficients dominates over any improvement in accuracy the covariance
would provide.  We assume that the uncertainties in the two-photon Bethe
logarithms are sufficiently large to account for higher-order P and D
state two-photon uncertainties as well.

For S states, higher-order terms have been estimated by
\citet{2005212} with an effective potential model.  They find that
the next term has a coefficient of $B_{72}$ and is state independent.
We thus assume that the uncertainty $u_0[B_{60}(n{\rm S})]$ is
sufficient to account for the uncertainty due to omitting such a term
and higher-order state-independent terms.  In addition, they find an
estimate for the state dependence of the next term, given by
\begin{eqnarray}
\Delta B_{71}(n{\rm S}) &=&
B_{71}(n{\rm S}) - B_{71}(1{\rm S}) =
\rmpi\left(\frac{427}{36} - \frac{16}{3}\,\ln{2}\right)
\nonumber\\
&\times&\left[\frac{3}{4}-\frac{1}{n}+\frac{1}{4n^2}
+\rmgamma +\psi(n)-\ln{n}\right] ,
\end{eqnarray}
with a relative uncertainty of 50\,\%.  We include this additional term, which
is listed in Table~\ref{tab:b60}, along with the estimated uncertainty
$u_n(B_{71}) = B_{71}/2$.

\paragraph{Three-photon corrections}
\label{par:thpc}

The three-photon contribution in powers of $Z\alpha$ is
\begin{eqnarray}
E^{(6)} &=& \left({\alpha\over\rmpi}\right)^3 {(Z\alpha)^4\over n^3} m_{\rm e}
c^2
\left[C_{40} + C_{50}(Z\alpha) + \cdots \right] \ .
\nonumber\\
\label{eq:total6}
\end{eqnarray}
The leading term $C_{40}$ is
\cite{2000019,1996060,1995220,1995158}
\begin{eqnarray}
C_{40} &=& \bigg[
-{{568\,{\rm a_4}}\over{9}}+{{85\,\zeta(5)}\over{24}}
\nonumber\\&&
-{{121\,\rmpi^{2}\,\zeta(3)}\over{72}}
-{{84\,071\,\zeta(3)}\over{2304}}
-{{71\,\ln ^{4}2}\over{27}}
\nonumber\\&&
-{{239\,\rmpi^{2}\,\ln^{2}2}\over{135}}
+{{4787\,\rmpi^{2}\,\ln 2}\over{108}}
+{{1591\,\rmpi^{4}}\over{3240}}
\nonumber\\&&
-{{252\,251\,\rmpi^{2}}\over{9720}}+{679\,441\over93\,312}
\bigg] \delta_{\ell0} \nonumber\\
&&+ \bigg[
-{{100\,{\rm a_4}}\over{3}}+{{215\,\zeta(5)}\over{24}}
\nonumber\\&&
-{{83\,\rmpi^{2}\,\zeta(3)}\over{72}}-{{139\,\zeta(3)}\over{18}}
-{{25\,\ln ^{4}2}\over{18}}
\nonumber\\&&
+{{25\,\rmpi^{2}\,\ln ^{2}2}\over{18}}+{{298\,\rmpi^{2}\,\ln 2}\over{9}}
+{{239\,\rmpi^{4}}\over{2160}}
\nonumber\\&&
-{{17\,101\,\rmpi^{2}}\over{810}}-{28\,259\over5184}
\bigg] {1 - \delta_{\ell0} \over \kappa(2\ell+1)} \ ,
\nonumber\\
\label{eq:c40}
\end{eqnarray}
where $a_4 = \sum_{n=1}^\infty 1/(2^n\,n^4) = 0.517\,479\,061\dots$ .
Partial results for $C_{50}$ have been calculated by
\citet{2004242,2007138}.  The uncertainty is taken to be $u_0(C_{50}) =
30\delta_{\ell0}$ and $u_n(C_{63}) = 1$, where $C_{63}$ would be the
coefficient of $(Z\alpha)^{2}\ln^{3}{(Z\alpha)^{-2}}$ in the square
brackets in Eq.~(\ref{eq:total6}).  The dominant effect of the finite
mass of the nucleus is taken into account by multiplying the term
proportional to $\delta_{\ell0}$ by the reduced-mass factor $(m_{\rm
r}/m_{\rm e})^3$ and the term proportional to $1/[\kappa(2\ell+1)]$, the
magnetic moment term, by the factor $(m_{\rm r}/m_{\rm e})^2$.

The contribution from four photons would be of order
\begin{eqnarray}
\left({\alpha\over\rmpi}\right)^4 {(Z\alpha)^4\over n^3} m_{\rm e}c^2 \ ,
\label{eq:fourphoton}
\end{eqnarray}
which is about 10 Hz for the 1S state and is negligible at the level of
uncertainty of current interest.

\paragraph{Finite nuclear size}
\label{par:nucsize}

In the nonrelativistic limit, the level shift due to the finite size of
the nucleus is
\begin{eqnarray}
E^{(0)}_{\rm NS} = {\cal E}_{\rm NS}
\delta_{\ell0}  \ ,
\label{eq:ens0}
\end{eqnarray}
where
\begin{eqnarray}
{\cal E}_{\rm NS}
= {2\over3}\left({m_{\rm r}\over m_{\rm e}}\right)^3
{(Z\alpha)^2\over n^3} \ m_{\rm e}c^2
\left({Z\alpha r_{\rm N}\over\lbar_{\rm C}}
\right)^2  \ ,
\label{eq:enscoef}
\end{eqnarray}
$r_{\rm N}$ is the bound-state root-mean-square (rms) charge radius of
the nucleus, and $\lbar_{\rm C}$ is the Compton wavelength of the
electron divided by $2\rmpi$.

Higher-order contributions have been examined by
\citet{1979028,1997168,1997022} [see also \citet{1979029,1983005}].  For
S states the leading and next-order corrections are given by
\begin{eqnarray}
E_{\rm NS} &=& {\cal E}_{\rm NS}
\Bigg\{1 - C_\eta{ m_{\rm r}\over m_{\rm e}}
{r_{\rm N}\over\lbar_{\rm C}}
Z\alpha
-\bigg[
\ln{\left({m_{\rm r}\over m_{\rm e}}
{r_{\rm N}\over \lbar_{\rm C}}{Z\alpha\over n}\right)}
\nonumber\\
&& + \psi(n) + \rmgamma - {(5n+9)(n-1)\over4n^2}
-C_\theta
\bigg](Z\alpha)^2
\Bigg\} \ ,
\nonumber\\
\label{eq:ensho}
\end{eqnarray}
where $C_\eta$ and $C_\theta$ are constants that depend on the charge
distribution in the nucleus with values
$C_\eta = 1.7(1)$ and $C_\theta = 0.47(4)$ for hydrogen or
$C_\eta = 2.0(1)$ and $C_\theta = 0.38(4)$ for deuterium.

For the P$_{1/2}$ states in hydrogen the leading term is
\begin{eqnarray}
E_{\rm NS} = {\cal E}_{\rm NS}
{(Z\alpha)^2(n^2-1)\over 4 n^2} \ .
\label{eq:enpho}
\end{eqnarray}
For P$_{3/2}$ states and higher-$\ell$ states the nuclear-size
contribution is negligible.

As mentioned in Sec.~\ref{par:dev}, in the 2010 adjustment, we do not
use an effective radius for the deuteron, but rather simply $r_{\rm d}$
which is defined by Eq.~(\ref{eq:enscoef}).  In CODATA-02, and
CODATA-06, the adjustment code used $r_{\rm d}$ as an adjusted variable
and that value was reported for the rms radius, rather than the value
for $R_{\rm d}$ defined by Eq.~(56) of CODATA-98, which differs from
$r_{\rm d}$ by less than 0.1\,\%.

\paragraph{Nuclear-size correction to self energy and
vacuum polarization}
\label{par:nssevp}

There is a correction from the finite size of the nucleus to the self
energy \cite{1993124,1997158,2002194,2003127},
\begin{eqnarray}
E_{\rm NSE} = \left(4\ln{2}-\frac{23}{4}\right)
\alpha(Z\alpha){\cal E}_{\rm NS}\delta_{\ell0} \, ,
\label{eq:nse}
\end{eqnarray}
and to the vacuum polarization \cite{1979031,1985047,1997158},
\begin{eqnarray}
E_{\rm NVP} = {3\over4}\alpha(Z\alpha){\cal E}_{\rm NS}\delta_{\ell0} \, .
\label{eq:nvp}
\end{eqnarray}

For the self-energy, higher-order size corrections have been calculated
for S states by \citet{2002194} and for P states by
\citet{2003118,2003138,2004144}.  \citet{2011021} calculated the finite
nuclear size corrections to the self energy and vacuum polarization
nonperturbatively in $Z\alpha$ and has extrapolated the values for the
1S state to $Z=1$.  The results are consistent with the higher-order
analytic results.  Pachucki in a private communication quoted by
\citet{2011021} notes that the coefficients of the leading log terms are
the same for the nuclear size correction to the self energy as they are
for the self-energy correction to the hyperfine splitting.  The latter
terms have been calculated by \citet{2010013}.  However, these
higher-order terms are negligible at the level of accuracy under
consideration.  Corrections for higher-$\ell$ states are also expected
to be negligible.

\paragraph{Radiative-recoil corrections}
\label{par:rrc}

Corrections to the self energy and vacuum polarization for the finite
mass of the nucleus, beyond the reduced-mass corrections already
included, are radiative-recoil effects given by
\citet{2001060,2001075,1995124,1995100,1999164,1999182}:
\begin{eqnarray}
E_{\rm RR} &=& {m_{\rm r}^3\over m_{\rm e}^2m_{\rm N}}
{\alpha(Z\alpha)^5\over \rmpi^2 \, n^3} m_{\rm e}c^2
\delta_{\ell0}
\nonumber\\&&\times\bigg[
6\,\zeta(3) -2\,\rmpi^2\ln{2} + {35\,\rmpi^2\over 36} - {448\over27}
\nonumber\\&& \qquad +
{2\over3}\rmpi(Z\alpha)\,\ln^2{(Z\alpha)^{-2}} + \cdots
\bigg] \ . \qquad
\label{eq:radrec}
\end{eqnarray}
The uncertainty is taken to be the term $(Z\alpha)\ln(Z\alpha)^{-2}$
relative to the square brackets with numerical coefficients 10 for $u_0$
and 1 for $u_n$.  Corrections for higher-$\ell$ states are expected to
be negligible.

\paragraph{Nucleus self energy}
\label{par:nse}

A correction due to the self energy of the nucleus is
\cite{1995100,2001057}
\begin{eqnarray}
E_{\rm SEN} &=& {4Z^2\alpha(Z\alpha)^4 \over 3 \rmpi n^3}
{m_{\rm r}^3\over m_{\rm N}^2}c^2 \nonumber\\
&& \times\left[
\ln{\left({m_{\rm N}\over m_{\rm r}(Z\alpha)^2}\right)}\delta_{\ell0}
-\ln k_0(n,\ell) \right] \, . \qquad
\label{eq:nucse}
\end{eqnarray}
For the uncertainty, we assign a value to $u_0$ corresponding to an additive
constant of 0.5 in the square brackets in Eq.~(\ref{eq:nucse}) for S states.
For higher-$\ell$ states, the correction is not included.

\paragraph{Total energy and uncertainty}
\label{par:teu}

The energy $E_X(n{\rm L}_j)$ of a level (where L = S, P, ... and $X$ = H,
D) is the sum of the various contributions listed in the preceding
sections plus an additive correction $\delta_X(n{\rm L}_j)$ that is zero
with an uncertainty that is the rms sum of the uncertainties of the
individual contributions
\begin{eqnarray}
u^2[\delta_X(n{\rm L}_j)]
= \sum_i {u_{0i}^2(X{\rm L}_j) + u_{ni}^2(X{\rm L}_j) \over n^6 } \ ,
\end{eqnarray}
where $u_{0i}(X{\rm L}_j)/n^3$ and $u_{ni}(X{\rm L}_j)/n^3$ are the components
of
uncertainty $u_0$ and $u_n$ of contribution $i$.  Uncertainties from the
fundamental constants are not explicitly included here, because they are
taken into account through the least-squares adjustment.

The covariance of any two $\delta$'s follows from Eq.~(F7)
of Appendix~F of CODATA-98. For a given isotope
\begin{eqnarray}
u\left[\delta_X(n_1{\rm L}_j),
\delta_X(n_2{\rm L}_j)\right]
= \sum_i{u^2_{0i}(X{\rm L}_j) \over
(n_1 n_2)^3} \ ,
\end{eqnarray}
which follows from the fact that $u(u_{0i}, u_{ni}) = 0$ and
$u(u_{n_1i},u_{n_2i}) = 0$ for $n_1 \ne n_2$.  We also assume that
\begin{eqnarray}
u\left[\delta_X(n_1{\rm L_1}_{j_1}),
\delta_X(n_2{\rm L_2}_{j_2})\right] = 0 \ ,
\end{eqnarray}
if ${\rm L}_1 \ne {\rm L}_2$ or $j_1 \ne j_2$.

For covariances between $\delta$'s for hydrogen and deuterium,
we have for states of the same $n$
\begin{eqnarray}
&&u\left[\delta_{\rm H}(n{\rm L}_j),\delta_{\rm D}(n{\rm L}_j)\right]
\nonumber\\
\ &&= \sum_{i = \{i_{\rm c}\}}{ u_{0i}({\rm HL}_j)u_{0i}({\rm DL}_j)
+ u_{ni}({\rm HL}_j)u_{ni}({\rm DL}_j) \over n^6} \, , \qquad\quad
\end{eqnarray}
and for $n_1\neq n_2$
\begin{eqnarray}
u\left[\delta_{\rm H}(n_1{\rm L}_j),\delta_{\rm D}(n_2{\rm L}_j)\right]
= \sum_{i = i_{\rm c}}{ u_{0i}({\rm HL}_j)u_{0i}({\rm DL}_j)
\over (n_1 n_2)^3} \, ,
\end{eqnarray}
where the summation is over the uncertainties common to hydrogen and
deuterium.  We assume
\begin{eqnarray}
u\left[\delta_{\rm H}(n_1{\rm L_1}_{j_1}),
\delta_{\rm D}(n_2{\rm L_2}_{j_2})\right] = 0 \, ,
\end{eqnarray}
if ${\rm L}_1 \ne {\rm L}_2$ or
$j_1 \ne j_2$.

The values of $u\left[\delta_X(n{\rm L}_j)\right]$ of interest for the
2010 adjustment are given in Table~\ref{tab:rdata} of Sec.~\ref{sec:ad},
and the non negligible covariances of the $\delta$'s are given as
correlation coefficients in Table~\ref{tab:rdcc} of that section.  These
coefficients are as large as 0.9999.

\paragraph{Transition frequencies between levels with $n = 2$ and the
fine-structure constant $\alpha$}
\label{par:trfreq}

To test the QED predictions, we calculate the values of the transition
frequencies between levels with $n=2$ in hydrogen.  This is done by
running the least-squares adjustment with the hydrogen and deuterium
spectroscopic data included, but excluding experimental values for the
transitions being calculated (items $A39$, $A40.1$, and $A40.2$ in
Table~\ref{tab:rdata}).  The necessary constants $A_{\rm r}$(e), $A_{\rm
r}$(p), $A_{\rm r}$(d), and $\alpha$, are assigned their 2010 adjusted
values.  The results are
\begin{eqnarray}
\nu_{\rm H}(2{\rm P}_{1/2} - 2{\rm S}_{1/2}) &=&
1\,057\,844.4(1.8) \ {\rm kHz} \ [ 1.7\times 10^{-6}],  \nonumber \\
\nu_{\rm H}(2{\rm S}_{1/2} - 2{\rm P}_{3/2}) &=&
9\,911\,197.1(1.8) \ {\rm kHz} \ [ 1.8\times 10^{-7}],  \nonumber \\
\nu_{\rm H}(2{\rm P}_{1/2} - 2{\rm P}_{3/2})&&
\nonumber\\&&\hbox to -65pt {} =
10\,969\,041.571(41) \ {\rm kHz} \ [ 3.7\times 10^{-9}] ,
\label{eq:neq2data}
\end{eqnarray}
which are consistent with the relevant experimental results given in
Table~\ref{tab:rdata}.  There is more than a factor of two reduction in
uncertainty in the first two frequencies compared to the corresponding
2006 theoretical values.

We obtain a value for the fine-structure constant $\alpha$ from the data
on the hydrogen and deuterium transitions.  This is done by running a
variation of the 2010 least-squares adjustment that includes all the
transition frequency data in Table~\ref{tab:rdata} and the 2010 adjusted
values of $A_{\rm r}$(e), $A_{\rm r}$(p), and $A_{\rm r}$(d).  This
yields
\begin{eqnarray}
\alpha^{-1} &=&  137.036\,003(41) \qquad [ 3.0\times 10^{-7}] \ ,
\label{eq:alphinvhd}
\end{eqnarray}
which is in excellent agreement with, but substantially less accurate
than, the 2010 recommended value, and is included in
Table~\ref{tab:alpha}.

\paragraph{Isotope shift and the deuteron-proton radius difference}
\label{par:isdprd}

A new experimental result for the hydrogen-deuterium isotope shift is
included in Table \ref{tab:rydfreq} \cite{2010083,2011058}.  In
\citet{2011058} there is a discussion of the theory of the isotope
shift, with the objective of extracting the difference of the squares of
the charge radii for the deuteron and proton.  The analysis in
\citet{2011058} is in general agreement with the review given in the
preceding sections of the present work, with a few differences in the
estimates of uncertainties.

As pointed out by \citet{2011058}, the isotope shift is roughly given by
\begin{eqnarray}
\Delta f_{\rm 1S-2S,d} - \Delta f_{\rm 1S-2S,p}
&\approx& -\frac{3}{4}\,R_\infty c
\left(\frac{m_{\rm e}}{m_{\rm d}}-\frac{m_{\rm e}}{m_{\rm p}}\right)
\nonumber\\&=&
\frac{3}{4}\,R_\infty c \,
\frac{m_{\rm e}\left({m_{\rm d}}-{m_{\rm p}}\right)}{{m_{\rm d}}{m_{\rm p}}} \,
,
\qquad
\label{eq:isos}
\end{eqnarray}
and from a comparison of experiment and theory, they obtain
\begin{eqnarray}
r_{\rm d}^2 - r_{\rm p}^2 &=& 3.820\,07(65) \mbox{ fm}^2
\label{eq:drmpq}
\end{eqnarray}
for the difference of the squares of the radii.  This can be compared to
the result of the 2010 adjustment given by
\begin{eqnarray}
r_{\rm d}^2 - r_{\rm p}^2 &=& 3.819\,89(42) \mbox{ fm}^2 \, ,
\label{eq:drcodata}
\end{eqnarray}
which is in good agreement.  (The difference of the squares of the
quoted 2010 recommended values of the radii gives 87 in the last two
digits of the difference, rather than 89, due to rounding.)  The
uncertainty follows from Eqs.~(F11) and (F12) of CODATA-98.  Here there
is a significant reduction in the uncertainty compared to the
uncertainties of the individual radii because of the large correlation
coefficient (physics.nist.gov/constants)
\begin{eqnarray}
r(r_{\rm d},r_{\rm p}) = 0.9989 \, .
\end{eqnarray}
Part of the reduction in uncertainty in Eq.~(\ref{eq:drcodata}) compared
to Eq.~(\ref{eq:drmpq}) is due to the fact that the correlation
coefficient takes into account the covariance of the electron-nucleon
mass ratios in Eq.~(\ref{eq:isos}).

\subsubsection{Experiments on hydrogen and deuterium}
\label{sssec:rydex}

The hydrogen and deuterium transition frequencies used in the 2010
adjustment for the determination of the Rydberg constant $R_{\infty}$
are given in Table~\ref{tab:rydfreq}. These are items $A26$ to $A48$ in
Table~\ref{tab:rdata}, Sec.~\ref{sec:ad}. There are only three
differences between Table~\ref{tab:rydfreq} and its counterpart, Table
XII, in CODATA-06.

First, the last two digits of the $1{\rm S}_{1/2}-2{\rm S}_{1/2}$
transition frequency obtained by the group at the Max-Planck-Institute
f\"ur Quantenoptik (MPQ), Garching, Germany have changed from $74$ to
$80$ as a result of the group's improved measurement of the $2{\rm S}$
hydrogen hyperfine splitting frequency (HFS). Their result is
\cite{2009096}

\begin{eqnarray}
\nu_{\rm HFS}({\rm H};2\rm S) &=&  177\,556\,834.3(6.7) \mbox{ Hz}
\quad [ 3.8\times 10^{-8}].
\nonumber\\
\label{eq:hhfs2s}
\end{eqnarray}
The reduction in the uncertainty of their previous value for this
frequency \cite{2004006} by a factor of $2.4$ was mainly due to the use
of a new ultra stable optical reference \cite{2008238} and a reanalysis
of the shift with pressure of the $2{\rm S}$ HFS frequency that showed
it was negligible in their apparatus. The $2{\rm S}$ HFS enters the
determination of the $1{\rm S}_{1/2}-2{\rm S}_{1/2}$ transition
frequency because the transition actually
measured is $(1{\rm S},F=1,m_{F}={\pm 1})\rightarrow(2{\rm
S},F^{\prime}=1,m_{F}^{\prime}={\pm 1})$ and the well known $1{\rm S}$
HFS \cite{1990021} and the $2{\rm S}$ HFS are required to convert the
measured frequency to the frequency of the hyperfine centroid.

For completeness, we note that the MPQ group has very recently reported
a new value for the $1{\rm S}_{1/2}-2{\rm S}_{1/2}$ transition frequency
that has an uncertainty of 10 Hz, corresponding to a relative standard
uncertainty of $4.2\times10^{-15}$, or about 30\,\% of the uncertainty
of the value in the table \cite{2011221}.

Second, the previous MPQ value \cite{1998002} for the hydrogen-deuterium
$1{\rm S}-2{\rm S}$ isotope shift, that is, the frequency difference
$\nu_{\rm D}(1{\rm S}_{1/2}-2{\rm S}_{1/2})-\nu_{\rm H}(1{\rm
S}_{1/2}-2{\rm S}_{1/2})$, has been replaced by their recent, much more
accurate value \cite{2010083}; its uncertainty of $15~{\rm Hz}$,
corresponding to a relative uncertainty of $2.2\times10^{-11}$, is a
factor of 10 smaller than the uncertainty of their previous result.
Many experimental advances enabled this significant uncertainty
reduction, not the least of which was the use of a fiber frequency comb
referenced to an active hydrogen maser steered by the Global Positioning
System (GPS) to measure laser frequencies. The principal uncertainty
components in the measurement are $11~{\rm Hz}$ due to density effects
in the atomic beam, $6~{\rm Hz}$ from second-order Doppler shift, and
$5.1~{\rm Hz}$ statistical.

Third, Table~\ref{tab:rydfreq} includes a new result from the group at
the Laboratoire Kastler-Brossel (LKB), \'Ecole Normale Sup\'erieure et
Universit\'e Pierre et Marie Curie, Paris, France.  These researchers
have extended their previous work and determined the $1{\rm
S}_{1/2}-3{\rm S}_{1/2}$ transition frequency in hydrogen using
Doppler-free two-photon spectroscopy with a relative uncertainty of
$4.4\times10^{-12}$ \cite{2010184}, the second smallest uncertainty for
a hydrogen or deuterium optical transition frequency ever obtained. The
transition occurs at a wavelength of $205~{\rm nm}$, and  light at this
wavelength was obtained by twice doubling the frequency of light emitted
by a titanium-sapphire laser of wavelength $820~{\rm nm}$ whose
frequency was measured using an optical frequency comb.

A significant problem in the experiment was the second-order Doppler
effect due to the velocity $v$ of the $1{\rm S}$ atomic beam which
causes an apparent shift of the transition frequency. The velocity was
measured by having the beam pass through a transverse magnetic field,
thereby inducing a motional electric field and hence a quadratic Stark
shift that varies as $v^2$.  The variation of this Stark shift with
field was used to determine $v$ and thus the correction for the
second-order Doppler effect.  The dominant $12.0~{\rm kHz}$ uncertainty
component in the LKB experiment is statistical, corresponding to a
relative uncertainty of $4.1\times10^{-12}$; the remaining components
together contribute an additional uncertainty of only $4.8~{\rm kHz}$.

As discussed in CODATA-98, some of the transition frequencies measured
in the same laboratory are correlated. Table~\ref{tab:rdcc},
Sec~\ref{sec:ad}, gives the relevant correlation coefficients.

\shortcites{2004111,1995159,1998002,1997001,1999072,1996001,
1995138,1994090,1986003,1979001,2010083,2010184}
\def\vsp{\vbox to 10pt{}}
\def\hsp{\hbox to 7.0 pt{}}
\begin{table*}
\caption{
Summary of measured transition frequencies $\nu$ considered in the present
work for the determination of the Rydberg constant $R_\infty$
(H is hydrogen and D is deuterium).
}
\label{tab:rydfreq}
\begin{tabular}{l@{\hsp}l@{\hsp}l@{\hsp}l@{\hsp}l}
\toprule
\noalign{\vbox to 5 pt {}}
Authors  & Laboratory$^1$
& \hbox to 23 pt {} Frequency interval(s) & Reported value
& Rel. stand. \\
& & \hbox to 10 pt {} &
\hbox to 10pt{} $\nu$/kHz & uncert. $u_{\rm r}$ \\
\noalign{\vbox to 5 pt {}}
\colrule
\noalign{\vbox to 5 pt {}}
\cite{2004111} & MPQ
& $\nu_{\rm H}({\rm 1S_{1/2}}-{\rm 2S_{1/2}})$
& $ 2\,466\,061\,413\,187.080(34)$ & $ 1.4\times 10^{-14}$ \\
\vsp\cite{1995159}  & MPQ
& $\nu_{\rm H}({\rm 2S_{1/2}}-{\rm 4S_{1/2}})
- \frac{1}{4}\nu_{\rm H}({\rm 1S_{1/2}}-{\rm 2S_{1/2}})$
& $ 4\,797\,338(10)$ & $ 2.1\times 10^{-6}$ \\
&
& $\nu_{\rm H}({\rm 2S_{1/2}}-{\rm 4D_{5/2}})
- {1\over4}\nu_{\rm H}({\rm 1S_{1/2}}-{\rm 2S_{1/2}})$
& $ 6\,490\,144(24)$ & $ 3.7\times 10^{-6}$ \\
&& $\nu_{\rm D}({\rm 2S_{1/2}}-{\rm 4S_{1/2}})
- {1\over4}\nu_{\rm D}({\rm 1S_{1/2}}-{\rm 2S_{1/2}})$
& $ 4\,801\,693(20)$ & $ 4.2\times 10^{-6}$ \\
&
& $\nu_{\rm D}({\rm 2S_{1/2}}-{\rm 4D_{5/2}})
- {1\over4}\nu_{\rm D}({\rm 1S_{1/2}}-{\rm 2S_{1/2}})$
& $ 6\,494\,841(41)$ & $ 6.3\times 10^{-6}$ \\
\vsp\cite{2010083} & MPQ
& $\nu_{\rm D}({\rm 1S_{1/2}} -{\rm 2S_{1/2}})
- \nu_{\rm H}({\rm 1S_{1/2}} - {\rm 2S_{1/2}})$
& $ 670\,994\,334.606(15)$ & $ 2.2\times 10^{-11}$ \\

\vsp\cite{1997001} & LKB/SYRTE
& $\nu_{\rm H}({\rm 2S_{1/2}}-{\rm 8S_{1/2}})$ &
$ 770\,649\,350\,012.0(8.6)$ & $ 1.1\times 10^{-11}$ \\
& & $\nu_{\rm H}({\rm 2S_{1/2}}-{\rm 8D_{3/2}})$ &
$ 770\,649\,504\,450.0(8.3)$ & $ 1.1\times 10^{-11}$ \\
& & $\nu_{\rm H}({\rm 2S_{1/2}}-{\rm 8D_{5/2}})$ &
$ 770\,649\,561\,584.2(6.4)$ & $ 8.3\times 10^{-12}$ \\
& & $\nu_{\rm D}({\rm 2S_{1/2}}-{\rm 8S_{1/2}})$ &
$ 770\,859\,041\,245.7(6.9)$ & $ 8.9\times 10^{-12}$ \\
& & $\nu_{\rm D}({\rm 2S_{1/2}}-{\rm 8D_{3/2}})$ &
$ 770\,859\,195\,701.8(6.3)$ & $ 8.2\times 10^{-12}$ \\
& & $\nu_{\rm D}({\rm 2S_{1/2}}-{\rm 8D_{5/2}})$ &
$ 770\,859\,252\,849.5(5.9)$ & $ 7.7\times 10^{-12}$ \\

\vsp\cite{1999072} & LKB/SYRTE
& $\nu_{\rm H}({\rm 2S_{1/2}}-{\rm 12D_{3/2}})$ &
$ 799\,191\,710\,472.7(9.4)$ & $ 1.2\times 10^{-11}$ \\
& & $\nu_{\rm H}({\rm 2S_{1/2}}-{\rm 12D_{5/2}})$ &
$ 799\,191\,727\,403.7(7.0)$ & $ 8.7\times 10^{-12}$ \\
& & $\nu_{\rm D}({\rm 2S_{1/2}}-{\rm 12D_{3/2}})$ &
$ 799\,409\,168\,038.0(8.6)$ & $ 1.1\times 10^{-11}$ \\
& & $\nu_{\rm D}({\rm 2S_{1/2}}-{\rm 12D_{5/2}})$ &
$ 799\,409\,184\,966.8(6.8)$ & $ 8.5\times 10^{-12}$ \\

\vsp\cite{2010184} & LKB
& $\nu_{\rm H}({\rm 1S_{1/2}}-{\rm 3S_{1/2}})$
& $ 2\,922\,743\,278\,678(13)$ & $ 4.4\times 10^{-12}$ \\

\vsp \cite{1996001} & LKB
& $\nu_{\rm H}({\rm 2S_{1/2}}-{\rm 6S_{1/2}})
- {1\over4}\nu_{\rm H}({\rm 1S_{1/2}}-{\rm 3S_{1/2}})$
& $ 4\,197\,604(21)$ & $ 4.9\times 10^{-6}$ \\
&& $\nu_{\rm H}({\rm 2S_{1/2}}-{\rm 6D_{5/2}})
- {1\over4}\nu_{\rm H}({\rm 1S_{1/2}}-{\rm 3S_{1/2}})$
& $ 4\,699\,099(10)$ & $ 2.2\times 10^{-6}$ \\

\vsp\cite{1995138} & Yale
& $\nu_{\rm H}({\rm 2S_{1/2}}-{\rm 4P_{1/2}})
- {1\over4}\nu_{\rm H}({\rm 1S_{1/2}}-{\rm 2S_{1/2}})$
& $ 4\,664\,269(15)$ & $ 3.2\times 10^{-6}$ \\
&& $\nu_{\rm H}({\rm 2S_{1/2}}-{\rm 4P_{3/2}})
- {1\over4}\nu_{\rm H}({\rm 1S_{1/2}}-{\rm 2S_{1/2}})$
& $ 6\,035\,373(10)$ & $ 1.7\times 10^{-6}$ \\

\vsp\cite{1994090} & Harvard
& $\nu_{\rm H}({\rm 2S_{1/2}}-{\rm 2P_{3/2}})$
& $ 9\,911\,200(12)$ & $ 1.2\times 10^{-6}$ \\

\vsp\cite{1986003} & Harvard
& $\nu_{\rm H}({\rm 2P_{1/2}}-{\rm 2S_{1/2}})$
& $ 1\,057\,845.0(9.0)$ & $ 8.5\times 10^{-6}$ \\

\vsp\cite{1979001} & U. Sussex
& $\nu_{\rm H}({\rm 2P_{1/2}}-{\rm 2S_{1/2}})$
& $ 1\,057\,862(20)$ & $ 1.9\times 10^{-5}$ \\
\botrule
\end{tabular}
$^1$MPQ: Max-Planck-Institut f\"ur Quantenoptik, Garching.
LKB: Laboratoire Kastler-Brossel, Paris.
SYRTE: Syst\`emes de r\'ef\'erence Temps Espace, Paris, formerly
Laboratoire Primaire du Temps et des Fr\'equences (LPTF).
\end{table*}

\subsubsection{Nuclear radii}
\label{sssec:nucrad}

Transition frequencies in hydrogen and deuterium depend on the rms
charge radius of the nucleus, denoted by $r_{\rm p}$ and $r_{\rm d}$
respectively.  The main difference between energy levels for a point
charge nucleus and for a nucleus with a finite charge radius is given by
Eq.~(\ref{eq:enscoef}).  These radii are treated as adjusted constants,
so the H and D experimental transition-frequency input data, together
with theory, provide adjusted values for them.

\paragraph{Electron scattering}

The radii can also be determined from elastic electron-proton (e-p)
scattering data in the case of $r_{\rm p}$, and from elastic
electron-deuteron (e-d) scattering data in the case of $r_{\rm d}$.
These independently determined values are used as additional input data
which, together with the H and D spectroscopic data and the theory,
determine the 2010 recommended values of the radii.  The experimental
electron-scattering values of $r_{\rm p}$ and $r_{\rm d}$ that we take
as input data in the 2010 adjustment are
\begin{eqnarray}
r_{\rm p} &=&  0.895(18) \ {\rm fm} \, ,
\label{eq:rpes03}\\
r_{\rm p} &=&  0.8791(79) \ {\rm fm} \, ,
\label{eq:rpbes10}\\
r_{\rm d} &=&  2.130(10) \ {\rm fm} \, .
\label{eq:rdes98}
\end{eqnarray}

The first result for $r_{\rm p}$, which was also used in the 2002 and
2006 adjustments, is due to \citet{2003222,2007137,2008275} and is based
on a reanalysis of the world e-p cross section and polarization transfer
data.  The value in Eq.~(\ref{eq:rpes03}) is consistent with the more
accurate result $r_{\rm p}=0.894(8)$ reported after the closing date of
the 2010 adjustment by \citet{2011093} using an improved method to treat
the proton's charge density at large radii. It is also consistent with
the very recent result $r_{\rm p}=0.886(8)$ calculated by \citet{pc12is}
that extends this method and is based in part on the data obtained by
\citet{2010199} in the experiment that yields the second result for
$r_{\rm p}$, which we now discuss. [Note that the recent paper of
\citet{pc12is} gives an overview of the problems associated with
determining a reliable value of $r_{\rm p}$ from e-p scattering data.
Indeed, \citet{pc12ca} find $r_{\rm p}=0.844(7)$ based on a reanalysis
of selected nucleon form-factor data; see also \citet{2007354}.]

The value of $r_{\rm p}$ given in Eq.~(\ref{eq:rpbes10}) was obtained at
the Mainz University, Germany, with the Mainz linear electron
accelerator MAMI. About 1400 elastic e-p scattering cross sections were
measured at six beam energies from 180~MeV to 855~MeV, covering the
range of four-momentum transfers squared from $Q^2=0.004$~(GeV/c)$^2$ to
1~(GeV/c)$^2$.  The value of $r_{\rm p}$ was extracted from the data using
spline fits or polynomial fits, and because the reason for the
comparatively small difference between the resulting values could not be
identified, \citet{2010199} give as their final result the average of
the two values with an added uncertainty equal to half the difference.
[Note that the value in Eq.~(\ref{eq:rpbes10}) contains extra digits
provided by \citet{pc10jb}. See also the exchange of comments of
\citet{2011180,2011179}.]

The result for $r_{\rm d}$ is that given by \citet{2008275} and is based
on an analysis of the world data on e-d scattering similar to that used
to determine the value of $r_{\rm p}$ in Eq.~(\ref{eq:rpes03}).

For completeness we note the recent e-p scattering result for $r_{\rm
p}$ based in part on new data obtained in the range $Q^2 = 0.3~({\rm
GeV}/c)^2$ to $0.7~({\rm GeV}/c)^2$ at the Thomas Jefferson National
Accelerator Facility, Newport News, Virginia, USA, often referred to as
simply JLab. The new data, acquired using a technique called
polarization transfer or recoil polarimetry, were combined with previous
cross section and polarization measurements to produce the result
$r_{\rm p}= 0.875(10)$~fm from an updated global fit in this range of
$Q^2$ \cite{2011222,2011242}. It is independent of and agrees with the
Mainz result in Eq.~(\ref{eq:rpbes10}), and it also agrees with the
result in Eq.~(\ref{eq:rpes03}) but the two are not independent since
the data used to obtain the latter result were included in the JLab fit.
This result became available after the 31 December 2010 closing date of
the 2010 adjustment.

\paragraph{Muonic hydrogen}
\label{par:muhrad}

A muonic hydrogen atom, $\rmmu^-$p, consists of a negative muon and a
proton. Since $m_{\rmssmu}/m_{\rm e} \approx 207$, the Bohr radius of
the muon is about 200 times smaller than the electron Bohr radius, so
the muon is more sensitive to the size of the nucleus.  Indeed, the
finite-size effect for the 2S state in ${\rmmu^-}$p is about 2\,\% of
the total Lamb shift, that is, the energy difference between the 2S and
2P states, which should make it an ideal system for measuring the size
of the proton. (Because of the large electron vacuum polarization effect
in muonic hydrogen, the $2{\rm S}_{1/2}$ level is well below both the
$2{\rm P}_{3/2}$ and $2{\rm P}_{1/2}$ levels.)

In a seminal experiment carried out using pulsed laser spectroscopy at a
specially built muon beam line at the proton accelerator of the Paul
Scherrer Institute (PSI), Villigen, Switzerland, \citet{2010108,2011079}
have measured the 206~meV (50~THz or 6~${\rmmu \rm m}$) ${\rmmu^-}$p
Lamb shift, in particular, the $2{\rm S}_{1/2}(F=1)-2{\rm P}_{3/2}(F=2)$
transition, with an impressive relative standard uncertainty of 15 parts
in $10^6$. The result, when combined with the theoretical expression for
the transition, leads to \cite{2011017}
\begin{eqnarray}
r_{\rm p} &=& 0.84169(66) \ {\rm fm} \, .
\label{eq:rpmuhuj}
\end{eqnarray}

The value given in Eq.~(\ref{eq:rpmuhuj}) is based on a review
and reanalysis of the theory by \citet{2011017,2011018} but is not
significantly different from the value first given by \citet{2010108}.
Because the muonic hydrogen value of $r_{\rm p}$ differs markedly from
the 2006 CODATA recommended value given in CODATA-06, its publication in
2010 has led to a significant number of papers that reexamine various
aspects of the theory or propose possible reasons for the disagreement;
see, for example, the recent review of \citet{2012024}.  If
Eq.~(\ref{eq:rpmuhuj}) is compared to the 2010 recommended value of
$0.8775(51)$~fm, the disagreement is $7\,\sigma$.  If it is compared to
the value $0.8758(77)$~fm based on only H and D spectroscopic data (see
Table~\ref{tab:adjustsa}), the disagreement is $4.4\,\sigma$.

The impact of including Eq.~(\ref{eq:rpmuhuj}) on the 2010 adjustment
and the reasons the Task Group decided not to include it are discussed
in Sec.~\ref{sssec:idd}. We also note the following fact.  If the
least-squares adjustment that leads to the value of $\alpha$ given in
Eq.~(\ref{eq:alphinvhd}) is carried out with the value in
Eq.~(\ref{eq:rpmuhuj}) added as an input datum, the result is
$\alpha^{-1} = 137.035\,881(35)~[2.6\times10^{-7}]$, which differs from
the 2010 recommended value by $3.4\,\sigma$.  The value of $R_\infty$
from this adjustment is $10\,973\,731.568\,016(49)$ m$^{-1}$.

\subsection{Antiprotonic helium transition frequencies and
$\bm{A_{\rm r}({\rm e})}$}
\label{ssec:aph}

Consisting of a $^4{\rm He}$ or a $^3{\rm He}$ nucleus, an antiproton,
and an electron, the antiprotonic helium atom is a three-body system
denoted by $\bar{\rm p}{\rm He}^+$. Because it is assumed that $CPT$ is
a valid symmetry, determination of the antiproton-electron mass ratio
from antiprotonic helium experiments can be interpreted as determination
of the proton-electron mass ratio.  Further, because the relative atomic
mass of the proton $A_{\rm r}(\rm p)$ is known with a significantly
smaller relative uncertainty from other data than is $A_{\rm r}(\rm e)$,
a value of the antiproton-electron mass ratio with a sufficiently small
uncertainty can provide a competitive value of $A_{\rm r}(\rm e)$.

Theoretical and experimental values of frequencies corresponding to
transitions between atomic levels of the antiprotons with large
principal quantum number $n$ and angular momentum quantum number $l$,
such that $n\approx{l+1}\approx38$, were used to obtain a value of
$A_{\rm r}(\rm e)$ in the 2006 adjustment.  Table~\ref{tab:aphe}
summarizes the relevant experimental and theoretical data. The first
column indicates the mass number of the helium nucleus of the
antiprotonic atom and the principal and angular momentum quantum numbers
of the energy levels involved in the transitions.  The second column
gives the experimentally measured values of the transition frequencies
while the third gives the theoretically calculated values.  The last two
columns give the values in the unit $2cR_\infty$ of quantities $a$ and
$b$ used in the observational equations that relate the experimental
values of the transition frequencies to their calculated values and
relevant adjusted constants, as discussed in the next section. Besides a
few comparatively minor changes in some of the calculated frequencies
and their uncertainties, the only significant difference between
Table~\ref{tab:aphe} and the corresponding Table~XIII in CODATA-06 is
the addition of recently acquired data on three two-photon transitions:
$(33,32)\rightarrow(31,30)$ and $(36,34)\rightarrow(34,32)$ for
$\bar{\rm p}^4{\rm He}^+$, and $(35,33)\rightarrow(33,31)$ for $\bar{\rm
p}^3{\rm He}^+$.

It is noteworthy that \citet{2011150}, who determined the experimental
values of these three frequencies (discussed further in
Sec.~\ref{sssec:aphelexh} below), have used the new experimental and
theoretical data to obtain an important new limit. With the aid of the
long-known result that the  absolute value of the charge-to-mass ratio
of $\rm p$ and $\bar {\rm p}$ are the same within at least $9$ parts in
$10^{11}$ \cite{2006037}, they showed that the charge and mass of $\rm
p$ and $\bar{\rm p}$ are the same within $7$ parts in $10^{10}$ at the
$90\%$ confidence level.

\subsubsection{Theory relevant to antiprotonic helium}
\label{sssec:aphelth}

The calculated transition frequencies in Table~\ref{tab:aphe} are due to
\citet{2008109,pc10vk} and are based on the 2002 recommended values of
the required fundamental constants with no uncertainties. Korobov's
publication updates some of the values and uncertainties of the
calculated transition frequencies used in the 2006 adjustment that he
provided directly to the Task Group \cite{pc06vk}, but it also includes
results for the $\bar{\rm p}^4{\rm He}^+$ and $\bar{\rm p}^3{\rm He}^+$
two-photon transition frequencies $(36,34)\rightarrow(34,32)$ and
$(35,33)\rightarrow(33,31)$. The calculated value for the $\bar{\rm
p}^4{\rm He}^+$ two-photon frequency $(33,32)\rightarrow(31,30)$ was
again provided directly to the Task Group by \citet{pc10vk}, as were
slightly updated values for the two other two-photon frequencies. The
same calculated values of the three two-photon frequencies are also
given in the paper by \citet{2011150} cited above.

The quantities $a\equiv a_{\bar{\rm p}\rm He}(n,l:{n^\prime},{l^\prime})$
and $b\equiv b_{\bar{\rm p}\rm He}(n,l:{n^\prime},{l^\prime})$ in
Table~\ref{tab:aphe}, also directly  provided to the Task Group by
\citet{pc10vk,pc06vk}, are actually the numerical values of derivatives
defined and used as follows (in these and other similar expressions in
this section, $\rm He$ is $^3{\rm He}$ or $^4{\rm He}$).

The theoretical values of the transition frequencies are functions of
the mass ratios $A_{\rm r}(\bar{\rm p})/A_{\rm r}(\rm e)$ and $A_{\rm
r}(N)/A_{\rm r}(\bar{\rm p})$, where $N$ is either $^4{\rm He}^{2+}$ or
$^3{\rm He}^{2+}$, that is, the alpha particle $\rm\alpha$ or helion
$\rm h$. If the transition frequencies as a function of these mass
ratios are denoted by $\nu_{\bar{\rm p}\rm
He}(n,l:{n^\prime},{l^\prime})$, and  the calculated values in
Table~\ref{tab:aphe} by $\nu^{(0)}_{\bar{\rm p}\rm
He}(n,l:{n^\prime},{l^\prime})$, we have
\begin{eqnarray}
a_{\rm \bar p\,{\rm He}}(n,l:n^\prime,l^\prime)
&=&
\left(\frac{A_{\rm r}({\rm \bar p})}{A_{\rm r}({\rm e})}\right)^{(0)}
\frac{\partial \Delta \nu_{\rm \bar p\,{\rm He}}(n,l:n^\prime,l^\prime)}
{\partial \left(\frac{A_{\rm r}({\rm \bar p})}{A_{\rm r}({\rm
e})}\right)}\, ,
\nonumber \\ \\
b_{\rm \bar p\,{\rm He}}(n,l:n^\prime,l^\prime)
&=&\left(\frac{A_{\rm r}({\rm He})}{A_{\rm r}({\rm \bar p})}\right)^{(0)}
\frac{\partial \Delta \nu_{\rm \bar p\,{\rm He}}(n,l:n^\prime,l^\prime)}
{\partial \left(\frac{A_{\rm r}({N})}{A_{\rm r}({\rm \bar p})}\right)}
\, ,
\nonumber \\
\end{eqnarray}
where the superscript $(0)$ denotes the fact that the 2002 CODATA values
of the relative atomic mass ratios were used by Korobov in his
calculations.  The zero-order frequencies, mass ratios, and the
derivatives $a$ and $b$ provide a first-order approximation to the
transition frequencies as a function of changes in the mass ratios:
\begin{eqnarray}
&&\hbox to -10 pt {} \nu_{\rm \bar p\,{\rm He}}\,(n,l:n^\prime,l^\prime)
= \nu_{\rm \bar p\,{\rm He}}^{(0)}(n,l:n^\prime,l^\prime)
\nonumber\\
&&+ a_{\rm \bar p\,{\rm He}}(n,l:n^\prime,l^\prime)
\left[\left(\frac{A_{\rm r}({\rm e})}{A_{\rm r}({\rm \bar p)}}\right)^{\!(0)}
\!\!
\left(\frac{A_{\rm r}({\rm \bar p}\,)}{A_{\rm r}({\rm e})}\right)-1 \right]
\label{eq:apheth}
\nonumber\\
&&+ b_{\bar{\rm p}\,{\rm He}}(n,l:n^\prime,l^\prime)
\left[\left(\frac{A_{\rm r}({\rm \bar p})}{A_{\rm r}({N)}}\right)^{\!(0)} \!\!
\left(\frac{A_{\rm r}({N})}{A_{\rm r}({\rm \bar p})}\right)-1 \right]
\nonumber\\ && + \dots \, .
\end{eqnarray}

This expression is the basis for the observational equations for the
measured and calculated transition frequencies as a function of the mass
ratios in the least-squares adjustment; see Table~\ref{tab:pobseqsc},
Sec.~\ref{sec:ad}. Although $A_{\rm r}(\rm e)$, $A_{\rm r}(\rm p)$ and
$A_{\rm r}(N)$ are adjusted constants, the principal effect of including
the antiprotonic helium transition frequencies in the adjustment is to
provide information about $A_{\rm r}(\rm e)$.  This is because
independent data in the adjustment provide values of $A_{\rm r}(\rm p)$
and $A_{\rm r}(N)$ with significantly smaller relative uncertainties
than the uncertainty of $A_{\rm r}(\rm e)$.

The uncertainties of the calculated transition frequencies  are taken
into account by including an additive constant $\delta_{\bar{\rm p}\rm
He}(n,l:{n^\prime},{l^\prime})$ in the observational equation for each
measured frequency; see Table~\ref{tab:adjconc} and $C13-C24$ in
Table~\ref{tab:pobseqsc}, Sec.~\ref{sec:ad}.  The additive constants are
adjusted constants and their assigned values are zero with the
uncertainties of the theoretical values.  They are data items $C1$ to
$C15$ in Table~\ref{tab:cdata}.  Moreover, the input data for the
additive constants are correlated; their correlation coefficients,
calculated from information provided by \citet{pc10vk}, are given in
Table~\ref{tab:cdcc}.  (In the 2006 adjustment, the correlations between
the $^4{\rm He}$ and $^3{\rm He}$ calculated frequencies were omitted.)

\subsubsection{Experiments on antiprotonic helium}
\label{sssec:aphelexh}

Recent reviews of the experimental work, which is carried out at CERN,
have been given by \citet{2011026} and by \citet{2010055}.  The first
seven $^4{\rm He}$ and the first five $^3{\rm He}$ experimental
transition frequencies in Table~\ref{tab:aphe}, obtained by
\citet{2006056}, were used in the 2006 adjustment and are discussed in
CODATA-06. The measurements were carried out with antiprotons from the
CERN Antiproton Decelerator and employed the technique of single-photon
precision laser-spectroscopy. The transition frequencies and their
uncertainties include an extra digit beyond those reported by
\citet{2006056} that were provided to the Task Group by \citet{pc06mh}
to reduce rounding errors.

During the past 4 years the CERN group has been able to improve their
experiment and, as noted above, \citet{2011150} have recently  reported
results for three transitions based on two-photon laser spectroscopy. In
this work $\bar{\rm p}^4{\rm He}^+$ or $\bar{\rm p}^3{\rm He}^+$ atoms
are irradiated by two counter-propagating laser beams that excite deep
ultraviolet, nonlinear, two-photon transitions of the type
$(n,l)\rightarrow(n-2,,l-2)$. This technique reduces thermal Doppler
broadening of the resonances of the antiprotonic atoms, thereby
producing narrower spectral lines and reducing the uncertainties of the
measured transition frequencies.

In normal two-photon spectroscopy the frequencies of the two counter
propagating laser beams are the same and equal to one-half the resonance
frequency. In consequence, to first order in the atom's velocity,
Doppler broadening is reduced to zero. However, normal two-photon
spectroscopy is difficult to do in antiprotonic helium because of the
small transition probabilities of the  nonlinear two-photon transitions.
The CERN group was able to mitigate this problem by using the fact that
the probability can be increased some five orders of magnitude if the
two beams have different frequencies $\nu_1$ and $\nu_2$ such that the
virtual state of the two-photon transition is within approximately
$10~{\rm GHz}$ of a real state with quantum numbers $(n-1,l-1)$
\cite{2010114}. In this case the first-order Doppler width of the
resonance is reduced by the factor $|\nu_1-\nu_2|/(\nu_1+\nu_2)$.

As for the earlier data, an extra digit, provided to the Task Group by
\citet{pc10mh}, has been added to the three new two-photon frequencies
and their uncertainties.  Further, as for the one-photon transitions
used in 2006, \citet{pc10mh} has provided the Task Group with a detailed
uncertainty budget for each of the new frequencies so that their
correlation coefficients could be properly evaluated. (There are no
correlations between the 12 older one-photon frequencies and the 3 new
two-photon frequencies.) As for the one-photon frequencies, the dominant
uncertainty component for the two-photon frequencies is statistical; it
varies from $3.0~{\rm MHz}$ to $6.6~{\rm MHz}$ compared to $3.2~{\rm
MHz}$ to $13.8~{\rm MHz}$ for the one-photon frequencies.  The 15
transition frequencies are data items $16$ to $30$ in
Table~\ref{tab:cdata}; all relevant correlation coefficients are given
in Table~\ref{tab:cdcc}.

\subsubsection{Inferred value of $A_{\rm r}(\rm e)$ from antiprotonic helium}
\label{sssec:apheare}

Use of the 2010 recommended values of $A_{\rm r}(\rm p)$, $A_{\rm r}(\rm
\alpha)$, and $A_{\rm r}(\rm h)$, the experimental and theoretical
values of the $15$ transition frequencies in Table~\ref{tab:aphe}, the
correlation coefficients in Table~\ref{tab:cdcc}, and the observational
equations in Table~\ref{tab:pobseqsc} derived as discussed above, yields
the following inferred value of the electron relative atomic mass:
\begin{eqnarray}
A_{\rm r}({\rm e}) &=&   0.000\,548\,579\,909\,14(75) \quad [ 1.4\times 10^{-9}]
\, . \qquad
\label{eq:areaphe}
\end{eqnarray}
The $\bar {\rm p} ^3{\rm He}$ data alone give a value of $A_{\rm r}({\rm
e})$ that has an uncertainty that is $1.7$ times as large as the
uncertainty of the value in Eq.~(\ref{eq:areaphe}); and it is smaller by
a factor $1.2$ times its uncertainty.  The combined result is consistent
and competitive with other values, as discussed in Sec.~\ref{sec:ad}.

\begin{table*}
\caption{
Summary of data related to the determination of $A_{\rm r}(\rm e)$ from
measurements of antiprotonic helium. The uncertainties of the 15
calculated values are the root-sum-square (rss) of the following 15
pairs of uncertainty components in MHz, where the first component
reflects the possible size of uncalculated terms of order
$R_{\infty}\alpha^5\ln\alpha$ and higher, and the second component
reflects the uncertainty of the numerical calculations: (0.8, 0.2);
(1.0, 0.3); (1.1, 0.3); (1.1, 0.3); (1.1, 0.4); (1.0, 0.8); (1.8, 0.4);
(1.6, 0.3); (2.1, 0.3); (0.9, 0.1); (1.1, 0.2); (1.1, 0.4); (1.1, 0.3);
(1.8, 0.3); (2.2, 0.2).
}
\label{tab:aphe}
\begin{tabular}{c@{\hbox to 40pt {}}D{.}{.}{13.11}D{.}{.}{13.11}D{.}{.}{6.6}D{.}
{.}{6.6}}
\toprule
\vbox to 10 pt {}
Transition   & \text{Experimental} \s{-17}& \text{Calculated}\s{-10}
& a\s{-10}  & b\s{-10} \\

$(n,l)\rightarrow(n^\prime,l^\prime)$& \text{Value (MHz)}\s{-15} &
\text{Value (MHz)}\s{-15} & (2cR_\infty)\s{-25} & (2cR_\infty)\s{-25}\\

\colrule
\vbox to 10 pt {}

$\bar{\rm p}^4$He$^+$: $(32,31) \rightarrow (31,30)$ &
1\,132\,609\,209(15)\s{-16} &  1\,132\,609\,223.50(82) &  0.2179 &  0.0437  \\
$\bar{\rm p}^4$He$^+$: $(35,33) \rightarrow (34,32)$ &  804\,633\,059.0(8.2)
&  804\,633\,058.0(1.0) &  0.1792 &  0.0360  \\
$\bar{\rm p}^4$He$^+$: $(36,34) \rightarrow (35,33)$ &  717\,474\,004(10)\s{-16}
&  717\,474\,001.1(1.1) &  0.1691 &  0.0340  \\
$\bar{\rm p}^4$He$^+$: $(37,34) \rightarrow (36,33)$ &  636\,878\,139.4(7.7)
&  636\,878\,151.7(1.1) &  0.1581 &  0.0317  \\
$\bar{\rm p}^4$He$^+$: $(39,35) \rightarrow (38,34)$ &  501\,948\,751.6(4.4)
&  501\,948\,755.6(1.2) &  0.1376 &  0.0276  \\
$\bar{\rm p}^4$He$^+$: $(40,35) \rightarrow (39,34)$ &  445\,608\,557.6(6.3)
&  445\,608\,569.3(1.3) &  0.1261 &  0.0253  \\
$\bar{\rm p}^4$He$^+$: $(37,35) \rightarrow (38,34)$ &  412\,885\,132.2(3.9)
&  412\,885\,132.8(1.8) &  -0.1640 &  -0.0329  \\
$\bar{\rm p}^4$He$^+$: $(33,32) \rightarrow (31,30)$ &  2\,145\,054\,858.2(5.1)
&  2\,145\,054\,857.9(1.6) &  0.4213 &  0.0846  \\
$\bar{\rm p}^4$He$^+$: $(36,34) \rightarrow (34,32)$ &  1\,522\,107\,061.8(3.5)
&  1\,522\,107\,058.9(2.1) &  0.3483 &  0.0699  \\
\colrule
\vbox to 10 pt {}
$\bar{\rm p}^3$He$^+$: $(32,31) \rightarrow (31,30)$ &
1\,043\,128\,608(13)\s{-16} &  1\,043\,128\,579.70(91) &  0.2098 &  0.0524  \\
$\bar{\rm p}^3$He$^+$: $(34,32) \rightarrow (33,31)$ &  822\,809\,190(12)\s{-16}
&  822\,809\,170.9(1.1) &  0.1841 &  0.0460  \\
$\bar{\rm p}^3$He$^+$: $(36,33) \rightarrow (35,32)$ &  646\,180\,434(12)\s{-16}
&  646\,180\,408.2(1.2) &  0.1618 &  0.0405  \\
$\bar{\rm p}^3$He$^+$: $(38,34) \rightarrow (37,33)$ &  505\,222\,295.7(8.2)
&  505\,222\,280.9(1.1) &  0.1398 &  0.0350  \\
$\bar{\rm p}^3$He$^+$: $(36,34) \rightarrow (37,33)$ &  414\,147\,507.8(4.0)
&  414\,147\,507.8(1.8) &  -0.1664 &  -0.0416  \\
$\bar{\rm p}^3$He$^+$: $(35,33) \rightarrow (33,31)$ &  1\,553\,643\,099.6(7.1)
&  1\,553\,643\,100.7(2.2) &  0.3575 &  0.0894  \\

\botrule
\end{tabular}
\end{table*}

\subsection{Hyperfine structure and fine structure}
\label{ssec:hfsfs}

During the past 4 years two highly accurate values of the
fine-structure constant $\rm\alpha$ from dramatically different
experiments have become available, one from the electron magnetic-moment
anomaly $a_{\rm e}$ and the other from $h/m(^{87}{\rm Rb})$ obtained by
atom recoil. They are consistent and have relative standard
uncertainties of $3.7\times 10^{-10}$ and $6.6\times 10^{-10}$,
respectively; see Table~\ref{tab:alpha}. These uncertainties imply that
for another value of $\rm\alpha$ to be competitive, its relative
uncertainty should be no more than about a factor of 10 larger.

By equating the experimentally measured ground-state hyperfine
transition frequency of a simple atom such as hydrogen, muonium
(${\rm\mu}^{+}{\rm e}^{-}$ atom), or positronium (${\rm e}^{+}{\rm
e}^{-}$ atom) to its  theoretical expression, one could in principle
obtain a value of $\rm\alpha$, since this frequency is proportional to
${\rm\alpha}^2{R_\infty}c$. Muonium is, however, still the only atom for
which both the measured value of the hyperfine frequency and its
theoretical expression have sufficiently small uncertainties to be of
possible interest, and even for this atom with a structureless nucleus
the resulting value of ${\rm\alpha}$ is no longer competitive; instead,
muonium  provides the most accurate value of the electron-muon mass
ratio, as discussed in Sec.~\ref{ssec:muhfs}.

Also proportional to ${\rm\alpha}^2{R_\infty}c$ are fine-structure
transition frequencies, and thus in principal these could provide a
useful value of $\rm\alpha$. However, even the most accurate
measurements of such frequencies in the relatively simple one-electron
atoms hydrogen and deuterium do not provide a competitive value; see
Table~\ref{tab:rydfreq} and Sec.~\ref{par:trfreq}, especially
Eq.~(\ref{eq:alphinvhd}). Rather, the experimental hydrogen fine-structure
transition frequencies given in that table are included in the 2010
adjustment, as in past adjustments, because of their influence on the
adjusted constant $R_{\infty}$.

The large natural line widths of the $2{\rm P}$ levels in H and D limit
the accuracy with which the fine-structure frequencies in these atoms
can be measured. By comparison, the $2^{3}{\rm P}_{J}$ states of $^4{\rm
He}$ are narrow ($1.6~{\rm MHz}$ vs. $100~{\rm MHz}$) because they
cannot decay to the ground $1^{1}{\rm S}_{0}$ state by allowed electric
dipole transitions. Since the energy differences between the three
$2^{3}{\rm P}$ levels and the corresponding transition frequencies can
be calculated and measured with reasonably small uncertainties, it has
long been hoped that the fine structure of $^4{\rm He}$ could one day
provide a competitive value of $\rm\alpha$. Although the past 4 years
has seen considerable progress toward this goal, it has not yet been
reached. In brief, the situation is as follows.

The fine structure of the $2^{3}{\rm P}_{J}$ triplet state of $^4{\rm
He}$ consists of three levels; they are, from highest to lowest,
$2^{3}{\rm P}_{0}$, $2^{3}{\rm P}_{1}$, and $2^{3}{\rm P}_{2}$.  The
three transition frequencies of interest are
${\rm\nu}_{01}\approx29.6~{\rm GHz}$, ${\rm\nu}_{12}\approx2.29~{\rm
GHz}$, and ${\rm\nu}_{02}\approx31.9~{\rm GHz}$. In a series of papers
\citet{2006053} and  \citet{2009166,2010012,2011029,2011072}, but see
also \citet{2010024} and \citet{2010047}, have significantly advanced
the theory of these transitions in both helium and light helium-like
ions. Based on this work, the theory is now complete to orders
$m\alpha^{7}$ and $m(m/M)\alpha^{6}$ ($m$ the electron mass and $m/M$
the electron-alpha particle mass ratio), previous disagreements among
calculations have been resolved, and an estimate of  uncertainty due to
the uncalculated $m\alpha^{8}$ term has been made. Indeed, the
uncertainty of the theoretical expression for the ${\rm\nu}_{02}$
transition, which is the most accurately known both theoretically and
experimentally, is estimated to be $1.7~{\rm kHz}$, corresponding to a
relative uncertainty of $5.3\times 10^{-8}$ or $2.7\times 10^{-8}$ for
${\alpha}$. Nevertheless, even if an experimental value of
${\rm\nu}_{02}$ with an uncertainty of just a few hertz were available,
the uncertainty in the value of $\alpha$ from helium fine structure
would still be too large to be included in the 2010 adjustment

In fact, the most accurate experimental value of ${\rm\nu}_{02}$ is that
measured by  \citet{2010147} with an uncertainty of $300~\rm Hz$,
corresponding to a relative uncertainty of $9.4\times 10^{-9}$ or
$4.7\times10^{-9}$ for ${\rm\alpha}$.  As given by \citet{2011029}, the
value of ${\rm\alpha}$ obtained by equating this experimental result and
the theoretical result is ${\alpha}^{-1}=137.035\,9996(37)$
$[2.7\times10^{-8}]$, which agrees well with the two most accurate
values mentioned at the start of this section but is not competitive
with them.

Another issue is that the agreement among different experimental values
of the various helium fine-structure transitions and their agreement
with theory is not completely satisfactory.  Besides the result of
\citet{2010147} for ${\rm\nu}_{02}$, there is the measurement of
${\rm\nu}_{12}$ by \citet{2009169}, all three frequencies by
\citet{2005131}, ${\rm\nu}_{01}$ by \citet{2005145}, ${\rm\nu}_{01}$ by
\citet{2001254}, ${\rm\nu}_{12}$ by  \citet{2000022}, and
${\rm\nu}_{02}$ by \citet{1995049}. Graphical comparisons of these data
among themselves and with theory may be found in the paper by
\citet{2010147}.

In summary, no $^4{\rm He}$ fine-structure datum is included in the 2010
adjustment, because the resulting value of $\rm\alpha$ has too large an
uncertainty compared to the uncertainties of the values from $a_{\rm e}$
and  $h/m(^{87}{\rm Rb})$.

\section{Magnetic moment anomalies and $\bm g$-factors}
\label{sec:mmagf}

As discussed in CODATA-06, the magnetic moment of any of the three
charged leptons $\ell={\rm e},\,\rmmu,\,\rmtau$ is
\begin{eqnarray}
\bm\mu_\ell = g_\ell { e \over 2m_\ell}\bm s \ ,
\label{eq:lgdef}
\end{eqnarray}
where $g_\ell$ is the $g$-factor of the particle, $m_\ell$ is its mass,
and $\bm s$ is its spin.  In Eq.~(\ref{eq:lgdef}), $e$ is the (positive)
elementary charge.  For the negatively charged leptons $\ell^{\,-}$,
$g_\ell$ is negative.  These leptons have eigenvalues of spin projection
$s_z = \pm \hbar/2$, so that
\begin{eqnarray}
\mu_{\rm \ell} = {g_{\ell}\over2} \,\frac{e\hbar}{2m_\ell} \ ,
\label{eq:gedef}
\end{eqnarray}
and $\hbar/2m_{\rm e} = \mu_{\rm B}$,
the Bohr magneton.
The magnetic moment anomaly $a_\ell$ is defined by
\begin{eqnarray}
|g_\ell| &=& 2(1+a_\ell) \ ,
\label{eq:adef}
\end{eqnarray}
where the free-electron Dirac equation gives $a_\ell = 0$.
In fact, the anomaly is not zero, but is given by
\begin{eqnarray}
a_\ell({\rm th}) = a_\ell({\rm QED}) + a_\ell({\rm weak})
+ a_\ell({\rm had}) \ ,
\label{eq:aeth}
\end{eqnarray}
where the terms denoted by QED, weak, and had account for the purely
quantum electrodynamic, predominantly electroweak, and predominantly
hadronic (that is, strong interaction) contributions to $a_\ell$,
respectively.

For a comprehensive review of the theory of $a_{\rm e}$, but particularly
of $a_{\rmssmu}$, see \citet{2009189}.  It has long been recognized, as
these authors duly note, that the comparison of experimental and
theoretical values of the electron and muon $g$-factors can test our
description of nature, in particular, the Standard Model of particle
physics, which is the theory of the electromagnetic, weak, and strong
interactions.  Nevertheless, our main purpose here is  not to test
physical theory critically, but to obtain ``best'' values of the
fundamental constants.  \vspace{.25 in}

\subsection{Electron magnetic moment anomaly $\bm{a_{\rm e}}$
and the fine-structure constant $\bm{\alpha}$}
\label{ssec:emma}

Comparison of theory and experiment for the electron magnetic moment
anomaly gives the value for the fine-structure constant $\alpha$ with the
smallest estimated uncertainty in the 2010 adjustment.

\subsubsection{Theory of $a_{\rm e}$}
\label{sssec:ath}

The QED contribution for the electron may be written as \cite{1990005}
\begin{eqnarray}
a_{\rm e}({\rm QED})&=& A_1 + A_2(m_{\rm e}/m_{\rmssmu})
+ A_2(m_{\rm e}/m_{\rmsstau})
\nonumber\\
&&+ A_3(m_{\rm e}/m_{\rmssmu},m_{\rm e}/m_{\rmsstau}) \ .
\label{eq:aqed}
\end{eqnarray}
The leading term $A_1$ is mass independent and the masses in the
denominators of the ratios in $A_2$ and $A_3$ correspond to particles in
vacuum polarization loops.

Each of the four terms on the right-hand side
of Eq.~(\ref{eq:aqed}) is expressed as a power series in the fine-structure
constant $\alpha$:
\begin{eqnarray}
A_i &=& A_i^{(2)}\left({\alpha\over\rmpi}\right)
+A_i^{(4)}\left({\alpha\over\rmpi}\right)^2
+A_i^{(6)}\left({\alpha\over\rmpi}\right)^3
\nonumber\\
&&+A_i^{(8)}\left({\alpha\over\rmpi}\right)^4
+A_i^{(10)}\left({\alpha\over\rmpi}\right)^5
+\cdots \ ,
\label{eq:aeqedsal}
\end{eqnarray}
where $A_2^{(2)}= A_3^{(2)} = A_3^{(4)}=0$.  Coefficients proportional
to $(\alpha/\rmpi)^n$ are of order $e^{2n}$ and are referred to as
2nth-order coefficients.  The second-order coefficient is known
exactly, and the fourth- and sixth-order coefficients are known
analytically in terms of readily evaluated functions:
\begin{eqnarray}
A_1^{(2)} &=& \fr{1}{2} \, ,
\label{eq:a12}
\\
A_1^{(4)} &=&  -0.328\,478\,965\,579\ldots \, ,
\label{eq:a14}
\\
A_1^{(6)} &=&  1.181\,241\,456\ldots \ .
\label{eq:a16}
\end{eqnarray}

The eighth-order coefficient $A_1^{(8)}$ arises from $891$ Feynman
diagrams of which only a few are known analytically. Evaluation of this
coefficient  numerically by Kinoshita and co-workers has been underway
for many years \cite{2010028}.  The value used in the 2006 adjustment is
$A_1^{(8)}= -1.7283(35)$ as reported by \citet{2006003}. However, and as
discussed in CODATA-06, well after the 31 December 2006 closing date of
the 2006 adjustment, as well as the date when the 2006 CODATA recommended
values of the constants were made public, it was discovered by
\citet{2007167} that a significant error had been made in the
calculation. In particular, 2 of the 47 integrals representing $518$
diagrams that had not been confirmed independently required a corrected
treatment of infrared divergences.  The error was identified by using
FORTRAN code generated by an automatic code generator.  The new value is
\cite{2007167}
\begin{eqnarray}  A_1^{(8)}&=&  -1.9144(35) \, ;
\label{eq:a18}
\end{eqnarray}
details of the calculation are given by \citet{2008092}.  In view of the
extensive effort made by these workers to ensure that the result in
Eq.~(\ref{eq:a18}) is reliable, the Task Group adopts both its value and
quoted uncertainty for use in the 2010 adjustment.

Independent work is in progress on analytic calculations of eighth-order
integrals.  See, for example,
\citet{2001339,2001106,2004210,2008233}.  Work is also in progress
on numerical calculations of the 12\,672 Feynman diagrams for the
tenth-order coefficient.  See \citet{2011199} and references cited
therein.

The evaluation of the contribution to the uncertainty of $a_{\rm e}({\rm
th})$ from the fact that $A_1^{(10)}$ is unknown follows the procedure
in CODATA-98 and yields $A_1^{(10)} = 0.0( 4.6)$, which
contributes a standard uncertainty component to $a_{\rm e}({\rm th})$ of
$ 2.7\times 10^{-10} \, a_{\rm e}$.  This uncertainty is larger than the
uncertainty attributed to $A_1^{(10)}$ in CODATA-06, because the
absolute value of $A_1^{(8)}$ has increased.  All higher-order
coefficients are assumed to be negligible.

The mass-dependent coefficients for the electron based on the 2010
recommended values of the mass ratios are
\begin{eqnarray}
A_2^{(4)}\!(m_{\rm e}/m_{\rmssmu}) &=&  5.197\,386\,68(26)\times 10^{-7}
\nonumber\\
&&\quad\rightarrow
 24.182\times 10^{-10}a_{\rm e} \, ,
\\
\nonumber\\
A_2^{(4)}\!(m_{\rm e}/m_{\rmsstau}) &=&  1.837\,98(33)\times 10^{-9}
\nonumber\\
&&\quad\rightarrow  0.086\times 10^{-10}a_{\rm e} \, ,
\\
\nonumber\\
A^{(6)}_2\!(m_{\rm e}/m_{\rmssmu}) &=&  -7.373\,941\,62(27)\times 10^{-6}
\nonumber\\
&&\quad\rightarrow  -0.797\times 10^{-10}a_{\rm e} \, ,
\label{eq:a26s}
\\
\nonumber\\
A^{(6)}_2\!(m_{\rm e}/m_{\rmsstau}) &=&  -6.5830(11)\times 10^{-8}
\nonumber\\
&&\quad\rightarrow  -0.007\times 10^{-10}a_{\rm e} \, ,
\end{eqnarray}
where the standard uncertainties of the coefficients are due to the
uncertainties of the mass ratios and are negligible.  The
contributions from $A_3^{(6)}\!(m_{\rm e}/m_{\rmssmu},m_{\rm
e}/m_{\rmsstau})$ and all higher-order mass-dependent terms are
also negligible.

The dependence on $\alpha$ of any contribution other
than $a_{\rm e}({\rm QED})$ is negligible, hence
the anomaly as a function of $\alpha$ is given by
combining QED terms that have like powers of $\alpha/\rmpi$:
\begin{eqnarray}
a_{\rm e}({\rm QED})
&=& C_{\rm e}^{(2)}\left({\alpha\over\rmpi}\right)
+ C_{\rm e}^{(4)}\left({\alpha\over\rmpi}\right)^2
+ C_{\rm e}^{(6)}\left({\alpha\over\rmpi}\right)^3
\nonumber\\
&&+ C_{\rm e}^{(8)}\left({\alpha\over\rmpi}\right)^4
+ C_{\rm e}^{(10)}\left({\alpha\over\rmpi}\right)^5
+ \cdots,
\label{eq:appaeqed}
\end{eqnarray}
with
\begin{eqnarray}
C_{\rm e}^{(2)} &=&
 0.5 \, ,
\nonumber\\
C_{\rm e}^{(4)} &=&
 -0.328\,478\,444\,00\, ,
\nonumber\\
C_{\rm e}^{(6)} &=&
 1.181\,234\,017\, ,
\nonumber\\
C_{\rm e}^{(8)} &=&
 -1.9144(35)\, ,
\nonumber\\
C_{\rm e}^{(10)} &=&
 0.0(4.6) \, .
\label{eq:appaecs}
\end{eqnarray}

The electroweak contribution, calculated as in CODATA-98 but with the
2010 values of $G_{\rm F}$ and ${\rm sin}^2\theta_{\rm W}$, is
\begin{eqnarray}
a_{\rm e}({\rm weak})
&=&  0.029\,73(52)\times 10^{-12}
\nonumber\\
&=&  0.2564(45)\times 10^{-10}a_{\rm e} \, .
\label{eq:aeweak}
\end{eqnarray}

The hadronic contribution can be written as
\begin{eqnarray}
a_{\rm e}({\rm had})=a_{\rm e}^{(4)}({\rm had})+a_{\rm e}^{(6a)}({\rm
had}) +a_{\rm e}^{(\rmssgamma\rmssgamma)}({\rm had})+\cdots,
\nonumber\\
\label{eq:ehadcontr}
\end{eqnarray}
where $a_{\rm e}^{(4)}{({\rm had})}$ and $a_{\rm e}^{(6a)}{({\rm had})}$
are due to hadronic vacuum polarization and are of order
$({\alpha/\rmpi})^2$ and $({\alpha/\rmpi})^3$, respectively; also of order
$({\alpha/\rmpi})^3$ is $a_{\rmssmu}^{(\rmssgamma\rmssgamma)}$, which is due to
light-by-light vacuum polarization. Its value,
\begin{eqnarray}
a_{\rm e}({\rm had}) &=&  1.685(22)\times 10^{-12}
\nonumber\\
                     &=&  1.453(19)\times 10^{-9}a_{\rm e} \, ,
\label{eq:aehad}
\end{eqnarray}
is the sum of the following three contributions: $a_{\rm e}^{(4)}({\rm
had}) =  1.875(18)\times 10^{-12}$ obtained by \citet{1998089}; $a_{\rm
e}^{(6a)}({\rm had}) =  -0.225(5)\times 10^{-12}$ given by
\citet{1997005}; and $a_{\rm e}^{(\rmssgamma \rmssgamma)}(\rm had)=
0.035(10)\times 10^{-12}$ as given by \citet{2010022}. In past
adjustments this contribution was calculated by assuming that $a_{\rm
e}^{(\rmssgamma \rmssgamma)} = (m_{\rm e}/m_{\rmssmu})^2 \,
a_{\rmssmu}^{(\rmssgamma \rmssgamma)}(\rm had)$.  However,
\citet{2010022} have shown that such scaling is not adequate for the
neutral pion exchange contribution to $a_{\rmssmu}^{(\rmssgamma
\rmssgamma)}(\rm had)$ and have taken this into account in obtaining their
above result for $a_{\rm e}^{(\rmssgamma \rmssgamma)}(\rm had)$ from
their muon value $a_{\rmssmu}^{(\rmssgamma \rmssgamma)}(\rm had)=
105(26)\times 10^{-11}$.

The theoretical prediction is
\begin{eqnarray}
a_{\rm e}({\rm th}) = a_{\rm e}({\rm QED}) + a_{\rm e}({\rm weak})
 + a_{\rm e}({\rm had}) \ .
\label{eq:appaeth}
\end{eqnarray}
The various contributions can be put into context by comparing them to the
most accurate experimental value of $a_{\rm e}$ currently available,
which has an uncertainty of $2.8\times10^{-10}a_{\rm e}$; see
Eq.~(\ref{eq:aeharv08}) below.

The standard uncertainty of $a_{\rm e}({\rm th})$ from the uncertainties
of the terms listed above is
\begin{eqnarray}
u[a_{\rm e}({\rm th)}] =  0.33\times 10^{-12} =  2.8\times 10^{-10}\, a_{\rm e},
\label{eq:uncaeth}
\end{eqnarray}
and is dominated by the uncertainty of the
coefficient $C_{\rm e}^{(10)}$.

For the purpose of the least-squares calculations carried out in
Sec.~\ref{sec:ad}, we include an additive correction $\delta_{\rm e}$
to $a_{\rm e}({\rm th)}$ to account for the uncertainty of $a_{\rm
e}({\rm th)}$ other than that due to $\alpha$, and hence the complete
theoretical expression in the observational equation for the electron
anomaly ($B13$ in Table~\ref{tab:pobseqsb1}) is
\begin{eqnarray}
a_{\rm e}(\alpha,\delta_{\rm e}) = a_{\rm e}({\rm th)}
+ \delta_{\rm e} \, .
\label{eq:aefth}
\end{eqnarray}
The input datum for $\delta_{\rm e}$ is zero with standard uncertainty
$u[a_{\rm e}({\rm th})]$, or $ 0.00(33)\times 10^{-12}$, which is data item
  $B12$
in Table~\ref{tab:pdata}.

\subsubsection{Measurements of $a_{\rm e}$}
\label{sssec:aemeas}

\paragraph{University of Washington.}
\label{par:aeuw}
The classic series of measurements of the electron and positron anomalies
  carried
out at the University of Washington by \citet{1987003}
yield the value
\begin{eqnarray}
a_{\rm e} =  1.159\,652\,1883(42)\times 10^{-3} ~~ [ 3.7\times 10^{-9}] \ ,
\label{eq:aeuwash}
\end{eqnarray}
as discussed in CODATA-98.  This result, which assumes that $CPT$
invariance holds for the electron-positron system, is data item $B13.1$
in Table~\ref{tab:pdata}.

\paragraph{Harvard University.}
\label{par:harvard}
In both the University of Washington and Harvard University, Cambridge
MA, USA experiments, the electron magnetic moment anomaly is essentially
determined from the relation $a_{\rm e}=f_{\rm a}/f_{\rm c}$ by
measuring in the same magnetic flux density $B\approx 5~{\rm T}$ the
anomaly difference frequency $f_{\rm a}=f_{\rm s}-f_{\rm c}$ and
cyclotron frequency $f_{\rm c}=eB/2{\rmpi}m_{\rm e}$, where $f_{\rm
s}=|g_{\rm e}|{\rmssmu}_{\rm B}B/h$ is the electron spin-flip (or precession)
frequency.

Because of its small relative standard uncertainty of
$7.6\times10^{-10}$, the then new result for $a_{e}$ obtained by
\citet{2006081} at Harvard using a cylindrical rather than a hyperbolic
Penning trap played the dominant role in determining the 2006
recommended value of $\rm\alpha$. This work continued with a number of
significant improvements and a new value of $a_{e}$ consistent with the
earlier one but with an uncertainty nearly a factor of three smaller was
reported by \citet{2008043}:
\begin{eqnarray}
a_{\rm e} &=&  1.159\,652\,180\,73(28)\times 10^{-3}\, .
\label{eq:aeharv08}
\end{eqnarray}
A paper that describes this measurement in detail was subsequently
published by \citet{2011102} (see also the review by \citet{2010026}).
As discussed by \citet{2011102}, the improvement that contributed most
to the reduction in uncertainty is a better understanding of the Penning
trap cavity frequency shifts of the radiation used to measure $f_{\rm
c}$.  A smaller reduction resulted from narrower linewidths of the
anomaly and cyclotron resonant frequencies.  Consequently,
\citet{2011102} state that their 2008 result should be viewed as
superseding the earlier Harvard result.  Therefore, only the value of
$a_{e}$ in Eq.~(\ref{eq:aeharv08}) is included as an input datum in the
2010 adjustment; it is data item $B13.2$ in Table~\ref{tab:pdata}.

\subsubsection{Values of $\alpha$ inferred from $a_{\rm e}$}
\label{sssec:alphaae}

Equating the theoretical expression with the two experimental values of
$a_{\rm e}$ given in Eqs.~(\ref{eq:aeuwash}) and (\ref{eq:aeharv08})
yields
\begin{eqnarray}
\alpha^{-1}(a_{\rm e}) =  137.035\,998\,19(50)
~~ [ 3.7\times 10^{-9}]
\label{eq:alphinvuwash87}
\end{eqnarray}
from the University of Washington result and
\begin{eqnarray}
\alpha^{-1}(a_{\rm e}) =  137.035\,999\,084(51)
~~ [ 3.7\times 10^{-10}]
\label{eq:alphinvharvu08}
\end{eqnarray}
from the Harvard University result.  The contribution of the uncertainty
in $a_{\rm e}({\rm th})$ to the relative uncertainty of either of these
results is $ 2.8\times 10^{-10}$.  The value in
Eq.~(\ref{eq:alphinvharvu08}) has the smallest uncertainty of any value
of alpha currently available.  The fact that the next most accurate
value of ${\rm\alpha}$, which has a relative standard uncertainty of
$6.6\times10^{-10}$ and is obtained from the quotient $h/m(^{87}{\rm
Rb})$ measured by atom recoil, is consistent with this value suggests
that the theory of $a_{\rm e}$ is well in hand; see Sec.~\ref{sec:ad}.

\subsection{Muon magnetic moment anomaly $\bm{a_{\rmssmu}}$}
\label{ssec:mmma}

The 2006 adjustment included data that provided both an experimental
value and a theoretical value for $a_{\rmssmu}$. Because of problems with
the theory, the uncertainty assigned to the theoretical value was over
three times larger than that of the experimental value.  Nevertheless,
the theoretical value with its increased uncertainty was included in the
adjustment, even if with a comparatively small weight.

For the 2010 adjustment, the Task Group decided not to include the
theoretical value for $a_{\rmssmu}$, with the result that the 2010
recommended value is based mainly on experiment.  This is consistent
with the fact that the value of $a_{\rmssmu}$ recommended by the
Particle Data Group in their biennial 2010 \emph{Review of Particle
Physics} \cite{2010129} is the experimental value. The current situation
is briefly summarized in the following sections.

\subsubsection{Theory of ${a_{\rmssmu}}$}
\label{sssec:amuth}

The mass-independent coefficients $A_1^{(n)}$ for the muon are the same
as for the electron.  Based on the 2010 recommended values of the
mass ratios, the relevant mass-dependent terms are
\begin{eqnarray}
A_2^{(4)}\!(m_{\rmssmu}/m_{\rm e}) &=&
 1.094\,258\,3118(81)
\nonumber\\
 &\rightarrow&  506\,386.4620(38)\times 10^{-8} a_{\rmssmu} \, ,
\\ \nonumber \\
A_2^{(4)}\!(m_{\rmssmu}/m_{\rmsstau}) &=&
 0.000\,078\,079(14)
\nonumber\\
&\rightarrow&  36.1325(65)\times 10^{-8} a_{\rmssmu} \, ,
\\ \nonumber \\
A^{(6)}_2\!(m_{\rmssmu}/m_{\rm e}) &=&  22.868\,380\,04(19)
\label{eq:a26lmume}
\nonumber\\ &\rightarrow&  24\,581.766\,56(20)\times 10^{-8} a_{\rmssmu} \, ,
\\ \nonumber\\
A^{(6)}_2\!(m_{\rmssmu}/m_{\rmsstau}) &=&  0.000\,360\,63(11)
\label{eq:a26smumtau}
\nonumber\\ &\rightarrow&  0.387\,65(12)\times 10^{-8} a_{\rmssmu} \, ,
\\ \nonumber\\
A_2^{(8)}\!(m_{\rmssmu}/m_{\rm e}) &=&  132.6823(72)
\label{eq:a28mume}
\nonumber\\ &\rightarrow&  331.288(18)\times 10^{-8} a_{\rmssmu} \, ,
\\ \nonumber\\
A_2^{(10)}\!(m_{\rmssmu}/m_{\rm e}) &=&  663(20)
\label{eq:a210mume}
\nonumber\\ &\rightarrow&  3.85(12)\times 10^{-8} a_{\rmssmu} \, ,
\end{eqnarray}
\begin{eqnarray}
A_3^{(6)}\!(m_{\rmssmu}/m_{\rm e},m_{\rmssmu}/m_{\rmsstau}) &=&
  0.000\,527\,762(94)
\label{eq:a36mumemumtau}
\nonumber\\ &\rightarrow&  0.567\,30(10)\times 10^{-8} a_{\rmssmu \, ,}
\\ \nonumber\\
A_3^{(8)}\!(m_{\rmssmu}/m_{\rm e},m_{\rmssmu}/m_{\rmsstau}) &=&  0.037\,594(83)
\label{eq:a38mumemumtau}
\nonumber\\ &\rightarrow&  0.093\,87(21)\times 10^{-8} a_{\rmssmu} \, .
\qquad
\end{eqnarray}

The QED contribution to the theory of $a_{\rmssmu}$, where terms that
have like powers of $\alpha/\rmpi$ are combined, is
\begin{eqnarray}
a_{\rmssmu}({\rm QED})
&=& C_{\rmssmu}^{(2)}\left({\alpha\over\rmpi}\right)
+ C_{\rmssmu}^{(4)}\left({\alpha\over\rmpi}\right)^2
+ C_{\rmssmu}^{(6)}\left({\alpha\over\rmpi}\right)^3
\nonumber\\
&&+ C_{\rmssmu}^{(8)}\left({\alpha\over\rmpi}\right)^4
+ C_{\rmssmu}^{(10)}\left({\alpha\over\rmpi}\right)^5
+ \cdots,
\label{eq:appamqed}
\end{eqnarray}
with
\begin{eqnarray}
C_{\rmssmu}^{(2)} &=&
 0.5\, ,
\nonumber\\
C_{\rmssmu}^{(4)} &=&
 0.765\,857\,426(16)\, ,
\nonumber\\
C_{\rmssmu}^{(6)} &=&
 24.050\,509\,88(28)\, ,
\nonumber\\
C_{\rmssmu}^{(8)} &=&
 130.8055(80)\, ,
\nonumber\\
C_{\rmssmu}^{(10)} &=&
 663(21) \, ,
\label{eq:appamcs}
\end{eqnarray}
which yields, using the 2010 recommended value of $\alpha$,
\begin{eqnarray}
a_{\rmssmu}(\rm QED) =  0.001\,165\,847\,1810(15) \quad [ 1.3\times 10^{-9}] \,
  .
\quad
\label{eq:amuqed}
\end{eqnarray}
In absolute terms, the uncertainty in $a_{\rmssmu}({\rm QED})$ is
$ 0.15\times 10^{-11}$.
\vspace{.25 in}

The current theoretical expression for the muon anomaly is of the same
form as for the electron:
\begin{eqnarray}
a_{\rmssmu}({\rm th}) = a_{\rmssmu}({\rm QED}) + a_{\rmssmu}({\rm weak})
 + a_{\rmssmu}({\rm had}) \, .
\label{eq:appamth}
\end{eqnarray}
The electroweak contribution, calculated by \citet{2003052}, is
$a_{\rmssmu}{({\rm weak})}=154(2)\times 10^{-11}$. In contrast to the
case of the electron, $a_{\rmssmu}({\rm weak})$ is a significant
contribution compared to $a_{\rmssmu}({\rm QED})$.

In a manner similar to that for the electron, the hadronic contribution
can be written as
\begin{eqnarray}
a_{\rmssmu}{({\rm had})} = a_{\rmssmu}^{(4)}{({\rm had})}
 + a_{\rmssmu}^{(6a)}({\rm had})
 + a_{\rmssmu}^{(\rmssgamma\rmssgamma)}({\rm had})+\cdots \ .
\nonumber\\
\label{eq:mhadcontr}
\end{eqnarray}
It is also of much greater importance for the muon than for the
electron.  Indeed, $a_{\rmssmu}(\rm had)$ is roughly $7000(50)\times
10^{-11}$, which should be compared with the $63\times10^{-11}$
uncertainty of the experimental value  $a_{\rmssmu}(\rm exp)$ discussed
in the next section.

For well over a decade a great deal of effort has been devoted by many
researchers to the improved evaluation of $a_{\rmssmu}({\rm had})$. The
standard method of calculating $a_{\rmssmu}^{(4)}{({\rm had})}$ and
$a_{\rmssmu}^{(6a)}({\rm had})$ is to evaluate dispersion integrals over
experimentally measured cross sections for the scattering of ${\rm
e}^+{\rm e}^-$ into hadrons.  However, in some calculations data on
decays of the $\rmtau$ into hadrons are used to replace the ${\rm
e}^+{\rm e}^-$ data in certain energy regions.  The results of three
evaluations which include the most recent data can be concisely
summarized as follows.

\citet{2011080} find that $a_{\rmssmu}({\rm exp})$ exceeds their
theoretically predicted value $a_{\rmssmu}({\rm th})$ by $3.6$ times the
combined standard uncertainty of the difference, or $3.6{\sigma}$, using
only ${\rm e}^+{\rm e}^-$ data, and by $2.4 \sigma$ if $\rmtau$ data are
included. On the other hand, \citet{2011095} find that by correcting the
$\rmtau$ data for the effect they term $\rmrho$\,-\,$\rmgamma$ mixing,
the values of $a_{\rmssmu}^{(4)}({\rm had})$ obtained from only ${\rm
e}^+{\rm e}^-$ data, and from ${\rm e}^+{\rm e}^-$ and $\rmtau$ data
together, are nearly identical and that the difference between
experiment and theory is $3.3{\sigma}$.  And \citet{2011169} find the
same $3.3{\sigma}$ difference using ${\rm e}^+{\rm e}^-$ data alone.
Finally, we note that in a very recent paper, \citet{2012010} obtain a
difference in the range $4.07\sigma$ to $4.65\sigma$, depending on the
assumptions made, using a ``hidden local symmetry'' model.

The disagreement between experiment and theory has long been known and
numerous theoretical papers have been published that attempt to explain
the discrepancy in terms of New Physics; see the review by
\citet{2010019}.  Although a contribution to $a_{\rmssmu}({\rm th})$
large enough to bring it into agreement with $a_{\rmssmu}({\rm exp})$
from physics beyond the Standard Model is possible, no outside
experimental evidence currently exists for such physics.  Thus, because
of the persistence of the discrepancy and its confirmation by the most
recent calculations, and because no known physics has yet been able to
eliminate it, the Task Group has decided to omit the theory of
$a_{\rmssmu}$ from the 2010 adjustment.

\subsubsection{Measurement of $a_{\rmssmu}$: Brookhaven}
\label{sssec:amb}

Experiment E821 at BNL has been discussed in the past three CODATA
reports. It involves the direct measurement of the anomaly difference
frequency $f_{\rm a}=f_{\rm s}-f_{\rm c}$, where $f_{\rm
s}=|g_{\rmssmu}|(e\hbar/2m_{\rmssmu})B/h$ is the muon spin-flip (or
precession) frequency in the applied magnetic flux density $B$ and
$f_{\rm c}=eB/2{\rmpi}m_{\rmssmu}$ is the corresponding muon cyclotron
frequency. However, in contrast to the case of the electron where both
$f_{\rm a}$ and $f_{\rm c}$ are measured directly and the electron
anomaly is calculated from $a_{\rm e}=f_{\rm a}/f_{\rm c}$, for the muon
$B$ is eliminated by determining its value from proton nuclear magnetic
resonance (NMR) measurements.  This means that the muon anomaly is
calculated from
\begin{eqnarray}
a_{\rmssmu}({\rm
exp})=\frac{\overline{R}}{|\mu_{\rmssmu}/{\mu_{p}}|-\overline{R}}\, ,
\label{eq:amurbar}
\end{eqnarray}
where $\overline{R}=f_{\rm a}/\overline{f}_{\rm p}$ and
$\overline{f}_{\rm p}$ is the free proton NMR frequency corresponding to
the average flux density $B$ seen by the muons in their orbits in the
muon storage ring.

The final value of $\overline{R}$ obtained in the E821 experiment is
\cite{2006132}
\begin{eqnarray}
\overline{R} =  0.003\,707\,2063(20)\, ,
\label{eq:rbar06}
\end{eqnarray} which is used as an input datum in
the 2010 adjustment and is data item $B14$ in Table~\ref{tab:pdata}.
[The last digit of this value is one less than that of the value used
in 2006, because the 2006 value was taken from Eq.~(57) in the paper by
\citet{2006132} but the correct value is that given in Table~XV
\cite{pc09br}.]  Based on this value of $\overline{R}$,
Eq.~(\ref{eq:amurbar}), and the 2010 recommended value of
$\mu_{\rmssmu}/{\mu_{\rm p}}$, whose uncertainty is negligible in this
context, the experimental value of the muon anomaly is
\begin{eqnarray}
a_{\rmssmu}({\rm exp}) =  1.165\,920\,91(63)\times 10^{-3}\, .
\label{eq:amu10}
\end{eqnarray}
Further, with the aid of Eq.~(\ref{eq:mumemump}), the equation for
$\overline{R}$ can be written as
\begin{eqnarray}
\overline{R} = -\frac{a_{\rmssmu}}
{1 + a_{\rm e}(\alpha,\delta_{\rm e})}\frac{m_{\rm e}
}{m_{\rmssmu }}\frac{\mu_{{\rm e}^-}}{\mu_{\rm p}} \ ,
\label{eq:rbarobs}
\end{eqnarray}
where use has been made of the relations $g_{\rm e} = -2(1+a_{\rm e})$,
$g_{\rmssmu}=-2(1+a_{\rmssmu})$, and $a_{\rm e}$ is replaced by the
theoretical expression $a_{\rm e}(\alpha,\delta_{\rm e})$ given in
Eq.~(\ref{eq:aeth}).  However, since the theory of $a_{\rmssmu}$ is
omitted from the 2010 adjustment, $a_{\rmssmu}$ is not replaced in
Eq.~(\ref{eq:rbarobs}) by a theoretical expression, rather it is made to
be an adjusted constant.

\subsection{Bound electron $\bm{g}$-factor
in $\bm{^{12}{\rm C}^{5+}}$ and
in $\bm{^{16}{\rm O}^{7+}}$ and $\bm{A_{\rm r}({\rm e})}$}
\label{ssec:ehco}

Competitive values of $A_{\rm r}$(e) can be obtained from precise
measurements and theoretical calculations of the $g$-factor of the
electron in hydrogenic $^{12}$C and $^{16}$O.

For a ground-state hydrogenic ion $^{A}X^{(Z-1)+}$ with
mass number $A$, atomic number (proton number) $Z$, nuclear spin quantum
number $i$ = 0, and $g$-factor
$g_{\rm e^-} (^{A}{X}^{(Z-1)+})$
in an applied magnetic flux density $B$, the ratio of the electron's spin-flip
(or precession) frequency $f_{\rm s}=|g_{{\rm e}^-}
(^{A}{X}^{(Z-1)+})| (e\hbar/2m_{\rm e})B/h$ to the cyclotron frequency of the
ion $f_{\rm c} = (Z-1) eB/2\rmpi m(^{A}{X}^{(Z-1)+})$
in the same magnetic flux
density is
\begin{eqnarray}
\frac{f_{\rm s} (^A{X}^{(Z-1)+})}{f_{\rm c} (
^A{X}^{(Z-1)+})}&=&-\frac{g_{{\rm e}^-}(
^A{X}^{(Z-1)+})}{2(Z-1)}\frac{A_{\rm r}(
^A{X}^{(Z-1)+})}{A_{\rm r} ({\rm e})} \ ,
\nonumber\\
\label{eq:fsfcgx}
\end{eqnarray}
where $A_{\rm r}(X)$ is the relative atomic mass of particle $X$.

This expression can be used to obtain a competitive result for $A_{\rm
r}(\rm e)$ if for a particular ion the quotient $f_{\rm s}/f_{\rm c}$,
its bound state $g$-factor, and the relative atomic mass of the ion can
be obtained with sufficiently small uncertainties. In fact, work
underway since the mid-1990s has been so successful that
Eq.~(\ref{eq:fsfcgx}) now provides the most accurate values of $A_{\rm
r}(\rm e)$.  Measurements of $f_{\rm s}/f_{\rm c}$ for $^{12}{\rm
C}^{5+}$ and $^{16}{\rm O}^{7+}$, performed at the Gesellschaft f\"{u}r
Schwerionenforschung, Darmstadt, Germany (GSI) by GSI and University of
Mainz researchers, are discussed in CODATA-06 and the results were
included in the 2006 adjustment. These data are recalled in
Sec.~\ref{sssec:bsgfexps} below, and the present status of the theoretical
expressions for the bound-state $g$-factors of the two ions are
discussed in the following section.

For completeness, we note that well after the closing date of the 2010
adjustment \citet{2011141} reported a value of $f_{\rm s}/f_{\rm c}$ for
the hydrogenic ion $^{28}{\rm Si}^{13+}$.  Using the 2006 recommended
value of $A_{\rm r}(\rm e)$ and the applicable version of
Eq.~(\ref{eq:fsfcgx}), they found good agreement between the theoretical
and experimental values of the $g$-factor of this ion, thereby
strengthening confidence in our understanding of bound-state QED theory.

\subsubsection{Theory of the bound electron $g$-factor}
\label{sssec:thbegf}

The energy of a free electron with spin projection
$s_z$ in a magnetic flux density $B$ in the $z$ direction is
\begin{eqnarray}
E &=& -\bm \mu \cdot \bm B = - g_{\rm e^-} {e\over2m_{\rm e}}s_z B\ ,
\label{eq:eemf}
\end{eqnarray}
and hence the spin-flip energy difference is
\begin{eqnarray}
\Delta E = - g_{\rm e^-} \mu_{\rm B} B \ .
\label{eq:deemf}
\end{eqnarray}
(In keeping with the definition of the $g$-factor in Sec.~\ref{sec:mmagf},
the quantity $g_{\rm e^-}$ is negative.)
The analogous expression for ions with no nuclear spin is
\begin{eqnarray}
\Delta E_{\rm b}(X)
&=& - g_{\rm e^-}(X) \mu_{\rm B} B \ ,
\label{eq:deebmf}
\end{eqnarray}
which defines the bound-state electron $g$-factor,
and where $X$ is either $^{12}{\rm C}^{5+}$ or
$^{16}${\rm O}$^{7+}$.

The theoretical expression for
$g_{\rm e^-}(X)$ is written as
\begin{eqnarray}
g_{\rm e^-}(X)
= g_{\rm D} + \Delta g_{\rm rad} + \Delta g_{\rm rec}
+ \Delta g_{\rm ns} + \cdots
\ ,
\label{eq:gsumdef}
\end{eqnarray}
where the individual terms are the Dirac value, the radiative
corrections, the recoil corrections, and the nuclear size corrections,
respectively.  Numerical results are summarized in
Tables~\ref{tab:gfactthc} and \ref{tab:gfacttho}.

\begin{table}[t]
\def\m{\phantom{-}}
\caption{Theoretical contributions and total for the $g$-factor
of the electron in
hydrogenic carbon 12 based on the 2010 recommended values of the
constants.}
\label{tab:gfactthc}
\begin{center}
\begin{tabular}{c@{\quad}l@{\quad}c}
\toprule
\vbox to 10 pt {}
Contribution   &  \hbox to .8 cm{} Value & Source \\
\colrule
\vbox to 10 pt {}
Dirac $g_{\rm D}$                 & $   -1.998\,721\,354\,390\,9(8)  $ &
  Eq.~(\ref{eq:diracg}) \\
 $\Delta g^{(2)}_{\rm SE}       $ & $   -0.002\,323\,672\,436(4)$ &
  Eq.~(\ref{eq:yerokgco}) \\
 $\Delta g^{(2)}_{\rm VP}       $ & $\m 0.000\,000\,008\,512(1)$ &
  Eq.~(\ref{eq:cvp2co})  \\
 $\Delta g^{(4)}                $ & $\m 0.000\,003\,545\,677(25)  $ &
  Eq.~(\ref{eq:c4co})  \\
 $\Delta g^{(6)}                $ & $   -0.000\,000\,029\,618  $ &
  Eq.~(\ref{eq:c6co})  \\
 $\Delta g^{(8)}                $ & $\m 0.000\,000\,000\,111  $ &
  Eq.~(\ref{eq:c8co})  \\
 $\Delta g^{(10)}                $ & $\m 0.000\,000\,000\,000(1)  $ &
  Eq.~(\ref{eq:c10co})  \\
 $\Delta g_{\rm rec}            $ & $   -0.000\,000\,087\,629 $ &
  Eqs.~(\ref{eq:grr0})-(\ref{eq:grr2}) \\
 $\Delta g_{\rm ns}             $ & $   -0.000\,000\,000\,408(1)  $ &
  Eq.~(\ref{eq:gns})  \\
 $g_{\rm e^-}(^{12}{\rm C}^{5+})$ & $   -2.001\,041\,590\,181(26) \vbox to .4 cm
  {}  $ & Eq.~(\ref{eq:gco}) \\
\botrule
\end{tabular}
\end{center}
\end{table}

\begin{table}[t]
\def\m{\phantom{-}}
\caption{Theoretical contributions and total for the $g$-factor
of the electron in
hydrogenic oxygen 16 based on the 2010 recommended values of the
constants.}
\label{tab:gfacttho}
\begin{center}
\begin{tabular}{c@{\qquad}l@{\qquad}c}
\toprule
\vbox to 10 pt {}
Contribution   &  \hbox to .8 cm{} Value & Source \\
\colrule
\vbox to 10 pt {}
Dirac $g_{\rm D}$                 & $   -1.997\,726\,003\,06  $ &
  Eq.~(\ref{eq:diracg}) \\
 $\Delta g^{(2)}_{\rm SE}       $ & $   -0.002\,324\,442\,14(1)$ &
  Eq.~(\ref{eq:yerokgco}) \\
 $\Delta g^{(2)}_{\rm VP}       $ & $\m 0.000\,000\,026\,38$ &
  Eq.~(\ref{eq:cvp2co})  \\
 $\Delta g^{(4)}                $ & $\m 0.000\,003\,546\,54(11)  $ &
  Eq.~(\ref{eq:c4co})  \\
 $\Delta g^{(6)}                $ & $   -0.000\,000\,029\,63  $ &
  Eq.~(\ref{eq:c6co})  \\
 $\Delta g^{(8)}                $ & $\m 0.000\,000\,000\,11  $ &
  Eq.~(\ref{eq:c8co})  \\
 $\Delta g^{(10)}                $ & $\m 0.000\,000\,000\,00  $ &
  Eq.~(\ref{eq:c10co})  \\
 $\Delta g_{\rm rec}            $ & $   -0.000\,000\,117\,00 $ &
  Eqs.~(\ref{eq:grr0})-(\ref{eq:grr2}) \\
 $\Delta g_{\rm ns}             $ & $   -0.000\,000\,001\,56(1)  $ &
  Eq.~(\ref{eq:gns})  \\
 $g_{\rm e^-}(^{16}{\rm O}^{7+})$ & $   -2.000\,047\,020\,35(11) \vbox to .4 cm
  {}  $ & Eq.~(\ref{eq:gco}) \\
\botrule
\end{tabular}
\end{center}
\end{table}

\citet{1928001} obtained the exact value
\begin{eqnarray}
g_{\rm D} &=& - \frac{2}{3}\left[1+2\sqrt{1-(Z\alpha)^2}\right]
\nonumber\\
&=& - 2\left[1-\frac{1}{3}(Z\alpha)^2 -\frac{1}{12}(Z\alpha)^4
-\frac{1}{24}(Z\alpha)^6
 + \cdots\right]
\nonumber\\
\label{eq:diracg}
\end{eqnarray}
from the Dirac equation for an electron in the field of a fixed point
charge of magnitude $Ze$, where the only uncertainty is that due to
the uncertainty in $\alpha$.

For the radiative corrections we have
\begin{eqnarray}
\Delta g_{\rm rad} &=&
- 2\left[C_{\rm e}^{(2)}(Z\alpha)\left({\alpha\over\rmpi}\right)
+ C_{\rm e}^{(4)}(Z\alpha)\left({\alpha\over\rmpi}\right)^2
+ \cdots\right] \ ,
\nonumber\\
\end{eqnarray}
where
\begin{eqnarray}
\lim_{Z\alpha\rightarrow 0}C_{\rm e}^{(2n)}(Z\alpha)
= C_{\rm e}^{(2n)} \, ,
\end{eqnarray}
and where the $C_{\rm e}^{(2n)}$ are given in Eq.~(\ref{eq:appaecs}).

For the coefficient $C_{\rm e}^{(2)}(Z\alpha)$, we have
\cite{1970020,1970019,1971019,2004078,2005079}
\begin{eqnarray}
C_{\rm e,SE}^{(2)}(Z\alpha)
&=& \frac{1}{2}\bigg\{1 + \frac{(Z\alpha)^2}{6}
+(Z\alpha)^4\left[
\frac{32}{9}\,\ln{(Z\alpha)^{-2}}
\right.
\nonumber\\ &&\left.
\quad + \frac{247}{216}
-\frac{8}{9}\,\ln{k_0}
- \frac{8}{3}\, \ln{k_3}
\right]
\nonumber\\ && \quad +
 (Z\alpha)^5\,R_{\rm SE}(Z\alpha) \bigg\} \ ,
\label{eq:pachetal}
\end{eqnarray}
where
\begin{eqnarray}
\ln{k_0} &=&  2.984\,128\,556 \, ,
\label{eq:lnk0}
\\
\ln{k_3} &=&  3.272\,806\,545 \, ,
\label{eq:lnk3}
\\
R_{\rm SE}(6\alpha) &=&  22.160(10) \, ,
\\
R_{\rm SE}(8\alpha) &=&  21.859(4) \, .
\end{eqnarray}
The quantity $\ln{k_0}$ is the Bethe logarithm for the 1S state (see
Table~\ref{tab:bethe}), $\ln{k_3}$ is a generalization of the Bethe
logarithm, and $R_{\rm SE}(Z\alpha)$ was obtained by extrapolation of
the results of numerical calculations at higher $Z$
\cite{2002126,2004078}.  Equation~(\ref{eq:pachetal}) yields
\begin{eqnarray}
C_{\rm e,SE}^{(2)}(6\alpha) &=&  0.500\,183\,606\,65(80) \, ,
\nonumber\\
C_{\rm e,SE}^{(2)}(8\alpha) &=&  0.500\,349\,2887(14) \, .
\label{eq:yerokgco}
\end{eqnarray}
The one loop self energy has been calculated directly at $Z = 6$ and $Z
= 8$ by \citet{2008088,2010017}.  The results are in agreement with, but
less accurate than the extrapolation from higher-$Z$.

The lowest-order vacuum-polarization correction consists of a
wave-function correction and a potential correction, each of which can
be separated into a lowest-order Uehling potential contribution and a
Wichmann-Kroll higher-contribution.  The wave-function correction is
\cite{2000111,2000037,2001061,2001218}
\begin{eqnarray}
C_{\rm e,VPwf}^{(2)}(6\alpha) &=&  -0.000\,001\,840\,3431(43) \, ,
\nonumber\\
C_{\rm e,VPwf}^{(2)}(8\alpha) &=&  -0.000\,005\,712\,028(26) \, .
\label{eq:gvpwfco}
\end{eqnarray}
For the potential correction, we have
\cite{2000111,2000153,2002139,2005191,2005090}
\begin{eqnarray}
C_{\rm e,VPp}^{(2)}(6\alpha) &=&  0.000\,000\,008\,08(12) \, ,
\nonumber\\
C_{\rm e,VPp}^{(2)}(8\alpha) &=&  0.000\,000\,033\,73(50) \, ,
\label{eq:gvppco}
\end{eqnarray}
which is the unweighted average of two slightly inconsistent results
with an uncertainty of half their difference.  The total one-photon
vacuum polarization coefficients are given by the sum of
Eqs.~(\ref{eq:gvpwfco}) and (\ref{eq:gvppco}):
\begin{eqnarray}
C_{\rm e,VP}^{(2)}(6\alpha) &=&
C_{\rm e,VPwf}^{(2)}(6\alpha)+
C_{\rm e,VPp}^{(2)}(6\alpha) \nonumber\\
&=& -0.000\,001\,832\,26(12) \, ,
\nonumber\\
C_{\rm e,VP}^{(2)}(8\alpha) &=&
C_{\rm e,VPwf}^{(2)}(8\alpha)+
C_{\rm e,VPp}^{(2)}(8\alpha) \nonumber\\
&=& -0.000\,005\,678\,30(50) \, .
\label{eq:cvp2co}
\end{eqnarray}

The total one-photon coefficient is the sum of Eqs.~(\ref{eq:yerokgco})
and (\ref{eq:cvp2co}):
\begin{eqnarray}
C_{\rm e}^{(2)}(6\alpha) &=&
C_{\rm e,SE}^{(2)}(6\alpha)+
C_{\rm e,VP}^{(2)}(6\alpha) \nonumber\\
&=&  0.500\,181\,774\,39(81)\, ,
\nonumber\\
C_{\rm e}^{(2)}(8\alpha) &=&
C_{\rm e,SE}^{(2)}(8\alpha)+
C_{\rm e,VP}^{(2)}(8\alpha) \nonumber\\
&=&  0.500\,343\,6104(14) \, ,
\label{eq:c2co}
\end{eqnarray}
and the total one-photon contribution is
\begin{eqnarray}
\Delta g^{(2)} &=&
- 2\,C_{\rm e}^{(2)}(Z\alpha)\left({\alpha\over\rmpi}\right)
\nonumber\\
&=&  -0.002\,323\,663\,924(4) \quad {\rm for}~Z = 6
\nonumber\\
&=&  -0.002\,324\,415\,756(7) \quad {\rm for}~Z = 8 \ .
\nonumber\\
\end{eqnarray}
Separate one-photon self energy and vacuum polarization
contributions to the $g$-factor are given in Tables~\ref{tab:gfactthc}
and \ref{tab:gfacttho}.

The leading binding correction to the higher-order coefficients is
\cite{1997162,2001004}
\begin{eqnarray}
C_{\rm e}^{(2n)}(Z\alpha) = C_{\rm e}^{(2n)}
\left(1 + {(Z\alpha)^2\over 6} + \cdots \right) \, .
\label{eq:egbinding}
\end{eqnarray}

The two-loop contribution of relative order $(Z\alpha)^4$ for the ground
S state is \cite{2005079,2006041}
\begin{eqnarray}
C_{\rm e}^{(4)}(Z\alpha) &=& C_{\rm e}^{(4)}
\left(1 + {(Z\alpha)^2\over 6} \right)
\nonumber\\[6 pt] && \hbox to -2cm {}+
(Z\alpha)^4\,\bigg[
\frac{14}{9}\,\ln{(Z\alpha)^{-2}} +
\frac{991343}{155520} - \frac{2}{9}\,\ln{k_0} - \frac{4}{3}\,\ln{k_3}
\nonumber\\[6 pt] && \hbox to -2cm {}+
\frac{679\,\rmpi^2}{12960} - \frac{1441\,\rmpi^2}{720}\,\ln{2} +
\frac{1441}{480}\,\zeta({3})
\bigg]
 + {\cal O}(Z\alpha)^5
\nonumber\\
\nonumber\\ &=&
 -0.328\,5778(23) \quad {\rm for}~Z = 6
\nonumber\\ &=&
 -0.328\,6578(97) \quad {\rm for}~Z = 8\, ,
\label{eq:c4co}
\end{eqnarray}
where $\ln{k_0}$ and $\ln{k_3}$ are given in Eqs.~(\ref{eq:lnk0}) and
(\ref{eq:lnk3}).  As in CODATA-06, the uncertainty due to uncalculated
terms is taken to be \cite{2005079}
\begin{eqnarray}
u\left[C_{\rm e}^{(4)}(Z\alpha)\right] &=&
2\,\left|
(Z\alpha)^5 \,
C_{\rm e}^{(4)}\,
R_{\rm SE}(Z\alpha)\right| .
\end{eqnarray}
\citet{2009089} has calculated a two-loop gauge-invariant set of vacuum
polarization diagrams to obtain a contribution of the same order in
$Z\alpha$ as the above uncertainty.  However, in general we do not
include partial results of a given order. \citeauthor{2009089} also
speculates that the complete term of that order could be somewhat larger
than our uncertainty.

The three- and four-photon terms are calculated with the leading
binding correction included:
\begin{eqnarray}
C_{\rm e}^{(6)}(Z\alpha) &=& C_{\rm e}^{(6)}
\left(1 + {(Z\alpha)^2\over 6} + \cdots \right)
\nonumber\\ &=&
 1.181\,611\dots \quad {\rm for}~Z = 6
\nonumber\\ &=&
 1.181\,905\dots \quad {\rm for}~Z = 8 \, ,
\label{eq:c6co}
\end{eqnarray}
where $C_{\rm e}^{(6)} = 1.181\,234\dots$~,
and
\begin{eqnarray}
C_{\rm e}^{(8)}(Z\alpha) &=& C_{\rm e}^{(8)}
\left(1 + {(Z\alpha)^2\over 6} + \cdots \right)
\nonumber\\ &=&
 -1.9150(35)\dots \quad {\rm for}~Z = 6
\nonumber\\ &=&
 -1.9155(35)\dots \quad {\rm for}~Z = 8 \, ,
\label{eq:c8co}
\end{eqnarray}
where $C_{\rm e}^{(8)} =  -1.9144(35)$.
An uncertainty estimate
\begin{eqnarray}
C_{\rm e}^{(10)}(Z\alpha) &\approx&  C_{\rm e}^{(10)} =  0.0(4.6)
\label{eq:c10co}
\end{eqnarray}
is included for the five-loop correction.

The recoil correction to the bound-state $g$-factor is $\Delta g_{\rm
rec} = \Delta g_{\rm rec}^{(0)} + \Delta g_{\rm rec}^{(2)} + \dots$
where the terms on the right are zero- and first-order in $\alpha/\rmpi$,
respectively.  We have
\begin{eqnarray}
\Delta g_{\rm rec}^{(0)}  &=&
\bigg\{- (Z\alpha)^2 + {(Z\alpha)^4\over
3[1+\sqrt{1-(Z\alpha)^2}]^2}
\nonumber\\
&& -(Z\alpha)^5\,P(Z\alpha)\bigg\}{m_{\rm e}\over m_{\rm N}}
+{\cal O}\left({m_{\rm e}\over m_{\rm N}}\right)^{\!2}
\nonumber\\ \nonumber \\
&=&  -0.000\,000\,087\,70 \dots ~ {\rm for}~Z=6
\nonumber \\
&=&  -0.000\,000\,117\,09 \dots ~ {\rm for}~Z=8 \, ,
\label{eq:grr0}
\end{eqnarray}
where $m_{\rm N}$ is the mass of the nucleus.  The mass ratios, obtained
from the 2010 adjustment, are
${m_{\rm e}/ m(^{12}{\rm C}^{6+})} =  0.000\,045\,727\,5\ldots$ and
${m_{\rm e}/ m(^{16}{\rm O}^{8+})} =  0.000\,034\,306\,5\ldots$.
The recoil terms are the same as in CODATA-02 and references to the
original calculations are given there.  An additional term of the order
of the mass ratio squared \cite{1997162,pc02me}
\begin{eqnarray}
(1+Z) (Z\alpha)^2\left({m_{\rm e}\over m_{\rm N}}\right)^{\!2}
\label{eq:msqgfcorr}
\end{eqnarray}
should also be included in the theory.  The validity of this term for a
nucleus of any spin has been reconfirmed by
\citet{2008303,2010145,2011074}.

For $\Delta g_{\rm rec}^{(2)}$, we have
\begin{eqnarray}
\Delta g_{\rm rec}^{(2)}  &=&
{\alpha\over\rmpi}{(Z\alpha)^2\over3}
{m_{\rm e}\over m_{\rm N}} +\cdots
\nonumber\\
&=&  0.000\,000\,000\,06\ldots ~ {\rm for}~Z=6
\nonumber\\
&=&  0.000\,000\,000\,09\ldots ~ {\rm for}~Z=8 \, .
\label{eq:grr2}
\end{eqnarray}

There is a small correction to the bound-state $g$-factor
due to the finite size of the nucleus, of order
\cite{2000037}
\begin{eqnarray}
\Delta g_{\rm ns} = - {8\over3}(Z\alpha)^4
\left({R_{\rm N}\over \lbar_{\rm C}}\right)^2 + \cdots \, ,
\label{eq:gnsg}
\end{eqnarray}
where $R_{\rm N}$ is the bound-state nuclear rms charge radius and
$\lbar_{\rm C}$ is the Compton wavelength of the electron divided by
$2\rmpi$.  This term is calculated by scaling the results of
\citet{2002058} with the squares of updated values for the nuclear
radii $R_{\rm N} =  2.4703(22)$ fm and $R_{\rm N} =  2.7013(55)$ from the
compilation of \citet{2004198} for $^{12}$C and $^{16}$O, respectively.
This yields the correction
\begin{eqnarray}
\Delta g_{\rm ns} &=&  -0.000\,000\,000\,408(1) \quad {\rm for}~^{12}{\rm C} \,
  ,
\nonumber\\
\Delta g_{\rm ns} &=&  -0.000\,000\,001\,56(1) \quad {\rm for}~^{16}{\rm O} \, .
\label{eq:gns}
\end{eqnarray}

The theoretical value for the $g$-factor of the electron in hydrogenic
carbon 12 or oxygen 16 is the sum of the
individual contributions discussed above
and summarized in Tables~\ref{tab:gfactthc} and \ref{tab:gfacttho}:
\begin{eqnarray}
g_{\rm e^-}(^{12}{\rm C}^{5+})
&=&  -2.001\,041\,590\,181(26)\, ,
\nonumber\\
g_{\rm e^-}(^{16}{\rm O}^{7+})
&=&  -2.000\,047\,020\,35(11) \, .
\nonumber\\
\label{eq:gco}
\end{eqnarray}

For the purpose of the least-squares calculations carried out in
Sec.~\ref{sec:ad}, we define $g_{\rm C}({\rm th})$ to be the sum of
$g_{\rm D}$ as given in Eq.~(\ref{eq:diracg}), the term
$-2(\alpha/\rmpi)C_{\rm e}^{(2)}$, and the numerical values of the
remaining terms in Eq.~(\ref{eq:gsumdef}) as given in
Table~\ref{tab:gfactthc}, where the standard uncertainty of these latter
terms is
\begin{eqnarray}
u[g_{\rm C}({\rm th})] &=&  0.3\times 10^{-10} =
 1.3\times 10^{-11}|g_{\rm C}({\rm th})| \, .
\nonumber\\
\end{eqnarray}
The uncertainty in $g_{\rm C}({\rm th})$ due to the uncertainty in
$\alpha$ enters the adjustment primarily through the functional
dependence of $g_{\rm D}$ and the term $-2(\alpha/\rmpi)C_{\rm e}^{(2)}$
on $\alpha$.  Therefore this particular component of uncertainty is not
explicitly included in $u[g_{\rm C}({\rm th})]$.  To take the
uncertainty $u[g_{\rm C}({\rm th})]$ into account we employ as the
theoretical expression for the $g$-factor ($B17$ in
Table~\ref{tab:pobseqsb1})
\begin{eqnarray}
g_{\rm C}(\alpha,\delta_{\rm C}) &=& g_{\rm C}({\rm th})
+ \delta_{\rm C} \, ,
\end{eqnarray}
where the input value of the additive correction $\delta_{\rm C}$ is
taken to be zero with standard uncertainty $u[g_{\rm C}({\rm th})]$, or
$ 0.00(26)\times 10^{-10}$, which is data item $B15$ in Table~\ref{tab:pdata}.
Analogous considerations apply for the $g$-factor in oxygen, where
\begin{eqnarray}
u[g_{\rm O}({\rm th})] &=&  1.1\times 10^{-10} =
 5.3\times 10^{-11}|g_{\rm O}({\rm th})| \nonumber\\
\end{eqnarray}
and ($B18$ in Table~\ref{tab:pobseqsb1})
\begin{eqnarray}
g_{\rm O}(\alpha,\delta_{\rm O}) &=& g_{\rm O}({\rm th})
+ \delta_{\rm O} \, .
\end{eqnarray}
The input value for $\delta_{\rm O}$ is $ 0.0(1.1)\times 10^{-10}$, which is
  data
item $B16$ in Table~\ref{tab:pdata}.

The covariance of the quantities $\delta_{\rm C}$ and $\delta_{\rm O}$
is
\begin{eqnarray}
u(\delta_{\rm C},\delta_{\rm O}) =  27\times 10^{-22} \ ,
\end{eqnarray}
which corresponds to a correlation coefficient of
$r(\delta_{\rm C},\delta_{\rm O})= 0.994$.

The theoretical value of the ratio of the two $g$-factors is
\begin{eqnarray}
{g_{\rm e^-}(^{12}{\rm C}^{5+}) \over
g_{\rm e^-}(^{16}{\rm O}^{7+})}
&=&  1.000\,497\,273\,224(40) \ ,
\label{eq:gcogo}
\end{eqnarray}
where the covariance of the two values is taken into account.

\subsubsection{Measurements of $g_{\rm e}(^{12}{\rm C}^{5+})$ and $g_{\rm e}
  (^{16}{\rm O}^{7+})$}
\label{sssec:bsgfexps}
\vspace{.25 in}

The experimental values of $f_{\rm s}/f_{\rm c}$ for $^{12}{\rm C}^{5+}$
and $^{16}{\rm O}^{7+}$ obtained at GSI using the double Penning trap
method are discussed in CODATA-02 and the slightly updated result for
the oxygen ion is discussed in CODATA-06. For $^{12}{\rm C}^{5+}$ we
have \cite{2002007,2003022,pc03gw}
\begin{eqnarray}
{f_{\rm s}\left(^{12}{\rm C}^{5+}\right)\over
 f_{\rm c}\left(^{12}{\rm C}^{5+}\right)} =
 4376.210\,4989(23) \ ,
\label{eq:rfsfcc02}
\end{eqnarray}
while for $^{16}{\rm O}^{7+}$ we have \cite{2002008,pc06jv}
\begin{eqnarray}
{f_{\rm s}\left(^{16}{\rm O}^{7+}\right)\over
 f_{\rm c}\left(^{16}{\rm O}^{7+}\right)} =
 4164.376\,1837(32)  \, .
\label{eq:rfsfco02}
\end{eqnarray}
The correlation coefficient of these two frequency ratios, which are
data items $B17$ and $B18$ in Table~\ref{tab:pdata}, is
$ 0.082$.

Equations~(\ref{eq:araxn}) and (\ref{eq:fsfcgx}) together yield
\begin{eqnarray}
{f_{\rm s}\left(^{12}{\rm C}^{5+}\right)\over
 f_{\rm c}\left(^{12}{\rm C}^{5+}\right)} &=&
-{g_{\rm e^-}\left(^{12}{\rm C}^{5+}\right)\over
10 A_{\rm r}({\rm e})}
\nonumber\\&&\hbox to -2 cm {} \times
\left[12-5A_{\rm r}({\rm e}) + {E_{\rm b}\left(^{12}{\rm C}\right)
-E_{\rm b}\left(^{12}{\rm C}^{5+}\right)\over m_{\rm u}c^2}\right] \, ,
\qquad
\label{eq:rfsfccoe}
\end{eqnarray}
which is the basis of the observational equation for the $^{12}{\rm
C}^{5+}$ frequency ratio input datum, Eq.~(\ref{eq:rfsfcc02}); see $B17$
in Table~\ref{tab:pobseqsb1}. In a similar manner we may write
\begin{eqnarray}
&&{f_{\rm s}\left(^{16}{\rm O}^{7+}\right)\over
 f_{\rm c}\left(^{16}{\rm O}^{7+}\right)} =
-{g_{\rm e^-}\left(^{16}{\rm O}^{7+}\right)\over
14 A_{\rm r}({\rm e})} \,
A_{\rm r}\!\left(^{16}{\rm O}^{7+}\right) , \qquad
\label{eq:rfsfcooe}
\end{eqnarray}
with
\begin{eqnarray}
A_{\rm r}\left(^{16}{\rm O}\right) &=&
A_{\rm r}\left(^{16}{\rm O}^{7+}\right)
+7A_{\rm r}({\rm e})
\nonumber\\ &&\qquad - {
E_{\rm b}\left(^{16}{\rm O}\right) -
E_{\rm b}\left(^{16}{\rm O}^{7+}\right)
\over m_{\rm u}c^2} \, , \qquad
\label{eq:aroobseq}
\end{eqnarray}
which are the basis for the observational equations for the oxygen
frequency ratio and $A_{\rm r}(^{16}{\rm O})$, respectively; see $B18$
and $B8$ in Table~\ref{tab:pobseqsb1}.

Evaluation of Eq.~(\ref{eq:rfsfccoe}) using the result for the carbon
frequency ratio in Eq.~(\ref{eq:rfsfcc02}), the theoretical result for
$g_{{\rm e}^-}(^{12}$C$^{5+})$ in Table~\ref{tab:gfactthc}, and the
relevant binding energies in Table~IV of CODATA-02, yields
\begin{eqnarray}
A_{\rm r}({\rm e}) =  0.000\,548\,579\,909\,32(29) \quad [ 5.2\times 10^{-10}]
  \, . \quad
\label{eq:arec02}
\end{eqnarray}
A similar calculation for oxygen using the value of $A_{\rm r}(^{16}$O) in
Table~\ref{tab:rmass10} yields
\begin{eqnarray}
A_{\rm r}({\rm e}) =  0.000\,548\,579\,909\,57(42) \quad [ 7.6\times 10^{-10}]
  \, . \quad
\label{eq:areo02}
\end{eqnarray}
These values of $A_{\rm r}({\rm e})$ are consistent with each other.

Finally, as a further consistency test, the experimental and theoretical
values of the ratio of $g_{{\rm e}^-}(^{12}$C$^{5+})$ to $g_{{\rm
e}^-}(^{16}$O$^{7+})$ can be compared \cite{2002168}.  The theoretical
value of the ratio is given in Eq.~(\ref{eq:gcogo}) and the experimental
value is
\begin{eqnarray}
\frac{g_{{\rm e}^-}(^{12}{\rm C}^{5+})}{g_{{\rm e}^-}(^{16}{\rm O}^{7+})}
&=&  1.000\,497\,273\,68(89) ~ [ 8.9\times 10^{-10}] \, ,\ \  \nonumber\\
\label{eq:gcogoex}
\end{eqnarray}
in agreement with the theoretical value.

\section{Magnetic moment ratios and the muon-electron mass ratio}
\label{sec:mmrmemr}

Magnetic moment ratios and the muon-electron mass ratio are determined by
experiments on bound states of the relevant particles and must be
corrected to determine the free particle moments.

For nucleons or nuclei with spin $\bm I$, the magnetic moment can be written as
\begin{eqnarray}
\bm\mu = g { e \over 2m_{\rm p}}\bm I \ ,
\label{eq:ngdef}
\end{eqnarray}
or
\begin{eqnarray}
\mu = g \mu_{\rm N} i \ .
\label{eq:ngdeff}
\end{eqnarray}
In Eq.~(\ref{eq:ngdeff}), $\mu_{\rm N} = e\hbar/2m_{\rm p}$ is the
nuclear magneton, defined in analogy with the Bohr magneton, and $i$ is
the spin quantum number of the nucleus defined by $\bm I^2 =
i(i+1)\hbar^2$ and $I_z = -i\hbar, ... , (i-1)\hbar, i\hbar$, where
$I_z$ is the spin projection.

Bound state $g$-factors for atoms with a non-zero nuclear spin are
defined by considering their interactions in an applied magnetic flux
density $\bm B$.  For hydrogen, in the Pauli approximation, we have
\begin{eqnarray}
{\cal H} &=& \beta({\rm H})\bm\mu_{\rm e^-}\cdot\bm\mu_{\rm p}
 -\bm\mu_{\rm e^-}({\rm H})\cdot \bm B
  -\bm\mu_{\rm p}({\rm H})\cdot \bm B
\nonumber\\
    &=& {2\rmpi\over\hbar}\Delta\nu_{\rm H}\bm s \cdot \bm I
     - g_{\rm e^-}({\rm H})\,{\mu_{\rm B}\over\hbar}\ \bm s \cdot \bm B
    - g_{\rm p}({\rm H})\,{\mu_{\rm N}\over\hbar}\ \bm I \cdot \bm B \ ,
\nonumber\\
\label{eq:gdefs}
\end{eqnarray}
where $\beta({\rm H})$ characterizes the strength of the hyperfine
interaction, $\Delta\nu_{\rm H}$ is the ground-state hyperfine
frequency, $\bm s$ is the spin of the electron, and $\bm I$ is the spin
of the nucleus.  Equation~(\ref{eq:gdefs}) defines the corresponding
bound-state $g$-factors $g_{\rm e^-}({\rm H})$ and $g_{\rm p}({\rm H})$.

\subsection{Magnetic moment ratios}
\label{ssec:mmr}

Theoretical binding corrections relate $g$-factors measured in the bound
state to the corresponding free-particle $g$-factors.  The corrections
are sufficiently small that the adjusted constants used to calculate
them are taken as exactly known.  These corrections and the references
for the relevant calculations are discussed in CODATA-98 and CODATA-02.

\subsubsection{Theoretical ratios of atomic bound-particle to
free-particle $g$-factors}
\label{sssec:thbfrats}

For the electron in hydrogen, we have
\begin{eqnarray}
{g_{\rm e^-}({\rm H}) \over g_{\rm e^-}}
&=& 1 -\fr{1}{3}(Z\alpha)^2 - \fr{1}{12}(Z\alpha)^4
+ \fr{1}{4}(Z\alpha)^2\left({\alpha\over\rmpi}\right)
\nonumber\\ &&
+ \fr{1}{2}(Z\alpha)^2{m_{\rm e}\over m_{\rm p}}
+ \fr{1}{2}\left(A_1^{(4)}-\fr{1}{4}\right)(Z\alpha)^2
\left(\alpha\over\rmpi\right)^2
\nonumber \\ &&
-\fr{5}{12}(Z\alpha)^2\left({\alpha\over\rmpi}\right)
{m_{\rm e}\over m_{\rm p}}
+ \cdots \ ,
\label{eq:ehgrat}
\end{eqnarray}
where $A_1^{(4)}$ is given in Eq.~(\ref{eq:a14}).
For the proton in hydrogen, we have
\begin{eqnarray}
{g_{\rm p}({\rm H}) \over g_{\rm p}} &=& 1 - \fr{1}{3}\alpha(Z\alpha)
- \fr{97}{108}\alpha(Z\alpha)^3
\nonumber\\
&&+\fr{1}{6} \alpha(Z\alpha) {m_{\rm e} \over m_{\rm p}}
{3+4a_{\rm p}\over 1+a_{\rm p}} + \cdots \ ,
\label{eq:phgrat}
\end{eqnarray}
where the proton magnetic moment anomaly $a_{\rm p}$ is defined by
\begin{eqnarray}
a_{\rm p} &=& {\mu_{\rm p} \over \left(e\hbar/2m_{\rm p}\right)}-1
\approx 1.793
\ .
\label{eq:apdef}
\end{eqnarray}

For deuterium, similar expressions apply for the electron
\begin{eqnarray}
{g_{\rm e^-}({\rm D}) \over g_{\rm e^-}}
&=& 1 -\fr{1}{3}(Z\alpha)^2 - \fr{1}{12}(Z\alpha)^4
+ \fr{1}{4}(Z\alpha)^2\left({\alpha\over\rmpi}\right)
\nonumber \\ &&
+ \fr{1}{2}(Z\alpha)^2{m_{\rm e}\over m_{\rm d}}
+ \fr{1}{2}\left(A_1^{(4)}-\fr{1}{4}\right)(Z\alpha)^2
\left(\alpha\over\rmpi\right)^2
\nonumber \\ &&
-\fr{5}{12}(Z\alpha)^2\left({\alpha\over\rmpi}\right)
{m_{\rm e}\over m_{\rm d}}
+ \cdots \ ,
\label{eq:edgrat}
\end{eqnarray}
and deuteron
\begin{eqnarray}
{g_{\rm d}({\rm D}) \over g_{\rm d}} &=& 1 - \fr{1}{3}\alpha(Z\alpha)
- \fr{97}{108}\alpha(Z\alpha)^3
\nonumber\\
&&+\fr{1}{6} \alpha(Z\alpha) {m_{\rm e} \over m_{\rm d}}
{3+4a_{\rm d}\over 1+a_{\rm d}} + \cdots \ ,
\label{eq:ddgrat}
\end{eqnarray}
where the deuteron magnetic moment anomaly $a_{\rm d}$ is defined by
\begin{eqnarray}
a_{\rm d} = {\mu_{\rm d}\over
 \left(e\hbar/ m_{\rm d}\right)} - 1 \approx -0.143 \ .
\end{eqnarray}

In the case of muonium Mu, some additional higher-order terms are
included.  For the electron in muonium, we have
\begin{eqnarray}
{g_{\rm e^-}({\rm Mu}) \over g_{\rm e^-}}
&=& 1 -\fr{1}{3}(Z\alpha)^2 - \fr{1}{12}(Z\alpha)^4
+ \fr{1}{4}(Z\alpha)^2\left({\alpha\over\rmpi}\right)
\nonumber \\ &&
+ \fr{1}{2}(Z\alpha)^2{m_{\rm e}\over m_{\rmssmu}}
+ \fr{1}{2}\left(A_1^{(4)}-\fr{1}{4}\right)(Z\alpha)^2
\left(\alpha\over\rmpi\right)^2
\nonumber \\ && \hbox to -0.5 cm {}
-\fr{5}{12}(Z\alpha)^2\left({\alpha\over\rmpi}\right)
{m_{\rm e}\over m_{\rmssmu}}
- \fr{1}{2}(1+Z)(Z\alpha)^2\left({m_{\rm e}\over m_{\rmssmu}}\right)^{\!2}
\nonumber \\ && \hbox to -0.5 cm {}
+ \cdots \ ,
\label{eq:emugrat}
\end{eqnarray}
and for the muon in muonium, the ratio is
\begin{eqnarray}
{g_{{\rmssmu}^+}({\rm Mu}) \over g_{{\rmssmu}^+}} &=&
1 - \fr{1}{3}\alpha(Z\alpha) - \fr{97}{108}\alpha(Z\alpha)^3
\nonumber\\ &&
+ \fr{1}{2} \alpha(Z\alpha) {m_{\rm e} \over m_{\rmssmu}}
+ \fr{1}{12} \alpha(Z\alpha)
\left({\alpha\over\rmpi}\right)
{m_{\rm e} \over m_{\rmssmu}}
\nonumber\\ &&
- \fr{1}{2}(1+Z)\alpha(Z\alpha) \left({m_{\rm e} \over m_{\rmssmu}}\right)^2
 + \cdots \ .
\nonumber\\
\label{eq:mumugrat}
\end{eqnarray}

The numerical values of the corrections in Eqs.~(\ref{eq:ehgrat}) to
(\ref{eq:mumugrat}), based on the 2010 adjusted values of the relevant
constants, are listed in Table \ref{tab:gfactrat}; uncertainties are
negligible here.  An additional term of order $\alpha(Z\alpha)^5$,
relevant to Eqs.~(\ref{eq:phgrat}), (\ref{eq:ddgrat}), and
(\ref{eq:mumugrat}) has been calculated by \citet{2009018}, but it is
negligible at the present level of uncertainty.

\begin{table}[t]
\caption{Theoretical values for various bound-particle to free-particle
$g$-factor ratios relevant to the 2010 adjustment
based on the 2010 recommended values of the
constants.}
\label{tab:gfactrat}
\begin{center}
\begin{tabular}{c@{\qquad}c}
\toprule
Ratio   &   Value \\
\colrule
 $g_{\rm e^-}({\rm H})/g_{\rm e^-}$      & $  1  -17.7054\times 10^{-6} $ \T
  \\[3 pt]
 $g_{\rm p}({\rm H})/g_{\rm p}$          & $  1  -17.7354\times 10^{-6} $ \\[3
  pt]
 $g_{\rm e^-}({\rm D})/g_{\rm e^-}$      & $  1  -17.7126\times 10^{-6} $ \\[3
  pt]
 $g_{\rm d}({\rm D})/g_{\rm d}$          & $  1  -17.7461\times 10^{-6} $ \\[3
  pt]
 $g_{\rm e^-}({\rm Mu})/g_{\rm e^-}$     & $ 1  -17.5926\times 10^{-6} $ \\[3
  pt]
 $g_{\rmssmu^+}({\rm Mu})/g_{\rmssmu^+}$ & $ 1  -17.6254\times 10^{-6} $ \B \\[1
  pt]
\botrule
\end{tabular}
\end{center}
\end{table}

\subsubsection{Bound helion to free helion magnetic moment ratio
$\mu_{\rm h}^\prime/\mu_{\rm h}$}
\label{sssec:bfhmmr}

The bound helion to free helion magnetic moment ratio correction
$\sigma_{\rm h}$, defined by
\begin{eqnarray}
\frac{\mu_{\rm h}^\prime}{\mu_{\rm h}} &=& 1 - \sigma_{\rm h} \, ,
\end{eqnarray}
has been calculated by \citet{2009161}, who obtain
\begin{eqnarray}
\sigma_{\rm h} &=&  59.967\,43(10)\times 10^{-6} \quad [ 1.7\times 10^{-6}] \, .
\end{eqnarray}
This provides a recommended value for the unshielded helion magnetic
moment, along with other related quantities.

\subsubsection{Ratio measurements}
\label{sssec:exps}

Since all of the experimental bound-state magnetic-moment ratios of
interest for the 2010 adjustment are discussed in one or more of the
previous three CODATA reports, only minimal information is given here.
The relevant input data are items $B19$-$B27$ of Table~\ref{tab:pdata}
and their respective observational equations are $B19$-$B27$ in
Table~\ref{tab:pobseqsb1}.  The adjusted constants in those equations
may be identified using Table~\ref{tab:adjconb}, and theoretical
bound-particle to free-particle $g$-factor ratios, which are taken to be
exact, are given in Table~\ref{tab:gfactrat}. The symbol $\mu_{\rm
p}^{\,\prime}$ denotes the magnetic moment of a proton in a spherical
sample of pure $\rm H_2{\rm O}$ at  25~$^\circ{\rm C}$ surrounded by
vacuum; and the symbol $\mu_{\rm h}^{\,\prime}$ denotes the magnetic
moment of a helion bound in a $^3{\rm He}$ atom. Although the exact
shape and temperature of the gaseous $^3{\rm He}$ sample is unimportant,
we assume that it is spherical, at  25~$^\circ{\rm C}$, and surrounded
by vacuum.

Item $B19$, labeled MIT-72, is the ratio $\mu_{\rm e^{-}}({\rm
H})/\mu_{\rm p}({\rm H})$ in the 1S state of hydrogen obtained at MIT by
\citet{1972028,pc97dk}; and $B20$, labeled MIT-84, is the ratio
$\mu_{\rm d}({\rm D})/\mu_{\rm e^{-}}({\rm D})$ in the 1S state of
deuterium also obtained at MIT \cite{pc84pkw}.

Item $B21$ with identification StPtrsb-03 is the magnetic moment ratio
$\mu_{\rm p}(\rm HD)/\mu_{\rm d}(\rm HD)$, and $B23$ with the same
identification is the ratio $\mu_{\rm t}(\rm HT)/\mu_{\rm p}(\rm HT)$,
both of which were determined from NMR measurements on the HD and HT
molecules (bound state of hydrogen and deuterium and of hydrogen and
tritium, respectively) by researchers working at institutes in St.
Petersburg, Russian Federation \cite{2003262,2005103}.  Here $\mu_{\rm
p}(\rm HD)$ and $\mu_{\rm d}(\rm HD)$ are the proton and the deuteron
magnetic moments in HD and $\mu_{\rm t}(\rm HT)$ and $\mu_{\rm p}(\rm
HT)$ are the triton and the proton magnetic moments in HT.  Item $B22$
and $B24$, also with the identifications StPtrsb-03 and due to
\citet{2003262} and \citet{2005103}, are defined according to
$\sigma_{\rm dp}\equiv\sigma_{\rm d}(\rm HD)-\sigma_{\rm p}(\rm HD)$ and
$\sigma_{\rm tp}\equiv\sigma_{\rm t}(\rm HT)-\sigma_{\rm p}(\rm HT)$,
where $\sigma_{\rm p}(\rm HD)$, $\sigma_{\rm d}(\rm HD)$, $\sigma_{\rm
t}(\rm HT)$, and $\sigma_{\rm p}(\rm HT)$ are the corresponding nuclear
magnetic shielding corrections, which are small: $\mu({\rm
bound})=(1-\sigma){\mu}({\rm free})$.

We note that after the 31 December 2010 closing date of the 2010
adjustment, \citet{2011216} reported a result for the ratio $\mu_{\rm
t}(\rm HT)/\mu_{\rm p}(\rm HT)$ with a relative standard uncertainty of
$7\times10^{-10}$ and which is consistent with data item $B23$.

Item $B25$, labeled MIT-77, is the ratio  $\mu_{\rm e^{-}}({\rm
H})/\mu_{\rm p}^{\,\prime}$ obtained at MIT by \citet{1977016}, where
the electron is in the 1S state of hydrogen. The results of
\citet{1984035} are used to correct the measured value of the ratio
based on a spherical $\rm H_2{\rm O}$ NMR sample at 34.7~$^\circ{\rm C}$
to the reference temperature 25~$^\circ{\rm C}$.

Item $B26$ with identification NPL-93 is the ratio $\mu_{\rm
h}^{\,\prime}/\mu_{\rm p}^{\,\prime}$ determined at the National
Physical Laboratory (NPL), Teddington, UK, by \citet{1993113}. And
$B27$, labeled ILL-79, is the neutron to shielded proton magnetic-moment
ratio $\mu_{\rm n}/\mu_{\rm p}^{\,\prime}$ determined at the Institut
Max von Laue-Paul Langevin (ILL) in Grenoble, France
\cite{1979014,1977020}.

\subsection{Muonium transition frequencies,
the muon-proton magnetic moment ratio
$\bm{\mu_{\rmssmu}/\mu_{\rm p}}$,
and muon-electron mass ratio $\bm{m_{\rmssmu}/m_{\rm e}}$}
\label{ssec:muhfs}

Experimental frequencies for transitions between Zeeman energy levels in
muonium ($\rmmu^+{\rm e}^-$ atom) provide measured values of
$\mu_{\rmssmu}/\mu_{\rm p}$ and the muonium ground-state hyperfine
splitting $\Delta \nu_{\rm Mu}$ that depend only on the commonly used
Breit-Rabi equation \cite{1931002}.

The theoretical expression for the hyperfine splitting
$\Delta\nu_{\rm Mu}({\rm th})$ is discussed
in the following section and
may be written as
\begin{eqnarray}
\Delta\nu_{\rm Mu}({\rm th}) &=& {16\over3} c R_\infty\alpha^2
{m_{\rm e}\over m_{\rmssmu}}\left(1+{m_{\rm e}\over m_{\rmssmu}}
\right)^{-3} {\cal F}
\hbox to -2pt {}\left(\alpha, {m_{\rm e}/ m_{\rmssmu}}\right)
\nonumber\\
&=&\Delta\nu_{\rm F}
       {\cal F}\hbox to -2pt {}
\left(\alpha, {m_{\rm e}/ m_{\rmssmu}}\right) \, ,
\label{eq:hfsth}
\end{eqnarray}
where the function ${\cal F}$ depends weakly on $\alpha$ and $m_{\rm e}/
m_{\rmssmu}$.

\subsubsection{Theory of the muonium ground-state hyperfine splitting}
\label{sssec:muhfs}

Presented here is a brief summary of the present theory of $\Delta
\nu_{\rm Mu}$.  Complete results of the relevant calculations are given
along with references to new work; references to the original literature
included in earlier CODATA reports are not repeated.

The hyperfine splitting is given mainly by the Fermi formula:
\begin{eqnarray}
\Delta\nu_{\rm F}
= {16\over 3} c R_\infty Z^3\alpha^2 {m_{\rm e}\over m_{\rmssmu}}
\left[ 1+{m_{\rm e}\over m_{\rmssmu}} \right]^{-3} \, .
\label{eq:dfermi}
\end{eqnarray}
In order to identify the source of the terms, some of the theoretical
expressions are for a muon with charge $Ze$ rather than $e$.

The general expression for the hyperfine splitting is
\begin{eqnarray}
\Delta \nu_{\rm Mu} (\hbox{th})
&=& \Delta \nu_{\rm D} + \Delta \nu_{\rm rad} + \Delta \nu_{\rm rec}
\nonumber\\
&&+ \Delta \nu_{\rm r\hbox{-}r}
+ \Delta \nu_{\rm weak} + \Delta \nu_{\rm had} \, ,
\label{eq:dmuth}
\end{eqnarray}
where the terms labeled D, rad, rec, r-r, weak, and had account for the
Dirac, radiative, recoil, radiative-recoil, electroweak, and hadronic
contributions to the hyperfine splitting, respectively.

The Dirac equation yields
\begin{eqnarray}
\Delta \nu_{\rm D} &=&
\Delta \nu_{\rm F}(1 + a_{\rmssmu})
\left[1+\fr{3}{2}(Z\alpha)^2 + \fr{17}{8}(Z\alpha)^4
+ \cdots \ \right] ,
\nonumber\\
\label{eq:dbreit}
\end{eqnarray}
where $a_{\rmssmu}$ is the muon magnetic moment anomaly.

The radiative corrections are
\begin{eqnarray}
\Delta \nu_{\rm rad} &=&
\Delta \nu_{\rm F}(1 + a_{\rmssmu})
\Big[
D^{(2)}\!(Z\alpha) \left({\alpha\over\rmpi}\right)
\nonumber \\
&& + D^{(4)}\!(Z\alpha) \left({\alpha\over\rmpi}\right)^2 +
 D^{(6)}\!(Z\alpha) \left({\alpha\over\rmpi}\right)^3 +
\cdots \Big] \ ,
\nonumber\\ &&
\label{eq:drad}
\end{eqnarray}
where the functions $D^{(2n)}\!(Z\alpha)$ are contributions from $n$
virtual photons.  The leading term is
\begin{eqnarray}
D^{(2)}\!(Z\alpha) &=&
A_{1}^{(2)} + \left(\ln 2 - \fr{5}{2}\right)\rmpi Z\alpha
\nonumber\\
&& + \Big[-\fr{2}{3}\ln^2 (Z\alpha)^{-2}
 + \left(\fr{281}{360}-\fr{8}{3}\ln 2\right)\ln(Z\alpha)^{-2}
\nonumber\\
&& + 16.9037\ldots\Big](Z\alpha)^2
\nonumber\\
&&+ \Big[\left(\fr{5}{2}\ln 2 - \fr{547}{96}\right)
\ln(Z\alpha)^{-2}\Big]\rmpi(Z\alpha)^3
\nonumber\\
&& + G(Z\alpha)(Z\alpha)^3 \ ,
\label{eq:mud2}
\end{eqnarray}
where $A_{1}^{(2)} = \fr{1}{2}$, as in Eq.~(\ref{eq:a12}).  The function
$G(Z\alpha)$ accounts for all higher-order contributions in powers of
$Z\alpha$; it can be divided into self-energy and vacuum polarization
contributions, $G(Z\alpha) = G_{\rm SE}(Z\alpha) + G_{\rm VP}(Z\alpha)$.
\citet{2008088,2010017} have calculated the one-loop self energy for the
muonium HFS with the result
\begin{eqnarray}
G_{\rm SE}(\alpha) = - 13.8308(43) \, ,
\end{eqnarray}
which agrees with the value $G_{\rm SE}(\alpha) = - 13.8(3)$ from an
earlier calculation by \citet{2005234}, as well as with other previous
estimates.  The vacuum polarization part is
\begin{eqnarray}
G_{\rm VP}(\alpha) = 7.227(9)\ .
\end{eqnarray}

For $D^{(4)}\!(Z\alpha)$, we have
\begin{eqnarray}
D^{(4)}\!(Z\alpha) &=&
A_{1}^{(4)} + 0.770\,99(2) \rmpi Z\alpha
+ \Big[-\fr{1}{3}\ln^2 (Z\alpha)^{-2}
\nonumber\\ &&
-0.6390 \ldots\times\ln (Z\alpha)^{-2}
+10(2.5) \Big](Z\alpha)^2
\nonumber\\ &&
+ \cdots \ ,
\label{eq:mud4}
\end{eqnarray}
where $A_1^{(4)}$ is given in Eq.~(\ref{eq:a14}), and the coefficient of
$\rmpi Z\alpha$ has been calculated by \citet{2010119}.

The next term is
\begin{eqnarray}
D^{(6)}\!(Z\alpha) = A_{1}^{(6)} + \cdots \ ,
\label{eq:mud6}
\end{eqnarray}
where the leading contribution $A_{1}^{(6)}$ is given in
Eq.~(\ref{eq:a16}), but only partial results of relative order $Z\alpha$
have been calculated \cite{2007138}.  Higher-order functions
$D^{(2n)}\!(Z\alpha)$ with $n>3$ are expected to be negligible.

The recoil contribution is
\begin{eqnarray}
\label{eq:hfsrec}
\Delta \nu_{\rm rec} &=&
\Delta\nu_{\rm F}{m_{\rm e}\over m_{\rmssmu}}
\Bigg(-{3\over1-\left(m_{\rm e}/ m_{\rmssmu}\right)^2}
\ln \Big({m_{\rmssmu}\over m_{\rm e}}\Big){Z\alpha\over\rmpi} \nonumber\\
&& +\, {1\over\left(1+m_{\rm e}/m_{\rmssmu}\right)^2}
\bigg\{\ln {(Z\alpha)^{-2}}-8\ln
2 + {65\over 18}
\nonumber\\&&
+\bigg[{9\over 2\rmpi^2}
\ln^2\left({m_{\rmssmu}\over m_{\rm e}}\right)
+\left({27\over 2\rmpi^2}-1\right)
\ln\left({m_{\rmssmu}\over m_{\rm e}}\right)
\nonumber\\&&
+{93\over4\rmpi^2} + {33\zeta(3)\over\rmpi^2}
-{13\over12}-12\ln2 \bigg]
{m_{\rm e}\over m_{\rmssmu}}
\bigg\}(Z\alpha)^2 \nonumber\\
&& +\, \bigg\{-{3\over 2} \ln{\Big({m_
{\rmssmu}\over m_{\rm e}}\Big)} \ln(Z\alpha)^{-2}
-{1\over6}\ln^2 {(Z\alpha)^{-2}}
\nonumber\\ &&
+\left({101\over18} - 10 \ln 2\right)\ln(Z\alpha)^{-2}
\nonumber\\&&\qquad\qquad
+40(10)\bigg\}
{(Z\alpha)^3\over\rmpi}
\Bigg)
 + \cdots \,,
\nonumber\\&&
\end{eqnarray}
\label{eq:drec}
as discussed in CODATA-02

The radiative-recoil contribution is
\begin{eqnarray}
\label{eq:hfsradrec}
\Delta \nu_{\rm r\hbox{-}r} &=& \Delta \nu_{\rm F}
\left({\alpha\over \rmpi}\right)^2 {m_{\rm e}\over m_{\rmssmu}}
\bigg\{\bigg[ -2\ln^2\Big({m_{\rmssmu}\over m_{\rm e}}\Big) +{13\over
12}\ln{\Big({m_{\rmssmu}\over m_{\rm e}}\Big)} \nonumber\\
&& +\, {21\over2}\zeta(3)+{\rmpi^2\over6}+{35\over9}\bigg]
+\bigg[\, {4\over3} \ln^2\alpha^{-2}
\nonumber\\
&&+\left({16\over3} \ln 2 - {341\over180}\right) \ln\alpha^{-2}
-40(10) \bigg]\rmpi\alpha \nonumber\\
&&+\bigg[-{4\over3}\ln^3 \Big({m_{\rmssmu}\over
m_{\rm e}}\Big)
+{4\over3}\ln^2 \Big({m_{\rmssmu}\over m_{\rm e}}\Big)
\bigg] {\alpha\over \rmpi}\bigg\}
\nonumber\\
&& - \nu_{\rm F}\alpha^2\!\left(m_{\rm e}\over m_{\rmssmu}\right)^{\!2}
\left(6\ln{2} + {13\over6}\right) + \cdots \ ,
\label{eq:dradrec}
\end{eqnarray}
where, for simplicity, the explicit dependence on $Z$ is not shown.
Partial radiative recoil results are given by
\citet{2009231,2009234,2010034}, and are summarized as
\begin{eqnarray}
\Delta \nu_{\rm ES} &=&
\Delta \nu_{\rm F}
\left(\frac{\alpha}{\rmpi}\right)^3
\frac{m_{\rm e}}{m_{\rmssmu}}
\bigg\{\left[3\zeta(3)-6\rmpi^2\ln{2}
+\rmpi^2-8 \right] \ln{\frac{m_{\rmssmu}}{m_{\rm e}}}
\nonumber\\&&\qquad
+ 63.127(2) \bigg\}
\label{eq:essum}
= -34.7 \mbox{ Hz} \, .
\end{eqnarray}

The electroweak contribution due to the exchange
of a Z$^0$ boson is \cite{1996059}
\begin{eqnarray}
\Delta \nu_{\rm weak} &=& -65 \ {\rm Hz} \ ,
\label{eq:dweak}
\end{eqnarray}
while for the hadronic vacuum polarization contribution
we have \cite{2002136}
\begin{eqnarray}
\Delta \nu_{\rm had} &=& 236(4) \ {\rm Hz}\, ,
\label{eq:dhad}
\end{eqnarray}
as in CODATA-06.  A negligible contribution ($\approx 0.0065$ Hz)
from the hadronic light-by-light correction has been given by
\citet{2008206}.  Tau vacuum polarization contributes 3 Hz, which is also
negligible at the present level of uncertainty \cite{1984001}.

The four principle sources of uncertainty in $\Delta\nu_{\rm Mu}(\rm
th)$ are $\Delta\nu_{\rm rad}$, $\Delta\nu_{\rm rec}$, $\Delta\nu_{\rm
r-r}$, and $\Delta\nu_{\rm had}$ in Eq.~(\ref{eq:dmuth}).  Based on the
discussion in CODATA-02, CODATA-06, and the new results above, the
current uncertainties from these contributions are 7~Hz, 74~Hz, 63~Hz,
and 4~Hz, respectively, for a total of 98~Hz.  Since this is only 3 \%
less than the value 101~Hz used in the 2006 adjustment, and in view of
the incomplete nature of the calculations, the Task Group has retained
the 101~Hz standard uncertainty of that adjustment:
\begin{eqnarray}
u[\Delta \nu_{\rm Mu} {\rm(th)}] =  101 \ {\rm Hz}
\quad [ 2.3\times 10^{-8}] \ .
\label{eq:runcmuhfs}
\end{eqnarray}
For the least-squares calculations, we use as the theoretical
expression for the hyperfine splitting
\begin{eqnarray}
\Delta \nu_{\rm Mu}\!\!\left(R_\infty,\alpha,{m_{\rm e}\over m_{\rmssmu}},
\delta_{\rmssmu},\delta_{\rm Mu}\right) =
\Delta \nu_{\rm Mu} \rm (th) + \delta_{\rm Mu} \ ,
\nonumber\\
\end{eqnarray}
where the input datum for the additive correction $\delta_{\rm Mu}$,
which accounts for the uncertainty of the theoretical expression and is
data item $B28$ in Table~\ref{tab:pdata}, is  0(101)~Hz.

The above theory yields
\begin{eqnarray}
\Delta \nu_{\rm Mu} =  4\,463\,302\,891(272) \ {\rm Hz}
\quad [ 6.1\times 10^{-8}]
\label{eq:dnmup}
\end{eqnarray}
using values of the constants obtained from the 2010 adjustment without
the two LAMPF measured values of $\Delta\nu_{\rm Mu}$ discussed in the
following section.  The main source of uncertainty in this value is the
mass ratio $m_{\rm e}/m_{\rmssmu}$.

\subsubsection{Measurements of muonium transition frequencies
and values of $\mu_{\rmssmu}/\mu_{\rm p}$ and $m_{\rmssmu}/m_{\rm e}$}
\label{sssec:mufreqs}

The two most precise determinations of muonium Zeeman transition frequencies
were carried out at the Clinton P. Anderson Meson Physics Facility at Los
Alamos (LAMPF), USA, and were reviewed in detail in CODATA-98.  The
results are as follows.

Data reported in 1982 by \citet{1982003, pc81mar} are
\begin{eqnarray}
\Delta \nu_{\rm Mu} =  4\,463\,302.88(16) \ {\rm kHz} \quad [ 3.6\times 10^{-8}]
\, ,
\label{eq:nupL82}
\end{eqnarray}
\begin{eqnarray}
\nu(f_{\rm p}) =  627\,994.77(14) \ {\rm kHz} \quad [ 2.2\times 10^{-7}]
\, ,
\label{eq:numL82}
\end{eqnarray}
\begin{eqnarray}
r[\Delta \nu_{\rm Mu},\nu(f_{\rm p})] =  0.227 \, ,
\label{eq:rnupm82}
\end{eqnarray}
where $f_{\rm p}$ is $57.972\,993$ MHz, corresponding to the magnetic
flux density of about $1.3616$ T used in the experiment, and $r[\Delta
\nu_{\rm Mu},\nu(f_{\rm p})]$ is the correlation coefficient of $\Delta
\nu_{\rm Mu}$ and $\nu(f_{\rm p})$.  The data reported in 1999 by
\citet{1999002} are
\begin{eqnarray}
\Delta \nu_{\rm Mu} =  4\,463\,302\,765(53) \ {\rm Hz} \quad [ 1.2\times 10^{-8}
  ]
\, ,
\label{eq:nupL99}
\end{eqnarray}
\begin{eqnarray}
\nu(f_{\rm p}) =  668\,223\,166(57) \ {\rm Hz} \quad [ 8.6\times 10^{-8}]
\, ,
\label{eq:numL99}
\end{eqnarray}
\begin{eqnarray}
r[\Delta \nu_{\rm Mu},\nu(f_{\rm p})] =  0.195
\, ,
\label{eq:rnupm99}
\end{eqnarray}
where $f_{\rm p}$ is $72.320\,000$ MHz, corresponding to the flux
density of approximately $1.7$ T used in the experiment, and $r[\Delta
\nu_{\rm Mu},\nu(f_{\rm p})]$ is the correlation coefficient of $\Delta
\nu_{\rm Mu}$ and $\nu(f_{\rm p})$.  The data in Eqs.~(\ref{eq:nupL82}),
(\ref{eq:numL82}), (\ref{eq:nupL99}), and (\ref{eq:numL99}) are data
items $B29.1$, $B30$, $B29.2$, and $B31$, respectively, in
Table~\ref{tab:pdata}.

The expression for the magnetic moment ratio is
\begin{eqnarray}
{\mu_{{\rmssmu}^+} \over \mu_{\rm p}} &=&
{\Delta \nu_{\rm Mu}^2 - \nu^2(f_{\rm p}) + 2 s_{\rm e}f_{\rm p}\,\nu(f_{\rm p})
\over 4 s_{\rm e} f_{\rm p}^2 - 2 f_{\rm p}\,\nu(f_{\rm p})}
\left({g_{{\rmssmu}^+}(\rm Mu})\over g_{{\rmssmu}^+}\right)^{-1} ,
\nonumber\\
\label{eq:murat}
\end{eqnarray}
where $\Delta \nu_{\rm Mu}$ and $\nu(f_{\rm p})$ are the sum and
difference of two measured transition frequencies, $f_{\rm p}$ is the
free proton NMR reference frequency corresponding to the flux
density used in the experiment, $g_{{\rmssmu}^+}({\rm Mu})/
g_{{\rmssmu}^+}$ is the bound-state correction for the muon in muonium
given in Table \ref{tab:gfactrat}, and
\begin{eqnarray}
s_{\rm e} = {\mu_{\rm e^-}\over\mu_{\rm p}}
{g_{\rm e^-}({\rm Mu})\over g_{\rm e^-}} \, ,
\label{eq:se}
\end{eqnarray}
where $g_{\rm e^-}({\rm Mu}) / g_{\rm e^-}$ is the
bound-state correction for the electron in muonium given in
the same table.

The muon to electron mass ratio $m_{\rmssmu}/m_{\rm e}$ and the muon
to proton magnetic moment ratio $\mu_{\rmssmu}/\mu_{\rm p}$ are
related by
\begin{eqnarray}
{m_{\rmssmu}\over m_{\rm e}} =
\left({\mu_{\rm e}\over \mu_{\rm p}}\right)
\left({\mu_{\rmssmu}\over \mu_{\rm p}}\right)^{-1}
\left({g_{\rmssmu}\over g_{\rm e}}\right) .
\label{eq:mumemump}
\end{eqnarray}

A least-squares adjustment using the LAMPF data, the 2010 recommended
values of $R_{\infty}$, $\mu_{\rm e}/\mu_{\rm p}$, $g_{\rm e}$, and
$g_{\rmssmu}$, together with Eq.~(\ref{eq:hfsth}) and
Eqs.~(\ref{eq:murat}) to (\ref{eq:mumemump}), yields
\begin{eqnarray}
{\mu_{{\rmssmu}^+}\over \mu_{\rm p}} &=&   3.183\,345\,24(37) \quad [ 1.2\times
  10^{-7}]
\, ,
\label{eq:mumupsmupL}
\\
{m_{\rmssmu}\over m_{\rm e}} &=&   206.768\,276(24) \quad [ 1.2\times 10^{-7}]
\, ,
\label{eq:mesmmuiL}
\\
\alpha^{-1} &=&   137.036\,0018(80) \quad [ 5.8\times 10^{-8}]
\, ,
\label{eq:alphiL}
\end{eqnarray}
where this value of $\alpha$ is denoted as $\alpha^{-1}(\Delta\nu_{\rm Mu})$.

The uncertainty of $m_{\rmssmu}/m_{\rm e}$ in Eq.~(\ref{eq:mesmmuiL}) is
nearly five times the uncertainty of the 2010 recommended value.  In
Eq.~(\ref{eq:mesmmuiL}), the value follows from Eqs.~(\ref{eq:murat}) to
(\ref{eq:mumemump}) with almost the same uncertainty as the moment ratio
in Eq.~(\ref{eq:mumupsmupL}).  Taken together, the experimental value of
and theoretical expression for the hyperfine splitting essentially
determine the value of the product $\alpha^2 m_{\rm e}/m_{\rmssmu}$, as
is evident from Eq.~(\ref{eq:hfsth}), with an uncertainty dominated by
the $ 2.3\times 10^{-8}$ relative uncertainty in the theory, and in this
limited least-squares adjustment $\alpha$ is otherwise unconstrained.
However, in the full adjustment the value of $\alpha$ is determined by
other data which in turn determines the value of $m_{\rmssmu}/m_{\rm e}$
with a significantly smaller uncertainty than that of
Eq.~(\ref{eq:mesmmuiL}).

\section{Quotient of Planck constant and particle mass $\bm{h/m(X)}$
and $\bm{\alpha}$}
\label{sec:pcpmq}

Measurements of $h/m(X)$ are of potential importance because the
relation $R_{\infty} = \alpha^2m_{\rm e}c/2h$ implies
\begin{eqnarray}
\alpha = \left[{2R_{\infty}\over c}{A_{\rm r}(X)\over
A_{\rm r}(\rm e)}{h\over m(X)}\right]^{1/2},
\label{eq:alhmx}
\end{eqnarray}
where $A_{\rm r}(X)$ is the relative atomic mass of particle $X$ with
mass $m(X)$ and $A_{\rm r}({\rm e})$ is the relative atomic mass of the
electron.  Because $c$ is exactly known, the relative standard
uncertainties of $R_\infty$ and $A_{\rm r}({\rm e})$ are $5.0
\times10^{-12}$ and $4.0\times10^{-10}$, respectively, and the
uncertainty of $A_{\rm r}(X)$ for many particles and atoms is less than
that of $A_{\rm r}({\rm e})$, Eq.~(\ref{eq:alhmx}) can provide a
competitive value of $\alpha$ if $h/m(X)$ is determined with a
sufficiently small uncertainty.  This section discusses measurements of
$h/m({\rm ^{133}Cs})$ and $h/m({\rm ^{87}Rb})$.

\subsection{Quotient $h/m({\rm ^{133}Cs})$}
\label{sssec:pccsmr}

\citet{2002223} determined $h/m({\rm^{133}Cs})$ by measuring the atomic
recoil frequency shift of photons absorbed and emitted by ${\rm
^{133}Cs}$ atoms using atom interferometry.  Carried out at Stanford
University, Stanford, California, USA, the experiment is discussed in
CODATA-06 and CODATA-02. Consequently, only the final result is given
here:
\begin{eqnarray}
{h\over m(^{133}{\rm Cs})} &=&  3.002\,369\,432(46)\times 10^{-9}~{\rm m}^2~{\rm
  s}^{-1}
\nonumber\\
\label{eq:homcs02}
&&\qquad\qquad[ 1.5\times 10^{-8}] \ .
\end{eqnarray}
The observational equation for this datum is, from Eq.~(\ref{eq:alhmx}),
\begin{eqnarray}
{h\over m(^{133}{\rm Cs})} = {A_{\rm r}({\rm e})\over A_{\rm r}({\rm ^{133}Cs})}
{c\,\alpha^2 \over 2 R_\infty} \ .
\label{eq:homcsoe}
\end{eqnarray}
The value of $\alpha$ inferred from this expression and
Eq.~(\ref{eq:homcs02}) is given in Table~\ref{tab:alpha},
Sec.~\ref{ssec:infv}.

The Stanford result for $h/m({\rm^{133}Cs})$ was not included as an
input datum in the final adjustment on which the 2006 recommended values
are based because of its low weight, and is omitted from the 2010 final
adjustment for the same reason. Nevertheless, it is included as an
initial input datum to provide a complete picture of the available data
that provide values of $\alpha$.

\subsection{Quotient $h/m({\rm ^{87}Rb})$}
\label{sssec:pcrbmr}

A value of $h/m({\rm^{87}Rb})$ with a relative standard uncertainty of
$1.3\times 10^{-8}$ obtained at LKB in Paris was taken as an input datum
in the 2006 adjustment and its uncertainty was sufficiently small for it
to be included in the 2006 final adjustment.  Reported by
\citet{2006276} and discussed in CODATA-06, $h/m({\rm^{87}Rb})$ was
determined by measuring the rubidium recoil velocity $v_{\rm r}=\hbar
k/m({\rm ^{87}Rb})$ when a rubidium atom absorbs or emits a photon of
wave vector $k = 2{\rmpi}/\lambda$, where $\lambda$ is the wavelength of
the photon and $\nu = c/\lambda$ is its frequency. The measurements were
based on Bloch oscillations in a moving standing wave.

A value of $h/m({\rm^{87}Rb})$ with a relative uncertainty of $9.2\times
10^{-9}$ and in agreement with the earlier result, obtained from a new
LKB experiment using combined Bloch oscillations and atom
interferometry, was subsequently reported by \citet{2008243}.  In this
approach Bloch oscillations are employed to transfer a large number of
photon momenta to rubidium atoms and an atom interferometer is used to
accurately determine the resulting variation in the velocity of the
atoms. Significant improvements incorporated into this version of the
experiment have now provided a newer value of $h/m({\rm^{87}Rb})$ that
not only agrees with the two previous values, but has an uncertainty
over 10 and 7 times smaller, respectively. As given by \citet{2011014},
the new LKB result is \begin{eqnarray} {h\over m(^{87}{\rm Rb})} &=&
 4.591\,359\,2729(57)\times 10^{-9}~{\rm m}^2~{\rm s}^{-1} \nonumber\\
  \label{eq:homrb11}
&&\qquad\qquad[ 1.2\times 10^{-9}] \ .  \end{eqnarray}

Because the LKB researchers informed the Task Group that this result
should be viewed as superseding the two earlier results \cite{pc11fb},
it is the only value of $h/m({\rm^{87}Rb})$ included as an
input datum in the 2010 adjustment . The observational equation for this
datum is, from Eq.~(\ref{eq:alhmx}),
\begin{eqnarray}
{h\over m(^{87}{\rm Rb})} = {A_{\rm r}({\rm e})\over A_{\rm r}({\rm ^{87}Rb})}
{c\,\alpha^2 \over 2 R_\infty} \ .
\label{eq:homrboe}
\end{eqnarray}
The value of $\alpha$ inferred from this expression and
Eq.~(\ref{eq:homrb11}) is given in Table~\ref{tab:alpha},
Sec.~\ref{ssec:infv}.

The experiment of the LKB group from which the result given in
Eq.~(\ref{eq:homrb11}) was obtained is described in the paper by
\citet{2011014}, the references cited therein; see also
\citet{2011164,2010142,2009148,2008243}.  It is worth noting, however,
that the reduction in uncertainty of the 2008 result by over a factor of
7 was achieved by reducing the uncertainties of a number of individual
components, especially those due to the alignment of beams, wave front
curvature and Gouy phase, and the second order Zeeman effect.  The total
fractional correction for systematic effects is $-26.4(5.9)\times
10^{-10}$ and the statistical or Type A uncertainty is 2 parts in
$10^{10}$.

\subsection{Other data}
\label{ssec:pcnmr}

A result for the quotient $h/m_{\rm n}d_{220}({\rm \scriptstyle W04})$
with a relative standard uncertainty of $4.1\times 10^{-8}$, where
$m_{\rm n}$ is the neutron mass and $d_{220}({\rm {\scriptstyle W04}})$
is the \{220\} lattice spacing of the crystal WASO~04, was included in
the past three CODATA adjustments, although its uncertainty was
increased by the multiplicative factor 1.5 in the 2006 final adjustment.
It was obtained by PTB researchers working at the ILL high-neutron-flux
reactor in Grenoble \cite{1999075}.

Since the result has a relative uncertainty of $4.1\times 10^{-8}$, the
value of $\alpha$ that can be inferred from it, even assuming that
$d_{220}({\rm \scriptstyle  W04})$ is exactly known, has an uncertainty
of about $2\times 10^{-8}$.  This is over 50 times larger than that of
$\alpha$ from $a_{\rm  e}$ and is not competitive.  Further, the
inferred value disagrees with the $a_{\rm e}$ value.

On the other hand, the very small uncertainty of the $a_{\rm e}$ value
of $\alpha$ means that the PTB result for $h/m_{\rm n}d_{220}({\rm
\scriptstyle W04})$ can provide an inferred value of $d_{220}({\rm
\scriptstyle W04})$ with the competitive relative uncertainty of about 4
parts in $10^8$.  However, this inferred lattice-spacing value,
reflecting the disagreement of the inferred value of alpha, is
inconsistent with the directly determined XROI value.  This discrepancy
could well be the result of the different effective lattice parameters
for the different experiments.  In the PTB measurement of $h/m_{\rm
n}d_{220}({\rm \scriptstyle W04})$, the de Broglie wavelength,
$\lambda\approx  0.25$ nm, of slow neutrons was determined using back
reflection from the surface of a silicon crystal.  As pointed out to the
Task Group by Peter \citet{pc11pb} of the PTB, the lattice spacings near
the surface of the crystal, which play a more critical role than in the
XROI measurements carried out using x-ray transmission, may be strained
and not the same as the spacings in the bulk of the crystal.

For these reasons, the Task Group decided not to consider this result
for inclusion in the 2010 adjustment.

\section{Electrical measurements}
\label{sec:elmeas}

This section focuses on 18 input data resulting from high-accuracy
electrical measurements, 16 of which were also available for the 2006
adjustment. The remaining two became available in the intervening 4
years.  Of the 16, 13 were not included in the final adjustment on which
the 2006 recommended values are based because of their low weight. These
same data and one of the two new values are omitted in the final 2010
adjustment for the same reason.  Nevertheless, all are initially
included as input data because of their usefulness in providing an
overall picture of the consistency of the data and in testing the
exactness of the Josephson and quantum Hall effect relations $K_{\rm
J}=2e/h$ and $R_{\rm K}=h/e^2$. As an aid, we begin with a concise
overview of the seven different types of electrical quantities of which
the 18 input data are particular examples.  \vspace{.25 in}

\subsection{Types of electrical quantities}
\label{ssec:eq}
\vspace{.25 in}

If microwave radiation of frequency $f$ is applied to a Josephson effect
device, quantized voltages $U_{\rm J}(n)= nf/K_{\rm J}$ are induced
across the device, where $n$, an integer, is the step number of the
voltage and $K_{\rm J}=2e/h$ is the Josephson constant.  Similarly, the
quantized Hall resistance of the $i$th resistance plateau of a quantum
Hall effect device carrying a current and in a magnetic field, $i$ an
integer, is given by $R_{\rm H}(i)=R_{\rm K}/i$, where $R_{\rm
K}=h/e^2=\mu_0c/2\alpha$ is the von Klitzing constant. Thus, measurement
of $K_{\rm J}$ in its SI unit Hz/V determines the quotient $2e/h$, and
since in the SI $c$ and $\mu_0$ are exactly known constants, measurement
of $R_{\rm K}$ in its SI unit $\Omega$ determines $\alpha$. Further,
since $K_{\rm J}^2R_{\rm K}=4/h$, a measurement of this product in its
SI unit $({\rm J}~{\rm s})^{-1}$ determines $h$.

The gyromagnetic ratio $\gamma_{x}$ of a bound particle $x$ of spin
quantum number $i$ and magnetic moment $\mu_{x}$ is given by
\begin{eqnarray}
\gamma_{x}=\frac{2\rmpi{f}}{B}=\frac{\omega}{B}=\frac{|\mu_{x}|}{i\hbar},
\label{eq:gammadef2}
\end{eqnarray}
where $f$ is the spin-flip (or precession) frequency and $\omega$ is the
angular precession frequency of the particle in the magnetic flux
density $B$. For a bound and shielded proton p and helion h
Eq.~(\ref{eq:gammadef2}) gives
\begin{eqnarray}
\gamma^\prime_{\rm p}=\frac{2\mu^\prime_{\rm p}}{\hbar},
\qquad\gamma^\prime_{\rm h}=\frac{2\mu^\prime_{\rm h}}{\hbar},
\label{eq:gampphe}
\end{eqnarray}
where the protons are in a spherical sample of pure $\rm H_2{\rm O}$ at
25~$^\circ{\rm C}$ surrounded by vacuum; and the helions are in a
spherical sample of low-pressure, pure $^3{\rm He}$ gas at
25~$^\circ{\rm C}$ surrounded by vacuum.

The shielded gyromagnetic ratio of a particle can be determined by two
methods but the quantities actually measured are different: the
low-field method determines $\gamma_{x}^\prime/K_{\rm J}R_{\rm K}$ while
the high-field method determines $\gamma_{x}^\prime K_{\rm J}R_{\rm K}$.
In both cases an electric current $I$  is measured using the Josephson
and quantum Hall effects with the conventional values of the Josephson
and von Klitzing constants. We have for the two methods
\begin{eqnarray}
\gamma_{x}^{\,\prime} &=&
{\it \Gamma}_{x-90}^{\,\prime}({\rm lo})
{K_{\rm J} \, R_{\rm K}\over K_{{\rm J}-90} \, R_{{\rm K}-90}} \ ,
\label{eq:sgammaslo90}
\\ \nonumber\\
\gamma_{x}^{\,\prime} &=&
{\it \Gamma}_{x-90}^{\,\prime}({\rm hi})
{K_{{\rm J}-90} \, R_{{\rm K}-90} \over
K_{\rm J} \, R_{\rm K}} \ ,
\label{eq:sgammashi90}
\end{eqnarray}
where ${\it \Gamma}^\prime_{x-90}{(\rm lo)}$ and
${\it{\Gamma}}^{\prime}_{x-90}{(\rm hi)}$ are the experimental values of
$\gamma_{x}^\prime$ in SI units that would result from low- and hi-
field experiments, respectively, if $K_{\rm J}$ and $R_{\rm K}$ had the
exactly known conventional values $K_{\rm J-90}$ and $R_{\rm K-90}$. The
actual input data used in the adjustment are
${\it{\Gamma}}^{\prime}_{x-90}{(\rm lo)}$ and
${\it{\Gamma}}^{\prime}_{x-90}{(\rm hi)}$ since these are the quantities
actually measured in the experiments, but their observational equations
(see Table~\ref{tab:pobseqsb2}) account for the fact that $K_{\rm
J-90}\neq K_{\rm J}$ and $R_{\rm K-90}\neq R_{\rm K}$.

Finally, for the Faraday constant $F$ we have
\begin{eqnarray}
F &=&
{\cal F}_{90}
{K_{{\rm J}-90} \, R_{{\rm K}-90} \over
K_{\rm J} \, R_{\rm K}} \ ,
\label{eq:ff90}
\end{eqnarray}
where ${\cal F}_{90}$ is the actual quantity experimentally measured.
Equation~(\ref{eq:ff90}) is similar to Eq.~(\ref{eq:sgammashi90})
because ${\cal F}_{90}$ depends on current in the same way as ${\it
\Gamma}_{x-90}^{\,\prime}({\rm hi})$, and the same comments apply.

\subsection{Electrical data}
\label{ssec:ed}
\vspace{.25 in}

The 18 electrical input data are data items $B32.1$ through $B38$ in
Table~\ref{tab:pdata}, Sec.~\ref{sec:ad}. Data items $B37.4$ and
$B37.5$, the two new input data mentioned above and which, like the
other three data in this category, are moving-coil watt balance results
for the product $K_{\rm J}^2R_{\rm K}$, are discussed in the next two
sections. Since the other 16 input data have been discussed in one or
more of the three previous CODATA reports, we provide only limited
information here.

$B32.1$ and $B32.2$, labeled NIST-89 and NIM-95, are values of ${\it
\Gamma}_{\rm p-90}^{\,\prime}({\rm lo})$ obtained at the National
Institute of Standards and Technology (NIST), Gaithersburg, MD, USA
\cite{1989008}, and at the National Institute of Metrology (NIM),
Beijing, PRC \cite{1995233}, respectively. $B33$, identified as KR/VN-98,
is a similar value of ${\it \Gamma}_{\rm h-90}^{\,\prime}({\rm lo})$
obtained at the Korea Research Institute of Standards and Science
(KRISS), Taedok Science Town, Republic of Korea in a collaborative
effort with researchers from the Mendeleyev All-Russian Research
Institute for Metrology (VNIIM), St. Petersburg, Russian Federation
\cite{1999153,1999083,1998117,1998132}.  $B34.1$ and $B34.2$ are values
of ${\it \Gamma}_{\rm p-90}^{\,\prime}({\rm hi})$ from NIM
\cite{1995233} and NPL \cite{1979012}, respectively, with
identifications NIM-95 and NPL-79.

$B35.1$-$B35.5$ are five calculable-capacitor determinations of $R_{\rm
K}$ from NIST \cite{1997033,1998049}, the National Metrology Institute
(NMI), Lindfield, Australia \cite{1997146}, NPL \cite{ccerknpl88}, NIM
\cite{1995237}, and 
Laboratoire national de m\'etrologie et d'essais (LNE), Trappes,
France \cite{2001026,2003191}, respectively, and are labeled NIST-97,
NMI-97, NPL-88, NIM-95, and LNE-01.

$B36.1$ with identification NMI-89 is the mercury electrometer result
for $K_{\rm J}$ from NMI \cite{1989050}; and $B36.2$, labeled PTB-91,
is the capacitor voltage balance result for  $K_{\rm J}$ from the
Physikalisch-Technische Bundesanstalt (PTB), Braunschweig, Germany
\cite{1991025,1986043,1985042}.

$B37.1$-$B37.3$, with identifications NPL-90, NIST-98, and NIST-07,
respectively, are moving-coil watt-balance results for $K_{\rm
J}^2R_{\rm K}$ from NPL \cite{1990057} and from NIST
\cite{1998071,2007085}.

The last electrical input datum, $B38$ and labeled NIST-80, is the
silver dissolution coulometer result for ${\cal F}_{90}$ from NIST
\cite{1980029}.

The correlation coefficients of these data, as appropriate, are given in
Table~\ref{tab:pdcc}, Sec.~\ref{sec:ad}; the observational equations for
the seven different types of electrical data of which the 18 input data
are particular examples are given in Table~\ref{tab:pobseqsb1} in the
same section and are $B32$-$B38$. Recalling that the relative standard
uncertainties of $R_{\infty}$, $\alpha$, $\mu_{\rm e^-}/{\mu^\prime_{\rm
p}}$, ${\mu^\prime_{\rm h}}/{\mu^\prime_{\rm p}}$, and $A_{\rm r}(\rm
e)$ are significantly smaller that those of the electrical input data,
inspection of these equations shows that measured values of ${\it
\Gamma}_{\rm p-90}^{\,\prime}({\rm lo})$, ${\it \Gamma}_{\rm
h-90}^{\,\prime}({\rm lo})$, ${\it \Gamma}_{\rm p-90}^{\,\prime}({\rm
hi})$, $R_{\rm K}$, $K_{\rm J}$, $K_{\rm J}^2R_{\rm K}$, and ${\cal
F}_{90}$ principally determine $\alpha$, $\alpha$, $h$, $\alpha$, $h$,
$h$, and $h$, respectively.  \vspace{.25 in}

\subsubsection{$K^{2}_{\rm J}R_{\rm K}$ and $h$: NPL watt balance}
\label{sssec:nplwb12}
\vspace{.25 in}

We consider here and in the following section the two new watt-balance
measurements of $K_{\rm J}^2R_{\rm K}=4/h$. For reviews of such
experiments, see, for example, the papers of
\citet{2012007,2011193,2009155}. The basic idea is to compare electrical
power measured in terms of the Josephson and quantum Hall effects to the
equivalent mechanical power measured in the SI unit W =
m$^2$~kg~s$^{-3}$.  The comparison employs an apparatus now called a
moving-coil watt balance, or simply a watt balance, first proposed by
\citet{1975027} at NPL. A watt balance experiment can be described by
the simple equation $m_{\rm s}gv = UI$, where, for example, $I$ is the
current in a circular coil in a radial magnetic flux density $B$ and the
force on the coil due to $I$ and $B$ is balanced by the weight $m_{\rm
s}g$ of a standard of mass $m_{\rm s}$; and $U$ is the voltage induced
across the terminals of the coil when it is moved vertically with a
velocity $v$ in the same flux density $B$.  Thus, a watt balance is
operated in two different modes: the weighing mode and the velocity
mode.

The NPL Mark II watt balance and its early history were briefly
discussed in CODATA-06, including the initial result obtained with it by
\citet{2007209}. Based on measurements carried out from October 2006 to
March 2007 and having a relative standard uncertainty of 66 parts in
$10^9$, this result became available only after the closing date of the
2006 adjustment.  Moreover, the NPL value of $K_{\rm J}^2R_{\rm K}$ was
308 parts in $10^9$ smaller than the NIST-07 value with a relative
uncertainty of 36 parts in $10^9$.

Significant modifications were subsequently made to the NPL apparatus in
order to identify previously unknown sources of error as well as to
reduce previously identified sources.  The modifications were completed
in November 2008, the apparatus was realigned in December 2008, and
measurements and error investigations were continued until June 2009.
From then to August 2009 the apparatus was dismantled, packed, and
shipped to the National Research Council (NRC), Ottawa, Canada.  A
lengthy, highly detailed preprint reporting the final Mark II result was
provided to the Task Group by I. A. Robinson of NPL prior to the 31
December 2010 closing date of the 2010 adjustment. This paper has now
been published and the reported value is
\cite{2012004}

\begin{eqnarray} h =
6.626\,07123(133) \times 10^{-34} \ {\rm J \ s} \quad [2.0\times 10^{-7}]
\, .
\qquad
\label{eq:hwbnpl12}
\end{eqnarray}
This corresponds to
\begin{eqnarray}
K_{\rm J}^2R_{\rm K} &=&  6.036\,7597(12)\times 10^{33} ~{\rm J}^{-1} \ {\rm s}
  ^{-1}
\nonumber\\&&\qquad\qquad\qquad\qquad[ 2.0\times 10^{-7}]
\label{eq:kj2rknpl12}
\end{eqnarray}
identified as NPL-12 and which is included as an input datum in the
current adjustment, data item $B37.4$.

The NPL final result is based on the initial data obtained from October
2006 to March 2007, data obtained during the first half of 2008, and
data obtained during the first half of 2009, the final period.  Many
variables were investigated to determine their possible influence on the
measured values of $K_{\rm J}^2R_{\rm K}$.  For example, several mass
standards with different masses and fabricated from different materials
were used during the course of the data taking. A comparison of the
uncertainty budgets for the 2007 data and the 2009 data shows
significant reductions in all categories, with the exception of the
calibration of the mass standards, resulting in the reduction of the
overall uncertainty from 66 parts in $10^9$ to 36 parts in $10^9$.

Nevertheless, during the week before the balance was to be dismantled, a
previously unrecognized possible systematic error in the weighing mode
of the experiment came to light. Although there was insufficient time to
derive a correction for the effect, Robinson obtained an uncertainty
estimate for it.  This additional uncertainty component, 197 parts in
$10^9$, when combined with the initially estimated overall uncertainty,
leads to the 200 parts in $10^9$ final uncertainty in
Eqs.~(\ref{eq:hwbnpl12}) and (\ref{eq:kj2rknpl12}).  Since the same
component applies to the initial Mark II result, its uncertainty is
increased from 66 parts in $10^9$ to $208$ parts in $10^9$.

Finally, there is a slight correlation between the final Mark II value
of $K_{\rm J}^2R_{\rm K}$, NPL-12, item $B37.4$ in
Table~\ref{tab:pdata}, and its 1990 predecessor, NPL-90, item $B37.1$ in
the same table.  Based on the paper by \citet{2012004}, the correlation
coefficient is 0.0025.

\subsubsection{$K^{2}_{\rm J}R_{\rm K}$ and $h$: METAS watt balance}
\label{sssec:metaswb11}
\vspace{.25 in}

The watt-balance experiment at the Federal Office of Metrology (METAS),
Bern-Wabern, Switzerland, was initiated in 1997, and progress reports
describing more than a decade of improvements and investigations of
possible systematic errors have been published and presented at
conferences \cite{1999087,2001039,2003041}. A detailed preprint giving
the final result of this effort,which is being continued with a new
apparatus, was provided to the Task Group by A. Eichenberger of METAS
prior to the 31 December 2010 closing date of the 2010 adjustment, and
was subsequently published by \citet{2011052}.  The METAS value for $h$
and the corresponding value for $K_{\rm J}^2R_{\rm K}$, identified as
METAS-11, input datum $B37.5$, are
\begin{eqnarray}
 h &=&  6.626\,0691(20)\times 10^{-34}  \ {\rm J \ s} 
 \quad [ 2.9\times 10^{-7}] \, ,
\qquad
\label{eq:hwbmetas11}
\end{eqnarray}
and
\begin{eqnarray}
K_{\rm J}^2R_{\rm K} &=& 6.036\,7617(18)\times 10^{33}  ~{\rm J}^{-1} \
{\rm s} ^{-1}\quad [ 2.9\times 10^{-7}] \, .  \qquad
\label{eq:kj2rkmetas11}
\end{eqnarray}

The METAS watt balance differs in a number of respects from those of
NIST and NPL.  For example, the METAS apparatus was designed to use a
100~g mass standard and a commercial mass comparator rather than a 1\,kg
standard and a specially designed and constructed balance in order to
reduce the size and complexity of the apparatus. Also, the velocity mode
was designed to be completely independent of the weighing mode. The use
of two separated measuring systems for the two modes in the same
apparatus make it possible to optimize each, but does require the
transfer of the coil between the two systems during the course of the
measurements.  Improvements in the apparatus over the last several years
of its operation focused on alignment, control of the coil position, and
reducing magnet hysteresis.

The METAS result is based on six sets of data acquired in 2010, each
containing at least 500 individual measurements which together represent
over 3400 hours of operation of the apparatus. The $7 \times 10^{-8}$
relative standard uncertainty of the mean of the means of the six data
sets is considered by \citet{2011052} to be a measure of the
reproducibility of the apparatus. The uncertainty budget from which the
$29 \times 10^{-8}$ relative uncertainty of the METAS value of $K_{\rm
J}^2R_{\rm K}$ is obtained contains nine components, but the dominant
contributions, totaling 20 parts in $10^8$, are associated with the
alignment of the apparatus.  \citet{2011052} point out that because of
the mechanical design of the current METAS watt balance, it is not
possible to reduce this source of uncertainty in a significant way.

\subsubsection{Inferred value of $K_{\rm J}$}
\label{sssec:kjfromkj2rk}
\vspace{.25 in}

As indicated in CODATA-06, a value of $K_{\rm J}$ with an uncertainty
significantly smaller than those of the two directly measured values
$B36.1$ and $B36.2$ can be obtained without assuming the validity of the
relations $K_{\rm J} = 2e/h$ and $R_{\rm K} = h/e^{2}$. Dividing the
weighted mean of the five directly measured watt-balance values of
$K^{2}_{\rm J}R_{\rm K}$, $B37.1$-$B37.5$, by the weighted mean of the
five directly measured calculable-capacitor values $R_{\rm K}$,
$B35.1$-$B35.5$, we have
\begin{eqnarray}
K_{\rm J} &=& K_{\rm J-90}
[1   -3.0(1.9)\times 10^{-8} ]
\nonumber\\
&=&  483\,597.8853(92) \  {\rm GHz/V}
 \quad  [ 1.9\times 10^{-8}] \, . \qquad
\label{eq:kjwbcc}
\end{eqnarray}
This result is consistent with the two directly measured values but has
an uncertainty that is smaller by more than an order of magnitude.
\vspace{.25 in}

\subsection{Josephson and quantum Hall effect relations}
\label{ssec:jqhrel}
\vspace{.25 in}

The theoretical and experimental evidence accumulated over the past 50
years for the Josephson effect and 30 years for the quantum Hall effect
that supports the exactness of the relations $K_{\rm J} = 2e/h$ and
$R_{\rm K} = h/e^{2}$ has been discussed in the three previous CODATA
reports and references cited therein. The vast majority of the
experimental evidence for both effects over the years comes from tests
of the universality of these relations; that is, their invariance with
experimental variables such as the material of which the Josephson
effect and quantum Hall effect devices are fabricated. However, in both
the 2002 and 2006 adjustments, the input data were used to test these
relations experimentally in an ``absolute'' sense, that is by
comparing the values of $2e/h$ and $h/e^2=\mu_0c/2\alpha$ implied by the
data assuming the relations are exact with those implied by the data
under the assumption that they are not exact. Indeed, such an analysis
is given in this report in Sec.~\ref{sssec:epstests}. Also briefly
discussed there is the ``metrology triangle.'' Here we discuss other
developments of interest that have occurred between the closing dates of
the 2006 and 2010 adjustments.

Noteworthy for the Josephson effect is the publication by
\citet{2009284} of ``A review of Josephson comparison results.'' These
authors examined a vast number of Josephson junction voltage comparisons
conducted over the past 30 years involving many different laboratories,
junction materials, types of junctions, operating frequencies, step
numbers, number of junctions in series, voltage level, and operating
temperature with some comparisons achieving a precision of a few parts
in $10^{11}$.  They find no evidence that the relation $K_{\rm J} =
2e/h$ is not universal.

There are three noteworthy developments for the quantum Hall effect.
First is the recent publication of a \emph{C. R. Physique} special issue
on the quantum Hall effect and metrology with a number of theoretical as
well as experimental papers that support the exactness of the relation
$R_{\rm K} = h/e^{2}$; see the Foreword to this issue by \citet{2011156}
and the papers contained therein, as well as the recent review article
by \citet{2011194}.

The second is the agreement found between the value of $R_{\rm K}$ in a
normal GaAs/AlGaAs heterostructure quantum Hall effect device and a
graphene (two dimensional graphite) device to within the 8.6 parts in
$10^{11}$ uncertainty of the experiment \cite{2011189}. This is an
extremely important result in support of the universality of the above
relation, because of the significant difference in the charge carriers
in graphene and the usual two dimensional semiconductor systems; see
\citet{2011161,2010154,2010038}.

The third is the theoretical paper by \citet{2009070}. This author's
calculations  appear to show that the relation $R_{\rm K}= h/e^2$ is not
exact but should be written as  $R_{\rm K}= (h/e^2)[1+C]$, where the
correction $C$ is due to vacuum polarization and is given by $C=
-(2/45)(\alpha/\rmpi)(B/B_0)^2$.  Here $B$ is the magnetic flux density
applied to the quantum Hall effect device and $B_0 =2{\rmpi}{c^2}{m^2_{\rm
e}}/he\approx 4.4\times 10^9$~T.  However, since $B$ is generally no
larger than 20~T, the correction, approximately $-2\times 10^{-21}$, is
vanishingly small and can be completely ignored.  Further,
\citet{2009070} argues that because of the topological nature of the
quantum Hall effect, there can be no other type of correction including
finite size effects.

\section{Measurements involving silicon crystals}
\label{sec:msc}

Experimental results obtained using nearly perfect single crystals of
natural silicon are discussed here, along with a new result for $N_{\rm
A}$ with a relative standard uncertainty of $3.0\times 10^{-8}$ obtained
using highly-enriched silicon.  For this material, ${x}(^{28}{\rm
Si})\approx0.999\,96$, compared to ${x}(^{28}{\rm Si})\approx0.92$, for
natural silicon, where $x(^A{\rm Si})$ is the amount-of-substance
fraction of the indicated isotope.

The new $N_{\rm A}$ result (see Sec.~\ref{ssec:naiac} below), as well as
much of the natural silicon data used in the current and previous CODATA
adjustments, were obtained as part of an extensive international effort
under way since the early 1990s to determine $N_{\rm A}$ with the
smallest possible uncertainty. This worldwide enterprise, which has many
participating laboratories and is called the International Avogadro
Coordination (IAC), carries out its work under the auspices of the
Consultative Committee for Mass and Related Quantities (CCM) of the
CIPM.

The eight natural silicon crystal samples of interest here are denoted
WASO~4.2a, WASO~04, WASO~17, NRLM3, NRLM4, MO*, ILL, and N, and the
\{220\} crystal lattice spacing of each, $d_{220}({\scriptstyle X})$, is
taken as an adjusted constant. For simplicity the shortened forms W4.2a,
W04, W17, NR3, and NR4 are used in quantity symbols for the first five
crystals. Note also that crystal labels actually denote the single
crystal ingot from which the crystal samples are taken, since no
distinction is made between different samples taken from the same ingot.

Silicon is a cubic crystal with $n=8$ atoms per face-centered cubic unit
cell of edge length (or lattice parameter) $a\approx 543~{\rm pm}$ with
\{220\} crystal lattice spacing $d_{220} = a/\sqrt{8} \approx 192~{\rm
pm}$. For practical purposes, it can be assumed that $a$, and thus
$d_{220}$, of an impurity free, crystallographically perfect or
``ideal'' silicon crystal at specified conditions of temperature $t$,
pressure $p$, and isotopic composition is an invariant of nature.  The
currently adopted reference conditions for natural silicon are
$t_{90}=\,\,$22.5~$^\circ{\rm C}$ and $p =0$ (vacuum), where $t_{90}$ is
Celsius temperature on the International Temperature Scale of 1990
(ITS-90). Reference values for $x(^A{\rm Si})$ have not been adopted,
because any variation of $d_{220}({\scriptstyle X})$ with the typical
isotopic composition variation observed for the natural silicon crystals
used is deemed negligible. To convert the lattice spacing
$d_{220}({\scriptstyle X})$ of a real crystal to the lattice spacing
$d_{220}$ of an ideal crystal requires the application of corrections
for impurities, mainly carbon, oxygen, and nitrogen.

Typical variation in the lattice spacing of different samples from the
same ingot is taken into account by including an additional relative
standard uncertainty component of $\sqrt{2}\times 10^{-8}$ for each
crystal in the uncertainty budget of any measurement result involving
one or more silicon lattice spacings. However, the component is
$(3/2)\sqrt{2}\times 10^{-8}$ in the case of crystal MO* because it is
known to contain a comparatively large amount of carbon. For simplicity,
we do not explicitly mention the inclusion of such components in the
following discussion.

\subsection{Measurements of $\bm{d_{220}({\scriptstyle X})}$ of
natural silicon}
\label{ssec:xroidx}

Measurements of $d_{220}({\scriptstyle X})$ are performed using a combined
x-ray and optical interferometer (XROI).  The interferometer has three
lamenae from a single crystal, one of which can be displaced and is
called the analyzer; see CODATA-98.  Also discussed there is the
measurement at PTB using an XROI with WASO~4.2a \cite{1981017}.  This
result, which was taken as an input datum in the past three adjustments,
is also used in the current adjustment; its value is
\begin{eqnarray}
d_{220}({\scriptstyle{\rm W4.2a}}) =  192\,015.563(12) \
{\rm fm} \quad [ 6.2\times 10^{-8}]\, , \qquad
\label{eq:dw42ptb81}
\end{eqnarray}
which is data item $B41.1$, labeled PTB-81, in Table~\ref{tab:pdata}.

The three other \{220\} natural silicon lattice spacings taken as input
data in the 2010 adjustment, determined at the Istituto Nazionale di
Ricerca Metrologica, (INRIM) Torino, Italy, using XROIs fabricated from
MO*, WASO~04, and WASO~4.2a, are much more recent results.
\citet{2008194} report
\begin{eqnarray}
d_{220}({\scriptstyle{\rm MO^*}})
&=&  192\,015.5508(42) \mbox{ fm} \quad [ 2.2\times 10^{-8}]\, ,
\label{eq:dmo4inrim08q}
\qquad
\end{eqnarray}
which is data item $B39$, labeled INRIM-08;
\citet{2009113} find
\begin{eqnarray} d_{220}({\scriptstyle{\rm WO4}})
&=&  192\,015.5702(29) \mbox{ fm} \quad [ 1.5\times 10^{-8}]\, ,
\qquad
\label{eq:dw04inrim09q}
\end{eqnarray}
which is data item $B40$, labeled INRIM-09; and
\citet{2009062} give
\begin{eqnarray} d_{220}({\scriptstyle{\rm W4.2a}})
&=&  192\,015.5691(29) \mbox{ fm} \quad [ 1.5\times 10^{-8}]\, ,
\qquad
\label{eq:dw42inrim09q}
\end{eqnarray}
which is data item $B41.2$, labeled INRIM-09.

The XROI used to obtain these three results is a new design with many
special features. The most significant advance over previous designs is
the capability to displace the analyzer by up to 5~cm. In the new
apparatus, laser interferometers and capacitive transducers sense
crystal displacement, parasitic rotations, and transverse motions, and
feedback loops provide positioning with picometer resolution, alignment
with nanometer resolution, and movement of the analyzer with nanometer
straightness. A number of fractional corrections for different effects,
such as laser wavelength, laser beam diffraction, laser beam alignment,
and temperature of the crystal, are applied in each determination; the
total correction for each of the three results, in parts in $10^9$, is
6.5, $-4.0$, and 3.7, respectively. The relative standard uncertainties
of the three lattice spacing measurements without the additional
uncertainty component for possible variation in the lattice spacing of
different samples from the same ingot, again in parts in $10^9$, are
6.1, 5.2, and 5.2.

The three INRIM lattice spacing values are correlated with one another,
as well as with the enriched silicon value of $N_{\rm A}$ discussed in
Sec.~\ref{ssec:naiac} below.  The latter correlation arises because the
\{220\} lattice spacing of the enriched silicon was determined at INRIM
by \citet{2011047} using the same XROI apparatus (relative standard
uncertainty of 3.5 parts in $10^9$ achieved).  The relevant correlation
coefficients for these data are given in Table~\ref{tab:pdcc} and
are calculated using information provided to the Task Group by
\citet{pc11gm}.

The many successful cross-checks of the performance of the new INRIM
combined x-ray and optical interferometer lend support to the
reliability of the results obtained with it. Indeed, \citet{2011039}
describe a highly successful test based on the comparison of the lattice
spacings of enriched and natural silicon determined using the new XROI.
Consequently, the IAC \cite{pc11gm}.  and the Task Group view the new
INRIM values for $d_{220}({\scriptstyle{\rm MO^*}})$ and
$d_{220}({\scriptstyle{\rm W04}})$ as superseding the earlier INRIM
values of these lattice spacings used in the 2006 adjustment.

\subsection{$\bm{d_{220}}$ difference measurements of natural silicon crystals}
\label{ssec:d220diff}
\vspace{.25 in}

Measurements of the fractional difference $\left[d_{220}({\scriptstyle
X}) - d_{220}({\rm ref})\right] /d_{220}({\rm ref})$ of the \{220\}
lattice spacing of a sample of a single crystal ingot $X$ and that of a
reference crystal ``ref'' enable the lattice spacings of crystals used
in various experiments to be related to one another.  Both NIST and PTB
have carried out such measurements, and the fractional differences from
these two laboratories that we take as input data in the 2010 adjustment
are data items $B42$ to $B53$ in Table~\ref{tab:pdata}, labeled
NIST-97, NIST-99, NIST-06, PTB-98, and PTB-03. Their relevant
correlation coefficients can be found in  Table~\ref{tab:pdcc}. For
details concerning the NIST and PTB difference measurements, see the
three previous CODATA reports. A discussion of item $B53$, the
fractional difference between the \{220\} lattice spacing of an ideal
natural silicon crystal $d_{220}$ and $d_{220}({\scriptstyle {\rm W04}})$,
is given in CODATA-06 following Eq.~(312).

\subsection{Gamma-ray determination of the
neutron relative atomic mass $\bm{A_{\rm r}({\rm n})}$}
\label{ssec:arn}

The value of $A_{\rm r}({\rm n})$ listed in Table~\ref{tab:rmass03} from
AME2003 is not used in the 2010 adjustment. Rather, $A_{\rm r}({\rm n})$
is obtained as described here so that the 2010 recommended value is
consistent with the current data on the \{220\} lattice spacing of
silicon.

The value of $A_{\rm r}$(n) is obtained from measurement of the
wavelength of the 2.2~MeV $\rmgamma$ ray in the reaction n + p
$\rightarrow$ d + $\rmgamma$.  The result obtained from Bragg-angle
measurements carried out at the high-flux reactor of ILL in a NIST and
ILL collaboration, is \cite{1999052}
\begin{eqnarray}
\frac{\lambda_{\rm meas}}{ d_{220}({\rm {\scriptstyle ILL}})}
=  0.002\,904\,302\,46(50) \qquad [ 1.7\times 10^{-7}] \ . \quad
\label{eq:ladill99}
\end{eqnarray}
Here $d_{220}({\rm {\scriptstyle ILL}})$ is the \{220\} lattice spacing
of the silicon crystals of the ILL GAMS4 spectrometer at $t_{90} =
22.5~^\circ$C and $p$ = 0 used in the measurements.  Relativistic
kinematics of the reaction yields the observational equation
\begin{eqnarray}
{\lambda_{\rm meas}\over d_{220}({\rm {\scriptstyle ILL}})}
&=& {\alpha^2 A_{\rm r}({\rm e})
\over R_\infty d_{220}({\rm {\scriptstyle ILL}})}
{A_{\rm r}({\rm n}) + A_{\rm r}({\rm p})
\over \left[A_{\rm r}({\rm n}) + A_{\rm r}({\rm p})\right]^2
- A_{\rm r}^2({\rm d})} \ ,
\nonumber\\
\label{eq:relkingarm}
\end{eqnarray}
where the quantities on the right-hand side are adjusted constants.

\subsection{Historic X-ray units}
\label{ssec:xru}

Units used in the past to express the wavelengths of
x-ray lines are the copper ${\rm K\rmalpha_1}$ x unit, symbol ${\rm
xu(CuK\rmalpha_1)}$, the molybdenum K${\rm \rmalpha}_1$ x unit, symbol
${\rm xu(MoK\rmalpha_1)}$, and the \aa ngstrom star, symbol \AA$^*$.
They are defined by assigning an exact, conventional value to the
wavelength of the ${\rm CuK\rmalpha_1}$, ${\rm MoK\rmalpha_1}$, and
${\rm WK\rmalpha_1}$ x-ray lines when each is expressed in its
corresponding unit:

\begin{eqnarray}
\lambda({\rm CuK\rmalpha_1}) &=& 1\,537.400 \
{\rm xu(CuK\rmalpha_1)} \, , \qquad\label{eq:cuxudef} \\[5 pt]
\lambda({\rm MoK\rmalpha_1}) &=& 707.831 \
{\rm xu(MoK\rmalpha_1)}  \, , \label{eq:moxudef} \\[5 pt]
\lambda({\rm WK\rmalpha_1}) &=& 0.209\,010\,0 \
{\rm \AA^*} \, . \label{eq:wxudef}
\end{eqnarray}

The data relevant to these units are (see CODATA-98)
\begin{eqnarray}
{\lambda({\rm CuK\rmalpha_1})\over d_{220}({\rm {\scriptstyle W4.2a}})} &=&
 0.802\,327\,11(24) \quad [ 3.0\times 10^{-7}]
\, ,
\label{eq:lcusdw42aptb}
\\
{\lambda({\rm WK\rmalpha_1})\over d_{220}({\rm {\scriptstyle N}})} &=&
 0.108\,852\,175(98) \quad [ 9.0\times 10^{-7}]
\, ,
\label{eq:lamnsdwnist}
\\
{\lambda({\rm MoK\rmalpha_1})\over d_{220}({\rm {\scriptstyle N}})} &=&
 0.369\,406\,04(19) \quad [ 5.3\times 10^{-7}]
\, ,
\label{eq:lmodn73}
\\
{\lambda({\rm CuK\rmalpha_1})\over d_{220}({\rm {\scriptstyle N}})} &=&
 0.802\,328\,04(77) \quad [ 9.6\times 10^{-7}]
\, , \qquad
\label{eq:lcudn73}
\end{eqnarray}
where ${d_{220}({\rm {\scriptstyle W4.2a}})}$ and ${d_{220}({\rm
{\scriptstyle N}})}$ denote the \{220\} lattice spacings, at the
standard reference conditions $p=0$ and $t_{90}=22.5~^\circ{\rm C}$, of
particular silicon crystals used in the measurements.  The result in
Eq.~(\ref{eq:lcusdw42aptb}) is from a collaboration between researchers
from Friedrich-Schiller University (FSUJ), Jena, Germany and the PTB
\cite{1991096}.

To obtain recommended values for ${\rm xu(CuK\rmalpha_1)}$, ${\rm
xu(MoK\rmalpha_1)}$, and ${\rm \AA^*}$, we take these units to be
adjusted constants.  The observational equations for the data of
Eqs.~(\ref{eq:lcusdw42aptb}) to (\ref{eq:lcudn73}) are
\begin{eqnarray}
\frac{\lambda({\rm CuK\rmalpha_1})}{d_{220}({\rm {\scriptstyle{W4.2a}})}} &=&
\frac{\rm 1\,537.400 ~xu(CuK\rmalpha_1)}{d_{220}({\rm {\scriptstyle{W4.2a}})}}
\, ,
\label{eq:lcud42a79obeq} \\
\frac{\lambda({\rm WK\rmalpha_1})}{d_{220}({\rm {\scriptstyle N}})} &=&
\frac{\rm 0.209\,010\,0 ~\AA^*}{d_{220}({\rm {\scriptstyle N}})}
\, ,
\label{eq:lwdn79obeq} \\
\frac{\lambda({\rm MoK\rmalpha_1})}{d_{220}({\rm {\scriptstyle N}})} &=&
\frac{\rm 707.831 ~xu(MoK\rmalpha_1)}{d_{220}({\rm {\scriptstyle N}})}
\, ,
\label{eq:lmodn73obeq} \\
\frac{\lambda({\rm CuK\rmalpha_1})}{d_{220}({\rm {\scriptstyle N}})} &=&
\frac{\rm 1\,537.400 ~xu(CuK\rmalpha_1)}{d_{220}({\rm {\scriptstyle N}})}
\, , \qquad
\label{eq:lcudn73obeq}
\end{eqnarray}
where $d_{220}({\rm {\scriptstyle N}})$ is taken to be an adjusted constant
and $d_{220}({\rm {\scriptstyle W17}})$ and $d_{220}({\rm {\scriptstyle{W4.2a}}}
  )$
are adjusted constants as well.

\subsection{Other data involving natural silicon crystals}
\label{ssec:othsidat}

Two input data used in the 2006 adjustment but not used in the 2010
adjustment at the request of the IAC \cite{pc10kf} are discussed in this
section.

The first is the NMIJ value of $d_{220}({\scriptstyle{\rm NR3}})$, the
\{220\} lattice spacing reported by \citet{2004001}.  The IAC formally
requested that the Task Group not consider this result for the 2010
adjustment, because of its questionable reliability due to
the problems discussed in Sec.~VIII.A.1.\emph{b} of CODATA-06.

The second is the molar volume of natural silicon $V_{\rm m}({\rm
Si})$ from which $N_{\rm A}$ can be determined.  The value used
in the 2006 adjustment is \cite{2005032}
$12.058\,8254(34)\times 10^{-6}~{\rm m^3~mol^{-1}}~~[2.8\times
10^{-7}]$.  The IAC requested that the Task Group no longer consider
this result, because of problems uncovered with the molar mass
measurements of natural silicon $M({\rm Si})$ at the Institute for
Reference Materials and Measurements (IRMM), Geel, Belgium.

One problem is associated with the experimental determination of the
calibration factors of the mass spectrometer used to measure the
amount-of-substance ratios (see following section) of the silicon
isotopes $^{28}{\rm Si}$, $^{29}{\rm Si}$, and $^{30}{\rm Si}$ in
various silicon crystals, as discussed by \citet{2011041}. The factors
are critical, because molar masses are calculated from these ratios and
the comparatively well-known relative atomic masses of the isotopes.
Another problem is the unexplained large scatter of $\pm 7$ parts in
$10^7$ in molar mass values among crystals taken from the same ingot, as
discussed by  \citet{2005032} in connection with their result for
$V_{\rm m}({\rm Si})$ given above.

More specifically, from 1994 to 2005 IRMM measured the molar masses of
natural silicon in terms of the molar mass of WASO17.2, which was
determined using the now suspect calibration factors \cite{2011041}.
Based on a new determination of the calibration factors, \citet{2011041}
report a value for the molar mass of WASO17.2 that has a relative
standard uncertainty of $2.4\times 10^{-7}$, compared to the $1.3\times
10^{-7}$ uncertainty of the value used since 1994, and which is
fractionally larger by $1.34 \times 10^{-6}$ than the earlier value.
(The recent paper by \citet{2012002} also points to a correction of the
same general magnitude.) This new result and the data and calculations
in \citet{2005032} yield the following revised value for the molar
volume of natural silicon:
\begin{eqnarray}
V_{\rm m}({\rm Si}) &=& 12.058\,8416(45)\times
10^{-6}~{\rm m^3~mol^{-1}}
\nonumber\\&&\qquad\qquad\qquad[3.7\times 10^{-7}]\, .
\qquad
\label{eq:vmsiliac11}
\end{eqnarray}

Although the IAC does not consider this result to be sufficiently
reliable for the Task Group to consider it for inclusion in the 2010
adjustment, we note that based on the 2010 recommended values of
$d_{220}$ and the molar Planck constant $N_{\rm A}h$,
Eq.~(\ref{eq:vmsiliac11}) implies
\begin{eqnarray}
N_{\rm A}
&=& 6.022\,1456(23)\times 10^{23}~{\rm mol^{-1}}
\quad[3.8\times 10^{-7}]
\, ,
\label{eq:navmsiliac11}
\nonumber\\
h &=& 6.626\,0649(25)\times 10^{-34}~{\rm J~s}~~[3.8\times 10^{-7}]\, .
\quad
\label{eq:hvmsiliac11}
\qquad
\end{eqnarray}
The difference between this value of $N_{\rm A}$ and the value with
relative standard uncertainty $3.0 \times 10^{-8}$ obtained from
enriched silicon discussed in the next section is $7.9(3.8)$ parts in
$10^7$, while the difference between the NIST 2007 watt-balance value of
$h$ with uncertainty $3.6 \times 10^{-8}$ and this value of $h$ is
$6.1(3.8)$ parts in $10^7$.  \vspace{.25 in}

\subsection{Determination of $\bm{N_{\rm A}}$ with enriched silicon}
\label{ssec:naiac}
\vspace{.25 in}

The IAC project to determine $N_{\rm A}$ using the XRCD method and
silicon crystals highly enriched with $^{28}{\rm Si}$ was formally
initiated in 2004, but its origin dates back two decades earlier. Its
initial result is discussed in detail in a \emph{Metrologia}
special issue; see the Foreword by \citet{2011036}, the 14 technical
papers in the issue, and the references cited therein.  The first paper,
by \citet{2011035}, provides an extensive overview of the entire
project.  The value  of the Avogadro constant obtained from this unique
international collaborative effort, identified as IAC-11, input datum
$B54$, is \cite{2011035}
\begin{eqnarray}
N_{\rm A} &=&  6.022\,140\,82(18)\times 10^{23} ~{\rm mol^{-1}}
\quad [ 3.0\times 10^{-8}] \, .
  \qquad
\label{eq:naiac11}
\end{eqnarray}
Note that this result differs slightly from the somewhat earlier result
reported by \citet{2011005} but is the preferred value \cite{pc11hb}.

The basic equation for the XRCD determination of $N_{\rm A}$ has been
discussed in previous CODATA reports. In brief,
\begin{eqnarray}
N_{\rm A} &=&
{A_{\rm r}({\rm Si})M_{\rm u}\over\sqrt{8}\,d_{220}^{\,3}\,\rho{\rm (Si)}} \ ,
\label{eq:naardrho}
\end{eqnarray}
which would apply to an impurity free, crystallographically perfect,
``ideal'' silicon crystal.  Here $A_{\rm r}$(Si) is the mean relative
atomic mass of the silicon atoms in such a crystal, and $\rho$(Si) is
the crystal's macroscopic mass density. Thus, to determine $N_{\rm A}$
from Eq.~(\ref{eq:naardrho}) requires determining the density
$\rho$(Si), the \{220\} lattice spacing $d_{220}$, and the
amount-of-substance ratios $R_{29/28}=n(^{29}\rm Si)$/$n(^{28}\rm Si)$
and $R_{30/28}= n(^{30}\rm Si)$/$n(^{28}\rm Si)$ so that $A_{\rm r}(\rm
Si)$ can be calculated using the well-known values of $A_{\rm r}(^{A}\rm
Si)$.  Equally important is the characterization of the material
properties of the crystals used, for example, impurity content,
non-impurity point defects, dislocations, and microscopic voids must be
considered.

The international effort to determine the Avogadro constant, as
described in the \emph{Metrologia} special issue, involved many tasks
including the following: enrichment and poly-crystal growth of silicon
in the Russian Federation; growth and purification of a 5 kg single
silicon crystal ingot in Germany; measurement of the isotopic
composition of the crystals at PTB; measurement of the lattice spacing
with the newly developed XROI described above at INRIM; grinding and
polishing of two spheres cut from the ingot to nearly perfect spherical
shape at NMI; optical interferometric measurement of the diameters
of the spheres at PTB and NMIJ; measurement of the masses of the spheres
in vacuum at PTB, NMIJ, and BIPM; and characterization of and correction
for the effect of the contaminants on the surfaces of the spheres at
various laboratories.

The uncertainty budget for the IAC value of $N_{\rm A}$ is dominated by
components associated with determining the volumes and the surface
properties of the spheres, followed by those related to measuring their
lattice spacings and their molar masses. These four components, in parts
in $10^9$, are 29, 15, 11, and 8 for the sphere designated AVO28-S5.

How this result compares with other data and its role in the 2010
adjustment is discussed in Sec.~\ref{sec:ad}.

\section{Thermal physical quantities}
\label{sec:tpq}

Table~\ref{tab:thermal} summarizes the eight results for the thermal
physical quantities $R$, $k$, and $k/h$, the molar gas constant, the
Boltzmann constant, and the quotient of the Boltzmann and Planck
constants, respectively, that are taken as input data in the 2010
adjustment.  They are data items $B58.1$ to $B60$ in
Table~\ref{tab:pdata} with  correlation coefficients as given in
Table~\ref{tab:pdcc} and observational equations as given in
Table~\ref{tab:pobseqsb2}.  Values of $k$ that can be inferred from
these data are given in Table~\ref{tab:k} and are graphically compared
in Fig.~\ref{fig:k}. The first two results, the NPL 1979 and NIST 1988
values of $R$, were included in the three previous CODATA adjustments,
but the other six became available during the 4 years between the
closing dates of the 2006 and 2010 adjustments.  (Note that not every
result in Table~\ref{tab:thermal} appears in the cited reference. For
some, additional digits have been provided to the Task Group to reduce
rounding errors; for others, the value of $R$ or $k$ actually determined
in the experiment is  recovered from the reported result using the
relation $R=kN_{\rm A}$ and the value of $N_{\rm A}$ used by the
researchers to obtain that result.) \vspace{.25 in}

\subsection{Acoustic gas thermometry}
\label{ssec:agt}
\vspace{.25 in}

As discussed in CODATA-98 and the references cited therein, measurement
of $R$ by the method of acoustic gas thermometry (AGT) is based on the
following expressions for the square of the speed of sound in a real gas
of atoms or molecules in thermal equilibrium at thermodynamic
temperature $T$ and pressure $p$ and occupying a volume $V$:
\begin{eqnarray}
c_{\rm a}^2{(T,p)} &=& A_0(T)+A_1(T)p
\nonumber\\&&+A_2(T)p^2+ A_3(T)p^3+\cdots\, . \qquad
\label{eq:ca2pt}
\end{eqnarray}

Here $A_1(T)$ is the first acoustic virial coefficient, $A_2(T)$ is the
second, etc.  In the limit $p \rightarrow0$, this becomes
\begin{eqnarray}
c_{\rm a}^2{(T,0)} = A_{0}(T)= {\gamma_{0}RT\over A_{\rm r}(X)M_{\rm u}}
\, ,
\label{eq:ca2p0t}
\end{eqnarray}
\noindent
where $\gamma_{0}=c_p/c_V$ is the ratio of the specific heat capacity of
the gas at constant pressure to that at constant volume and is 5/3 for
an ideal monotonic gas. The basic experimental approach to determining
the speed of sound of a gas, usually argon or helium, is to measure the
acoustic resonant frequencies of a cavity at or near the triple point of
water, $T_{\rm TPW}=273.16~{\rm K}$, and at various pressures and
extrapolating to $p=0$.  The cavities are either cylindrical of fixed or
variable length, or spherical, but most commonly quasispherical in the
form of a triaxial ellipsoid. This shape removes the degeneracy of the
microwave resonances used to measure the volume of the resonator in
order to calculate $c_{\rm a}^2(T,p)$ from the measured acoustic
frequencies and the corresponding acoustic resonator eigenvalues known
from theory.  The cavities are formed by carefully joining hemispherical
cavities.

In practice, the determination of $R$ by AGT with a relative standard
uncertainty of order one part in $10^6$ is complex; the application of
numerous corrections is required as well as the investigation of many
possible sources of error.  For a review of the advances made in AGT in
the past 20 years, see \citet{2009302}.

\subsubsection{NPL 1979 and NIST 1988 values of $\bm R$}
\label{sssec:NP79NI88}

Both the NPL and NIST experiments are discussed in detail in CODATA-98.
We only note here that the NPL measurement used argon in a vertical,
variable-path-length, 30~mm inner diameter cylindrical acoustic
resonator operated at a fixed frequency, and the displacement of the
acoustic reflector that formed the top of the resonator was measured
using optical interferometry.  The NIST experiment also used argon, and
the volume of the stainless steel spherical acoustic resonator, of
approximate inside diameter 180~mm, was determined from the mass of
mercury of known density required to fill it.  The 1986 CODATA
recommended value of $R$ is the NPL result while the 1998, 2002, and
2006 CODATA recommended values are the weighted means of the NPL and
NIST results.

\subsubsection{LNE 2009 and 2011 values of $\bm R$}
\label{sssec:LN0911}

\citet{2009333,pc11lp} obtained the LNE 2009 result using a copper
quasisphere of about 100~mm inner diameter and helium gas. The principal
advantage of helium is that its thermophysical properties are well-known
based on ab initio theoretical calculations; the principal disadvantage
is that because of its comparatively low mass, impurities have a larger
effect on the speed of sound. This problem is mitigated by passing the
helium gas through a liquid helium trap and having a continuous flow of
helium through the resonator, thereby reducing the effect of outgassing
from the walls of the resonator.  In calculating the molar mass of the
helium \citet{2009333} assumed that the only remaining impurity is
$^3$He and that the ratio of $^3$He to $^4$He is less than
$1.3\times10^{-6}$.

The critically important volume of the resonator was determined from
measurements of its electromagnetic (EM) resonances together with
relevant theory of the eigenvalues.  The dimensions of the
quasihemispheres were also measured using a coordinate measuring machine
(CMM). The volumes so obtained agreed, but the $17 \times 10^{-6}$
relative standard uncertainty of the CMM determination far exceeded the
$0.85 \times 10^{-6}$ relative uncertainty of the EM determination. The
principal uncertainty components that contribute to the 2.7 parts in
$10^6$ uncertainty of the final result are, in parts in $10^6$, 1.8,
1.0, 1.5, and 0.8 due, respectively, to measurement of the volume of the
quasisphere (including various corrections), its temperature relative to
$T_{\rm TPW}$, extrapolation of $c_{\rm a}^2(T_{\rm TPW},p)$ to $p=0$,
and the reproducibility of the result, based on two runs using different
purities of helium and different acoustic transducers \cite{pc11lp}.

The 2011 LNE result for $R$, which has the smallest uncertainty of any
reported to date, is described in great detail by \citet{2011207}. It
was obtained using the same quasispherical resonator employed in the
2009 experiment, but with argon in place of helium.  The reduction in
uncertainty by more than a factor of two  was achieved by improving all
aspects of the experiment \cite{2011207}. The volume of the resonator
was again determined from measurements of its EM resonances and cross
checked with CMM dimensional measurements of the quasispheres carried
out at NPL \cite{2010158}.  As usual in AGT, the square of the
speed-of-sound was determined from measurements of the quasisphere's
acoustic resonant frequencies at different pressures (50 kPa to 700 kPa
in this case) and extrapolation to $p=0$.  The isotopic composition of
the argon and its impurity content was determined at IRMM
\cite{2010064}.

The five uncertainty components of the final 1.24 parts in $10^6$
uncertainty of the result, with each component itself being composed of
a number of subcomponents, are, in parts in $10^6$, the following: 0.30
from temperature measurements (the nominal temperature of the
quasisphere was $T_{\rm TPW})$; 0.57 from the EM measurement of the
quasisphere's volume; 0.84 from the determination of $c_{\rm a}^2(T_{\rm
TPW},0)$; 0.60 associated with the argon molar mass and its impurities;
and 0.25 for experimental repeatability based on the results from two
series of measurements carried out in May and July of 2009.

Because the LNE 2009 and 2011 results are from experiments in which some
of the  equipment and measuring techniques are the same or similar, they
are correlated. Indeed, for the same reason, there are non-negligible
correlations among the four recent AGT determinations of $R$, that is,
LNE-09, NPL-10, INRIM-10, and LNE-11. These correlations are given in
Table~\ref{tab:pdcc} and have been calculated using information provided
to the Task Group by researchers involved in the experiments
\cite{pc11gpp}.

\subsubsection{NPL 2010 value of \bm $R$}
\label{sssec:NP10}

This result was obtained at NPL by \citet{2010177,pc11mp} at $T_{\rm
TPW}$ using a thin-walled copper quasispherical resonator of about
100~mm inner diameter on loan from LNE and argon as the working gas. The
internal surfaces of the quasihemispheres were machined using diamond
turning techniques.  The 5 mm wall thickness of the quasisphere, about
one-half that of the usual AGT resonators, was specially chosen to allow
improved study of the effect of resonator shell vibrations on acoustic
resonances. The volume of the quasisphere was determined from
measurements of EM resonances and checked with CMM dimensional
measurements of the quasihemispheres before assembly \cite{2010158}. Two
series of measurements were carried out, each lasting several days: one
with the quasisphere rigidly attached to a fixed stainless steel post
and one with it freely suspended by three wires attached to its equator.
Pressures ranged from 50~kPa to 650~kPa and were measured with
commercial pressure meters. The isotopic composition of the argon and
its impurity content were again determined at IRMM \cite{2010064}.

The final result is the average of the value obtained from each run. The
3.78 parts in $10^6$ difference between the molar mass of the argon used
in the fixed and hanging quasisphere runs is to a large extent canceled
by the $-2.77$ parts in $10^6$ difference between the values of $c_{\rm
a}^2(T_{\rm TPW},0)$ for the two runs, so the two values of $R$ agree
within 1.01 parts in $10^6$. The largest uncertainty components in parts
in $10^6$ contributing to the final uncertainty of 3.1 parts in $10^6$
are, respectively \cite{2010177,2010158}, 2, 1.1, 0.9, 1, and 1.4
arising from the difference between the acoustic and microwave volumes
of the resonator, temperature calibration, temperature measurement,
argon gas impurities, and correction for the layer of gas near the wall
of the resonator (thermal boundary layer correction).

\subsubsection{INRIM 2010 value of \bm $R$}
\label{sssec:IN10}

The INRIM determination of $R$ by \citet{2010080,pc11rg} employed a
stainless steel spherical resonator of about 182~mm inner diameter and
non-flowing helium gas. Although the measurements were performed with
the resonator very near $T_{\rm TPW}$ as in the other AGT
molar-gas-constant determinations, two important aspects of the INRIM
experiment are quite different.  First, the speed of sound was measured
at only one pressure, namely, 410~kPa, and the extrapolation to $p=0$
was implemented using the comparatively well-known theoretical values of
the required $^4{\rm He}$ equation-of-state and acoustic virial
coefficients.  Second, the radius of the resonator was determined using
the theoretical value of the $^4{\rm He}$ index of refraction together
with eight measured EM resonance frequencies and the corresponding
predicted eigenvalues. The speed of sound was then calculated from this
value of the radius and measured acoustic resonant frequencies.
\citet{2010080} calculated the molar mass of their He sample assuming
the known atmospheric abundance of $^3$He represents an upper limit.

The two uncertainty components that are by far the largest contributors
to the 7.5 parts in $10^6$ final uncertainty of the experiment are, in
parts in $10^6$, 4.2 from fitting the shape of the eight measured
microwave modes and 4.8 from the scatter of the squared frequencies of
the six measured radial acoustic modes used to determine $c_{\rm
a}^2(T_{\rm TPW},p =410~{\rm kPa})$.

\subsection{Boltzmann constant $\bm{k}$ and quotient $\bm{k/h}$}
\label{ssec:bckoh}

The following two sections discuss the two NIST experiments that have
yielded the last two entries of Table~\ref{tab:thermal}.

\subsubsection{NIST 2007 value of \bm $k$}
\label{sssec:NI07}

This result was obtained by \citet{2007117} using the technique of
refractive index gas thermometry (RIGT), an approach similar to that of
dielectric constant gas thermometry (DCGT) discussed in CODATA-98, and
to a lesser extent in CODATA-02 and CODATA-06. The starting point of
both DCGT and RIGT is the virial expansion of the equation of state for
a real gas of amount of substance $n$ in a volume $V$ \cite{2007117},
\begin{eqnarray}
p =\rho RT\left[1+\rho b(T)+\rho^2c(T)+\rho^3d(T)+\cdots \right],
\label{eq:vpvnrt}
\end{eqnarray}
where $\rho = n/V$ is the amount of substance density of the gas at
thermodynamic temperature $T$, and $b(T)$ is the first virial
coefficient, $c(T)$ is the second, etc.; and the Clausius-Mossotti
equation
\begin{eqnarray}
{\epsilon_{\rm r}- 1\over \epsilon_{\rm r}+ 2}&=&\rho A_{\epsilon}
\bigg[1+\rho B_{\epsilon}(T)
\nonumber\\&&\qquad+\rho^2C_{\epsilon}(T)+\rho^3D_{\epsilon}(T)+\cdots \bigg],
\qquad
\label{eq:cmeq}
\end{eqnarray}
where $\epsilon_{\rm r} = \epsilon/\epsilon_0$ is the relative
dielectric constant (relative permittivity) of the gas, $\epsilon$ is
its dielectric constant, $\epsilon_{0}$ is the exactly known electric
constant, $A_{\epsilon}$ is the molar polarizability of the atoms, and
$B_{\epsilon}(T)$, $C_{\epsilon}(T)$, etc. are the dielectric virial
coefficients. The static electric polarizability of a gas atom
$\alpha_0$, $A_{\epsilon}$, $R$, and $k$ are related by $A_{\epsilon}/R
= \alpha_0/3{\epsilon_0}k$, which shows that if $\alpha_0$ is known
sufficiently well from theory, which it currently is for $^4{\rm He}$
\cite{2004107,2011233,2011065}, then a competitive value of $k$ can be
obtained if the quotient $A_{\epsilon}/R$ can be measured with a
sufficiently small uncertainty.

In fact, by appropriately combining Eqs.~(\ref{eq:vpvnrt}) and
(\ref{eq:cmeq}), an expression is obtained from which $A_{\epsilon}/R$
can be experimentally determined by measuring $\epsilon_{\rm r}$ at a
known constant temperature such as $T_{\rm TPW}$ and at different pressures
and extrapolating to zero pressure.  This is done in practice by
measuring the fractional change in capacitance of a specially
constructed capacitor, first without helium gas and then with helium gas
at a known pressure.  This is the DCGT technique.

In the RIGT technique of \citet{2007117}, $A_{\epsilon}/R$ is
determined, and hence $k$, from measurements of $n^2(T,p)\equiv
\epsilon_{\rm r}\mu _{\rm r}$ of a gas of helium, where $n(T,p)$ is the
index of refraction of the  gas, $\mu_{\rm r} = \mu/\mu_0$ is the
relative magnetic permeability of the gas, $\mu$ is its magnetic
permeability, and $\mu_{0}$ is the exactly known magnetic constant.
Because $^4{\rm He}$ is slightly diamagnetic, the quantity actually
determined is $(A_{\epsilon}+A_{\mu})/R$, where
$A_{\mu}=4\rmpi\chi_{0}/3$ and $\chi_{0}$ is the diamagnetic
susceptibility of a $^4{\rm He}$ atom. The latter quantity is known from
theory and the theoretical value of $A_{\mu}$ was used to obtain
$A_{\epsilon}/R$ from the determined quantity.

\citet{2007117} obtained $n(T,p)$ by measuring the microwave resonant
frequencies from  2.7~GHz to 7.6~GHz of a quasispherical copper plated
resonator, either evacuated or filled with He at pressures of 0.1~MPa to
6.3~MPa.  The temperature of the resonator was within a few millikelvin
of $T_{\rm TPW}$. A network analyzer was used to measure the resonant
frequencies and a calibrated pressure balance to measure $p$. The
extrapolation to $p = 0$ employed  both theoretical and experimental
values of the virial coefficients $B, C, D, b$, and $c$ taken from the
literature. The uncertainties of these coefficients and of the pressure
and temperature measurements, and the uncertainty of the isothermal
compressibility of the resonator, are the largest components in the
uncertainty budget.

\subsubsection{NIST 2011 value of \bm $k/h$}
\label{sssec:NI11}

As discussed in CODATA-98, the Nyquist theorem predicts, with a
fractional error of less than one part in $10^6$ at frequencies less
than 10~MHz and temperatures greater than 250~K, that
\begin{eqnarray}
\langle U^2 \rangle = 4kTR_{\rm s}\Delta f \, .
\label{eq:nyt}
\end{eqnarray}
Here $\langle U^2 \rangle$ is the mean-square-voltage, or Johnson noise
voltage, in a measurement bandwidth of frequency $\Delta f$ across the
terminals of a resistor of resistance $R_{\rm s}$ in thermal equilibrium
at thermodynamic temperature $T$.  If $\langle U^2 \rangle$ is measured
in terms of the Josephson constant $K_{\rm J} = 2e/h$ and $R_{\rm s}$ in
terms of the von Klitzing constant $R_{\rm K} = h/e^2$, then this
experiment yields a value of $k/h$.

Such an experiment has been carried out at NIST, yielding the result in
Table~\ref{tab:thermal}; see the paper by \citet{2011051} and references
therein.  In that work, digitally synthesized pseudo-noise voltages are
generated by means of a pulse-biased Josephson junction array. These
known voltages are compared to the unknown thermal-noise voltages
generated by a specially designed 100~$\Omega$ resistor in a well
regulated thermal cell at or near $T_{\rm TPW}$. Since the spectral
density of the noise voltage of a 100~$\Omega$ resistor at 273.16~K is
only 1.23~nV$\sqrt{\rm Hz}\,$, it is measured using a low-noise,
two-channel, cross-correlation technique that enables the resistor
signal to be extracted from uncorrelated amplifier noise of comparable
amplitude and spectral density. The bandwidths range from 10~kHz to
650~kHz. The final result is based on two data runs, each of about 117
hours duration, separated in time by about three months.

The dominant uncertainty component of the 12.1 parts in $10^6$ total
uncertainty is the 12.0 parts in $10^6$ component due to the measurement
of the ratio $\langle V^{2}_{\rm R}/ V^{2}_{\rm Q} \rangle$, where
$V_{\rm R}$ is the resistor noise voltage and $V_{\rm Q}$ is the
synthesized voltage. The main uncertainty component contributing to the
uncertainty of the ratio is 10.4 parts in $10^6$ due to spectral
aberrations, that is, effects that lead to variations of the ratio with
bandwidth.

\subsection{Other data}
\label{ssec:rkod}

We note for completeness the following three results, each of which
agrees with its corresponding 2010 recommended value. The first has a
non-competitive uncertainty but is of interest because it is obtained
from a relatively new method that could yield a value with a competitive
uncertainty in the future.  The other two became available only after
the 31 December 2010 closing date of the 2010 adjustment.

\citet{2011166} find $R=8.314\,80(42)~{\rm J~m^{-1}~K^{-1}}~[50 \times
10^{-6}]$ determined by the method of Doppler spectroscopy, in
particular, by measuring near the ice point $T=273.15~{\rm K}$ the
absorption profile of a rovibrational line at $\nu= 30~{\rm THz}$ of
ammonia molecules in an ammonium gas in thermal equilibrium. The width
of the line is mainly determined by the Doppler width due to the
velocity distribution of the $^4{\rm NH}_3$ molecules along the
direction of the incident laser beam. The relevant expression is
\begin{eqnarray}
{\Delta\omega_{\rm D}\over\omega _{0}}= \left({2kT\over
m(^4{\rm NH}_3)c^2}\right)^{1/2}= \left({2RT\over A_{\rm r}(^4{\rm
NH}_3)M_{\rm u}c^2}\right)^{1/2},
\nonumber\\
\label{eq:dwovwkr}
\end{eqnarray}
where $\Delta\omega_{\rm D}$ is the e-fold angular frequency half-width
of the Doppler profile of the ammonium line at temperature $T$,
$\omega_0$ is its angular frequency, and $m(^4{\rm NH}_3)$ and $A_{\rm
r}(^4{\rm NH}_3)$ are the mass and relative atomic mass of the ammonium
molecule.

\citet{2011165,pc11jz} obtain $R=8.314\,474(66)~{\rm J~m^{-1}~K^{-1}}
~[7.9 \times 10^{-6}]$ using acoustic gas thermometry with argon gas,
more specifically, by measuring resonant frequencies of a
fixed-path-length cylindrical acoustic resonator at $T_{\rm TPW}$; its
approximate 129 mm length is measured by two-color optical
interferometry.

\citet{2012003,2011183} give $k=1.380\,655(11) \times 10^{-23}~{\rm
J/K}~[7.9 \times 10^{-6}]$ measured using dielectric gas thermometry
(see Sec.~\ref{sssec:NI07} above) and helium gas at $T_{\rm TPW}$ and
also at  temperatures in the range 21~K to 27~K surrounding the
triple point of neon at $T\approx25~{\rm K}$.

\subsection{Stefan-Boltzmann constant \boldmath $\sigma$}
\label{sssec:sfc}

The Stefan-Boltzmann constant is related to $c$, $h$, and  $k$ by
$\sigma = 2\rmpi^5 k^4/{15h^3 c^2}$, which, with the aid of the
relations $k = R/N_{\rm A}$ and $N_{\rm A}h = cA_{\rm r}({\rm e})M_{\rm
u}\alpha^2/2R_\infty$, can be expressed in terms of the molar gas
constant and other adjusted constants as

\begin{eqnarray}
\sigma = \frac{32\rmpi^5 h}{15c^6} \left(\frac{R_\infty R}{A_{\rm r}({\rm e})
M_{\rm u} \alpha^2}\right)^4  .
\label{eq:shcr4}
\end{eqnarray}

Since no competitive directly measured value of $\sigma$ is available
for the 2010 adjustment, the 2010 recommended value is obtained from
this equation.

\def\vsp{\vbox to 8 pt{}}
\def\hsp{\hbox to 15 pt{}}
\begin{table*}[t]
\caption{
Summary of thermal physical measurements relevant to the 2010 adjustment
(see text for details). AGT: acoustic gas thermometry; RIGT: refractive
index gas thermometry; JNT: Johnson noise thermometry; cylindrical,
spherical, quasispherical: shape of resonator used; JE and QHE: Josephson
effect voltage and quantum Hall effect resistance standards.
}
\label{tab:thermal}
\begin{tabular}{l@{\hsp}cc@{\hsp}l@{\hsp}l@{\hsp}l}
\toprule
\vbox to 10 pt {}
Source & Ident.$^\text{a}$ & Quant. & Method  & Value   & Rel. stand. \\
       &                     &        & & & uncert $u_{\rm r}$ \\
\noalign{\vbox to 2 pt {}}
\colrule
\noalign{\vbox to 5 pt {}}

\vsp  \citet{1979024}  & NPL-79  & $R$    & AGT, cylindrical, argon      & $
  8.314\,504(70)$~J mol$^{-1}$ K$^{-1}$     & $ 8.4\times 10^{-6}$  \\
                       &         &        &                              &
  \\

\vsp  \citet{1988027}  & NIST-88  & $R$   & AGT, spherical, argon        & $
  8.314\,471(15)$~J mol$^{-1}$ K$^{-1}$      & $ 1.8\times 10^{-6}$  \\
                       &          &       &                              &
  \\

\vsp  \citet{2009333}  & LNE-09   & $R$   & AGT, quasispherical, helium  & $
  8.314\,467(22)$~J mol$^{-1}$ K$^{-1}$       & $ 2.7\times 10^{-6}$  \\
                       &          &       &                              &
  \\
\vsp  \citet{2010177}  & NPL-10   & $R$   & AGT, quasispherical, argon   & $
  8.314\,468(26)$~J mol$^{-1}$ K$^{-1}$     &    $ 3.1\times 10^{-6}$  \\
                       &          &                                      &
  &                     \\

\vsp  \citet{2010080}  & INRIM-10 & $R$   & AGT, spherical, helium       & $
  8.314\,412(63)$~J mol$^{-1}$ K$^{-1}$       & $ 7.5\times 10^{-6}$  \\
                       &          &       &                              &
  \\

\vsp  \citet{2011207}  & LNE-11   & $R$   & AGT, quasispherical, argon   & $
  8.314\,456(10)$~J mol$^{-1}$ K$^{-1}$      & $ 1.2\times 10^{-6}$  \\
                       &          &       &                              &
  \\

\vsp  \citet{2007117}  & NIST-07  & $k$   & RIGT, quasispherical, helium & $
  1.380\,653(13)\times 10^{-23}$~J K$^{-1}$    & $ 9.1\times 10^{-6}$  \\
                       &          &       &                             &
  \\

\vsp  \citet{2011051}  & NIST-11  & $k/h$ & JNT, JE and QHE             & $
  2.083\,666(25)\times 10^{10}$~Hz K$^{-1}$    & $ 1.2\times 10^{-5}$  \\
                       &          &       &                             &
  \\

\botrule
\end{tabular}
\\ \vbox to 1 pt {}
$^\text{a}$NPL: National Physical Laboratory, Teddington, UK;
NIST: National Institute of Standards and Technology, Gaithersburg, MD, and
  Boulder, CO, USA;
LNE: Laboritoire commun de m\'etrologie (LCM), Saint-Denis, France, of
the Laboratoire national de m\'etrologie et d'essais (LNE);
INRIM: Istituto Nazionale di Ricerca Metrologica, Torino, Italy.
\end{table*}

\section{Newtonian constant of gravitation  $\bm{G}$}
\label{sec:ncg}

\def\vsp{\vbox to 8 pt{}}
\def\hsp{\hbox to 14 pt{}}
\begin{table*}[t]
\caption{
Summary of the results of measurements of the Newtonian
constant of gravitation relevant to the 2010 adjustment.
}
\label{tab:bg}
\begin{tabular}{l@{\hsp}ll@{\hsp}l@{~~}l}
\toprule
\vbox to 10 pt {}
Source & Identification$^\text{a}$ & Method & \hbox to 10 pt {} $10^{11} \, G$
  & Rel. stand. \\
       &                     &        & $\overline{{\rm m}^3 \  {\rm kg}^{-1} \
  {\rm s}^{-2}}$  & uncert $u_{\rm r}$ \\
\noalign{\vbox to 2 pt {}}
\colrule
\noalign{\vbox to 5 pt {}}

\vsp  \citet{1982013}  & NIST-82  & Fiber torsion balance,    & $ 6.672\,48(43)$
  & $ 6.4\times 10^{-5}$  \\
                         &           & dynamic mode              &
  &                     \\

\vsp  \citet{1996199}  & TR\&D-96  & Fiber torsion balance,    & $ 6.672\,9(5)$
  & $ 7.5\times 10^{-5}$  \\
                         &           & dynamic mode              &
  &                     \\

\vsp  \citet{1997025}      & LANL-97   & Fiber torsion balance,    & $
  6.673\,98(70)$     & $ 1.0\times 10^{-4}$  \\
                             &           & dynamic mode              &
  &                     \\

\vsp  \citet{2000088,pc02gm} & UWash-00    & Fiber torsion
balance,            & $ 6.674\,255(92)$   & $ 1.4\times 10^{-5}$  \\
                    &           & dynamic compensation      &                 &
  \\

\vsp  \citet{2001089}      & BIPM-01   & Strip torsion balance,    & $
  6.675\,59(27)$     & $ 4.0\times 10^{-5}$  \\
                             &    & compensation mode, static deflection
  &        &                     \\

\vsp  \citet{02kleinevoss,pc02kmph} &  UWup-02   & Suspended body,           & $
  6.674\,22(98)$    & $ 1.5\times 10^{-4}$  \\
                                      &           & displacement              &
  &                     \\

\vsp  \citet{2003219}      & MSL-03    & Strip torsion balance,    &  $
  6.673\,87(27)$     &  $ 4.0\times 10^{-5}$ \\
                             &           & compensation mode         &
  &                     \\

\vsp  \citet{2005292} & HUST-05  & Fiber torsion balance, & $ 6.672\,28(87)$& $
  1.3\times 10^{-4}$  \\
                        &           & dynamic mode              &
  &                     \\

\vsp  \citet{2006238}      & UZur-06   & Stationary body,          & $
  6.674\,25(12)$    & $ 1.9\times 10^{-5}$  \\
                             &           & weight change             &
  &                     \\

\vsp  \citet{2009099,2010122}   & HUST-09   & Fiber torsion balance, & $
  6.673\,49(18)$    & $ 2.7\times 10^{-5}$  \\
                             &           & dynamic mode             &
  &                     \\

\vsp  \citet{2010144}   & JILA-10   & Suspended body, & $ 6.672\,34(14)$    & $
  2.1\times 10^{-5}$  \\
                             &           & displacement             &
  &                     \\

\botrule
\end{tabular}
\\ \vbox to 0 pt {}
$^\text{a}$NIST: National Institute of Standards and Technology,
Gaithersburg, MD, USA;
TR\&D: Tribotech Research and Development Company, Moscow, Russian Federation;
LANL: Los Alamos National Laboratory, Los Alamos, New Mexico, USA;
UWash: University of Washington, Seattle, Washington, USA;
BIPM: International Bureau of Weights and Measures, S\`{e}vres, France;
UWup: University of Wuppertal, Wuppertal, Germany;
MSL: Measurement Standards Laboratory, Lower Hutt, New Zeland;
HUST: Huazhong University of Science and Technology, Wuhan, PRC;
UZur: University of Zurich, Zurich, Switzerland;
JILA: JILA, University of Colorado and National Institute of Standards
and Technology, Boulder, Colorado, USA.
\end{table*}

Table~\ref{tab:bg} summarizes the 11 values of the Newtonian constant of
gravitation $G$ of interest in the 2010 adjustment.  Because they are
independent of the other data relevant to the current adjustment, and
because there is no known quantitative theoretical relationship between
$G$ and other fundamental constants, they contribute only to the
determination of the 2010 recommended value of $G$. The calculation of
this value is discussed in Sec.~\ref{sssec:calcncg}.

The inconsistencies between different measurements of $G$ as discussed
in the reports of previous CODATA adjustments demonstrate the
historic difficulty of determining this most important constant.
Unfortunately, this difficulty has been demonstrated anew with the
publication of two new competitive results for $G$ during the past 4
years. The first is an improved value from the group at the Huazhong
University of Science and Technology (HUST), PRC, identified as
HUST-09 \cite{2009099,2010122}; the second is a completely new value from
researchers at JILA, Boulder, Colorado, USA, identified as JILA-10
\cite{2010144}.  (JILA, formerly known as the Joint Institute for
Laboratory Astrophysics, is a joint institute of NIST and the University
of Colorado and is located on the University of Colorado campus,
Boulder, Colorado.)

The publication of the JILA value has led the Task Group to re-examine
and modify two earlier results.  The first is that obtained at NIST
(then known as the National Bureau of Standards) by \citet{1982013} in a
NIST-University of Virginia (UVa) collaboration, labeled NIST-82.  This
value was the basis for the CODATA 1986 recommended value \cite{1987004}
and was taken into account in determining the CODATA 1998 value
\cite{2000035}, but played no role in either the 2002 or 2006
adjustments.  The second is the Los Alamos National Laboratory (LANL),
Los Alamos, USA, result of \citet{1997025}, labeled LANL-97; it was
first  included in the 1998 CODATA adjustment and in all subsequent
adjustments. Details of the modifications to NIST-82 and LANL-97 (quite
minor for the latter), the reasons for including NIST-82 in the 2010
adjustment, and discussions of the new values HUST-09 and JILA-10 are
given below.  The 11 available values of $G$, which are data items
$G1$-$G11$ in Table~\ref{tab:bigg}, Sec.~\ref{sec:ad}, are the same as
in 2006 with the exception of NIST-82, slightly modified LANL-97, and
the two new values.  Thus, in keeping with our approach in this report,
there is no discussion of the other seven values since they have been
covered in one or more of the previous reports.

For simplicity, in the following text, we write $G$ as a numerical
factor multiplying $G_0$, where
\begin{eqnarray}
G_0 = 10^{-11}~{\rm  m^{3}~kg^{-1}~s^{-2}} \ .
\end{eqnarray}

\subsection{Updated values}
\label{ssec:uv}

\subsubsection{National Institute of Standards and Technology
and University of Virginia}
\label{sssec:nist82}

As discussed in CODATA-98, the experiment of \citet{1982013} used a
fiber-based torsion balance operated in the dynamic mode and the
time-of-swing method, thereby requiring measurement of a small change in
the long oscillation period of the balance.  Ideally, the torsional
spring constant of the fiber should be independent of frequency at very
low frequencies, for example, at 3 mHz.

Long after the publication of the NIST-UVa result, \citet{1995147} [see
also  \citet{1998053} and \citet{1999058}] pointed out that the
anelasticity of such fibers is sufficiently large to cause the value of
$G$ determined in this way to be biased.  If $Q$ is the quality factor
of the main torsional mode of the fiber and it is assumed that the
damping of the torsion balance is solely due to the losses in the fiber,
then the unbiased value of $G$ is related to the experimentally observed
value $G({\rm obs})$ by \cite{1995147}
\begin{eqnarray}
G = {G({\rm obs})\over 1 + \rmpi Q \,}\, .
\label{eq:gqgobs}
\end{eqnarray}

Although the exact value of the $Q$ of the fiber used in the NIST-UVa
experiment is unknown, one of the researchers \cite{pc10gl} has provided
an estimate, based on data obtained during the course of the experiment,
of no less than 10\,000 and no greater than 30\,000.  Assuming a
rectangular probability density function for $Q$ with these lower and
upper limits then leads to $Q = 2\times 10^{4}$ with a relative standard
uncertainty of $4.6\times 10^{-6}$. Using these values, the result
$G({\rm obs}) = 6.672~59(43)G_0$   $[64\times10^{-6}] $
\cite{1982013}, \cite{pc86gl}, and Eq.~(\ref{eq:gqgobs}) we obtain

\begin{eqnarray}
G =  6.672\,48(43) G_0 \quad [ 6.4\times 10^{-5}] \, .
\label{eq:bgnbs82}
\end{eqnarray}
In this case the correction $1/(1 + \rmpi Q)$ reduced $G({\rm
obs})$  by the fractional amount $15.9(4.6)\times 10^{-6}$, but
increased its $64\times 10^{-6}$ relative standard uncertainty by a
negligible amount.

The Task Group decided to include the value given in
Eq.~(\ref{eq:bgnbs82}) as an input datum in the 2010 adjustment even
though it was not included in the 2002 and 2006 adjustments, because
information provided by \citet{pc10gl} allows the original result to be
corrected for the Kuroda effect.  Further, although there were plans to
continue the NIST-UVa experiment \cite{1982013}, recent conversations
with \citet{pc10gl} made clear that the measurements on which the result
is based were thorough and complete.

\subsubsection{Los Alamos National Laboratory}

\label{sssec:lanl97}

The experiment of \citet{1997025}, also described in detail in
CODATA-98, is similar to the NIST-UVa experiment of \citet{1982013}, and
in fact used some of the same components including the tungsten source
masses. Its purpose was not only to determine $G$, but also to test the
Kuroda hypothesis by using two different fibers, one with $Q = 950$ and
the other with $Q = 490$. Because the value of $G$ resulting from this
experiment is correlated with the NIST-UVa value and both values are now
being included in the adjustment, we evaluated the correlation
coefficient of the two results.  This was done with information from
\citet{pc10cb}, \citet{pc10gl}, and the Ph.D.~thesis of \citet{pc96cb}.
We take into account the uncertainties of the two $Q$ values (2\,\%) and
the correlation coefficient of the two values of $G$ obtained from the
two fibers (0.147) when computing their weighted mean.  The final result
is
\begin{eqnarray}
G =  6.673\,98(70) \quad [ 1.0\times 10^{-4}]\, ,
\label{eq:bglanl97}
\end{eqnarray}
which in fact is essentially the same as the value used in the 2002 and
2006 adjustments.  The correlation coefficient of the NIST-UVa and LANL
values of $G$ is 0.351.

\subsection{New values}
\label{ssec:nv}

\subsubsection{Huazhong University of Science and Technology}
\label{sssec:hust09}

The improved HUST-09 result for $G$ was first reported by
\citet{2009099} and subsequently described in detail by \citet{2010122};
it represents a reduction in uncertainty, compared to the previous
Huazhong University result HUST-05, of about a factor of five. As
pointed out by \citet{2010122}, a number of changes in the earlier
experiment contributed to this uncertainty reduction, including (i)
replacement of the two stainless steel cylindrical source masses by
spherical source masses with a more homogeneous density; (ii) use of a
rectangular quartz block as the principal portion of the torsion
balance's pendulum, thereby improving the stability of the period of the
balance and reducing the uncertainty of the pendulum's moment of
inertia; (iii) a single vacuum chamber for the source masses and
pendulum leading to a reduction of the uncertainty of their relative
positions; (iv) a remotely operated stepper motor to change the
positions of the source masses, thereby reducing environmental changes;
and (v) measurement of the anelasticity of the torsion fiber with the
aid of a high-$Q$ quartz fiber.

The final result is the average of two values of $G$ that differ by 9
parts in $10^6$ obtained from two partially correlated determinations
using the same apparatus.  The dominant components of uncertainty, in
parts in $10^6$, are 19 from the measurement of the fiber's
anelasticity, 14 (statistical) from the measurement of the change in the
square of the angular frequency of the pendulum when the source masses
are in their near and far positions, and 10 from the measured distance
between the geometric centers of the source masses. Although the
uncertainty of HUST-05 is five times larger than that of HUST-09,
\citet{pc10jl} and co-workers do not believe that HUST-09 supersedes
HUST-05. Thus, both are considered for inclusion in the 2010 adjustment.
Based on information provided to the Task Group by the researchers
\cite{pc10jl}, their correlation coefficient is estimated to be 0.234
and is used in the calculations of Sec.~\ref{sec:ad}. The extra digits
for the value and uncertainty of HUST-05 were also provided by
\citet{pc11jl}.

\subsubsection{JILA}

\label{sssec:jila10}

As can be seen from Table~\ref{tab:bg}, the $21\times 10^{-6}$ relative
standard uncertainty of the value of $G$ identified as JILA-10 and
obtained at JILA by \citet{2010144} has the third smallest estimated
uncertainty of the values listed and is the second smallest of those
values. It differs from the value with the smallest uncertainty,
identified as UWash-00, by $287(25)$ parts in $10^6$, which is $11$
times the standard uncertainty of their difference $u_{\rm diff}$, or
``$11\sigma$.'' This disagreement is an example of the ``historic
difficulty'' referred to at the very beginning of this section. The data
on which the JILA researchers based their result was taken in 2004, but
being well aware of this inconsistency they hesitated to publish it
until they checked and rechecked their work \cite{2010144}. With this
done, they decided it was time to report their value for $G$.

The apparatus used in the JILA experiment of \citet{2010144} consisted
of two $780~{\rm g}$ copper test  masses (or ``pendulum bobs'')
separated by $34~{\rm cm}$, each of which was suspended from a
supporting bar by four wires and together they formed a Fabry-Perot
cavity.  When the four $120~{\rm kg}$ cylindrical tungsten source
masses, two pairs with each member of the pair on either side of the
laser beam traversing the cavity, were periodically moved parallel to
the laser beam from their inner and outer positions (they remained
stationary for $80~{\rm s}$ in each position), the separation between
the bobs changed by about $90~{\rm nm}$. This change was observed as a
$125~{\rm MHz}$ beat frequency between the laser locked to the pendulum
cavity and the laser locked to a reference cavity that was part of the
supporting bar. The geometry of the experiment reduces the most
difficult aspect of determining the gravitational field of the source
masses to six one dimensional measurements: the distance between
opposite source mass pairs in the inner and outer positions and the
distances between adjacent source masses in the inner position.  The
most important relative standard uncertainty components contributing to
the uncertainty of $G$ are, in parts in $10^6$ \cite{2010144}, the six
critical dimension measurements, 14; all other dimension measurements
and source mass density inhomogeneities, 8 each; pendulum spring
constants, 7; and total mass measurement and interferometer
misalignment, 6 each.

As already noted, we leave the calculation of the 2010 recommended value
of $G$ to Sec.~\ref{sssec:calcncg}.

\section{Electroweak quantities}
\label{sec:xeq}
\shortcites{2002170}

As in previous adjustments, there are a few cases in the 2010 adjustment
where an inexact constant that is used in the analysis of input data is
not treated as an adjusted quantity, because the adjustment has a
negligible effect on its value.  Three such constants, used in the
calculation of the theoretical expression for the electron magnetic
moment anomaly $a_{\rm e}$, are the mass of the tau lepton
$m_{\rmsstau}$, the Fermi coupling constant $G_{\rm F}$, and sine
squared of the weak mixing angle sin$^{2}{\theta}_{\rm W}$; they are
obtained from the most recent report of the Particle Data Group
\cite{2010129}:
\begin {eqnarray}
m_{\rmsstau}c^{2} &=&  1776.82(16) \ {\rm MeV}
\qquad [ 9.0\times 10^{-5}]\, ,
\label{eq:mtaumev}
 \\[10 pt]
{G_{\rm F}\over(\hbar c)^{3}} &=&  1.166\,364(5)\times 10^{-5} \ {\rm GeV}^{-2}
\quad [ 4.3\times 10^{-6}]\, ,  \nonumber\\
\label{eq:gf}
 \\
{\rm sin}^{2}{\theta}_{\rm W} &=&  0.2223(21)
\qquad [ 9.5\times 10^{-3}] \, .
\label{eq:sin2thw}
\end{eqnarray}
The value for $G_{\rm F}/(\hbar c)^{3}$ is taken from p.~127 of
\citet{2010129}.  We use the definition sin$^{2}{\theta}_{\rm W} = 1 -
(m_{\rm W}/m_{\rm Z})^{2}$, where $m_{\rm W}$ and $m_{\rm Z}$ are,
respectively, the masses of the ${\rm W}^{\pm}$ and ${\rm Z}^{0}$
bosons, because it is employed in the calculation of the electroweak
contributions to $a_{\rm e}$ \cite{1996033}.  The Particle Data
Group's recommended value for the mass ratio of these bosons is $m_{\rm
W}/m_{\rm Z} =  0.8819(12)$, which leads to the value of
sin$^{2}{\theta}_{\rm W}$ given above.

\section{Analysis of Data}
\label{sec:ad}

We examine in this section the input data discussed in the previous
sections and, based upon that examination, select the data to be used in
the least-squares adjustment that determines the 2010 CODATA recommended
values of the constants. Tables~\ref{tab:rdata}, \ref{tab:pdata},
\ref{tab:cdata}, and \ref{tab:bigg} give the input data, including the
$\delta$'s, which are corrections added to theoretical expressions to
account for the uncertainties of those expressions.  The covariances of
the data are given as correlation coefficients in Tables~\ref{tab:rdcc},
\ref{tab:pdcc}, \ref{tab:cdcc}, and \ref{tab:bigg}.  There are 14 types
of input data for which there are two or more experiments, and the
data of the same type generally agree.

\setcounter{topnumber}{3}

\def\m{\phantom{-}}
\def\fixh{\vbox to 9pt {}}
\def\vsp{\vbox to 10pt{}}
\begin{table*}[t]
\caption{Summary of principal input data for the
determination of the 2010 recommended value of the Rydberg constant $R_\infty$.
}
\def\s{\hbox to 2 pt {}}
\label{tab:rdata}
\begin{tabular}{l@{\s}l@{\s}l@{\s}l@{\s}l@{\s}l}
\toprule
Item \vbox to 10 pt {}
 &  Input datum  \hbox to 125 pt {}  &  \hbox to 12pt {} Value
\hbox to 100 pt {} & \hbox to -20pt {} Relative standard
& Identification & Sec. \\
number& & & \hbox to -5pt {}uncertainty$^1$
 $u_{\rm r}$ &  \\
\colrule
 $A1   $ \vsp & $\delta_{\rm H}({\rm 1S_{1/2}})$ & $ 0.0(2.5)$  kHz & $[
  7.5\times 10^{-13}]$ & theory & \ref{par:teu}  \\
 $A2   $ & $\delta_{\rm H}({\rm 2S_{1/2}})$ & $ 0.00(31)$  kHz & $[ 3.8\times
  10^{-13}]$ & theory & \ref{par:teu}  \\
 $A3   $ & $\delta_{\rm H}({\rm 3S_{1/2}})$ & $ 0.000(91)$  kHz & $[ 2.5\times
  10^{-13}]$ & theory & \ref{par:teu}  \\
 $A4   $ & $\delta_{\rm H}({\rm 4S_{1/2}})$ & $ 0.000(39)$  kHz & $[ 1.9\times
  10^{-13}]$ & theory & \ref{par:teu}  \\
 $A5   $ & $\delta_{\rm H}({\rm 6S_{1/2}})$ & $ 0.000(15)$  kHz & $[ 1.6\times
  10^{-13}]$ & theory & \ref{par:teu}  \\
 $A6   $ & $\delta_{\rm H}({\rm 8S_{1/2}})$ & $ 0.0000(63)$  kHz & $[ 1.2\times
  10^{-13}]$ & theory & \ref{par:teu}  \\
\botrule
\end{tabular}
\\
$^1$ The values in brackets are relative to the frequency equivalent
of the binding energy of the indicated level.
\end{table*}

\def\m{\phantom{-}}
\def\fixh{\vbox to 9pt {}}
\def\vsp{\vbox to 10pt{}}
\def\hsp{\hbox to 51pt{}}
\begin{table*}[t]
\caption{Correlation coefficients
$r(x_i,x_j) \ge 0.0001$
of the input data related to $R_\infty$ in Table~\ref{tab:rdata}.
For simplicity, the two items of data to which a particular correlation
  coefficient
corresponds are identified by their item numbers in
Table~\ref{tab:rdata}.}
\label{tab:rdcc}
\begin{tabular}{ r @{~} l @{\hsp} r @{~} l @{\hsp} r @{~} l @{\hsp} r @{~} l }
\toprule
\vbox to 10 pt {}
$r$($A1$, $A2$)& = $ 0.9905 $ & $r$($A6$, $A19$)& = $ 0.7404 $ & $r$($A28$,
  $A29$)& = $ 0.3478 $ & $r$($A31$, $A45$)& = $ 0.1136 $ \\
$r$($A1$, $A3$)& = $ 0.9900 $ & $r$($A6$, $A20$)& = $ 0.9851 $ & $r$($A28$,
  $A30$)& = $ 0.4532 $ & $r$($A32$, $A35$)& = $ 0.0278 $ \\
$r$($A1$, $A4$)& = $ 0.9873 $ & $r$($A7$, $A8$)& = $ 0.0237 $ & $r$($A28$,
  $A31$)& = $ 0.0899 $ & $r$($A32$, $A36$)& = $ 0.0553 $ \\
$r$($A1$, $A5$)& = $ 0.7640 $ & $r$($A9$, $A10$)& = $ 0.0237 $ & $r$($A28$,
  $A32$)& = $ 0.1206 $ & $r$($A32$, $A41$)& = $ 0.1512 $ \\
$r$($A1$, $A6$)& = $ 0.7627 $ & $r$($A11$, $A12$)& = $ 0.0006 $ & $r$($A28$,
  $A35$)& = $ 0.0225 $ & $r$($A32$, $A42$)& = $ 0.1647 $ \\
$r$($A1$, $A17$)& = $ 0.9754 $ & $r$($A11$, $A21$)& = $ 0.9999 $ & $r$($A28$,
  $A36$)& = $ 0.0448 $ & $r$($A32$, $A43$)& = $ 0.1750 $ \\
$r$($A1$, $A18$)& = $ 0.9656 $ & $r$($A11$, $A22$)& = $ 0.0003 $ & $r$($A28$,
  $A41$)& = $ 0.1225 $ & $r$($A32$, $A44$)& = $ 0.1209 $ \\
$r$($A1$, $A19$)& = $ 0.9619 $ & $r$($A12$, $A21$)& = $ 0.0003 $ & $r$($A28$,
  $A42$)& = $ 0.1335 $ & $r$($A32$, $A45$)& = $ 0.1524 $ \\
\botrule
\end{tabular}
\end{table*}

\newpage
\null
\newpage
\null

\def\m{\phantom{-}}
\def\fixh{\vbox to 9pt {}}
\begin{longtable*}{ll@{\qquad}llll}
\caption{Summary of principal input data for the
determination of the 2010
recommended values of the fundamental constants ($R_\infty$ and $G$ excepted).
\label{tab:pdata}} \\
\toprule
\vbox to 10 pt {}
Item & Input datum  &  \hbox to 15pt {} Value
& \hbox to -20pt {} Relative standard
& Identification & Sec. and Eq. \\
number& & & \hbox to -5pt {}uncertainty$^1$
$u_{\rm r}$ &  \\
\colrule
\endfirsthead

\caption{{\it (Continued).}
Summary of principal input data for the
determination of the 2006
recommended values of the fundamental constants ($R_\infty$ and $G$ excepted).}
  \\
\colrule
\vbox to 10 pt {}
Item & Input datum  &  \hbox to 15pt {} Value
& \hbox to -20pt {} Relative standard
& Identification & Sec. and Eq. \\
number& & & \hbox to -5pt {}uncertainty$^1$
$u_{\rm r}$ &  \\
\colrule
\endhead
\colrule
\endfoot
\endlastfoot
 $B1   $ &\fixh $A_{\rm r}(^{1}{\rm H})$ \vbox to 10pt{}    &
 $\phantom{-}  1.007\,825\,032\,07(10)$ &  $ 1.0\times 10^{-10}$ & AMDC-03 &
  \ref{ssec:rama} \\
 $B2.1 $ &\fixh $A_{\rm r}(^{2}{\rm H})$     & $\phantom{-}
  2.014\,101\,777\,85(36)$ & $ 1.8\times 10^{-10}$ & AMDC-03 & \ref{ssec:rama}
  \\
 $B2.2 $ &\fixh $A_{\rm r}(^{2}{\rm H})$     & $\phantom{-}
  2.014\,101\,778\,040(80)$ & $ 4.0\times 10^{-11}$ & UWash-06 & \ref{ssec:rama}
  \\
 $B3   $ &\fixh $A_{\rm r}(E_{\rm av}) $     & $\phantom{-}  0.794(79)$ & $
  1.0\times 10^{-1}$ & StockU-08 & \ref{ssec:smtr} \ (\ref{eq:bengh2ps})\\
 $B4   $ &\fixh $f_{\rm c}({\rm H}_2^{+*})/f_{\rm c}({\rm d})$  & $\phantom{-}
  0.999\,231\,659\,33(17)$ & $ 1.7\times 10^{-10}$ & StockU-08 & \ref{ssec:smtr}
  \ (\ref{eq:rSMTp08}) \\
 $B5   $ &\fixh $f_{\rm c}({\rm t})/f_{\rm c}({\rm H}_2^{+*})$  & $\phantom{-}
  0.668\,247\,726\,86(55)$ & $ 8.2\times 10^{-10}$ & StockU-06 & \ref{ssec:smtr}
  \ (\ref{eq:rSMTt06}) \\
 $B6   $ &\fixh $f_{\rm c}(^3{\rm He}^+)/f_{\rm c}({\rm H}_2^{+*})$  &
  $\phantom{-}  0.668\,252\,146\,82(55)$ & $ 8.2\times 10^{-10}$ & StockU-06 &
  \ref{ssec:smtr}  \ (\ref{eq:rSMTh06}) \\
 $B7   $ &\fixh $A_{\rm r}(^{4}{\rm He})$    & $\phantom{-}
  4.002\,603\,254\,131(62)$ & $ 1.5\times 10^{-11}$ & UWash-06 & \ref{ssec:rama}
  \\
\end{longtable*}

\begin{table*}
\caption{Non-negligible correlation coefficients $r(x_i,x_j)$ of the
input data in Table~\ref{tab:pdata}.  For simplicity, the two items of
data to which a particular correlation coefficient corresponds are
identified by their item numbers in Table~\ref{tab:pdata}.}
\def\hsp{\hbox to 35.5pt{}}
\label{tab:pdcc}
\begin{tabular}{ r @{~} l @{\hsp} r @{~} l @{\hsp} r @{~} l @{\hsp} r @{~} l }
\toprule
\vbox to 10 pt {}
$r$($B1$, $B2.1$)& = $ 0.073 $ & $r$($B39$, $B40$)& = $ 0.023 $ & $r$($B43$,
  $B45$)& = $ 0.516 $ & $r$($B47$, $B48$)& = $ 0.509 $ \\
$r$($B2.2$, $B7$)& = $ 0.127 $ & $r$($B39$, $B41.2$)& = $ 0.023 $ & $r$($B43$,
  $B46$)& = $ 0.065 $ & $r$($B49$, $B50$)& = $ 0.469 $ \\
$r$($B2.2$, $B8$)& = $ 0.089 $ & $r$($B39$, $B54$)& = $ -0.026 $ & $r$($B43$,
  $B47$)& = $ 0.065 $ & $r$($B49$, $B51$)& = $ 0.372 $ \\
$r$($B5$, $B6$)& = $ 0.876 $ & $r$($B40$, $B41.2$)& = $ 0.027 $ & $r$($B43$,
  $B48$)& = $ -0.367 $ & $r$($B49$, $B52$)& = $ 0.502 $ \\
$r$($B7$, $B8$)& = $ 0.181 $ & $r$($B40$, $B54$)& = $ -0.029 $ & $r$($B44$,
  $B45$)& = $ 0.421 $ & $r$($B50$, $B51$)& = $ 0.347 $ \\
$r$($B15$, $B16$)& = $ 0.994 $ & $r$($B41.2$, $B54$)& = $ -0.029 $ & $r$($B44$,
  $B46$)& = $ 0.053 $ & $r$($B50$, $B52$)& = $ 0.469 $ \\
$r$($B17$, $B18$)& = $ 0.082 $ & $r$($B42$, $B43$)& = $ -0.288 $ & $r$($B44$,
  $B47$)& = $ 0.053 $ & $r$($B51$, $B52$)& = $ 0.372 $ \\
$r$($B29.1$, $B30$)& = $ 0.227 $ & $r$($B42$, $B44$)& = $ 0.096 $ & $r$($B44$,
  $B48$)& = $ 0.053 $ & $r$($B58.3$, $B58.4$)& = $ 0.002 $ \\
$r$($B29.2$, $B31$)& = $ 0.195 $ & $r$($B42$, $B45$)& = $ 0.117 $ & $r$($B45$,
  $B46$)& = $ -0.367 $ & $r$($B58.3$, $B58.5$)& = $ 0.001 $ \\
$r$($B32.2$, $B34.1$)& = $ -0.014 $ & $r$($B42$, $B46$)& = $ 0.066 $ &
  $r$($B45$, $B47$)& = $ 0.065 $ & $r$($B58.3$, $B58.6$)& = $ 0.032 $ \\
$r$($B36.1$, $B58.2$)& = $ 0.068 $ & $r$($B42$, $B47$)& = $ 0.066 $ & $r$($B45$,
  $B48$)& = $ 0.065 $ & $r$($B58.4$, $B58.6$)& = $ 0.012 $ \\
$r$($B37.1$, $B37.4$)& = $ 0.003 $ & $r$($B42$, $B48$)& = $ 0.504 $ & $r$($B46$,
  $B47$)& = $ 0.509 $ & &  \\
$r$($B37.2$, $B37.3$)& = $ 0.140 $ & $r$($B43$, $B44$)& = $ 0.421 $ & $r$($B46$,
  $B48$)& = $ 0.509 $ & &  \\
\botrule
\end{tabular}
\end{table*}

\def\m{\phantom{-}}
\def\fixh{\vbox to 9pt {}}
\def\vsp{\vbox to 10pt{}}
\def\hsp{\hbox to 36pt{}}
\begin{table*}[t]
\caption{Summary of principal input data for the determination of the
relative atomic mass of the electron from antiprotonic helium
transitions.  The numbers in parentheses $(n,l:n^\prime,l^\prime)$
denote the transition $(n,l)\rightarrow(n^\prime,l^\prime)$.}
\label{tab:cdata}
\begin{tabular}{l@{\hsp}l@{\hsp}l@{\hsp}l@{\hsp}ll}
\toprule
Item &  Input Datum    &  \hbox to 15pt {} Value
& \hbox to -20pt {} Relative standard
& Identification & Sec. \\
number& & & \hbox to -5pt {}uncertainty$^1$
 $u_{\rm r}$ &  \\
\colrule
 $C1   $ \vsp & $\delta_{\bar{\rm p}^4{\rm He}^+}(32,31:31,30)$ & $ 0.00(82)$
  MHz & $[ 7.3\times 10^{-10}]$ & JINR-06 & \ref{sssec:aphelth}  \\
 $C2   $ & $\delta_{\bar{\rm p}^4{\rm He}^+}(35,33:34,32)$ & $ 0.0(1.0)$  MHz &
  $[ 1.3\times 10^{-9}]$ & JINR-06 & \ref{sssec:aphelth}  \\
 $C3   $ & $\delta_{\bar{\rm p}^4{\rm He}^+}(36,34:35,33)$ & $ 0.0(1.1)$  MHz &
  $[ 1.6\times 10^{-9}]$ & JINR-06 & \ref{sssec:aphelth}  \\
 $C4   $ & $\delta_{\bar{\rm p}^4{\rm He}^+}(37,34:36,33)$ & $ 0.0(1.1)$  MHz &
  $[ 1.8\times 10^{-9}]$ & JINR-06 & \ref{sssec:aphelth}  \\
 $C5   $ & $\delta_{\bar{\rm p}^4{\rm He}^+}(39,35:38,34)$ & $ 0.0(1.2)$  MHz &
  $[ 2.3\times 10^{-9}]$ & JINR-06 & \ref{sssec:aphelth}  \\
 $C6   $ & $\delta_{\bar{\rm p}^4{\rm He}^+}(40,35:39,34)$ & $ 0.0(1.3)$  MHz &
  $[ 2.9\times 10^{-9}]$ & JINR-06 & \ref{sssec:aphelth}  \\
\botrule
\end{tabular}
\\
$^1$ The values in brackets are relative to the corresponding
transition frequency.
\end{table*}

\begin{table*}
\caption{Non-negligible correlation coefficients $r(x_i,x_j)$ of the
input data in Table~\ref{tab:cdata}.  For simplicity, the two items of
data to which a particular correlation coefficient corresponds are
identified by their item numbers in Table~\ref{tab:cdata}.}
\label{tab:cdcc}
\def\hsp{\hbox to 52pt{}}
\begin{tabular}{ r @{~} l @{\hsp} r @{~} l @{\hsp} r @{~} l @{\hsp} r @{~} l }
\toprule
\vbox to 10 pt {}
$r$($C1$, $C2$)& = $ 0.929 $ & $r$($C4$, $C10$)& = $ 0.959 $ & $r$($C9$, $C14$)&
  = $ -0.976 $ & $r$($C18$, $C27$)& = $ 0.141 $ \\
$r$($C1$, $C3$)& = $ 0.936 $ & $r$($C4$, $C11$)& = $ 0.949 $ & $r$($C9$, $C15$)&
  = $ 0.986 $ & $r$($C18$, $C28$)& = $ 0.106 $ \\
$r$($C1$, $C4$)& = $ 0.936 $ & $r$($C4$, $C12$)& = $ 0.907 $ & $r$($C10$,
  $C11$)& = $ 0.978 $ & $r$($C18$, $C29$)& = $ 0.217 $ \\
$r$($C1$, $C5$)& = $ 0.912 $ & $r$($C4$, $C13$)& = $ 0.931 $ & $r$($C10$,
  $C12$)& = $ 0.934 $ & $r$($C19$, $C20$)& = $ 0.268 $ \\
$r$($C1$, $C6$)& = $ 0.758 $ & $r$($C4$, $C14$)& = $ -0.952 $ & $r$($C10$,
  $C13$)& = $ 0.959 $ & $r$($C19$, $C21$)& = $ 0.193 $ \\
$r$($C1$, $C7$)& = $ -0.947 $ & $r$($C4$, $C15$)& = $ 0.961 $ & $r$($C10$,
  $C14$)& = $ -0.980 $ & $r$($C19$, $C22$)& = $ 0.302 $ \\
$r$($C1$, $C8$)& = $ 0.954 $ & $r$($C5$, $C6$)& = $ 0.734 $ & $r$($C10$, $C15$)&
  = $ 0.990 $ & $r$($C19$, $C25$)& = $ 0.172 $ \\
$r$($C1$, $C9$)& = $ 0.960 $ & $r$($C5$, $C7$)& = $ -0.917 $ & $r$($C11$,
  $C12$)& = $ 0.925 $ & $r$($C19$, $C26$)& = $ 0.190 $ \\
\botrule
\end{tabular}
\end{table*}

\begin{table*}
\caption{Summary of values of $G$ used to determine the 2010 recommended
value (see also Table~\ref{tab:bg}, Sec.~\ref{sec:ncg}).}
\label{tab:bigg}
\def\hsp{\hbox to 25pt{}}
\begin{tabular}{l@{\hsp}l@{\hsp}l@{\hsp}l}
\toprule
\vbox to 10 pt {}
 &  & Relative &  \\
Item & Value$^1$ & standard &  \\
number & ($10^{-11}$ m$^3$ kg$^{-1}$ s$^{-2}$) & uncertainty $u_{\rm r}$ &
  Identification \\
\colrule

$G1$\vbox to 10 pt {} & $ 6.672\,48(43)$ & $ 6.4\times 10^{-5}$ & NIST-82   \\
$G2 $ & $ 6.672\,9(5)$  & $ 7.5\times 10^{-5}$ &  TR\&D-96  \\
$G3 $ & $ 6.673\,98(70)$ & $ 1.0\times 10^{-4}$ &  LANL-97  \\
$G4 $ & $ 6.674\,255(92)$& $ 1.4\times 10^{-5}$ &  UWash-00  \\
$G5 $ & $ 6.675\,59(27)$ & $ 4.0\times 10^{-5}$ &  BIPM-01  \\
$G6 $ & $ 6.674\,22(98)$ & $ 1.5\times 10^{-4}$ &  UWup-02  \\
$G7 $ & $ 6.673\,87(27)$  & $ 4.0\times 10^{-5}$ &  MSL-03  \\
$G8 $ & $ 6.672\,28(87)$ & $ 1.3\times 10^{-4}$ &  HUST-05  \\
$G9 $ & $ 6.674\,25(12)$ & $ 1.9\times 10^{-5}$ &  UZur-06  \\
$G10$ & $ 6.673\,49(18)$ & $ 2.7\times 10^{-5}$ &  HUST-09  \\
$G11$ & $ 6.672\,34(14)$ & $ 2.1\times 10^{-5}$ &  JILA-10  \\

\botrule
\end{tabular}
\hbox to 10 cm {$^1$Correlation coefficients: $r(G1,G3)=0.351$;
$r(G8,G10)=0.234$}.
\end{table*}

\subsection{Comparison of data through inferred values of $\bm{\alpha}$,
$\bm h$,
$\bm k$ and  $\bm{A_{\rm r}({\rm e})}$}
\label{ssec:infv}
\vspace{.25 in}

Here the level of consistency of the data is shown by comparing values
of $\alpha$, $h$, $k$ and  $A_{\rm r}({\rm e})$ that can be inferred
from different types of experiments.  Note, however, that the inferred
value is for comparison purposes only; the datum from which it is
obtained, not the inferred value, is used as the input datum in the
least-squares calculations.

\def\m{\phantom{-}}
\def\fixh{\vbox to 10pt {}}
\def\h{\hbox to 24pt {}}
\begin{table*}
\caption{Inferred values of the fine-structure constant $\alpha$ in
order of increasing standard uncertainty obtained from the indicated
experimental data in Table~\ref{tab:pdata}.}
\label{tab:alpha}
\begin{tabular}{l@{\h}l@{\h}l@{\h}l@{\h}l@{\h}l@{\h}l}
\toprule
Primary & Item & Identification \fixh
& Sec. and Eq. & \hbox to 25pt{} $\alpha^{-1}$ & \hbox to -5pt {} Relative
  standard \\
source & number &
& & & \hbox to -5pt {} uncertainty $u_{\rm r}$  \\
\colrule

$a_{\rm e}$ & \fixh $B13.2$ &
HarvU-08 & \ref{sssec:alphaae} \ (\ref{eq:alphinvharvu08}) &
$ 137.035\,999\,084(51)$& $ 3.7\times 10^{-10}$\\

$h/m(^{87}{\rm Rb})$ & \fixh $B57$ &
LKB-11 & \ref{sssec:pcrbmr} &
$ 137.035\,999\,049(90)$& $ 6.6\times 10^{-10}$\\

$a_{\rm e}$ & \fixh $B11$ &
UWash-87 & \ref{sssec:alphaae} \ (\ref{eq:alphinvuwash87}) &
$ 137.035\,998\,19(50)$& $ 3.7\times 10^{-9}$\\

$h/m(^{133}{\rm Cs})$ & \fixh $B56$ &
StanfU-02 & \ref{sssec:pccsmr} &
$ 137.036\,0000(11)$& $ 7.7\times 10^{-9}$\\

$R_{\rm K}$ & \fixh $B35.1$ &
NIST-97 & \ref{ssec:ed} &
$ 137.036\,0037(33)$&$ 2.4\times 10^{-8}$ \\

${\it\Gamma}_{\rm p-90}^{\,\prime}({\rm lo})$ & \fixh $B32.1$ &
NIST-89 & \ref{ssec:ed} &
$ 137.035\,9879(51)$&$ 3.7\times 10^{-8}$ \\

$R_{\rm K}$ & \fixh $B35.2$ &
NMI-97 & \ref{ssec:ed} &
$ 137.035\,9973(61)$&$ 4.4\times 10^{-8}$ \\

$R_{\rm K}$ & \fixh $B35.5$ &
LNE-01 & \ref{ssec:ed} &
$ 137.036\,0023(73)$&$ 5.3\times 10^{-8}$ \\

$R_{\rm K}$ & \fixh $B35.3$ &
NPL-88 & \ref{ssec:ed} &
$ 137.036\,0083(73)$&$ 5.4\times 10^{-8}$ \\

$\Delta\nu_{\rm Mu}$ & \fixh $B29.1,B29.2$ &
LAMPF & \ref{sssec:mufreqs} \ (\ref{eq:alphiL}) &
$ 137.036\,0018(80)$ & $ 5.8\times 10^{-8}$ \\

${\it\Gamma}_{\rm h-90}^{\,\prime}({\rm lo})$ & \fixh $B33$ &
KR/VN-98 & \ref{ssec:ed} &
$ 137.035\,9852(82)$&$ 6.0\times 10^{-8}$ \\

$R_{\rm K}$ & \fixh $B35.4$ &
NIM-95 & \ref{ssec:ed} &
$ 137.036\,004(18)$&$ 1.3\times 10^{-7}$ \\

${\it\Gamma}_{\rm p-90}^{\,\prime}({\rm lo})$ & \fixh $B32.2$ &
NIM-95 & \ref{ssec:ed} &
$ 137.036\,006(30)$&$ 2.2\times 10^{-7}$ \\

$\nu_{\rm H},\nu_{\rm D}$ &&& \ref{par:trfreq} \ (\ref{eq:alphinvhd}) &
$ 137.036\,003(41)$&$ 3.0\times 10^{-7}$ \\

\botrule
\end{tabular}
\end{table*}

\begin{figure}
\rotatebox{-90}{\resizebox{!}{4.4 in}{
\includegraphics[clip,trim=10 40 40 10]{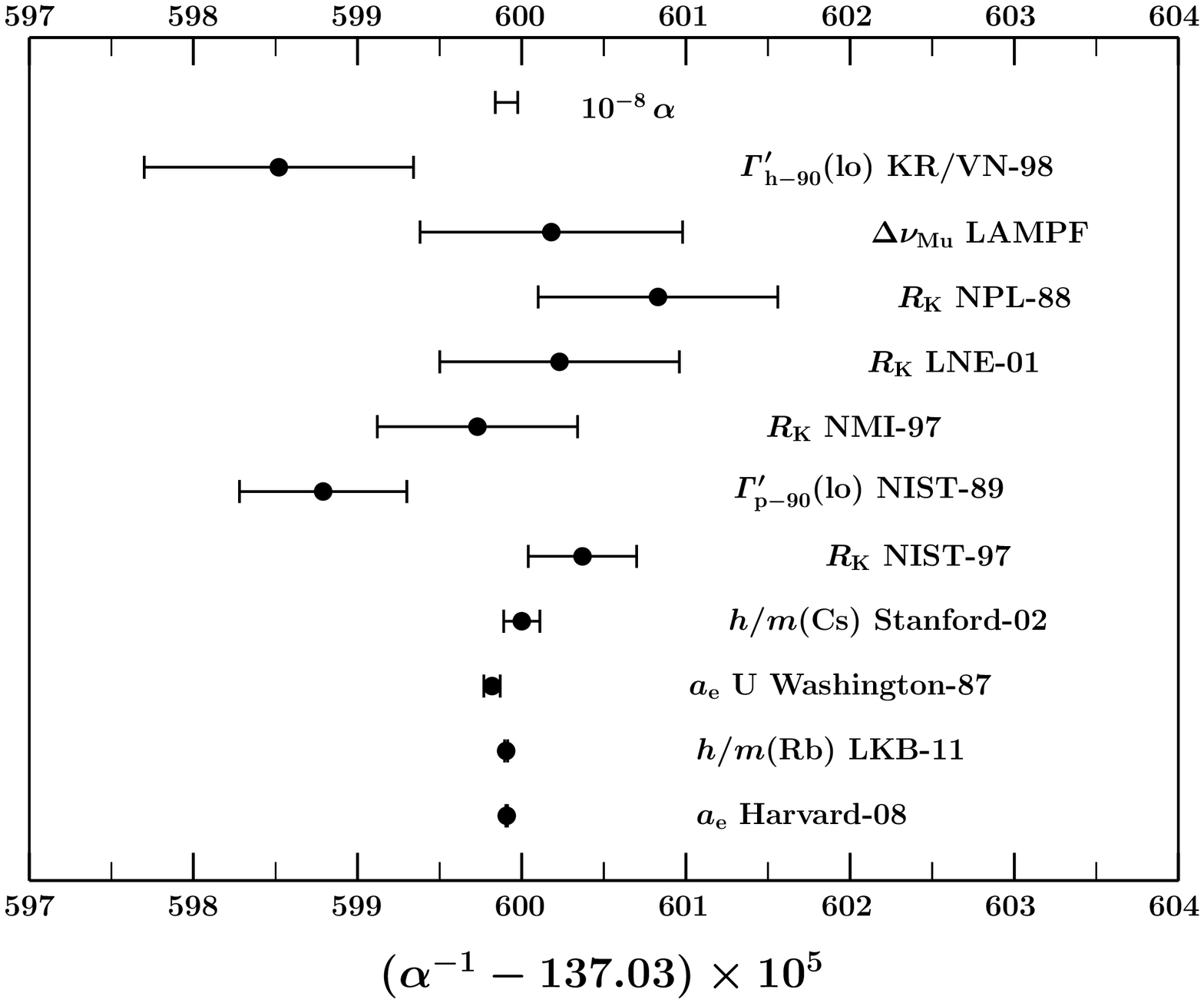} }}
\caption { Values
of the fine-structure constant $\alpha$ with $u_{\rm r}<10^{-7}$ implied by the
input data in Table~\ref{tab:pdata}, in order of decreasing uncertainty from
top to bottom (see Table \ref{tab:alpha}).}
\label{fig:aall}
\end{figure}

\begin{figure}
\rotatebox{-90}{\resizebox{!}{4.4in}{
\includegraphics[clip,trim=10 40 40 10]{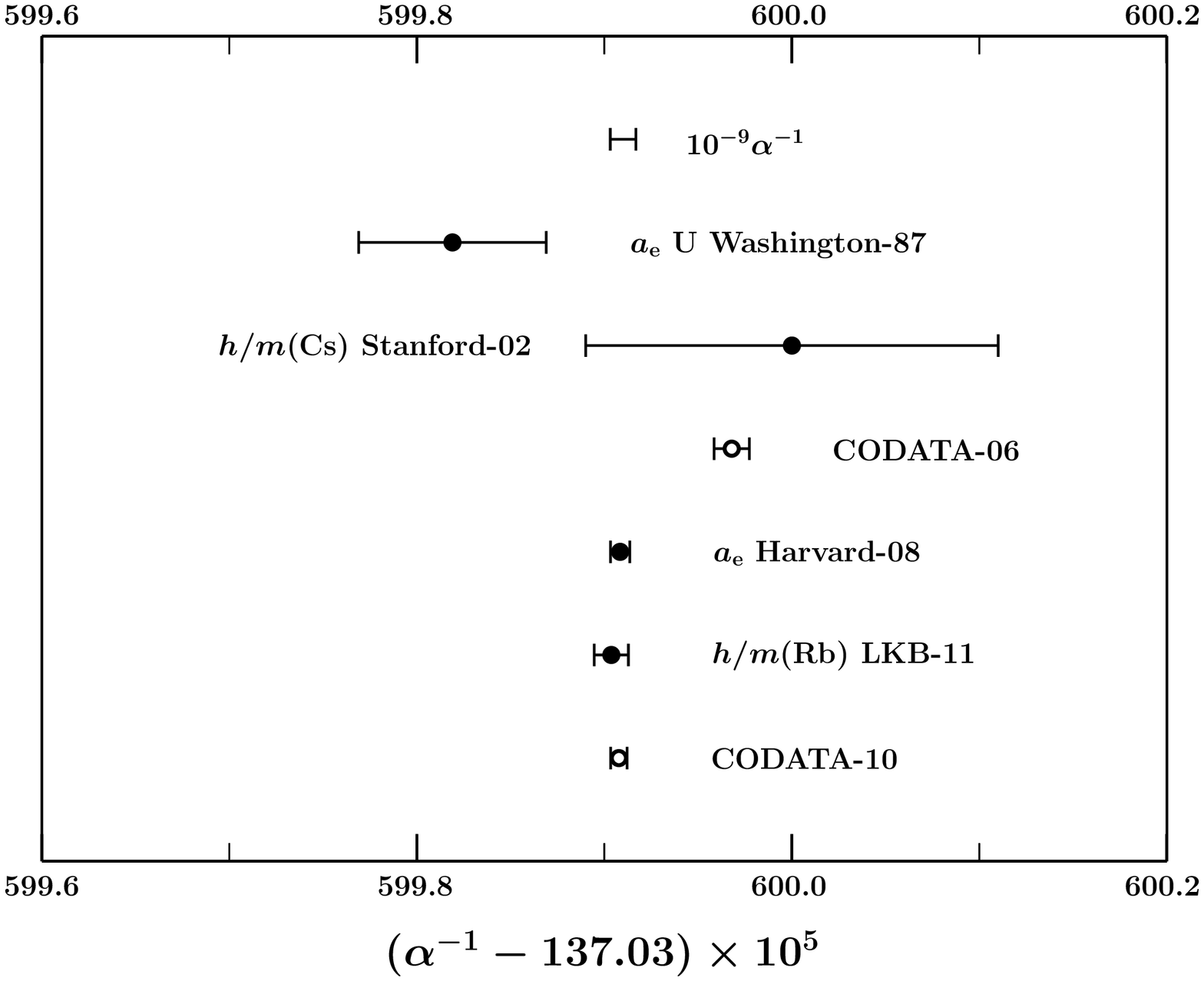} }}
\caption {
Values of the fine-structure constant $\alpha$ with $u_{\rm r}<10^{-8}$
implied by the input data in Table~\ref{tab:pdata} and the 2006 and 2010
CODATA recommended values in chronological order from top to bottom
(see Table~\ref{tab:alpha}).}
\label{fig:alpha}
\end{figure}

Table~\ref{tab:alpha} and Figs.~\ref{fig:aall} and \ref{fig:alpha}
compare values of $\alpha$ obtained from the indicated input data.
These values are calculated using the appropriate observational equation
for each input datum as given in Table~\ref{tab:pobseqsb1} and the 2010
recommended values of the constants other than $\alpha$ that enter that
equation.  (Some inferred values have also been given in the portion of
the paper where the relevant datum is discussed.) Inspection of the
Table and figures shows that there is agreement among the vast majority
of the various values of $\alpha$, and hence the data from which they
are obtained, to the extent that the difference between any two values
of $\alpha$ is less than $2u_{\rm diff}$, the standard uncertainty of
the difference.

The two exceptions are the values of $\alpha$ from the NIST-89 result
for ${\it \Gamma}_{\rm p-90}^{\,\prime}({\rm lo})$ and, to a lesser
extent, the KR/VN-98 result for ${\it \Gamma}_{\rm h-90}^{\,\prime}({\rm
lo})$; of the 91 differences, six involving $\alpha$ from NIST-89 and
two involving $\alpha$ from KR/VN-98 are greater than $2u_{\rm diff}$.
The inconsistency of these data has in fact been discussed in previous
CODATA reports but, as in 2006, because their self-sensitivity
coefficients $S_{\rm c}$ (see Sec.~\ref{ssec:mada} below) are less than
$0.01$, they are not included in the final adjustment on which the 2010
recommended values are based. Hence, their disagreement is not a serious
issue.  Examination of the table and figures also shows that even if all
of the data from which these values of $\alpha$ have been inferred were
to be included in the final adjustment, the recommended value of
$\alpha$ would still be determined mainly by the HarvU-08 $a_{\rm e}$
and LKB-10 $h/m(^{87}{\rm Rb})$ data.  Indeed, the comparatively large
uncertainties of some of the values of $\alpha$ means that the data from
which they are obtained will have values of $S_{\rm c} < 0.01$ and will
not be included in the final adjustment.

\def\m{\phantom{-}}
\def\fixh{\vbox to 9pt {}}
\def\hsp{\hbox to 25pt{}}
\begin{table*}
\caption{Inferred values of the Planck constant $h$ in order of
increasing standard uncertainty obtained from the indicated experimental
data in Table~\ref{tab:pdata}.}
\label{tab:plancks}
\begin{tabular}{l@{\hsp}l@{\hsp}l@{\hsp}l@{\hsp}l@{\hsp}l}
\toprule
\fixh Primary & Item & Identification
& Sec. and Eq. & \hbox to 20pt{}$h/({\rm J \ s})$
& Relative standard \\
source & number &
& & & \hbox to -5pt {} uncertainty $u_{\rm r}$  \\
\colrule

$N_{\rm A}$($^{28}$Si) & \fixh $B54$ &
IAC-11 & \ref{ssec:naiac} \ (\ref{eq:naiac11}) &
$ 6.626\,070\,09(20)\times 10^{-34}$ & $ 3.0\times 10^{-8}$ \\

$K_{\rm J}^2R_{\rm K}$ & \fixh $B37.3$ &
NIST-07 & \ref{ssec:ed} &
$ 6.626\,068\,91(24)\times 10^{-34}$ & $ 3.6\times 10^{-8}$ \\

$K_{\rm J}^2R_{\rm K}$ & \fixh $B37.2$ &
NIST-98 & \ref{ssec:ed} &
$ 6.626\,068\,91(58)\times 10^{-34}$ & $ 8.7\times 10^{-8}$ \\

$K_{\rm J}^2R_{\rm K}$ & \fixh $B37.1$ &
NPL-90 & \ref{ssec:ed} &
$ 6.626\,0682(13)\times 10^{-34}$ & $ 2.0\times 10^{-7}$ \\

$K_{\rm J}^2R_{\rm K}$ & \fixh $B37.4$ &
NPL-12 & \ref{sssec:nplwb12} \ (\ref{eq:hwbnpl12}) &
$ 6.626\,0712(13)\times 10^{-34}$ & $ 2.0\times 10^{-7}$ \\

$K_{\rm J}^2R_{\rm K}$ & \fixh $B37.5$ &
METAS-11 & \ref{sssec:metaswb11} \ (\ref{eq:hwbmetas11}) &
$ 6.626\,0691(20)\times 10^{-34}$ & $ 2.9\times 10^{-7}$ \\

$K_{\rm J}$ & \fixh $B36.1$ &
NMI-89 & \ref{ssec:ed} &
$ 6.626\,0684(36)\times 10^{-34}$ & $ 5.4\times 10^{-7}$ \\

$K_{\rm J}$ & \fixh $B36.2$ &
PTB-91 & \ref{ssec:ed} &
$ 6.626\,0670(42)\times 10^{-34}$ & $ 6.3\times 10^{-7}$ \\

${\it\Gamma}_{\rm p-90}^{\,\prime}({\rm hi})$ & \fixh $B34.2$ &
NPL-79 & \ref{ssec:ed} &
$ 6.626\,0730(67)\times 10^{-34}$&$ 1.0\times 10^{-6}$ \\

${\cal F}_{90}$ & \fixh $B38$ &
NIST-80 & \ref{ssec:ed} &
$ 6.626\,0657(88)\times 10^{-34}$ & $ 1.3\times 10^{-6}$ \\

${\it\Gamma}_{\rm p-90}^{\,\prime}({\rm hi})$ & \fixh $B34.1$ &
NIM-95 & \ref{ssec:ed} &
$ 6.626\,071(11)\times 10^{-34}$&$ 1.6\times 10^{-6}$ \\

\botrule
\end{tabular}
\end{table*}

\begin{figure}
\rotatebox{-90}{\resizebox{!}{4.2in}{
\includegraphics[clip,trim= 10 40 40 10]{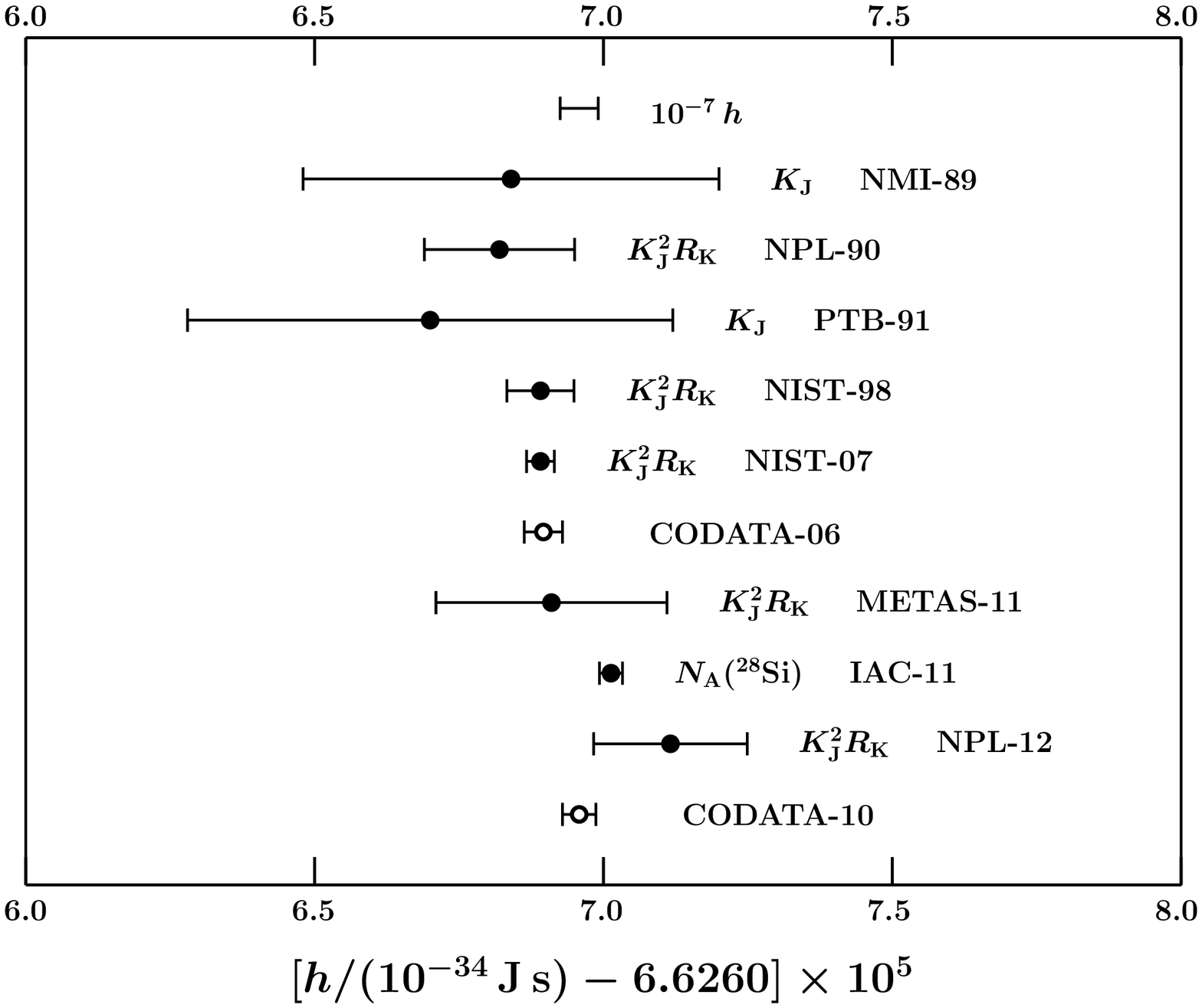}
}}
\caption{
Values of the Planck constant $h$ with $u_{{\rm r}}<10^{-6}$
implied by the input data in
Table~\ref{tab:pdata} and the 2006 and 2010 CODATA recommended values in
chronological order from top to bottom (see Table~\ref{tab:plancks}).}
\label{fig:h}
\end{figure}

Table~\ref{tab:plancks} and Fig.~\ref{fig:h} compare values of $h$
obtained from the indicated input data. The various values of $h$, and
hence the data from which they are calculated, agree to the extent that
the 55 differences between any two values of $h$ is less than $2u_{\rm
diff}$, except for the difference between the NIST-07 and IAC-11 values.
In this case, the difference is $3.8u_{\rm diff}$.

Because the uncertainties of these two values of $h$ are smaller than
other values and are comparable, they play the dominant role in the
determination of the recommended value of $h$.  This discrepancy is
dealt with before carrying out the final adjustment.  The relatively
large uncertainties of many of the other values of $h$ means that the
data from which they are calculated will not be included in the final
adjustment.

\def\m{\phantom{-}}
\def\fixh{\vbox to 9pt {}}
\def\fw{\hbox to 15pt {}}
\begin{table*}
\caption{Inferred values of the Boltzmann constant $k$ in
order of increasing standard uncertainty obtained from the indicated
experimental data in Table~\ref{tab:pdata}.}
\label{tab:k}
\begin{tabular}{l@{\fw}l@{\fw}l@{\fw}l@{\fw}l@{\fw}l}
\toprule
\fixh Primary & Item & Identification
& Section & \hbox to 13 pt {} $k/({\rm J \ K^{-1}})$
& Relative standard \\
source & number &
& & & \hbox to -5pt {} uncertainty $u_{\rm r}$  \\
\colrule

$R$ & \fixh $B58.6$ &
LNE-11 & \ref{sssec:LN0911} &
$ 1.380\,6477(17)\times 10^{-23}$ & $ 1.2\times 10^{-6}$ \\

$R$ & \fixh $B58.2$ &
NIST-88 & \ref{sssec:NP79NI88} &
$ 1.380\,6503(25)\times 10^{-23}$ & $ 1.8\times 10^{-6}$ \\

$R$ & \fixh $B58.3$ &
LNE-09 & \ref{sssec:LN0911} &
$ 1.380\,6495(37)\times 10^{-23}$ & $ 2.7\times 10^{-6}$ \\

$R$ & \fixh $B58.4$ &
NPL-10 & \ref{sssec:NP10} &
$ 1.380\,6496(43)\times 10^{-23}$ & $ 3.1\times 10^{-6}$ \\

$R$ & \fixh $B58.5$ &
INRIM-10 & \ref{sssec:IN10} &
$ 1.380\,640(10)\times 10^{-23}$ & $ 7.5\times 10^{-6}$ \\

$R$ & \fixh $B58.1$ &
NPL-79 & \ref{sssec:NP79NI88} &
$ 1.380\,656(12)\times 10^{-23}$ & $ 8.4\times 10^{-6}$ \\

$k$ & \fixh $B59$ &
NIST-07 & \ref{sssec:NI07} &
$ 1.380\,653(13)\times 10^{-23}$ & $ 9.1\times 10^{-6}$ \\

$k/h$ & \fixh $B60$ &
NIST-11 & \ref{sssec:NI11} &
$ 1.380\,652(17)\times 10^{-23}$ & $ 1.2\times 10^{-5}$ \\

\botrule
\end{tabular}
\end{table*}

\begin{figure}
\rotatebox{-90}{\resizebox{!}{4.2in}{
\includegraphics[clip,trim= 10 40 40 10]{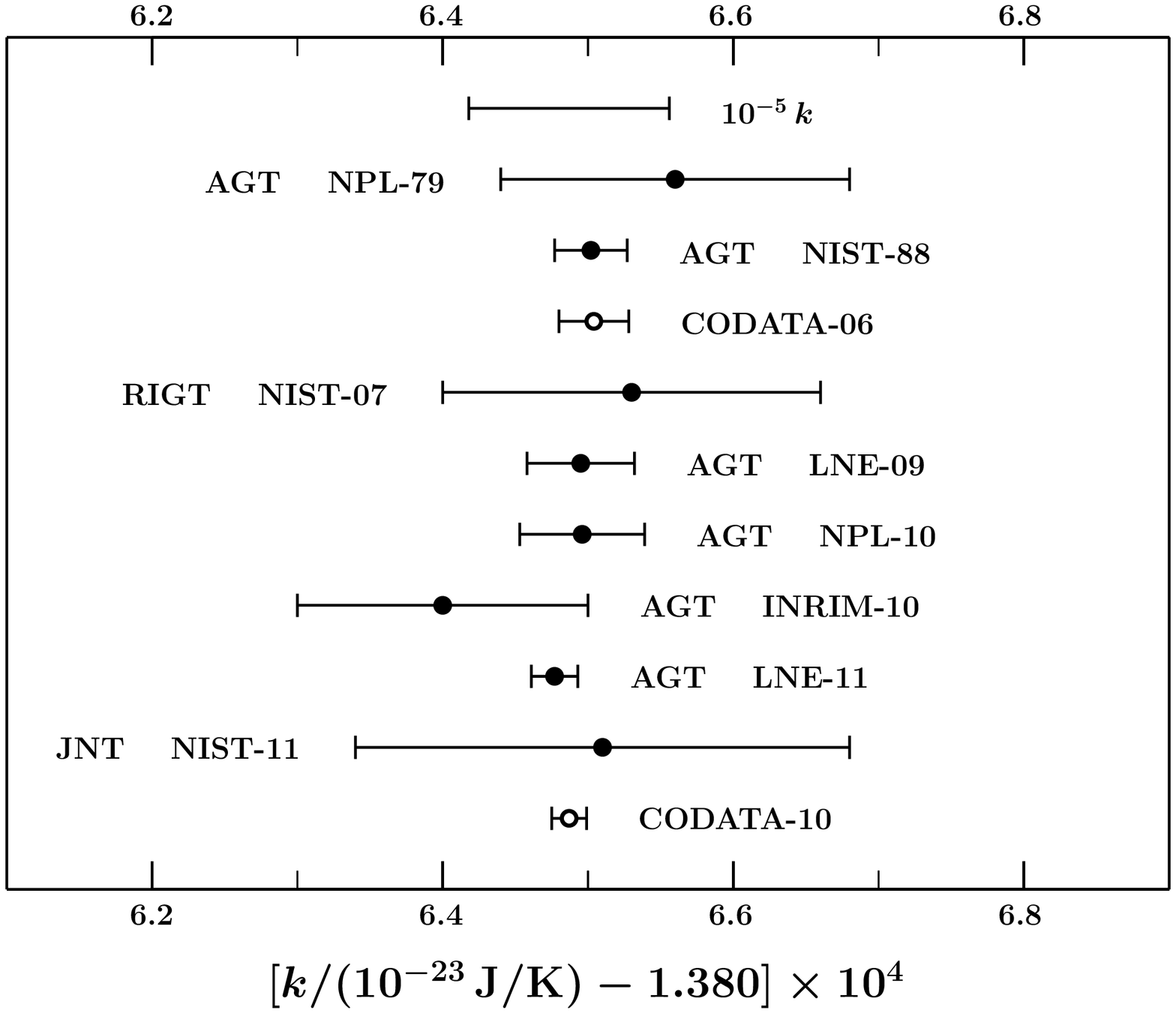}
}}
\caption{
Values of the Boltzmann constant $k$ implied by the input data in
Table~\ref{tab:pdata} and the 2006 and 2010 CODATA recommended values in
chronological order from top to bottom (see Table~\ref{tab:k}). AGT:
acoustic gas thermometry; RIGT: refractive index gas thermometry; JNT:
Johnson noise thermometry.}
\label{fig:k}
\end{figure}

Table~\ref{tab:k} and Fig.~\ref{fig:k} compare values of $k$ obtained
from the indicated input data.  Although most of the source data are
values of $R$, values of $k=R/N_{\rm A}$ are compared, because that is
the constant used to define the kelvin in the ``New'' SI; see, for
example, \citet{2011191}.  All of these values are in general agreement,
with none of the 28 differences exceeding $2u_{\rm diff}$.  However,
some of the input data from which they are calculated have uncertainties
so large that they will not be included in the final adjustment.

\def\m{\phantom{-}}
\def\fixh{\vbox to 10pt {}}
\def\hsp{\hbox to 22pt{}}
\begin{table*}
\caption{Inferred values of the electron relative atomic mass
$A_{\rm r}({\rm e})$ in order of increasing standard uncertainty obtained from
the indicated experimental data in Table~\ref{tab:pdata}.}
\label{tab:are}
\begin{tabular}{ll@{\hsp}l@{\hsp}l@{\hsp}l@{\hsp}l}
\toprule
Primary & Item & Identification \fixh
& Sec. and Eq. & \hbox to 25pt{} $A_{\rm r}({\rm e})$ & \hbox to -5pt {}
  Relative standard \\
source & number &
& & & \hbox to -5pt {} uncertainty $u_{\rm r}$  \\
\colrule

$f_{\rm s}({\rm C})/f_{\rm c}({\rm C})$ \hbox to 20pt{} & $B17$ \fixh  &
GSI-02 & \ref{sssec:bsgfexps} \ (\ref{eq:arec02}) &
$ 0.000\,548\,579\,909\,32(29)$ \hbox to 20pt {} & $ 5.2\times 10^{-10}$\\

$f_{\rm s}({\rm O})/f_{\rm c}({\rm O})$ & $B18$ \fixh  &
GSI-02 & \ref{sssec:bsgfexps} \ (\ref{eq:areo02}) &
$ 0.000\,548\,579\,909\,57(42)$& $ 7.6\times 10^{-10}$\\

$\Delta \nu_{\bar{\rm p}\,{\rm He^+}} $ & \fixh $C16-C30$ &
CERN-06/10 & \ref{sssec:apheare} \ (\ref{eq:areaphe}) &
$ 0.000\,548\,579\,909\,14(75)$& $ 1.4\times 10^{-9}$ \\

$A_{\rm r}({\rm e})$ & $B11$ \fixh  &
UWash-95 & \ref{ssec:ptmare} \ (\ref{eq:arexp}) &
$ 0.000\,548\,579\,9111(12)$& $ 2.1\times 10^{-9}$ \rule[-6pt]{0pt}{6pt}\\

\botrule
\end{tabular}
\end{table*}

\begin{figure}
\rotatebox{-90}{\resizebox{!}{4.2in}{
\includegraphics[clip,trim= 10 40 40 10]{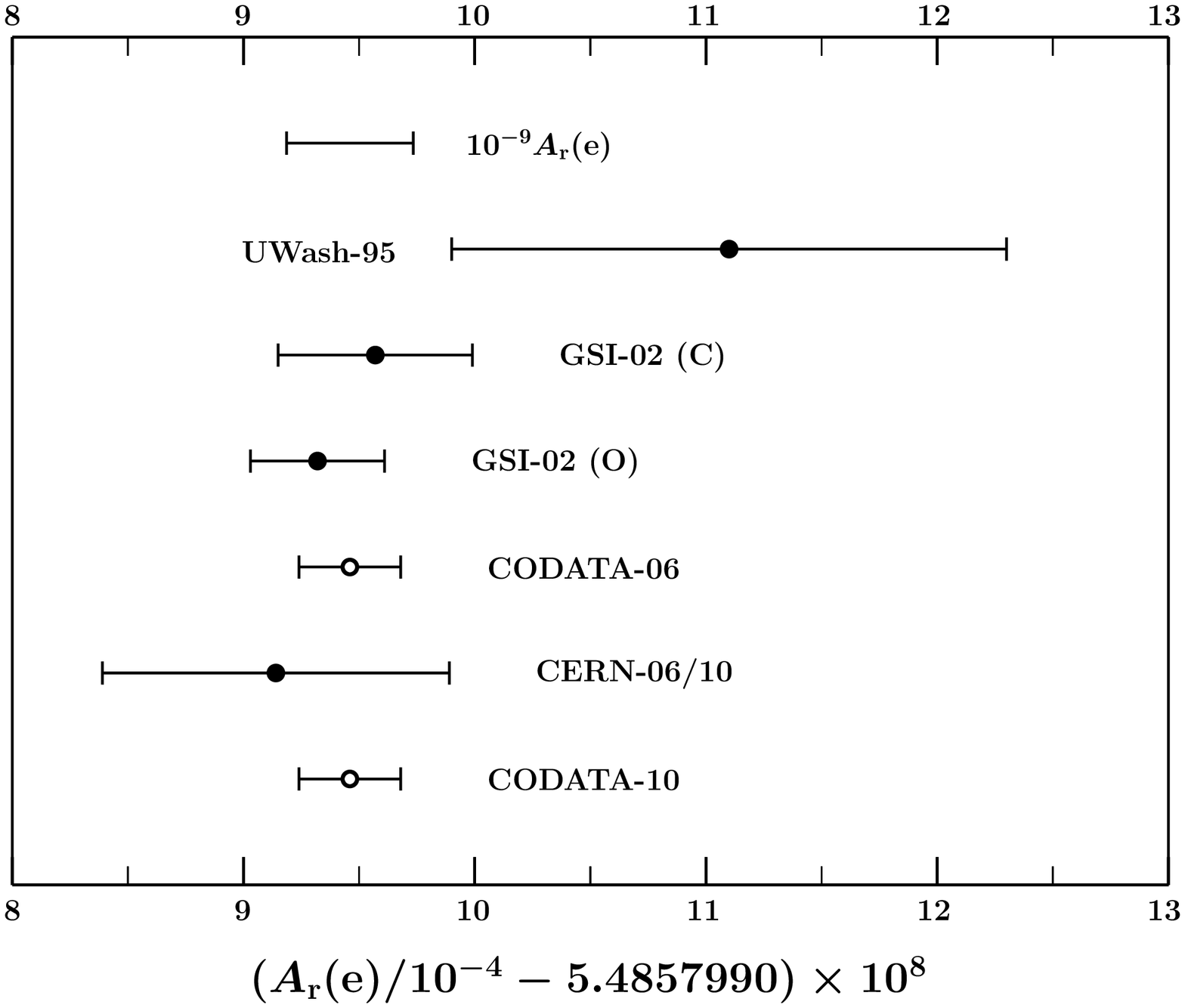}
}}
\caption{ Values of the electron relative atomic mass $A_{\rm r}({\rm e})$
implied by the input data in Tables~\ref{tab:pdata} and \ref{tab:cdata}
and the 2006 and 2010 CODATA recommended values in chronological order
from top to bottom (see Table~\ref{tab:are}).}
\label{fig:are}
\end{figure}

Finally, in Table~\ref{tab:are} and Fig.~\ref{fig:are} we compare four
values of $A_{\rm r}({\rm e})$ calculated from different input data as
indicated. They are in agreement, with all six differences less than
$2u_{\rm diff}$.  Further, since the four uncertainties are comparable,
all four of the source data are included in the final adjustment.

\subsection{Multivariate analysis of data}
\label{ssec:mada}
\vspace{.25 in}

Our multivariate analysis of the data employs a well known least-squares
method that takes correlations among the input data into account. Used
in the three previous adjustments, it is described in Appendix E of
CODATA-98 and references cited therein. We recall from that appendix
that a least-squares adjustment is characterized by the number of input
data $N$, number of variables or adjusted constants $M$, degrees of
freedom $\nu = N-M$, measure $\chi^2$, probability $p\,({\chi^{2}|\nu})$
of obtaining an observed value of $\chi^2$ that large or larger for the
given value of $\nu$, Birge ratio $R_{\rm B}=\sqrt{\chi^{2}/{\nu}}$, and
normalized residual of the $i$th input datum $r_i = (x_i-\langle x_i
\rangle)/u_i$, where $x_i$ is the input datum, $\langle x_i \rangle$ its
adjusted value, and $u_i$ its standard uncertainty.

The observational equations for the input data are given in
Tables~\ref{tab:pobseqsa}, \ref{tab:pobseqsb1}, and \ref{tab:pobseqsc}.
These equations are written in terms of a particular independent subset
of constants (broadly interpreted) called, as already noted,
\emph{adjusted constants}. These are the variables (or unknowns) of the
adjustment. The least-squares calculation yields values of the adjusted
constants that predict values of the input data through their
observational equations that best agree with the data themselves in the
least squares sense. The adjusted constants used in the 2010
calculations are  given in Tables~\ref{tab:adjcona}, \ref{tab:adjconb},
and \ref{tab:adjconc}.

The symbol $\doteq$ in an observational equation indicates that an input
datum of the type on the left-hand side is ideally given by the
expression on the right-hand side containing adjusted constants. But
because the equation is one of an overdetermined set that relates a
datum to adjusted constants, the two sides are not necessarily equal.
The best estimate of the value of an input datum is its observational
equation evaluated with the least-squares adjusted values of the
adjusted constants on which its observational equation depends. For some
input data such as $\delta_{\rm e}$ and $R$, the observational equation
is simply $\delta_{\rm e} \doteq \delta_{\rm e}$ and $R \doteq R$.

The binding energies $E_{\rm b}(X)/m_{\rm u}c^2$ in the observational
equations of Table~\ref{tab:pobseqsb1} are treated as fixed quantities
with negligible uncertainties, as are the bound-state $g$-factor ratios.
The frequency $f_{\rm p}$ is not an adjusted constant but is included in
the equation for data items $B30$ and $B31$ to indicate that they are
functions of $f_{\rm p}$. Finally, the observational equations for items
$B30$ and $B31$, which are based on
Eqs.~(\ref{eq:murat})-(\ref{eq:mumemump}) of Sec.~\ref{sssec:mufreqs},
include the function $a_{\rm e}(\alpha,\delta_{\rm e})$, as well as the
theoretical expression for input data of type $B29$, $\Delta\nu_{\rm
Mu}$. The latter expression is discussed in Sec.~\ref{sssec:muhfs} and
is a function of $R_\infty$, $\alpha$, $m_{\rm e}/m_{\rmssmu}$, and
$a_{\rmssmu}$.

The self-sensitivity coefficient $S_{\rm c}$ for an input datum is a
measure of the influence of a particular item of data on its
corresponding adjusted value.  As in previous adjustments, in general,
for an input datum to be included in the final adjustment on which the
2010 recommended values are based, its value of $S_{\rm c}$ must be
greater than 0.01, or $1\,\%$, which means that its uncertainty must be
no more than about a factor of 10 larger than the uncertainty of the
adjusted value of that quantity; see Sec.~I.D of CODATA-98 for the
justification of this $1\,\%$ cutoff.  However, the exclusion of a datum
is not followed if, for example, a datum with $S_{\rm c}< 0.01$ is part
of a group of data obtained in a given experiment, or series of
experiments, where most of the other data have self-sensitivity
coefficients greater than $0.01$.  It is also not followed for $G$,
because in this case there is substantial disagreement of some of the
data with the smallest uncertainties and hence relatively greater
significance of the data with larger uncertainties.

In summary, there is one major discrepancy among the data discussed in
this section: the disagreement of the NIST-07 watt balance value of
$K_{\rm J}^2R_{\rm K}$ and the IAC-11 enriched $^{28}$Si XRCD value of
$N_{\rm A}$, items $B37.3$ and $B54$ of Table~\ref{tab:pdata}.

\subsubsection{Data related to the Newtonian constant of gravitation $G$}
\label{sssec:calcncg}
\vspace{.25 in}

\begin{figure}
\rotatebox{-90}{\resizebox{!}{4.2in}{
\includegraphics[clip,trim=10 40 40 10]{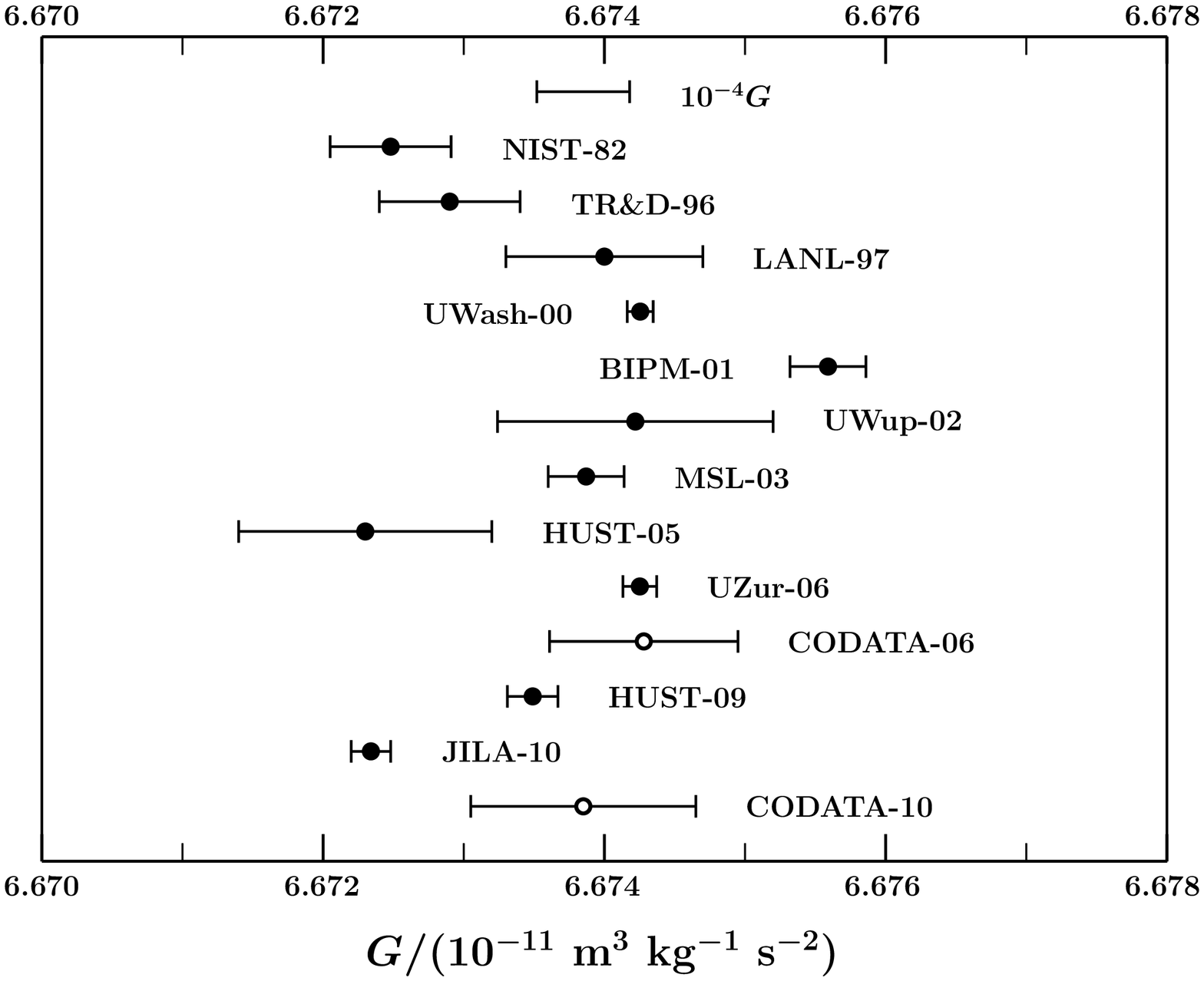}
}}
\caption{
Values of the Newtonian constant of gravitation $G$ in
Table~\ref{tab:bigg} and the 2006 and 2010 CODATA recommended values in
chronological order from top to bottom}
\label{fig:bigg}
\end{figure}

Our least-squares analysis of the input data begins with the 11 values
of $G$ in Table~\ref{tab:bigg}, which are graphically compared in
Fig.~\ref{fig:bigg}. (Because the $G$ data are independent of all other
data, they can be treated separately.) As discussed in
Secs.~\ref{sssec:lanl97} and \ref{sssec:hust09}, there are two
correlation coefficients associated with these data: $r(G1,G3) = 0.351$
and $r(G8,G10) = 0.234$. It is clear from both the table and figure that
the data are highly inconsistent.  Of the 55 differences among the 11
values, the three largest, $11.4u_{\rm diff}$, $10.7u_{\rm diff}$, and
$10.2u_{\rm diff}$ are between JILA-10 and UWash-00, BIPM-01, and
UZur-06, respectively. Further, eight range from $4u_{\rm diff}$ to
$7u_{\rm diff}$. The weighted mean of the 11 values has a relative
standard uncertainty of $8.6\times10^{-6}$. For this calculation, with
$\nu = 11-1=10$, we have $\chi^2=209.6$, $p\,(209.6|10)\approx0$, and
$R_{\rm B}= 4.58$. (Recall that a multivariate least-squares calculation
with only one variable is a weighted mean with covariances.) Five data
have normalized residuals $r_i > 2.0$: JILA-10, BIPM-01, UWash-00,
NIST-82, and UZur-06; their respective values are $-10.8$, $6.4$, $4.4$,
$-3.2$ and $3.2$.

Repeating the calculation using only the six values of $G$ with relative
uncertainties $\leq 4.0\times 10^{-5}$, namely, UWash-00, BIPM-01,
MSL-03, UZur-06, HUST-09, and JILA-10, has little impact: the value of
$G$ increases by the fractional amount $5.0\times10^{-6}$ and the
relative uncertainty increases to $8.8\times10^{-6}$; for this
calculation $\nu = 6-1=5$, $\chi^2 = 191.4$, $p\,(191.4|5)\approx0$, and
$R_{\rm B}= 6.19$; the values of $r_i$ are $4.0$, $6.3$, $-0.05$, $3.0$,
$-2.2$, and $-11.0$, respectively.

Taking into account the historic difficulty in measuring $G$ and the
fact that all $11$ values of $G$ have no apparent issue besides the
disagreement among them, the Task Group decided to take as the 2010
recommended value the weighted mean of the 11 values in
Table~\ref{tab:bigg} after each of their uncertainties
is multiplied by the factor 14. This yields
\begin{eqnarray}
G =  6.673\,84(80)~\mbox{m}^{3}~\mbox{kg}^{-1}~\mbox{s}^{-2}
\quad[ 1.2\times 10^{-4}] \, .
\label{eq:bg10}
\end{eqnarray}
The largest normalized residual, that of JILA-10, is now 0.77, and the
largest difference between values of $G$, that between JILA-10 and
UWash-00, is $0.82u_{\rm diff}$. For the calculation yielding the
recommended value, $\nu = 11-1=10$, $\chi^2 = 1.07$, $p\,(1.07|10) =
1.00$, and $R_{\rm B}= 0.33$.  In view of the significant scatter of the
measured values of $G$, the factor of 14 was chosen so that the smallest
and largest values would differ from the recommended value by about
twice its uncertainty; see Fig.~\ref{fig:bigg}.  The 2010 recommended
value represents a fractional decrease in the 2006 value of $0.66\times
10^{-4}$ and an increase in uncertainty of $20\,\%$.

\subsubsection{Data related to all other constants}
\label{sssec:idd}
\vspace{.25 in}

Tables~\ref{tab:adjustsall}, \ref{tab:adjres}, and \ref{tab:adjustsa}
summarize 12 least-squares analyses, discussed in the following
paragraphs, of the input data and correlation coefficients in
Tables~\ref{tab:rdata} to \ref{tab:cdcc}.  Because the adjusted value of
$R_\infty $ is essentially the same for all five adjustments summarized
in Table~\ref{tab:adjustsall} and equal to that of adjustment 3 of
Table~\ref{tab:adjustsa}, the values are not listed in
Table~\ref{tab:adjustsall}. (Note that adjustment 3 in
Tables~\ref{tab:adjustsall} and \ref{tab:adjustsa} is the same
adjustment.)

\emph{Adjustment 1}. The initial adjustment includes all of the input
data, three of which have normalized residuals whose absolute magnitudes
are problematically greater than 2; see Table~\ref{tab:adjres}.  They
are the 2007 NIST watt-balance result for $K^2_{\rm J}R_{\rm K}$, the
2011 IAC enriched silicon XRCD result for $N_{\rm A}$, and the 1989 NIST
result for ${\it\Gamma}^\prime_{\rm p-90}$(lo). All other input data
have values of $|r_i|$ less than 2, except those for two antiprotonic
$^3{\rm He}$ transitions, data items $C25$ and $C27$ in
Table~\ref{tab:cdata}, for which $r_{25}= 2.12$ and $r_{27} = 2.10$.
However, the fact that their normalized residuals are somewhat greater
than 2 is not a major concern, because their self-sensitivity
coefficients $S_{\rm c}$ are considerably less than 0.01. In this
regard, we see from Table~\ref{tab:adjres} that two of the three
inconsistent data have values of $S_{\rm c}$ considerably larger than
0.01; the exception is ${\it\Gamma}^\prime_{\rm p-90}$(lo) with $S_{\rm
c}= 0.0096$, which is rounded to 0.010 in the table.

\emph{Adjustment 2}. The difference in the IAC-11 and NIST-07 values of
$h$ (see first two lines of Table~\ref{tab:plancks}) is 3.8$u_{\rm
diff}$, where as before $u_{\rm diff}$ is the standard uncertainty of
the difference. To reduce the difference between these two highly
credible results to an acceptable level, that is, to 2$u_{\rm diff}$ or
slightly below, the Task Group decided that the uncertainties used in
the adjustment for these data would be those in the
Table~\ref{tab:pdata} multiplied by a factor of two.  It was also
decided to apply the same factor to the uncertainties of all the data
that contribute in a significant way to the determination of $h$, so
that the relative weights of this set of data are unchanged.  (Recall
that if the difference between two values of the same quantity is
$au_{\rm diff}$ and the uncertainty of each is increased by a factor
$b$, the difference is reduced to $(a/b)u_{\rm diff}$.)  Thus,
adjustment 2 differs from adjustment 1 in that the uncertainties of data
items $B36.1$, $B36.2$, $B37.1$ to $B37.5$, and $B54$ in
Table~\ref{tab:pdata}, which are the two values of $K_{\rm J}$, the five
values of $K^2_{\rm J}R_{\rm K}$, and the value of $N_{\rm A}$, are
increased by a factor of 2. (Although items $B31.1$, $B31.2$, and $B38$,
the two values of ${\it\Gamma}^\prime_{\rm p-90}$(hi) and ${\cal
F}_{90}$, also contribute to the determination of $h$, their
contribution is small and no multiplicative factor is applied.)

From Tables~\ref{tab:adjustsall} and \ref{tab:adjres} we see that the
values of $\alpha$ and $h$ from adjustment 2 are very nearly the same as
from adjustment 1, that $|r_i|$ for both $B37.3$ and $B54$ have been
reduced to below 1.4, and that the residual for ${\it\Gamma}^\prime_{\rm
p-90}$(lo) is unchanged.

\emph{Adjustment 3}. Adjustment 3 is the adjustment on which the 2010
CODATA recommended values are based, and as such it is referred to as
the ``final adjustment.''.  It differs from adjustment 2 in that,
following the prescription described above, 18 input data with values of
$S_{\rm c}$ less than 0.01 are deleted.  These are data items $B13.1$,
$B32.1$ to $B36.2$, $B37.5$, $B38$, $B56$, $B59$, and $B60$ in
Table~\ref{tab:pdata}. (The range in values of $S_{\rm c}$ for the
deleted data is 0.0003 to 0.0097, and no datum with a value of $S_{\rm
c} > 1$ was ``converted'' to a value with $S_{\rm c} < 1$ due to the
multiplicative factor.) Further, because $h/m(^{133}{\rm Cs})$, item
$B56$, is deleted as an input datum due to its low weight, the two
values of $A_{\rm r}(^{133}{\rm Cs})$, items $B10.1$ and $10.2$, which
are not relevant to any other input datum, are also deleted and $A_{\rm
r}(^{133}{\rm Cs})$ is omitted as an adjusted constant. This brings the
total number of omitted data items to 20.  Table~\ref{tab:adjustsall}
shows that deleting them has virtually no impact on the values of
$\alpha$ and $h$ and Birge ratio $R_{\rm B}$.  The data for the final
adjustment are quite consistent, as demonstrated by the value of
$\chi^2$: $p\,(58.1|67) = 0.77$.

\emph{Adjustments 4 and 5}. The purpose of these adjustments is to test
the robustness of the 2010 recommended values of $\alpha$ and $h$ by
omitting the most accurate data relevant to these constants.  Adjustment
4 differs from adjustment 2 in that the four data that provide values of
$\alpha$ with the smallest uncertainties are deleted, namely, items
$B13.1$, $B13.2$, $B56$ and $B57$, the two values of $a_{\rm e}$ and the
values of $h/m(^{133}{\rm Cs})$ and $h/m(^{87}{\rm Rb})$; see the first
four entries of Table~\ref{tab:alpha}. (For the same reason as in
adjustment 3, in adjustment 4 the two values of $A_{\rm r}(^{133}{\rm
Cs})$ are also deleted as input data and $A_{\rm r}(^{133}{\rm Cs})$ is
omitted as an adjusted constant; the same applies to $A_{\rm
r}(^{87}{\rm Rb})$.) Adjustment 5 differs from adjustment 1 in that the
three data that provide values of $h$ with the smallest uncertainties
are deleted, namely, items $B37.2$, $B37.3$, and $B54$, the two NIST
values of $K^2_{\rm J}R_{\rm K}$ and the IAC value of $N_{\rm A}$; see
the first three entries of Table~\ref{tab:plancks}.  Also deleted are
the data with $S_{\rm c} < 0.01$ that contribute in a minimal way to the
determination of $\alpha$ and are deleted in the final adjustment.
Table~\ref{tab:adjustsall} shows that the value of $\alpha$ from the
less accurate $\alpha$-related data used in adjustment 4, and the value
of $h$ from the less accurate $h$-related data used in adjustment 5,
agree with the corresponding recommended values from adjustment 3.  This
agreement provides a consistency check on the 2010 recommended values.

\emph{Adjustments 6 to 12}. The aim of the seven adjustments summarized
in Table~\ref{tab:adjustsa} is to investigate the data that determine the
recommended values of $R_{\infty}$, $r_{\rm p}$, and $r_{\rm d}$.
Results from adjustment 3, the final adjustment, are included in the
table for reference purposes. We begin with a discussion of adjustments
6 to 10, which are derived from adjustment 3 by deleting selected input
data. We then discuss adjustments 11 and 12, which examine the impact of
the value of the proton rms charge radius derived from the measurement
of the Lamb shift in muonic hydrogen discussed in Sec.~\ref{par:muhrad}
and given in Eq.~(\ref{eq:rpmuhuj}). Note that the
value of $R_{\infty}$ depends only weakly on the data in
Tables~\ref{tab:pdata} and \ref{tab:cdata}.

In adjustment 6, the electron scattering values of $r_{\rm p}$ and
$r_{\rm d}$, data items $A49.1$, $A49.2$, and $A50$ in
Table~\ref{tab:rdata}, are not included.  Thus, the values of these two
quantities from adjustment 6 are based solely on H and D spectroscopic
data. It is evident from a comparison of the results of this adjustment
and adjustment 3 that the scattering values of the radii play a smaller
role than the spectroscopic data in determining the 2010 recommended
values of $R_{\infty}$, $r_{\rm p}$ and $r_{\rm d}$.

Adjustment 7 is based on only hydrogen data, including the two
scattering values of $r_{\rm p}$ but not the difference between the
$1{\rm S_{1/2}}-2{\rm S_{1/2}}$ transition frequencies in H and D, item
$A48$ in Table~\ref{tab:rdata}, hereafter  referred to as the ``isotope
shift.''  Adjustment 8 differs from adjustment 7 in that the two
scattering values of $r_{\rm p}$ are deleted.  Adjustments 9 and 10 are
similar to 7 and 8 but are based on only deuterium data; that is,
adjustment 9 includes the scattering value of $r_{\rm d}$ but not the
isotope shift, while for adjustment 10 the scattering value is deleted.
The results of these four adjustments show the dominant role of the
hydrogen data and the importance of the isotope shift in determining the
recommended value of $r_{\rm d}$.  Further, the four values of
$R_\infty$ from these adjustments agree with the 2010 recommended value,
and the two values of $r_{\rm p}$ and of $r_{\rm d}$ also agree with
their respective recommended values: the largest difference from the
recommended value for the eight results is $1.4u_{\rm diff}$.

Adjustment 11 differs from adjustment 3 in that it includes the muonic
hydrogen value  $r_{\rm p}= 0.841\,69(66)~{\rm fm}$, and adjustment 12
differs from  adjustment 11 in that the three scattering values of the
nuclear radii are deleted.  Because the muonic hydrogen value is
significantly smaller and has a significantly smaller uncertainty than
the purely spectroscopic value of adjustment 6 and the two scattering
values, it has a major impact on the results of adjustments 11 and 12,
as can be seen from Table~\ref{tab:adjustsa}: for both adjustments the
value of $R_\infty$ shifts down by over 6 standard deviations and its
uncertainty is reduced by a factor of 4.6. Moreover, and not
surprisingly, the values of $r_{\rm p}$ and of $r_{\rm d}$ from both
adjustments are significantly smaller than the recommended values and
have significantly smaller uncertainties. The inconsistencies between
the muonic hydrogen result for $r_{\rm p}$ and the spectroscopic and
scattering results is demonstrated by the large value and low
probability of $\chi^2$ for adjustment 11; $p\,(104.9|68) = 0.0027$.

The impact of the muonic hydrogen value of $r_{\rm p}$ can also be seen
by examining for adjustments 3, 11, and 12 the normalized residuals and
self- sensitivity coefficients of the principal experimental data that
determine $R_\infty$, namely, items $A26$ to $A50$ of
Table~\ref{tab:rdata}. In brief, $|r_i|$ for these data in the final
adjustment range from near 0 to 1.24 for item $A50$, the $r_{\rm d}$
scattering result, with the vast majority being less than 1. For the
three greater than 1, $|r_i|$ is 1.03, 1.08, and 1.04. The value of
$S_{\rm c}$ is 1.00 for items $A26$ and $A48$, the hydrogen $1{\rm
S}_{1/2}-2{\rm S}_{1/2}$ transition frequency and the H-D isotope shift;
and 0.42 for item $A49.2$, which is the more accurate of the two
scattering values of $r_{\rm p}$. Most others are a few percent,
although some values of $S_{\rm c}$ are near 0.  The situation is
markedly different for adjustment 12. First, $|r_i|$ for item $A30$, the
hydrogen transition frequency involving the $8{\rm D}_{5/2}$ state, is
3.06 compared to 0.87 in adjustment 3; and items $A41$, $A42$, and
$A43$, deuterium transitions involving the $8{\rm S}_{1/2}$,  $8{\rm
D}_{3/2}$, and $8{\rm D}_{5/2}$ states, are now 2.5, 2.4, and 3.0,
respectively, compared to 0.40, 0.17, and 0.68. Further, ten other
transitions have residuals in the range 1.02 to 1.76.  As a result, with
this proton radius, the predictions of the theory for hydrogen and
deuterium transition frequencies are not generally consistent with the
experiments.  Equally noteworthy is the fact that although $S_{\rm c}$
for items $A26$ and $A48$ remain equal to 1.00, for all other transition
frequencies $S_{\rm c}$ is less than 0.01, which means that they play an
inconsequential role in determining $R_\infty$.  The results for
adjustment 11, which includes the scattering values of the nuclear radii
as well as the muonic hydrogen value, are similar.

In view of the impact of the latter value on the internal consistency of
the $R_\infty$ data and its disagreement with the spectroscopic and
scattering values, the Task Group decided that it was premature to
include it as an input datum in the 2010 CODATA adjustment; it was
deemed more prudent to wait to see if further research can resolve the
discrepancy.  See Sec.~\ref{par:muhrad} for additional discussion.

\subsubsection{Test of the Josephson and quantum Hall effect relations}
\label{sssec:epstests}
\vspace{.25 in}

As in CODATA-02 and CODATA-06, the exactness of the relations $K_{\rm J}
= 2e/h$ and $R_{\rm K} = h/e^2$ is investigated by writing
\begin{eqnarray}
K_{\rm J} &=&\frac{2e}{h}\left( {1+\varepsilon _{\rm J} } \right)=\left(
{\frac{8\alpha }{\mu _0 ch}} \right)^{1/2}\left( {1+\varepsilon _{\rm J} }
\right),
\label{eq:kjeps}
\\
R_{\rm K} &=&\frac{h}{e^2}\left( {1+\varepsilon_{\rm K} } \right)=\frac{\mu
_0 c}{2\alpha }\left( {1+\varepsilon_{\rm K} } \right),
\label{eq:rkeps}
\end{eqnarray}
where $\varepsilon_{\rm J}$ and $\varepsilon_{\rm K}$ are unknown
correction factors taken to be additional adjusted constants. Replacing
the relations $K_{\rm J} = 2e/h$ and $R_{\rm K} =h/e^2$ in the analysis
leading to the observational equations in Table~\ref{tab:pobseqsb1} with
the generalizations in Eqs.~(\ref{eq:kjeps}) and (\ref{eq:rkeps}) leads
to the modified observational equations given in
Table~\ref{tab:pobseqseps}.

Although the NIST value of $k/h$, item $B60$, was obtained using the
Josephson and quantum Hall effects, it is not included in the tests of
the relations $K_{\rm J} = 2e/h$ and $R_{\rm K} = h/e^2$, because of its
large uncertainty.

The results of seven different adjustments are summarized in
Table~\ref{tab:epsilons}. An entry of 0 in the $\varepsilon_{\rm K}$
column means that it is assumed that $R_{\rm K} = h/e^2$ in the
corresponding adjustment; similarly, an entry of 0 in the
$\varepsilon_{\rm J}$ column means that it is assumed that $K_{\rm J} =
2e/h$ in the corresponding adjustment. The following comments apply to
the adjustments of Table~\ref{tab:epsilons}.

Adjustment (i) uses all of the data and thus differs from adjustment 1
of Table~\ref{tab:adjustsall} discussed in the previous section only in
that the assumption $K_{\rm J} = 2e/h$ and $R_{\rm K} = h/e^2$ is
relaxed. For this adjustment, $\nu=86$, $\chi ^2 = 78.1$, and $R_{\rm
B}=1.02$. The normalized residuals $r_i$ for the three inconsistent data
items in Table~\ref{tab:adjres}, the companion table to
Table~\ref{tab:adjustsall}, are $0.75$, $-0.56$, and $2.88$.
Examination of Table~\ref{tab:epsilons} shows that $\epsilon _{\rm K}$
is consistent with 0 within 1.2 times its uncertainty of $1.8\times
10^{-8}$, while $\epsilon _{\rm J}$ is consistent with 0 within 2.4
times its uncertainty of $5.7\times 10^{-8}$.

\begin{figure}
\rotatebox{-90}{\resizebox{!}{4.2in}{
\includegraphics[clip,trim= 10 40 40 10]{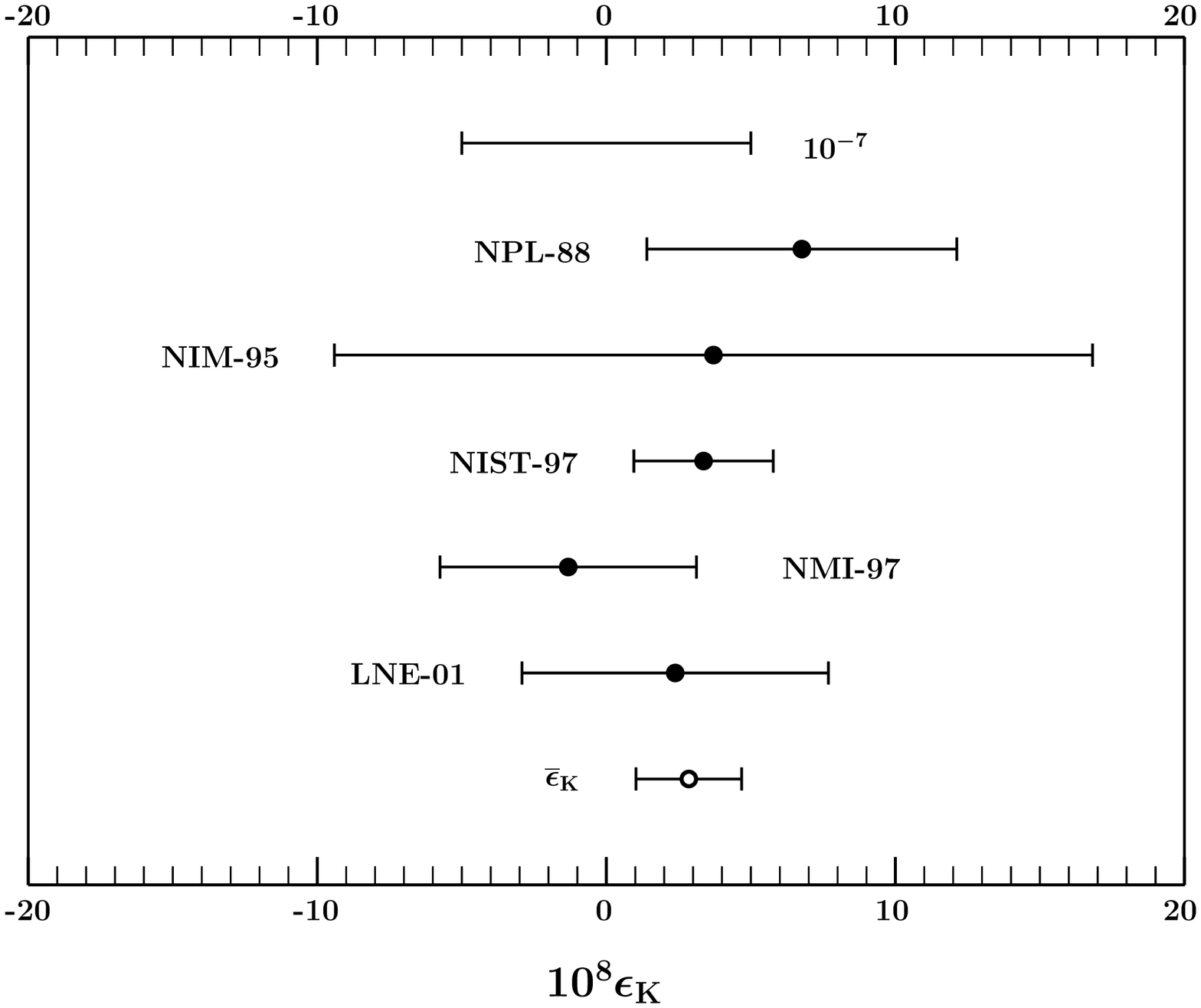}
}}
\caption{
Comparison of the five individual values of $\epsilon _{\rm K}$ obtained
from the five values of $R_{\rm K}$, data items $B35.1$ to $B35.5$, and
the combined value (open circle) from adjustment (ii) given in
Table~\ref{tab:epsilons}. The applicable observational equation in
Table~\ref{tab:pobseqseps} is $B35^*$.
}
\label{fig:epskr}
\end{figure}

\begin{figure}
\rotatebox{-90}{\resizebox{!}{4.2in}{
\includegraphics[clip,trim= 10 40 40 10]{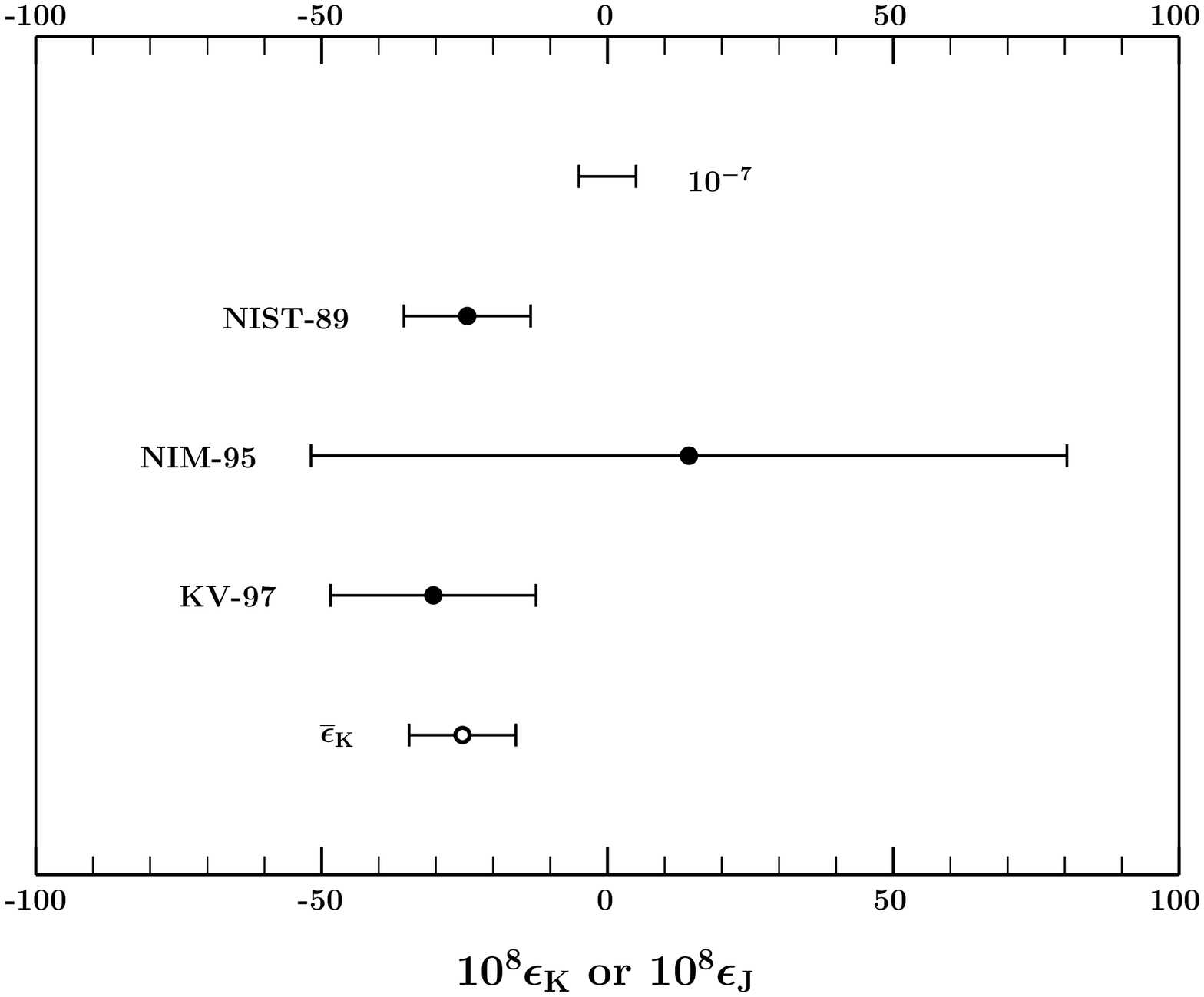}
}}
\caption{
Comparison of the three individual values of $\epsilon _{\rm K}$
obtained from the three low-field gyromagnetic ratios, data items
$B32.1$, $B32.2$, and $B33$, and the combined value (open circle) from
adjustment (iii) given in Table~\ref{tab:epsilons}.  The applicable
observational equations in Table~\ref{tab:pobseqseps} are $B32^*$ and
$B33^*$. Because of the form of these equations, the value of $\epsilon
_{\rm K}$ when $\epsilon _{\rm J} =0$ is identical to the value of
$\epsilon _{\rm J}$ when $\epsilon _{\rm K} =0$, hence the label at the
bottom of the figure.}
\label{fig:epskg}
\end{figure}

It is important to recognize that any conclusions that can be drawn from
the values of $\varepsilon_{\rm K}$ and $\varepsilon_{\rm J}$ of
adjustment (i) must be tempered, because not all of the individual
values of $\varepsilon_{\rm K}$ and $\varepsilon_{\rm J}$ that
contribute to their determination are consistent.  This is demonstrated
by adjustments (ii) to (vii) and Figs.~\ref{fig:epskr} and
\ref{fig:epskg}.  (Because of their comparatively small uncertainties,
it is possible in these adjustments to take the 2010 recommended values
for the constants $a_{\rm e}$, $\alpha$, $R_{\infty}$, and $A_{\rm
r}(\rm e)$, which appear in the observational equations of
Table~\ref{tab:pobseqseps}, and assume that they are exactly known.)

Adjustments (ii) and (iii) focus on $\epsilon _{\rm K}$: $\epsilon _{\rm
J}$ is set equal to 0 and values of $\epsilon _{\rm K}$ are obtained
from data whose observational equations are independent of $h$.  These
data are the five values of $R_{\rm K}$, items $B35.1$ to $B35.5$; and
the three low-field gyromagnetic ratios,  items $B32.1$, $B32.2$, and
$B33$. We see from Table~\ref{tab:epsilons} that the two values of
$\epsilon _{\rm K}$ resulting from the two adjustments not only have
opposite signs but their difference is $3.0u_{\rm diff}$.
Figure~\ref{fig:epskr} compares the combined value of $\epsilon _{\rm
K}$ obtained from the five values of $R_{\rm K}$ with the five
individual values, while Fig.~\ref{fig:epskg} does the same for the
results obtained from the three gyromagnetic ratios.

Adjustments (iv) to (vii) focus on $\epsilon _{\rm J}$: $\epsilon _{\rm
K}$ is set equal to 0 and values of $\epsilon _{\rm J}$ are, with the
exception of adjustment (iv), obtained from data whose observational
equations are dependent on $h$. Examination of Table~\ref{tab:epsilons}
shows that although the values of $\epsilon _{\rm J}$ from adjustments
(iv) and (v) are of opposite sign, their difference of $49.1\times
10^{-8}$ is less than the $72.0\times 10^{-8}$ uncertainty of the
difference.  However, the difference between the values of $\epsilon
_{\rm J}$ from adjustments (iv) and (vi) is $3.6u_{\rm diff}$, and is
$3.3u_{\rm diff}$ even for the value of $\epsilon _{\rm J}$ from
adjustment (vii), in which the uncertainties of the most accurate data
have been increased by the factor 2. (The multiplicative factor 2 is
that used in adjustment 2 and the final adjustment; see
Tables~\ref{tab:adjres}, \ref{tab:adjustsall}, and their associated
text.) On the other hand, we see that the value of $\epsilon _{\rm J}$
from adjustment (vi) is consistent with 0 only to within 3.9 times its
uncertainty, but that this is reduced to 2.0 for the value of $\epsilon
_{\rm J}$ from adjustment (vii) which uses expanded uncertainties.

The results of the adjustments discussed above reflect the disagreement
of the NIST-07 watt-balance value for $K_{\rm J}^2R_{\rm K}$, and to a
lesser extent that of the similar NIST-98 value, items $B37.2$ and
$B37.3$, with the IAC-11 enriched silicon value of $N_{\rm A}$, item
$B54$; and the disagreement of the NIST-89 result for  ${\it
\Gamma}^\prime_{p-90}$(lo), and to a lesser extent the KR/VN-98 result
for ${\it \Gamma}_{\rm h-90}^{\,\prime}({\rm lo})$, items $B32.1$ and
$B33$, with the highly accurate values of $\alpha$. If adjustment 1 is
repeated with these five data deleted, we find $\varepsilon_{\rm K} =
2.8(1.8)\times 10^{-8}$ and $\varepsilon_{\rm J} = 15(49)\times
10^{-8}$.  These values can be interpreted as confirming that
$\varepsilon_{\rm K}$ is consistent with 0 to within 1.6 times its
uncertainty of $1.8\times10^{-8}$ and that $\varepsilon_{\rm J}$ is
consistent with 0 well within its uncertainty of $49\times10^{-8}$.

We conclude this section by briefly discussing recent efforts to close
what is called the ``metrology triangle.''  Although there are variants,
the basic idea is to use a single electron tunneling (SET) device that
generates a quantized current $I = ef$ when an alternating voltage of
frequency $f$ is applied to it, where as usual $e$ is the elementary
charge. The current $I$ is then compared to a current derived from
Josephson and quantum Hall effect devices.  In view of quantization of
charge in units of $e$ and conservation of charge, the equality of the
currents shows that $K_{\rm J}R_{\rm K}e = 2$, as expected, within the
uncertainty of the measurements \cite{2008009,2008104,2009143}.
Although there is no indication from the results reported to date that
this relation is not valid, the uncertainties of the results are at best
at the 1 to 2 parts in $10^6$ level
\cite{2007276,2008104,2011149,2012001}.

\def\fixh{\vbox to 12pt {}}
\def\sp#1{\hbox to #1 pt {}}
\def\t{@{\sp{20}}}
\begin{table*}
\caption{
Summary of the results of several least-squares adjustments to
investigate the relations $K_{\rm J} = (2e/h)(1 + \epsilon _{\rm J})$
and $R_{\rm K} = (h/e^2)(1 + \epsilon _{\rm K})$.  See the text for an
explanation and discussion of each adjustment, but in brief, adjustment
(i) uses all the data, (ii) assumes $K_{\rm J} = 2e/h$ (that is,
$\epsilon _{\rm J} =0$) and obtains $\epsilon _{\rm K}$ from the five
measured values of $R_{\rm K}$, (iii) is based on the same assumption
and obtains $\epsilon _{\rm K}$ from the two values of the proton
gyromagnetic ratio and one value of the helion gyromagnetic ratio, (iv)
is (iii) but assumes $R_{\rm K} = h/e^2$ (that is, $\epsilon _{\rm K}
=0$) and obtains $\epsilon _{\rm J}$ in place of $\epsilon _{\rm K}$,
(v) to (vii) are based on the same assumption and obtain $\epsilon _{\rm
J}$ from all the measured values given in Table~\ref{tab:pdata} for the
quantities indicated.
}
\label{tab:epsilons}
\begin{tabular}{c @{\sp{20}} c @{\sp{20}} c @{\sp{20}} c }\\
\toprule
\fixh Adj. &  Data included$^1$ & $10^8\varepsilon_{\rm K}$ &
  $10^8\varepsilon_{\rm J}$ \\
\colrule
\fixh \ (i)   & All                                  & $2.2(1.8)$ & $5.7(2.4)$
  \\
\fixh \ (ii)  & $R_{\rm K}$                          & $2.6(1.8)$   & $ 0 $ \\
\fixh \ (iii) & ${\it \Gamma}^\prime_{\rm p,h-90}({\rm lo})$  & $-25.4(9.3)$ &
  $0$ \\
\fixh \ (iv)  & ${\it \Gamma}^\prime_{\rm p,h-90}({\rm lo})$  & $0$ &
  $-25.4(9.3) $\\
\fixh \ (v)   & ${\it \Gamma}^\prime_{\rm p-90}({\rm hi}),K_{\rm J},K_{\rm
J}^{\rm 2}\!R_{\rm K},{\cal F}_{{\rm 90}} $   & $0$ & $23.7(72.0)$\\
\fixh \ (vi)  & ${\it \Gamma}^\prime_{\rm p-90}({\rm hi}),K_{\rm J},K_{\rm
J}^{\rm 2}\!R_{\rm K},{\cal F}_{{\rm 90}},\!N_{\rm A} $   & $0$ &
$8.6(2.2)$\\
\fixh \ (vii) & ${\it \Gamma}^\prime_{\rm p-90}({\rm hi}),[K_{\rm J}],[K_{\rm
J}^{\rm 2}\!R_{\rm K}],{\cal F}_{{\rm 90}},\![N_{\rm A}] $& $0$ &
$8.6(4.4)$\\
\botrule
\end{tabular}
\hbox to 10 cm {}
\hbox to 10 cm {$^1$The data items in brackets have their uncertainties
expanded by a factor of two.}
\end{table*}

\def\fixh{\vbox to 9pt {}}
\begin{table}
\caption{The 28 adjusted constants
(variables) used in the least-squares multivariate analysis of
the Rydberg-constant data given in Table~\ref{tab:rdata}.
These adjusted constants appear as arguments
of the functions on the right-hand side of
the observational equations of Table~\ref{tab:pobseqsa}.
}
\label{tab:adjcona}
\begin{tabular}{l@{\hbox to 18 pt{}}l}
\toprule
Adjusted constant & Symbol \\
\colrule
 \fixh  Rydberg constant & $R_\infty$ \\
 \fixh  bound-state proton rms charge radius & $r_{\rm p}$ \\
 \fixh  bound-state deuteron rms charge radius~~~~~ & $r_{\rm d}$ \\
 \fixh  additive correction to $E_{\rm H}(1{\rm S}_{1/2})/h$ & $\delta_{\rm H}
  (1{\rm S}_{1/2})$ \\
 \fixh  additive correction to $E_{\rm H}(2{\rm S}_{1/2})/h$ & $\delta_{\rm H}
  (2{\rm S}_{1/2})$ \\
 \fixh  additive correction to $E_{\rm H}(3{\rm S}_{1/2})/h$ & $\delta_{\rm H}
  (3{\rm S}_{1/2})$ \\
 \fixh  additive correction to $E_{\rm H}(4{\rm S}_{1/2})/h$ & $\delta_{\rm H}
  (4{\rm S}_{1/2})$ \\
 \fixh  additive correction to $E_{\rm H}(6{\rm S}_{1/2})/h$ & $\delta_{\rm H}
  (6{\rm S}_{1/2})$ \\
 \fixh  additive correction to $E_{\rm H}(8{\rm S}_{1/2})/h$ & $\delta_{\rm H}
  (8{\rm S}_{1/2})$ \\
 \fixh  additive correction to $E_{\rm H}(2{\rm P}_{1/2})/h$ & $\delta_{\rm H}
  (2{\rm P}_{1/2})$ \\
 \fixh  additive correction to $E_{\rm H}(4{\rm P}_{1/2})/h$ & $\delta_{\rm H}
  (4{\rm P}_{1/2})$ \\
 \fixh  additive correction to $E_{\rm H}(2{\rm P}_{3/2})/h$ & $\delta_{\rm H}
  (2{\rm P}_{3/2})$ \\
 \fixh  additive correction to $E_{\rm H}(4{\rm P}_{3/2})/h$ & $\delta_{\rm H}
  (4{\rm P}_{3/2})$ \\
 \fixh  additive correction to $E_{\rm H}(8{\rm D}_{3/2})/h$ & $\delta_{\rm H}
  (8{\rm D}_{3/2})$ \\
 \fixh  additive correction to $E_{\rm H}(12{\rm D}_{3/2})/h$ & $\delta_{\rm H}
  (12{\rm D}_{3/2})$ \\
 \fixh  additive correction to $E_{\rm H}(4{\rm D}_{5/2})/h$ & $\delta_{\rm H}
  (4{\rm D}_{5/2})$ \\
 \fixh  additive correction to $E_{\rm H}(6{\rm D}_{5/2})/h$ & $\delta_{\rm H}
  (6{\rm D}_{5/2})$ \\
 \fixh  additive correction to $E_{\rm H}(8{\rm D}_{5/2})/h$ & $\delta_{\rm H}
  (8{\rm D}_{5/2})$ \\
 \fixh  additive correction to $E_{\rm H}(12{\rm D}_{5/2})/h$ & $\delta_{\rm H}
  (12{\rm D}_{5/2})$ \\
 \fixh  additive correction to $E_{\rm D}(1{\rm S}_{1/2})/h$ & $\delta_{\rm D}
  (1{\rm S}_{1/2})$ \\
 \fixh  additive correction to $E_{\rm D}(2{\rm S}_{1/2})/h$ & $\delta_{\rm D}
  (2{\rm S}_{1/2})$ \\
 \fixh  additive correction to $E_{\rm D}(4{\rm S}_{1/2})/h$ & $\delta_{\rm D}
  (4{\rm S}_{1/2})$ \\
 \fixh  additive correction to $E_{\rm D}(8{\rm S}_{1/2})/h$ & $\delta_{\rm D}
  (8{\rm S}_{1/2})$ \\
 \fixh  additive correction to $E_{\rm D}(8{\rm D}_{3/2})/h$ & $\delta_{\rm D}
  (8{\rm D}_{3/2})$ \\
 \fixh  additive correction to $E_{\rm D}(12{\rm D}_{3/2})/h$ & $\delta_{\rm D}
  (12{\rm D}_{3/2})$ \\
 \fixh  additive correction to $E_{\rm D}(4{\rm D}_{5/2})/h$ & $\delta_{\rm D}
  (4{\rm D}_{5/2})$ \\
 \fixh  additive correction to $E_{\rm D}(8{\rm D}_{5/2})/h$ & $\delta_{\rm D}
  (8{\rm D}_{5/2})$ \\
 \fixh  additive correction to $E_{\rm D}(12{\rm D}_{5/2})/h$ & $\delta_{\rm D}
  (12{\rm D}_{5/2})$ \\
\botrule
\end{tabular}
\end{table}

\def\tfrac#1#2{{\vbox to 9pt {}\phantom{_I}\textstyle{#1}\over\vbox to 9pt {}
  \textstyle{#2}}}
\def\vh{\vbox to 15pt {}}
\def\vhh{\vbox to 10pt {}}
\def\hsp{\hbox to 18pt{}}
\begin{table*}
\caption{Observational equations that express the input data related to
$R_\infty$ in Table~\ref{tab:rdata} as functions of the adjusted
constants in Table~\ref{tab:adjcona}.  The numbers in the first column
correspond to the numbers in the first column of Table~\ref{tab:rdata}.
Energy levels of hydrogenic atoms are discussed in Sec.~\ref{ssec:ryd}.
As pointed out at the beginning of that section, $E_{X}(n{\rm L}_j)/h$
is in fact proportional to $cR_\infty$ and independent of $h$, hence $h$
is not an adjusted constant in these equations.  See
Sec.~\ref{ssec:mada} for an explanation of the symbol $\doteq$.}
\label{tab:pobseqsa}
\begin{tabular}{l@{\hsp}rcll}
\toprule
Type of input & &
\multicolumn{2}{l}{Observational equation}& \\
datum      & & &  \\
\colrule

\vh$A1$--$A16 \quad $&$ \delta_{\rm H}(n{\rm L}_j) $&$\doteq$&$ \delta_{\rm H}
  (n{\rm L}_j) $ \\

\vh$A17$--$A25 \quad $&$ \delta_{\rm D}(n{\rm L}_j) $&$\doteq$&$ \delta_{\rm D}
  (n{\rm L}_j) $ \\

\vh$A26$--$A31 \quad $
&$ \nu_{\rm H}(n_1{\rm L_1}_{j_1} - n_2{\rm L_2}_{j_2}) $&$\doteq$&$
\big[E_{\rm H}\big(n_2{\rm L_2}_{j_2};R_\infty,\alpha,A_{\rm r}({\rm e}),A_{\rm
  r}({\rm p}),r_{\rm p},\delta_{\rm H}(n_2{\rm L_2}_{j_2})\big) $  \\
\vhh$A38,A39$&&&$-E_{\rm H}\big(n_1{\rm L_1}_{j_1};R_\infty,\alpha,A_{\rm r}
  ({\rm e}),A_{\rm r}({\rm p}),r_{\rm p},\delta_{\rm H}(n_1{\rm L_1}_{j_1}
  )\big)\big]/h $  \\

\vh$A32$--$A37 \quad $
&$ \nu_{\rm H}(n_1{\rm L_1}_{j_1} - n_2{\rm L_2}_{j_2})
-\fr{1}{4}\nu_{\rm H}(n_3{\rm L_3}_{j_3} - n_4{\rm L_4}_{j_4})$&$\doteq$&$
\Big\{E_{\rm H}\big(n_2{\rm L_2}_{j_2};R_\infty,\alpha,A_{\rm r}({\rm e}),A_{\rm
  r}({\rm p}),r_{\rm p},\delta_{\rm H}(n_2{\rm L_2}_{j_2})\big) $  \\
\vhh&&&$ \ -E_{\rm H}\big(n_1{\rm L_1}_{j_1};R_\infty,\alpha,A_{\rm r}({\rm e}
  ),A_{\rm r}({\rm p}),r_{\rm p},\delta_{\rm H}(n_1{\rm L_1}_{j_1})\big) $  \\
\vhh
&&&$  \ -\fr{1}{4} \big[E_{\rm H}\big(n_4{\rm L_4}_{j_4};R_\infty,\alpha,A_{\rm
  r}({\rm e}),A_{\rm r}({\rm p}),r_{\rm p},\delta_{\rm H}(n_4{\rm L_4}_{j_4}
  )\big) $  \\
\vhh&&&$ \quad -E_{\rm H}\big(n_3{\rm L_3}_{j_3};R_\infty,\alpha,A_{\rm r}({\rm
  e}),A_{\rm r}({\rm p}),r_{\rm p},\delta_{\rm H}(n_3{\rm L_3}_{j_3}
  )\big)\big]\Big\}/h $  \\

\vh$A40$--$A44 \quad $
&$ \nu_{\rm D}(n_1{\rm L_1}_{j_1} - n_2{\rm L_2}_{j_2}) $&$\doteq$&$
\big[E_{\rm D}\big(n_2{\rm L_2}_{j_2};R_\infty,\alpha,A_{\rm r}({\rm e}),A_{\rm
  r}({\rm d}),r_{\rm d},\delta_{\rm D}(n_2{\rm L_2}_{j_2})\big) $  \\
\vhh&&&$-E_{\rm D}\big(n_1{\rm L_1}_{j_1};R_\infty,\alpha,A_{\rm r}({\rm e}
  ),A_{\rm r}({\rm d}),r_{\rm d},\delta_{\rm D}(n_1{\rm L_1}_{j_1})\big)\big]/h
  $  \\

\vh$A45$--$A46 \quad $
&$ \nu_{\rm D}(n_1{\rm L_1}_{j_1} - n_2{\rm L_2}_{j_2})
-\fr{1}{4}\nu_{\rm D}(n_3{\rm L_3}_{j_3} - n_4{\rm L_4}_{j_4})$&$\doteq$&$
\Big\{E_{\rm D}\big(n_2{\rm L_2}_{j_2};R_\infty,\alpha,A_{\rm r}({\rm e}),A_{\rm
  r}({\rm d}),r_{\rm d},\delta_{\rm D}(n_2{\rm L_2}_{j_2})\big) $  \\
\vhh&&&$ \ -E_{\rm D}\big(n_1{\rm L_1}_{j_1};R_\infty,\alpha,A_{\rm r}({\rm e}
  ),A_{\rm r}({\rm d}),r_{\rm d},\delta_{\rm D}(n_1{\rm L_1}_{j_1})\big) $  \\
\vhh
&&&$  \ -\fr{1}{4} \big[E_{\rm D}\big(n_4{\rm L_4}_{j_4};R_\infty,\alpha,A_{\rm
  r}({\rm e}),A_{\rm r}({\rm d}),r_{\rm d},\delta_{\rm D}(n_4{\rm L_4}_{j_4}
  )\big) $  \\
\vhh&&&$ \quad -E_{\rm D}\big(n_3{\rm L_3}_{j_3};R_\infty,\alpha,A_{\rm r}({\rm
  e}),A_{\rm r}({\rm d}),r_{\rm d},\delta_{\rm D}(n_3{\rm L_3}_{j_3}
  )\big)\big]\Big\}/h $  \\

\vh$A47 \quad $
&$ \nu_{\rm D}(1{\rm S}_{1/2} - 2{\rm S}_{1/2})
-\nu_{\rm H}(1{\rm S}_{1/2} - 2{\rm S}_{1/2})$&$\doteq$&$
\Big\{E_{\rm D}\big(2{\rm S}_{1/2};R_\infty,\alpha,A_{\rm r}({\rm e}),A_{\rm r}
  ({\rm d}),r_{\rm d},\delta_{\rm D}(2{\rm S}_{1/2})\big) $  \\
\vhh&&&$ \ -E_{\rm D}\big(1{\rm S}_{1/2};R_\infty,\alpha,A_{\rm r}({\rm e}
  ),A_{\rm r}({\rm d}),r_{\rm d},\delta_{\rm D}(1{\rm S}_{1/2})\big) $  \\
\vhh
&&&$  \ - \big[E_{\rm H}\big(2{\rm S}_{1/2};R_\infty,\alpha,A_{\rm r}({\rm e}
  ),A_{\rm r}({\rm p}),r_{\rm p},\delta_{\rm H}(2{\rm S}_{1/2})\big) $  \\
\vhh&&&$ \quad -E_{\rm H}\big(1{\rm S}_{1/2};R_\infty,\alpha,A_{\rm r}({\rm e}
  ),A_{\rm r}({\rm p}),r_{\rm p},\delta_{\rm H}(1{\rm S}_{1/2})\big)\big]\Big\}
  /h $  \\

\vh$A48 \quad $&$ r_{\rm p} $&$\doteq$&$ r_{\rm p} $  \\

\vh$A49 \quad $&$ r_{\rm d} $&$\doteq$&$ r_{\rm d} $  \\

\botrule
\end{tabular}
\end{table*}

\def\fixh{\vbox to 9pt {}}
\def\hsp{\hbox to 23pt {}}
\begin{table}
\caption{The 39 adjusted constants (variables) used in the least-squares
multivariate analysis of the input data in Table~\ref{tab:pdata}.  These
adjusted constants appear as arguments of the functions on the
right-hand side of the observational equations of
Table~\ref{tab:pobseqsb1}.  }
\label{tab:adjconb}
\begin{tabular}{l@{\hsp}l}
\toprule
\vbox to 10 pt {}
Adjusted constant & Symbol \\
\colrule

 \fixh  electron relative atomic mass & $A_{\rm r}({\rm e})$ \\
 \fixh  proton relative atomic mass & $A_{\rm r}({\rm p})$ \\
 \fixh  neutron relative atomic mass & $A_{\rm r}({\rm n})$ \\
 \fixh  deuteron relative atomic mass & $A_{\rm r}({\rm d})$ \\
 \fixh  triton relative atomic mass & $A_{\rm r}({\rm t})$ \\
 \fixh  helion relative atomic mass & $A_{\rm r}({\rm h})$ \\
 \fixh  alpha particle relative atomic mass & $A_{\rm r}(\rmalpha)$ \\
 \fixh  $^{16}$O$^{7+}$ relative atomic mass & $A_{\rm r}(^{16}{\rm O}^{7+})$ \\
 \fixh  $^{87}$Rb relative atomic mass  & $A_{\rm r}(^{87}{\rm Rb})$ \\
 \fixh  $^{133}$Cs relative atomic mass & $A_{\rm r}(^{133}{\rm Cs})$ \\
 \fixh  average vibrational excitation energy & $A_{\rm r}(E_{\rm av})$ \\
 \fixh  fine-structure constant & $\alpha$ \\
 \fixh  additive correction to $a_{\rm e}$(th) & $\delta_{\rm e}$ \\
 \fixh  muon magnetic moment anomaly & $a_{\rmssmu}$ \\
 \fixh  additive correction to $g_{\rm C}$(th) & $\delta_{\rm C}$ \\
 \fixh  additive correction to $g_{\rm O}$(th) & $\delta_{\rm O}$ \\
 \fixh  electron-proton magnetic moment ratio & $\mu_{\rm e^-}/\mu_{\rm p}$ \\
 \fixh  deuteron-electron magnetic moment ratio \qquad& $\mu_{\rm d}/\mu_{\rm
  e^-}$ \\
 \fixh  triton-proton magnetic moment ratio & $\mu_{\rm t}/\mu_{\rm p}$ \\
 \fixh  shielding difference of d and p in HD & $\sigma_{\rm dp}$ \\
 \fixh  shielding difference of t and p in HT & $\sigma_{\rm tp}$ \\
 \fixh  electron to shielded proton & \\
 \fixh  \ \ magnetic moment ratio & $\mu_{\rm e^-}/\mu^\prime_{\rm p}$ \\
 \fixh  shielded helion to shielded proton  & \\
 \fixh  \ \ magnetic moment ratio & $\mu^\prime_{\rm h}/\mu^\prime_{\rm p}$ \\
 \fixh  neutron to shielded proton & \\
 \fixh  \ \ magnetic moment ratio & $\mu_{\rm n}/\mu_{\rm p}^\prime$ \\
 \fixh  electron-muon mass ratio & $m_{\rm e}/m_{\rmssmu}$ \\
 \fixh  additive correction to $\Delta\nu_{\rm Mu}({\rm th})$ & $\delta_{\rm Mu}
  $ \\
 \fixh  Planck constant & $ h $ \\
 \fixh  molar gas constant & $R$ \\
 \fixh  copper K${\rm \alpha}_1$ x unit & xu(CuK${\rm \alpha}_1)$ \\
 \fixh  molybdenum K${\rm \alpha}_1$ x unit & xu(MoK${\rm \alpha}_1)$ \\
 \fixh  \aa ngstrom star & \AA$^*$  \\
 \fixh  $d_{220}$ of Si crystal ILL & $d_{220}({\rm {\scriptstyle ILL}})$ \\
 \fixh  $d_{220}$ of Si crystal N & $d_{220}({\rm {\scriptstyle N}})$ \\
 \fixh  $d_{220}$ of Si crystal WASO 17 & $d_{220}({\rm {\scriptstyle W17}})$ \\
 \fixh  $d_{220}$ of Si crystal WASO 04 & $d_{220}({\rm {\scriptstyle W04}})$ \\
 \fixh  $d_{220}$ of Si crystal WASO 4.2a & $d_{220}({\rm {\scriptstyle W4.2a}}
  )$ \\
 \fixh  $d_{220}$ of Si crystal MO$^*$ & $d_{220}({\rm {\scriptstyle MO^*}})$ \\
 \fixh  $d_{220}$ of Si crystal NR3 & $d_{220}({\rm {\scriptstyle NR3}})$ \\
 \fixh  $d_{220}$ of Si crystal NR4 & $d_{220}({\rm {\scriptstyle NR4}})$ \\
 \fixh  $d_{220}$ of an ideal Si crystal & $d_{220}$ \\
\botrule
\end{tabular}
\end{table}

\newpage
\def\tfrac#1#2{{\vbox to 9pt {}\phantom{_I}\textstyle{#1}\over\vbox to 9pt {}
  \textstyle{#2}}}
\def\vh{\vbox to 15pt {}}
\def\vhh{\vbox to 20pt {}}
\def\vhhh{\vbox to 22pt {}}
\def\hsp{\hbox to 30pt{}}
\begin{table*}
\caption{ Observational equations that express the input data in
Table~\ref{tab:pdata} as functions of the adjusted constants in
Table~\ref{tab:adjconb}.  The numbers in the first column correspond to
the numbers in the first column of Table~\ref{tab:pdata}.  For
simplicity, the lengthier functions are not explicitly given.  See
Sec.~\ref{ssec:mada} for an explanation of the symbol $\doteq$.}
\label{tab:pobseqsb1}
\begin{tabular}{l@{\hsp}rcl@{\hsp}l}
\toprule
Type of input & &\multicolumn{2}{l}{Observational equation}& Sec. \hbox to
35pt{}\\
datum      &\hbox to 10pt{} & & \hbox to 240pt{} & \\
\colrule
\vh$B1 \quad $&$ A_{\rm r}(^1{\rm H})$ &$\doteq$&$ A_{\rm r}({\rm p}) + A_{\rm
  r}({\rm e}) - E_{\rm b}(^1{\rm H})/m_{\rm u}c^2$  & \ref{ssec:ramnuc} \\

\vh$B2 \quad $&$ A_{\rm r}(^2{\rm H}) $&$\doteq$&$ A_{\rm r}({\rm d}) + A_{\rm
  r}({\rm e}) - E_{\rm b}(^2{\rm H})/m_{\rm u}c^2$ & \ref{ssec:ramnuc} \\

\vh$B3 \quad $&$ A_{\rm r}(E_{\rm av}) $&$\doteq$&$ A_{\rm r}(E_{\rm av}) $&
  \ref{ssec:smtr} \\[3 pt]

\vh$B4 \quad $&$ \tfrac{f_{\rm c}({\rm H}_2^{+*})}{f_{\rm c}({\rm d})}
  $&$\doteq$&$ \tfrac{A_{\rm r}({\rm d})}{2A_{\rm r}({\rm p}) + A_{\rm r}({\rm
  e}) - \left[\,2E_{\rm I}({\rm H}) + E_{\rm B}({\rm H}_2) - E_{\rm I}({\rm H}
  _2) - E_{\rm av}\,\right]/m_{\rm u} c^2} $\hbox to 20 pt {} & \ref{ssec:smtr}
  \\[10 pt]

\vh$B5 \quad $&$ \tfrac{f_{\rm c}({\rm t})}{f_{\rm c}({\rm H}_2^{+*})}
  $&$\doteq$&$ \tfrac{2A_{\rm r}({\rm p}) + A_{\rm r}({\rm e}) - \left[\,2E_{\rm
  I}({\rm H}) + E_{\rm B}({\rm H}_2) - E_{\rm I}({\rm H}_2) - E_{\rm av}
  \,\right]/m_{\rm u} c^2}{A_{\rm r}({\rm t})} $\hbox to 20 pt {} &
  \ref{ssec:smtr} \\[10 pt]

\vh$B6 \quad $&$ \tfrac{f_{\rm c}(^3{\rm He}^+)}{f_{\rm c}({\rm H}_2^{+*})}
  $&$\doteq$&$ \tfrac{2A_{\rm r}({\rm p}) + A_{\rm r}({\rm e}) - \left[\,2E_{\rm
  I}({\rm H}) + E_{\rm B}({\rm H}_2) - E_{\rm I}({\rm H}_2) - E_{\rm av}
  \,\right]/m_{\rm u} c^2}{A_{\rm r}({\rm h})+A_{\rm r}({\rm e})-E_{\rm I}
  (^3{\rm He}^+)/m_{\rm u} c^2} $\hbox to 20 pt {} & \ref{ssec:smtr} \\

\vh$B7 \quad $&$ A_{\rm r}(^4{\rm He}) $&$\doteq$&$ A_{\rm r}({\rmalpha}) + 2
  A_{\rm r}({\rm e}) - E_{\rm b}(^4{\rm He})/m_{\rm u}c^2$ & \ref{ssec:ramnuc}
  \\

\vh$B8 \quad $&$ A_{\rm r}(^{16}{\rm O}) $&$\doteq$&$ A_{\rm r}(^{16}{\rm O}
  ^{7+}) + 7 A_{\rm r}({\rm e}) - \left[E_{\rm b}(^{16}{\rm O})-E_{\rm b}(^{16}
  {\rm O}^{7+})\right]\!/m_{\rm u}c^2$ & \ref{ssec:ramnuc} \\

\vh$B9 \quad $&$ A_{\rm r}(^{87}{\rm Rb}) $&$\doteq$&$ A_{\rm r}(^{87}{\rm Rb})$
  & \\

\vh$B10 \quad $&$ A_{\rm r}(^{133}{\rm Cs}) $&$\doteq$&$ A_{\rm r}(^{133}{\rm
  Cs})$ & \\

\vh$B11 \quad $&$ A_{\rm r}({\rm e}) $&$\doteq$&$ A_{\rm r}({\rm e})$ & \\

\vh$B12 \quad $&$ \delta_{\rm e}$&$\doteq$&$ \delta_{\rm e}$ \\

\vh$B13 \quad $&$ a_{\rm e} $&$\doteq$&$ a_{\rm e}(\alpha,\delta_e)$ &
  \ref{sssec:ath} \\

\vhh$B14 \quad $&$ \overline{R}$&$\doteq$&$ -\tfrac{a_{\rmssmu}} {1+a_{\rm e}
  (\alpha,\delta_{\rm e})} \tfrac{m_{\rm e}}{m_{\rmssmu}} \tfrac{\mu_{\rm e^-}}
  {\mu_{\rm p}} $ &\ref{sssec:amb} \\

\vh$B15 \quad $&$ \delta_{\rm C}$&$\doteq$&$ \delta_{\rm C}$ \\

\vh$B16 \quad $&$ \delta_{\rm O}$&$\doteq$&$ \delta_{\rm O}$ \\

\vhh$B17 \quad $&$ \tfrac{f_{\rm s}\left(^{12}{\rm C}^{5+}\right)}{f_{\rm c}
  \left(^{12}{\rm C}^{5+}\right)} $&$\doteq$&$ -\tfrac{g_{\rm C}
  (\alpha,\delta_{\rm C})}{10 A_{\rm r}({\rm e})} \left[12-5A_{\rm r}({\rm e}) +
  \tfrac{E_{\rm b}\left(^{12}{\rm C}\right) -E_{\rm b}\left(^{12}{\rm C}^{5+}
  \right)}{ m_{\rm u}c^2}\right] $ & \ref{sssec:bsgfexps} \\

\vhh$B18 \quad $&$ \tfrac{f_{\rm s}\left(^{16}{\rm O}^{7+}\right)}{f_{\rm c}
  \left(^{16}{\rm O}^{7+}\right)} $&$\doteq$&$ -\tfrac{g_{\rm O}
  (\alpha,\delta_{\rm O})}{14 A_{\rm r}({\rm e})} A_{\rm r}(^{16}{\rm O}^{7+}) $
  & \ref{sssec:bsgfexps} \\

\vhh$B19 \quad $&$ \tfrac{\mu_{\rm e^-}({\rm H})}{\mu_{\rm p}({\rm H})}
  $&$\doteq$&$ \tfrac{g_{\rm e^-}({\rm H})}{g_{\rm e^-}} \left(\tfrac{g_{\rm p}
  ({\rm H})}{g_{\rm p}}\right)^{-1} \tfrac{\mu_{\rm e^-}}{\mu_{\rm p}} $ & \\

\vhhh$B20 \quad $&$ \tfrac{\mu_{\rm d}({\rm D})}{\mu_{\rm e^-}({\rm D})}
  $&$\doteq$&$ \tfrac{g_{\rm d}({\rm D})}{{g_{\rm d}}} \left(\tfrac{g_{\rm e^-}
  ({\rm D})}{g_{\rm e^-}}\right)^{-1} \tfrac{\mu_{\rm d}}{\mu_{\rm e^-}} $ & \\

\vhhh$B21 \quad $&$ \tfrac{\mu_{\rm p}({\rm HD})}{\mu_{\rm d}({\rm HD})}
  $&$\doteq$&$ \left[1 + \sigma_{\rm dp} \right] \tfrac{\mu_{\rm p}}{\mu_{\rm
  e^-}} \tfrac{\mu_{\rm e^-}}{\mu_{\rm d}} $ & \\

\vh$B22 \quad $&$ \sigma_{\rm dp}$&$\doteq$&$ \sigma_{\rm dp}$ \\

\vhhh$B23 \quad $&$ \tfrac{\mu_{\rm t}({\rm HT})}{\mu_{\rm p}({\rm HT})}
  $&$\doteq$&$ \left[1 - \sigma_{\rm tp} \right] \tfrac{\mu_{\rm t}}{\mu_{\rm p}
  } $ & \\

\vh$B24 \quad $&$ \sigma_{\rm tp}$&$\doteq$&$ \sigma_{\rm tp}$ \\

\vhhh$B25 \quad $&$ \tfrac{\mu_{\rm e^-}({\rm H})}{\mu_{\rm p}^\prime}
  $&$\doteq$&$ \tfrac{g_{\rm e^-}({\rm H})}{g_{\rm e^-}} \tfrac{\mu_{\rm e^-}}
  {\mu_{\rm p}^\prime} $ & \\

\vhhh$B26 \quad $&$ \tfrac{\mu_{\rm h}^\prime}{\mu_{\rm p}^\prime} $&$\doteq$&$
  \tfrac{\mu_{\rm h}^\prime}{\mu_{\rm p}^\prime} $\\

\botrule
\end{tabular}
\end{table*}

\addtocounter{table}{-1}
\def\tfrac#1#2{{\vbox to 9pt {}\phantom{_I}\textstyle{#1}\over\vbox to 9pt {}
  \textstyle{#2}}}
\def\vh{\vbox to 15pt {}}
\def\vhh{\vbox to 20pt {}}
\def\vhhh{\vbox to 22pt {}}
\def\hsp{\hbox to 60pt{}}
\begin{table*}
\caption{{\it (Continued).} Observational equations that express the
input data in Table~\ref{tab:pdata} as functions of the adjusted
constants in Table~\ref{tab:adjconb}.  The numbers in the first column
correspond to the numbers in the first column of Table~\ref{tab:pdata}.
For simplicity, the lengthier functions are not explicitly given.  See
Sec.~\ref{ssec:mada} for an explanation of the symbol $\doteq$.}
\label{tab:pobseqsb2}
\begin{tabular}{l@{\hsp}rcl@{\hsp}l}
\toprule
Type of input & &\multicolumn{2}{l}{Observational equation}& \hbox to 73pt{}\\
datum      &\hbox to 10pt{} & & \hbox to 25pt{} & Sec. \\
\colrule
\vh$B27 \quad $&$ \tfrac{\mu_{\rm n}}{\mu_{\rm p}^\prime} $&$\doteq$&$
  \tfrac{\mu_{\rm n}}{\mu_{\rm p}^\prime} $\\

\vh$B28 \quad $&$ \delta_{\rm Mu}$&$\doteq$&$ \delta_{\rm Mu}$ \\

\vhhh$B29 \quad $&$ \Delta \nu_{\rm Mu} $&$\doteq$&$ \Delta \nu_{\rm Mu}
  \!\!\left(R_\infty,\alpha,\tfrac{m_{\rm e}}{m_{\rmssmu}}, \delta_{\rm Mu}
  \right)$ & \ref{sssec:muhfs} \\

\vhhh$B30,B31 \quad $&$ \nu(f_{\rm p}) $&$\doteq$&$ \nu\!\left(f_{\rm p}
  ;R_\infty,\alpha,\tfrac{m_{\rm e}}{m_{\rmssmu}},\tfrac{\mu_{\rm e^-}}{\mu_{\rm
  p}}, \delta_{\rm e}, \delta_{\rm Mu}\right) $ &\ref{sssec:mufreqs} \\

\vhh$B32 \quad $&$ {\it\Gamma}_{\rm p-90}^{\,\prime}({\rm lo}) $&$\doteq$& $
  -\tfrac{ K_{\rm J-90}R_{\rm K-90}[1+a_{\rm e}(\alpha,\delta_{\rm e})]\alpha^3
  } { 2\mu_0 R_\infty  } \left(\tfrac{\mu_{\rm e^-}}{\mu_{\rm p}^\prime}
  \right)^{-1} $ & \\

\vhhh$B33 \quad $&$ {\it\Gamma}_{\rm h-90}^{\,\prime}({\rm lo}) $&$\doteq$& $
  \tfrac{ K_{\rm J-90}R_{\rm K-90}[1+a_{\rm e}(\alpha,\delta_{\rm e})]\alpha^3 }
  { 2\mu_0 R_\infty  } \left(\tfrac{\mu_{\rm e^-}}{\mu_{\rm p}^\prime}
  \right)^{-1} \tfrac{\mu_{\rm h}^\prime}{\mu_{\rm p}^\prime} $ & \\

\vhhh$B34 \quad $&$ {\it\Gamma}_{\rm p-90}^{\,\prime}({\rm hi}) $&$\doteq$& $
  -\tfrac{ c [1+a_{\rm e}(\alpha,\delta_{\rm e})]\alpha^2 } { K_{\rm J-90}R_{\rm
  K-90} R_\infty h } \left(\tfrac{\mu_{\rm e^-}}{\mu_{\rm p}^\prime}\right)^{-1}
  $ & \\

\vh$B35 \quad $&$ R_{\rm K} $&$\doteq$&$ \tfrac{\mu_0c}{2\alpha} $ & \\

\vhhh$B36 \quad $&$ K_{\rm J} $&$\doteq$&$ \left(\tfrac{8\alpha}{\mu_0ch}
  \right)^{1/2} $ & \\

\vhh$B37 \quad $&$ K_{\rm J}^2R_{\rm K} $&$\doteq$&$ \tfrac{4}{h} $ & \\

\vhhh$B38 \quad $&$ {\cal F}_{90} $&$\doteq$& $ \tfrac{ c M_{\rm u} A_{\rm r}
  ({\rm e})  \alpha^2 } { K_{\rm J-90}R_{\rm K-90} R_\infty h } $&  \\

\vh$B39$-$B41  \quad $&$ d_{220}({{\scriptstyle X}}) $&$\doteq$&$ d_{220}
  ({{\scriptstyle X}}) $ \\

\vhh$B42$-$B53 \quad $&$ \tfrac{d_{220}({{\scriptstyle X}}) }{ d_{220}
  ({{\scriptstyle Y}})} - 1 $&$\doteq$& $\tfrac{d_{220}({{\scriptstyle X}}) }{
  d_{220}({{\scriptstyle Y}})}-1 $ & \\

\vhh$B54 \quad $&$ N_{\rm A} $&$\doteq$&$ \tfrac{cM_{\rm u}A_{\rm r}({\rm e}
  )\alpha^{2}}{ 2R_{\infty}h} $ & \\

\vhh$B55 \quad $&$\tfrac{\lambda_{\rm meas} }{ d_{220}({\rm {\scriptstyle ILL}}
  )} $&$\doteq$& $ \tfrac{\alpha^2 A_{\rm r}({\rm e}) }{ R_\infty d_{220}({\rm
  {\scriptstyle ILL}})} \tfrac{A_{\rm r}({\rm n}) + A_{\rm r}({\rm p}) }{
  \left[A_{\rm r}({\rm n}) + A_{\rm r}({\rm p})\right]^2 - A_{\rm r}^2({\rm d})}
  $& \ref{ssec:arn} \\

\vhhh$B56,B57 \quad $&$ \tfrac{h}{m(X)} $&$\doteq$& $ \tfrac{ A_{\rm r}({\rm e})
  } {  A_{\rm r}(X)}  $ $ \tfrac{ c\alpha^2 } { 2 R_\infty }  $
  &\ref{sssec:pccsmr} \\

\vh$B58 \quad $&$ R $&$\doteq$&$ R $ &\\

\vhh$B59 \quad $&$ k $&$\doteq$&$ \tfrac{2R_{\infty}hR}{cM_{\rm u}A_{\rm r}({\rm
  e})\alpha^{2}} $ &  \\

\vhh$B60 \quad $&$ \tfrac{k}{h} $&$\doteq$&$ \tfrac{2R_{\infty}R}{cM_{\rm u}
  A_{\rm r}({\rm e})\alpha^{2}} $ &  \\

\vhh$B61,B64 \quad $&$\tfrac{\lambda({\rm CuK\rmalpha_1})}{d_{220}
  ({{\scriptstyle X}})} $&$\doteq$&$ \tfrac{\rm 1\,537.400 ~xu(CuK\rmalpha_1)}
  {d_{220}({{\scriptstyle X}})}$ &\ref{ssec:xru} \\

\vhhh$B62 \quad $&$\tfrac{\lambda({\rm WK\rmalpha_1})}{d_{220}({\rm
  {\scriptstyle N}})} $&$\doteq$&$ \tfrac{\rm 0.209\,010\,0 ~\AA^*}{d_{220}({\rm
  {\scriptstyle N}})} $ & \ref{ssec:xru} \\

\vhhh$B63 \quad $&$\tfrac{\lambda({\rm MoK\rmalpha_1})}{d_{220}({\rm
  {\scriptstyle N}})} $&$\doteq$&$ \tfrac{\rm 707.831 ~xu(MoK\rmalpha_1)}
  {d_{220}({\rm {\scriptstyle N}})} $ & \ref{ssec:xru} \\

\botrule
\end{tabular}
\end{table*}

\def\fixh{\vbox to 9pt {}}
\def\hsp{\hbox to 50pt {}}
\begin{table}
\caption{The 15 adjusted constants relevant to the antiprotonic helium
data given in Table~\ref{tab:cdata}.  These adjusted constants appear as
arguments of the theoretical expressions on the right-hand side of the
observational equations of Table~\ref{tab:pobseqsc}.}
\label{tab:adjconc}
\begin{tabular}{@{~}l@{\hsp}l}
\toprule
\qquad Transition & Adjusted constant \\
\colrule

\fixh $\bar{\rm p}^4$He$^+$: $(32,31) \rightarrow (31,30)$ & $\delta_{\bar{\rm
  p}^4{\rm He}^+}(32,31\!\!:\!31,30)$ \\
\fixh $\bar{\rm p}^4$He$^+$: $(35,33) \rightarrow (34,32)$ & $\delta_{\bar{\rm
  p}^4{\rm He}^+}(35,33\!\!:\!34,32)$ \\
\fixh $\bar{\rm p}^4$He$^+$: $(36,34) \rightarrow (35,33)$ & $\delta_{\bar{\rm
  p}^4{\rm He}^+}(36,34\!\!:\!35,33)$ \\
\fixh $\bar{\rm p}^4$He$^+$: $(37,34) \rightarrow (36,33)$ & $\delta_{\bar{\rm
  p}^4{\rm He}^+}(37,34\!\!:\!36,33)$ \\
\fixh $\bar{\rm p}^4$He$^+$: $(39,35) \rightarrow (38,34)$ & $\delta_{\bar{\rm
  p}^4{\rm He}^+}(39,35\!\!:\!38,34)$ \\
\fixh $\bar{\rm p}^4$He$^+$: $(40,35) \rightarrow (39,34)$ & $\delta_{\bar{\rm
  p}^4{\rm He}^+}(40,35\!\!:\!39,34)$ \\
\fixh $\bar{\rm p}^4$He$^+$: $(37,35) \rightarrow (38,34)$ & $\delta_{\bar{\rm
  p}^4{\rm He}^+}(37,35\!\!:\!38,34)$ \\
\fixh $\bar{\rm p}^4$He$^+$: $(33,32) \rightarrow (31,30)$ & $\delta_{\bar{\rm
  p}^4{\rm He}^+}(33,32\!\!:\!31,30)$ \\
\fixh $\bar{\rm p}^4$He$^+$: $(36,34) \rightarrow (34,32)$ & $\delta_{\bar{\rm
  p}^4{\rm He}^+}(36,34\!\!:\!34,32)$ \\
&\\
\fixh $\bar{\rm p}^3$He$^+$: $(32,31) \rightarrow (31,30)$ & $\delta_{\bar{\rm
  p}^3{\rm He}^+}(32,31\!\!:\!31,30)$ \\
\fixh $\bar{\rm p}^3$He$^+$: $(34,32) \rightarrow (33,31)$ & $\delta_{\bar{\rm
  p}^3{\rm He}^+}(34,32\!\!:\!33,31)$ \\
\fixh $\bar{\rm p}^3$He$^+$: $(36,33) \rightarrow (35,32)$ & $\delta_{\bar{\rm
  p}^3{\rm He}^+}(36,33\!\!:\!35,32)$ \\
\fixh $\bar{\rm p}^3$He$^+$: $(38,34) \rightarrow (37,33)$ & $\delta_{\bar{\rm
  p}^3{\rm He}^+}(38,34\!\!:\!37,33)$ \\
\fixh $\bar{\rm p}^3$He$^+$: $(36,34) \rightarrow (37,33)$ & $\delta_{\bar{\rm
  p}^3{\rm He}^+}(36,34\!\!:\!37,33)$ \\
\fixh $\bar{\rm p}^3$He$^+$: $(35,33) \rightarrow (33,31)$ & $\delta_{\bar{\rm
  p}^3{\rm He}^+}(35,33\!\!:\!33,31)$ \\

\botrule
\end{tabular}
\end{table}

\def\tfrac#1#2{{\vbox to 9pt {}\phantom{_I}\textstyle{#1}\over\vbox to 9pt {}
  \textstyle{#2}}}
\def\vh{\vbox to 15pt {}}
\def\vhh{\vbox to 10pt {}}
\begin{table*}
\caption{Observational equations that express the input data related to
antiprotonic helium in Table~\ref{tab:cdata} as functions of adjusted
constants in Tables~\ref{tab:adjconb} and \ref{tab:adjconc}.  The
numbers in the first column correspond to the numbers in the first
column of Table~\ref{tab:cdata}.  Definitions of the symbols and values
of the parameters in these equations are given in Sec.~\ref{ssec:aph}.
See Sec.~\ref{ssec:mada} for an explanation of the symbol $\doteq$.  }
\label{tab:pobseqsc}
\begin{tabular}{l@{\hbox to 65 pt{}}rcll}
\toprule
Type of input & &
\multicolumn{2}{l}{Observational equation}& \\
datum      & & &  \\
\colrule

\vh$C1$--$C7 \quad $&$ \delta_{\bar{\rm p}{\rm ^4He^+}}(n,l:n^\prime,l^\prime)
  $&$\doteq$&$ \delta_{\bar{\rm p}{\rm ^4He^+}}(n,l:n^\prime,l^\prime) $ \\
\vh$C8$--$C12 \quad $&$ \delta_{\bar{\rm p}{\rm ^3He^+}}(n,l:n^\prime,l^\prime)
  $&$\doteq$&$ \delta_{\bar{\rm p}{\rm ^3He^+}}(n,l:n^\prime,l^\prime) $ \\

\vh$C13$--$C19 \quad $&$ \Delta \nu_{\bar{\rm p}{\rm ^4He^+}}
  (n,l:n^\prime,l^\prime) $&$
\doteq$&$ \Delta\nu_{\bar{\rm p}{\rm ^4He^+}}^{(0)}(n,l:n^\prime,l^\prime)
+ a_{\bar{\rm p}{\rm ^4He^+}}(n,l:n^\prime,l^\prime)\left[\left(\tfrac{A_{\rm r}
  ({\rm e})}{A_{\rm r}({\rm p)}}\right)^{\!(0)} \!\!
\left(\tfrac{A_{\rm r}({\rm p})}{A_{\rm r}({\rm e})}\right)-1 \right]$
\\ \vbox to 20 pt {}&&&
$+ b_{\bar{\rm p}{\rm ^4He^+}}(n,l:n^\prime,l^\prime)\left[\left(\tfrac{A_{\rm
  r}({\rm e})}{A_{\rm r}({\rmalpha)}}\right)^{\!(0)} \!\!
\left(\tfrac{A_{\rm r}({\rmalpha})}{A_{\rm r}({\rm e})}\right)-1 \right]
+\delta_{\bar{\rm p}{\rm ^4He^+}}(n,l:n^\prime,l^\prime) $ \\

\\

\vh$C20$--$C24 \quad $&$ \Delta \nu_{\bar{\rm p}{\rm ^3He^+}}
  (n,l:n^\prime,l^\prime) $&$
\doteq$&$ \Delta\nu_{\bar{\rm p}{\rm ^3He^+}}^{(0)}(n,l:n^\prime,l^\prime)
+ a_{\bar{\rm p}{\rm ^3He^+}}(n,l:n^\prime,l^\prime)\left[\left(\tfrac{A_{\rm r}
  ({\rm e})}{A_{\rm r}({\rm p)}}\right)^{\!(0)} \!\!
\left(\tfrac{A_{\rm r}({\rm p})}{A_{\rm r}({\rm e})}\right)-1 \right]$
\\ \vbox to 20 pt {}&&&
$+ b_{\bar{\rm p}{\rm ^3He^+}}(n,l:n^\prime,l^\prime)\left[\left(\tfrac{A_{\rm
  r}({\rm e})}{A_{\rm r}({\rm h)}}\right)^{\!(0)} \!\!
\left(\tfrac{A_{\rm r}({\rm h})}{A_{\rm r}({\rm e})}\right)-1 \right]
+\delta_{\bar{\rm p}{\rm ^3He^+}}(n,l:n^\prime,l^\prime) $   \\
\\

\botrule
\end{tabular}
\end{table*}

\def\fixh{\vbox to 12pt {}}
\def\hsp{\hbox to 17.3 pt {}}
\def\t{@{\sp{20}}}
\begin{table*}
\caption{
Summary of the results of some of the least-squares adjustments used to
analyze the input data given in Tables~\ref{tab:rdata}-\ref{tab:cdcc}.
The values of $\alpha$ and $h$ are those obtained in the adjustment, $N$
is the number of input data, $M$ is the number of adjusted constants,
$\nu=N-M$ is the degrees of freedom, and $R_{\rm B}=\sqrt{{\it
\chi}^2/\nu}$ is the Birge ratio.  See the text for an explanation and
discussion of each adjustment, but in brief, adjustment 1 is all the
data; 2 is the same as 1 except with the uncertainties of the key data
that determine $h$ multiplied by 2; 3 is 2 with the low-weight input
data deleted and is the adjustment on which the 2010 recommended values
are based; 4 is 2 with the input data that provide the most accurate
values of alpha deleted; and 5 is 1 with the input data that provide the
most accurate values of $h$ deleted.}
\label{tab:adjustsall}
\begin{tabular}{c c @{\hsp} c @{\hsp} c @{\hsp} c @{\hsp} c @{\hsp} c @{\hsp} c
  @{\hsp} c @{\hsp} c @{\hsp} c}
\toprule
\fixh Adj. &  $N$& $M$& $\nu$ & ${\it \chi}^2$& $R_{\rm B}$& $\alpha^{-1}$ &
  $u_{\rm r}(\alpha^{-1})$ & $h$/(J s) & $u_{\rm r}(h)$ \\
\colrule
\fixh \ 1 & 169& 83& 86 & 89.3& 1.02  & $137.035\,999\,075(44)$           &
  $3.2\times10^{-10}$ & $6.626\,069\,58(15)   \times 10^{-34}$   &
  $2.2\times10^{-8}$  \\

\fixh \ 2 & 169& 83& 86 & 75.7& 0.94  & $137.035\,999\,073(44)$           &
  $3.2\times10^{-10}$ & $6.626\,069\,57(29)   \times 10^{-34}$   &
  $4.4\times10^{-8}$  \\

\fixh \ 3 & 149& 82& 67 & 58.1& 0.93  & $137.035\,999\,074(44)$           &
  $3.2\times10^{-10}$ & $6.626\,069\,57(29)   \times 10^{-34}$   &
  $4.4\times10^{-8}$ \\

\fixh \ 4 & 161& 81& 80 & 69.4& 0.93  & $137.036\,0005(20)\phantom{\,44}$ &
  $1.4\times10^{-8~}$ & $6.626\,069\,50(31)   \times 10^{-34}$   &
  $4.7\times10^{-8}$  \\

\fixh \ 5 & 154& 82& 72 & 57.2& 0.89  & $137.035\,999\,074(44)$           &
  $3.2\times10^{-10}$ & $6.626\,069\,48(80)   \times 10^{-34}$   &
  $1.2\times10^{-7}$  \\
\botrule
\end{tabular}
\end{table*}

\def\sp{\hbox to 5 pt {}}
\def\sm{\phantom{-}}
\begin{table*}
\caption{
Normalized residuals $r_i $ and self-sensitivity coefficients $S_{\rm c}
$ that result from the five least-squares adjustments summarized in
Table~\ref{tab:adjustsall}  for the three input data with the largest
absolute values of $r_i $ in adjustment 1.  $S_{\rm c} $ is a measure of
how the least-squares estimated value of a given type of input datum
depends on a particular measured or calculated value of that type of
datum; see Appendix E of CODATA-98.  See the text for an explanation and
discussion of each adjustment; brief explanations are given at the end
of the caption to the previous table.}
\label{tab:adjres}
\begin{tabular}{|lcc|c|c|c|c|c|c|}
\toprule
 Item   & Input     & Identification &     Adj.~1     &     Adj.~2     &
  Adj.~3     & Adj.~4     &     Adj.~5     \\
number  & quantity  &                & $~r_i~~~~S_{\rm c}$ & $~r_i~~~~S_{\rm c}$
  & $~r_i~~~~S_{\rm c}$ & $~r_i~~~~S_{\rm c}$ & $~r_i~~~~S_{\rm c}$ \\
\colrule
 $B37.3$                      &\fixh $K_{\rm J}^2R_{\rm K}$
  & NIST-07  & $\sm2.83~~0.367~$ & $\sm1.39~~0.367~$ & $\sm1.39~~0.371~$ &
  $\sm1.23~~0.413~$ &    Deleted  \\
 $B54$                        &\fixh $N_{\rm A}$
  & IAC-11   & $  -2.57~~0.555~$ & $  -1.32~~0.539~$ & $  -1.31~~0.546~$ & $
  -1.16~~0.587~$ &    Deleted  \\
 $B32.1$\rule[-6pt]{0pt}{6pt} &\fixh ${\it \Gamma}^\prime_{\rm p-90}({\rm lo})$
  & NIST-89  & $\sm2.19~~0.010~$ & $\sm2.19~~0.010~$ &     Deleted       &
  $\sm2.46~~0.158~$ &    Deleted  \\
\botrule
\end{tabular}
\end{table*}

\def\fixh{\vbox to 12pt {}}
\def\hsp{\hbox to 20 pt {}}
\def\sd{\phantom{3}}
\begin{table*}
\caption{Summary of the results of some of the least-squares adjustments
used to analyze the input data related to $R_\infty$.  The values of
$R_\infty$, $r_{\rm p}$, and $r_{\rm d}$ are those obtained in the
indicated adjustment, $N$ is the number of input data, $M$ is the number
of adjusted constants, $\nu=N-M$ is the degrees of freedom, and $R_{\rm
B}=\sqrt{{\it \chi}^2/\nu}$ is the Birge ratio.  See the text for an
explanation and discussion of each adjustment, but in brief, adjustment
~6 is 3, but the scattering data for the nuclear radii are omitted;
~7 is 3, but with only the hydrogen data included (but not the isotope
shift); 8 is 7 with the $r_{\rm p}$ data deleted; 9 and 10 are similar
to 7 and 8, but for the deuterium data;  11 is 3 with the muonic
Lamb-shift value of $r_{\rm p}$ included; and 12 is 11, but without the
scattering values of $r_{\rm p}$ and $r_{\rm d}$.
\\}
\label{tab:adjustsa}
\begin{tabular}{c@{\hsp}c@{\hsp}l@{\hsp}l@{\hsp}l@{\hsp} c@{\hsp}l@{\hsp}
  l@{\hsp}l@{\hsp}l}
\toprule
\fixh \sp Adj. & $N$& $M$& $\nu$ & \quad ${\it \chi}^2$& $R_{\rm B}$ &
\quad\qquad $R_\infty/{\rm m}^{-1}$ & \sp$u_{\rm r}(R_\infty)$ & \sp $r_{\rm p}
  $/fm & \sp $r_{\rm d}$/fm \\
\colrule
\fixh  $\sd 3$ & $149$ & $82$ & 67 & \sd58.1& 0.93&  $10\,973\,731.568\,539(55)$
  & $5.0\times 10^{-12}$ &  $0.8775(51)$   & $2.1424(21)$ \\
\fixh  $\sd 6$ & $146$ & $82$ & 64 & \sd55.5& 0.93&  $10\,973\,731.568\,521(82)$
  & $7.4\times 10^{-12}$ &  $0.8758(77)$   & $2.1417(31)$ \\
\fixh  $\sd 7$ & $131$ & $72$ & 59 & \sd53.4& 0.95&  $10\,973\,731.568\,561(60)$
  & $5.5\times 10^{-12}$ &  $0.8796(56)$   &              \\
\fixh  $\sd 8$ & $129$ & $72$ & 57 & \sd52.5& 0.96&  $10\,973\,731.568\,528(94)$
  & $8.6\times 10^{-12}$ &  $0.8764(89)$   &              \\
\fixh  $\sd 9$ & $114$ & $65$ & 49 & \sd46.9& 0.98&  $10\,973\,731.568\,37(13)$&
  $1.1\times 10^{-11}$ &                   & $2.1288(93)$ \\
\fixh  $10$ &    $113$ & $65$ & 48 & \sd46.8& 0.99&  $10\,973\,731.568\,28(30)$&
  $2.7\times 10^{-11}$ &                   & $2.121(25)$ \\
\fixh  $11$ &    $150$ & $82$ & 68 &   104.9& 1.24&  $10\,973\,731.568\,175(12)$
  & $1.1\times 10^{-12}$ & $0.842\,25(65)$ & $2.128\,24(28)$\\
\fixh  $12$ &    $147$ & $82$ & 65 & \sd74.3& 1.07&  $10\,973\,731.568\,171(12)$
  & $1.1\times 10^{-12}$ & $0.841\,93(66)$ & $2.128\,11(28)$ \\
\botrule
\end{tabular}
\end{table*}

\def\tfrac#1#2{{\vbox to 9pt {}\phantom{_I}\textstyle{#1}\over\vbox to 9pt {}
  \textstyle{#2}}}
\def\vh{\vbox to 15pt {}}
\def\vhh{\vbox to 20pt {}}
\def\vhhh{\vbox to 22pt {}}

\begin{table*}
\caption{
Generalized observational equations that express input data $B32$-$B38$
in Table~\ref{tab:pdata} as functions of the adjusted constants in
Tables~\ref{tab:adjconb} and \ref{tab:adjcona} with the additional
adjusted constants $\varepsilon_{\rm J}$ and $\varepsilon_{\rm K}$ as
given in Eqs.~(\ref{eq:kjeps}) and (\ref{eq:rkeps}).  The numbers in the
first column correspond to the numbers in the first column of
Table~\ref{tab:pdata}.  For simplicity, the lengthier functions are not
explicitly given.  See Sec.~\ref{ssec:mada} for an explanation of the
symbol $\doteq$.}
\label{tab:pobseqseps}
\begin{tabular}{lrcl}
\toprule
Type of input & &\multicolumn{2}{l}{Generalized observational equation}\\
datum      &\hbox to 10pt{} & & \hbox to 80pt{} \\
\colrule

\vhh$B32^* \quad $&$ {\it\Gamma}_{\rm p-90}^{\,\prime}({\rm lo}) $&$\doteq$&
$ -\tfrac{ K_{\rm J-90}R_{\rm K-90}[1+a_{\rm e}(\alpha,\delta_{\rm e})]\alpha^3
  }
{ 2\mu_0 R_\infty (1+\varepsilon_{\rm J})(1+\varepsilon_{\rm K})
 } \left(\tfrac{\mu_{\rm e^-}}{\mu_{\rm p}^\prime}\right)^{-1} $
\\

\vhhh$B33^* \quad $&$ {\it\Gamma}_{\rm h-90}^{\,\prime}({\rm lo}) $&$\doteq$&
$ \tfrac{ K_{\rm J-90}R_{\rm K-90}[1+a_{\rm e}(\alpha,\delta_{\rm e})]\alpha^3 }
{ 2\mu_0 R_\infty  (1+\varepsilon_{\rm J})(1+\varepsilon_{\rm K})
} \left(\tfrac{\mu_{\rm e^-}}{\mu_{\rm p}^\prime}\right)^{-1}
\tfrac{\mu_{\rm h}^\prime}{\mu_{\rm p}^\prime} $
\\

\vhhh$B34^* \quad $&$ {\it\Gamma}_{\rm p-90}^{\,\prime}({\rm hi}) $&$\doteq$&
$ -\tfrac{ c [1+a_{\rm e}(\alpha,\delta_{\rm e})]\alpha^2 }
{ K_{\rm J-90}R_{\rm K-90} R_\infty h }
(1+\varepsilon_{\rm J})(1+\varepsilon_{\rm K})
\left(\tfrac{\mu_{\rm e^-}}{\mu_{\rm p}^\prime}\right)^{-1} $
\\

\vh$B35^* \quad $&$ R_{\rm K} $&$\doteq$&$ \tfrac{\mu_0c}{2\alpha}
(1+\varepsilon_{\rm K})$
\\

\vhhh$B36^* \quad $&$ K_{\rm J} $&$\doteq$&$
\left(\tfrac{8\alpha}{\mu_0ch}\right)^{1/2} (1+\varepsilon_{\rm J})$
\\

\vhh$B37^* \quad $&$ K_{\rm J}^2R_{\rm K} $&$\doteq$&$ \tfrac{4}{h}
(1+\varepsilon_{\rm J})^2(1+\varepsilon_{\rm K})$
\\

\vhhh$B38^* \quad $&$ {\cal F}_{90} $&$\doteq$&
$ \tfrac{ c M_{\rm u} A_{\rm r}({\rm e})  \alpha^2 }
{ K_{\rm J-90}R_{\rm K-90} R_\infty h } (1+\varepsilon_{\rm J}
  )(1+\varepsilon_{\rm K})$
\\

\botrule
\end{tabular}
\end{table*}

\section{The 2010 CODATA recommended values}
\label{sec:2010crv}

\subsection{Calculational details}
\label{ssec:cd}

The 168 input data and their correlation coefficients initially
considered for inclusion in the 2010 CODATA adjustment of the values of
the constants are given in Tables~\ref{tab:rdata} to \ref{tab:cdcc}. The
2010 recommended values are based on adjustment 3, called the final
adjustment, summarized in Tables~\ref{tab:adjustsall} to
\ref{tab:adjustsa} and discussed in the associated text. Adjustment 3
omits 20 of the 168 initially considered input data, namely, items
$B10.1$, $B10.2$, $B13.1$, $B32.1$ to $B36.2$, $B37.5$, $B38$, $B56$,
$B59$, and $B56$, because of their low weight (self sensitivity
coefficient $S_{\rm c} < 0.01$).  However, because the observational
equation for $h/m(^{133}{\rm Cs})$, item $B56$, depends on $A_{\rm
r}(^{133}{\rm Cs})$ but item $B56$ is deleted because of its low weight,
the two values of $A_{\rm r}(^{133}{\rm Cs})$, items $B10.1$ and
$B10.2$, are also deleted and $A_{\rm r}(^{133}{\rm Cs})$ itself is
deleted as an adjusted constant. Further, the initial uncertainties of
five input data, items $B37.1$ to $B37.4$ and $B56$, are multiplied by
the factor 2, with the result that the absolute values of the normalized
residuals $|r_i|$ of the five data are less than 1.4 and their
disagreement is reduced to an acceptable level.

Each input datum in this final adjustment has a self sensitivity
coefficient $S_{c}$ greater than 0.01, or is a subset of the data of an
experiment or series of experiments that provide an input datum or input
data with $S_{\rm c} > 0.01$.  Not counting such input data with $S_{\rm
c} < 0.01$, the seven  data with $|r_i|>1.2$ are $A50$, $B11$, $B37.3$,
$B54$, $C19$, $C21$, and $C28$;  their values of $r_i$ are $-1.24$,
$1.43$, $1.39$, $-1.31$, $-1.60$, $-1.83$, and $1.76$, respectively.

As discussed in Sec.~\ref{sssec:calcncg}, the 2010 recommended value of
$G$ is the weighted mean of the 11 measured values in
Table~\ref{tab:bigg} after the uncertainty of each is multiplied by the
factor 14. Although these data can be treated separately because they
are independent of all of the other data, they could have been included
with the other data.  For example, if the 11 values of $G$ with expanded
uncertainties are added to the 148 input data of adjustment 3,  $G$ is
taken as an additional adjusted constant so that these 11 values can be
included in a new adjustment using the observational equation $G \doteq
G$, and the so-modified adjustment 3 is repeated, then we find for
this ``grand final adjustment'' that $N = 160$, $M = 83$, $\nu = 77$,
$\chi^2 = 59.1$, $p(59.1|77) = 0.94$, and $R_{\rm B} = 0.88$. Of course,
the resulting values of the adjusted constants, and of the normalized
residuals and self sensitivity coefficients of the input data, are
exactly the same as those from adjustment 3 and the weighted mean of the
11 measured values of $G$ with expanded uncertainties.

In any event, the 2010 recommended values are calculated from the set of
best estimated values, in the least-squares sense, of 82 adjusted
constants, including $G$, and their variances and covariances, together
with (i) those constants that have exact values such as $\mu_0$ and $c$;
and (ii) the values of $m_{\rmsstau}$, $G_{\rm F}$, and $\sin^2
\theta_{\rm W}$ given in Sec.~\ref{sec:xeq}.  See Sec.~V.B of CODATA-98
for details.

\subsection{Tables of values}
\label{ssec:tov}

Tables~\ref{tab:abbr} to \ref{tab:enconv2} give the 2010 CODATA
recommended values of the basic constants and conversion factors of
physics and chemistry and related quantities.  Although very similar in
form and content to their 2006 counterparts, several new recommended
values have been included in the 2010 tables and a few have been
deleted.  The values of the four new constants, $m_{\rm n} - m_{\rm p}$
in kg and u, and $(m_{\rm n} - m_{\rm p})c^2$ in J and MeV, are given in
Table~\ref{tab:constants} under the heading ``Neutron, n''; and the
values of the four new constants $\mu_{\rm h}$, $\mu_{\rm h}/\mu_{\rm
B}$,  $\mu_{\rm h}/\mu_{\rm N}$, and $g_{\rm h}$ are given in the same
table under the heading ``Helion, h.''  The three constants deleted,
$\mu_{\rm t}/\mu_{\rm e}$, $\mu_{\rm t}/\mu_{\rm p}$, and $\mu_{\rm
t}/\mu_{\rm n}$, were in the 2006 version of Table~\ref{tab:constants}
under the heading ``Triton, t.''  It was decided that these constants
were of limited interest and the values can be calculated from other
constants in the table.

The values of the four new helion-related constants are calculated from
the adjusted constant $\mu_{\rm h}^\prime/\mu_{\rm p}^\prime$ and the
theoretically predicted shielding correction $\sigma_{\rm h}=
59.967\,43(10)\times 10^{-6}$ due to \citet{2009161} using the relation
$\mu_{\rm h}^\prime = \mu_{\rm h}(1 - \sigma_{\rm h})$; see
Sec.~\ref{sssec:bfhmmr}.

\def\s#1{\hbox to #1pt{}}
\def\hsp{\hbox to 15 pt{}}
\def\b{\hbox to 10 pt{}}
\begin{table*}[h]
\caption{An abbreviated list of the CODATA recommended values of the
fundamental constants of physics and chemistry based on the
2010 adjustment.}
\label{tab:abbr}
\begin{tabular}{l@{\hsp}l@{\hsp}l@{\hsp}l@{\hsp}l}
\toprule
& & & & Relative std. \\
\s{35}Quantity & \s{-10}Symbol & \s{15}Numerical value & \s{2}Unit
& uncert. $u_{\rm r}$ \\
\colrule
speed of light in vacuum & $ c,c_0 $ & 299\,792\,458 & m~s$^{-1}$ & exact
\vbox to 12 pt {} \\
magnetic constant & $\mu_0$ & $ 4\rmpi\times10^{-7}$ & N~A$^{-2}$ & \\
& & $=12.566\,370\,614...\times10^{-7}$ & N~A$^{-2}$ & exact \\
electric constant 1/$\mu_0c^{2}$ & $\epsilon_0$ &
$8.854\,187\,817...\times 10^{-12}$ & F~m$^{-1}$ & exact \\
Newtonian constant
of gravitation~~ & $ G $ & $ 6.673\,84(80)\times 10^{-11}$ & m$^{3}$~kg$^{-1}
  $~s$^{-2}$ & $ 1.2\times 10^{-4}$ \\
Planck constant & $ h $ & $ 6.626\,069\,57(29)\times 10^{-34}$ & J~s & $
  4.4\times 10^{-8}$ \\
\b $h/2\rmpi$ & $\hbar$ & $ 1.054\,571\,726(47)\times 10^{-34}$ & J~s & $
  4.4\times 10^{-8}$ \\
elementary charge & $ e $ & $ 1.602\,176\,565(35)\times 10^{-19}$ & C & $
  2.2\times 10^{-8}$ \\
magnetic flux quantum $h$/2$e$ & ${\it \Phi}_0$ & $ 2.067\,833\,758(46)\times
  10^{-15}$ & Wb & $ 2.2\times 10^{-8}$ \\
conductance quantum $2e^2\!/h$ & $G_0$ & $ 7.748\,091\,7346(25)\times 10^{-5}$ &
  S & $ 3.2\times 10^{-10}$ \\
electron mass & $ m_{\rm e}$ & $ 9.109\,382\,91(40)\times 10^{-31}$ & kg & $
  4.4\times 10^{-8}$ \\
proton mass & $ m_{\rm p}$ & $ 1.672\,621\,777(74)\times 10^{-27}$ & kg & $
  4.4\times 10^{-8}$ \\
proton-electron mass ratio & $m_{\rm p}$/$m_{\rm e}$ & $ 1836.152\,672\,45(75)$
  & & $ 4.1\times 10^{-10}$ \\
fine-structure constant $e^2\!/4\rmpi\epsilon_0 \hbar c$ & $\alpha$ & $
  7.297\,352\,5698(24)\times 10^{-3}$ & & $ 3.2\times 10^{-10}$ \\
\b inverse fine-structure constant & $\alpha^{-1}$ & $ 137.035\,999\,074(44)$ &
  & $ 3.2\times 10^{-10}$ \\
Rydberg constant $\alpha^2m_{\rm e}c/2h$ & $ R_\infty$ & $
  10\,973\,731.568\,539(55)$ & m$^{-1}$ & $ 5.0\times 10^{-12}$ \\
Avogadro constant & $N_{\rm A},L$ & $ 6.022\,141\,29(27)\times 10^{23}$ &
  mol$^{-1}$ & $ 4.4\times 10^{-8}$ \\
Faraday constant $N_{\rm A}e$ & $ F $ & $ 96\,485.3365(21)$ & C~mol$^{-1}$ & $
  2.2\times 10^{-8}$ \\
molar gas constant & $ R $ & $ 8.314\,4621(75)$ & J~mol$^{-1}$~K$^{-1}$ & $
  9.1\times 10^{-7}$ \\
Boltzmann constant $R$/$N_{\rm A}$ & $k$ & $ 1.380\,6488(13)\times 10^{-23}$ &
  J~K$^{-1}$ & $ 9.1\times 10^{-7}$ \\
Stefan-Boltzmann constant & & & & \\
\, ($\rmpi^2$/60)$k^4\!/\hbar^3c^2$ & $\sigma$ & $ 5.670\,373(21)\times 10^{-8}$
  & W~m$^{-2}$~K$^{-4}$ & $ 3.6\times 10^{-6}$ \\
\multicolumn {5} {c} { \vbox to 12 pt {}
Non-SI units accepted for use with the SI} \\
electron volt ($e$/{\rm C}) {\rm J} & eV & $ 1.602\,176\,565(35)\times 10^{-19}$
  & J & $ 2.2\times 10^{-8}$ \\
(unified) atomic mass unit ${1\over12}m(^{12}$C)~~ & u & $
  1.660\,538\,921(73)\times 10^{-27}$ & kg & $ 4.4\times 10^{-8}$
  \phantom{\Big|}\\
\botrule
\end{tabular}
\end{table*}

Table~\ref{tab:abbr} is a highly-abbreviated list of the values of the
constants and conversion factors most commonly used.
Table~\ref{tab:constants} is a much more extensive list of values
categorized as follows: UNIVERSAL; ELECTROMAGNETIC; ATOMIC AND NUCLEAR;
and PHYSICOCHEMICAL.  The ATOMIC AND NUCLEAR category is subdivided into
11 subcategories: General; Electroweak; Electron, ${\rm e}^{-}$; Muon,
${\rmmu}^{-}$; Tau, ${\rmtau}^{-}$; Proton, ${\rm p}$; Neutron, ${\rm
n}$; Deuteron, ${\rm d}$; Triton, ${\rm t}$; Helion, ${\rm h}$; and
Alpha particle, ${\rmalpha}$.  Table~\ref{tab:varmatrix} gives the
variances, covariances, and correlation coefficients of a selected group
of constants.  (Use of the covariance matrix is discussed in Appendix~E
of CODATA-98.)  Table~\ref{tab:adopted} gives the internationally
adopted values of various quantities; Table~\ref{tab:xrayvalues} lists
the values of a number of x-ray related quantities;
Table~\ref{tab:units} lists the values of various non-SI units; and
Tables~\ref{tab:enconv1} and \ref{tab:enconv2} give the values of
various energy equivalents.

All of the values given in Tables~\ref{tab:abbr} to \ref{tab:enconv2}
are available on the Web pages of the Fundamental Constants Data Center
of the NIST Physical Measurement Laboratory at
physics.nist.gov/constants.  This electronic version of the 2010 CODATA
recommended values of the constants also includes a much more extensive
correlation coefficient matrix.  In fact, the correlation coefficient of
any two constants listed in the tables is accessible on the Web site, as
well as the automatic conversion of the value of an energy-related
quantity expressed in one unit to the corresponding value expressed in
another unit (in essence, an automated version of
Tables~\ref{tab:enconv1} and \ref{tab:enconv2}).

\clearpage

\def\lbar{\lambda\hskip-4.5pt\vrule height4.6pt depth-4.3pt width4pt}
\def\b{\hbox to 12pt{}}
\def\s#1{\hbox to #1pt{}}
\shortcites{2010129}
\begin{longtable*}{lllll}
\caption{The CODATA recommended values of the
fundamental constants of physics and chemistry based on the
2010 adjustment.\label{tab:constants}} \\
\toprule
& & & & Relative std. \\
\s{35}Quantity & \s{-15} Symbol & \s{15} Numerical value & \s{5}Unit & uncert.
  $u_{\rm r}$ \\
\colrule
\endfirsthead

\caption{{\it (Continued).}} \\
\colrule
& & & & Relative std. \\
\s{35}Quantity & \s{-15} Symbol & \s{15} Numerical value & \s{5}Unit & uncert.
  $u_{\rm r}$ \\
\colrule
\endhead
\colrule
\endfoot
\endlastfoot
\multicolumn {5} {c} { \vbox to 12 pt {}
UNIVERSAL} \\
speed of light in vacuum & $ c,c_0 $ & $299\,792\,$458 & m~s$^{-1}$ & exact \\
magnetic constant & $\mu_0$ & 4$\rmpi\times10^{-7}$ & N~A$^{-2}$ & \\
& & $=12.566\,370\,614...\times10^{-7}$ & N~A$^{-2}$ & exact \\
electric constant 1/$\mu_0c^2$ & $\epsilon_0$ &
  $8.854\,187\,817...\times10^{-12}$ & F~m$^{-1}$ & exact\\
characteristic impedance of vacuum $\mu_0c$ & $Z_0$ & $376.730\,313\,461...$ &
  ${\rm \Omega}$ & exact\\
Newtonian constant of gravitation & $ G $ & $ 6.673\,84(80)\times 10^{-11}$ &
  m$^3$~kg$^{-1}$~s$^{-2}$ & $ 1.2\times 10^{-4}$ \\
& $G/\hbar c $ & $ 6.708\,37(80)\times 10^{-39}$ & $ ({\rm GeV}/c^2)^{-2}$ & $
  1.2\times 10^{-4}$ \\
\botrule
\end{longtable*}

\clearpage

\begin{table*}
\caption{The variances, covariances, and correlation coefficients of the values
  of a
selected group of constants based on the 2010 CODATA adjustment.  The numbers in
  bold
above the main diagonal are $10^{16}$ times the numerical values of the relative
  covariances;
the numbers in bold on the main diagonal are $10^{16}$ times the numerical
  values of the
relative variances; and the numbers in italics below the main diagonal are the
  correlation
coefficients.$^1$}
\label{tab:varmatrix}
\def\sp{\hbox to 30.7 pt{}}
\hbox to 6.8 in{
\begin{tabular} {c@{\sp}r@{\sp}r@{\sp}r@{\sp}r@{\sp}r@{\sp}r@{\sp}r}
\toprule
& $\alpha$~~~ & $h$~~~  & $e$~~~  & $m_{\rm e}$~~~ & $N_{\rm A}$~~~  & $m_{\rm
  e}/m_{\rm \mu}$  & $F$~~~ \\
\colrule
$\alpha$ & ${\bf  0.0010}$ & ${\bf  0.0010}$ & ${\bf
 0.0010}$ & ${\bf  -0.0011}$ & ${\bf  0.0009}$ &
${\bf  -0.0021}$ & ${\bf  0.0019}$ \vbox to 10 pt {}\\
$h$ & ${\it  0.0072}$ & ${\bf  19.4939}$ & ${\bf  9.7475}$ & ${\bf  19.4918}$ &
  ${\bf  -19.4912}$ & ${\bf  -0.0020}$ & ${\bf  -9.7437}$ \\
$e$ & ${\it  0.0145}$ & ${\it  1.0000}$ & ${\bf  4.8742}$ & ${\bf  9.7454}$ &
  ${\bf  -9.7452}$ & ${\bf  -0.0020}$ & ${\bf  -4.8709}$ \\
$m_{\rm e}$ & ${\it  -0.0075}$ & ${\it  0.9999}$ & ${\it  0.9998}$ & ${\bf
  19.4940}$ & ${\bf  -19.4929}$ & ${\bf  0.0021}$ & ${\bf  -9.7475}$ \\
$N_{\rm A}$ & ${\it  0.0060}$ & ${\it  -0.9999}$ & ${\it  -0.9997}$ & ${\it
  -1.0000}$ & ${\bf  19.4934}$ & ${\bf  -0.0017}$ & ${\bf  9.7483}$ \\
$m_{\rm e}/m_{\rm \mu}$ & ${\it  -0.0251}$ & ${\it  -0.0002}$ & ${\it  -0.0004}$
  & ${\it  0.0002}$ & ${\it  -0.0002}$ & ${\bf  6.3872}$ & ${\bf  -0.0037}$ \\
$F$ & ${\it  0.0265}$ & ${\it  -0.9993}$ & ${\it  -0.9990}$ & ${\it  -0.9997}$ &
  ${\it  0.9997}$ & ${\it  -0.0007}$ & ${\bf  4.8774}$ \\
\botrule
\end{tabular}
}
$^1$ The relative covariance is
$u_{\rm r}(x_i,x_j) = u(x_i,x_j)/(x_ix_j)$, where $u(x_i,x_j)$ is the covariance
  of
$x_i$ and $x_j$; the relative variance is
$u_{\rm r}^2(x_i) = u_{\rm r}(x_i,x_i)$:
and the correlation coefficient is
$r(x_i,x_j) = u(x_i,x_j)/[u(x_i)u(x_j)]$.
\end{table*}

\begin{table*}
\caption{Internationally adopted values of various quantities.}
\label{tab:adopted}
\def\hsp{\hbox to 22pt{}}
\begin{tabular}{l@{\hsp}l@{\hsp}l@{\hsp}l@{\hsp}l}
\toprule
& & & & Relative std. \\
~~~~~~Quantity & Symbol~~~~~ & Numerical value~~~~~ & Unit & uncert. $u_{\rm r}$
  \\
\colrule
relative atomic mass$^1$ of $^{12}$C & $A_{\rm r}(^{12}$C) & $12$ &  & ~~~exact
  \vbox to 12 pt {} \\
molar mass constant  & $M_{\rm u}$ & $1 \times 10 ^{-3}$ & kg mol$^{-1}$ &
  ~~~exact \\
molar mass$^2$ of $^{12}$C & $M(^{12}$C) & $12 \times 10 ^{-3}$ & kg mol$^{-1}$ &
  ~~~exact \\
conventional value of Josephson constant$^3$ & $K_{\rm J-90}$ & 483\,597.9 & GHz
  V$^{-1}$ & ~~~exact \\
conventional value of von Klitzing constant$^4$ ~~~~~ & $R_{\rm K-90}$ &
  25\,812.807 & ${\rm \Omega}$ & ~~~exact \\
standard-state pressure & & $100$ & kPa & ~~~exact \\
standard atmosphere & & $101.325$ & kPa & ~~~exact \\
\botrule
\end{tabular}\\
$^1$ The relative atomic mass $A_{\rm r}(X)$ of particle $X$ with
 mass $m(X)$ is defined by $A_{\rm r}(X) = m(X) /m_{\rm u}$, where
$m_{\rm u} = m(^{12}{\rm C})/12 = M_{\rm u}/N_{\rm A} = 1~{\rm u}$ is the
atomic mass constant, $M_{\rm u}$ is the molar mass constant,
$N_{\rm A}$ is the Avogadro constant, and u is the unified
atomic mass unit.  Thus the mass of particle $X$ is $m(X) = A_{\rm r}(X)$~u
and the molar mass of $X$ is $M(X) = A_{\rm r}(X)M_{\rm u}$.
\\$^2$ Value fixed by the SI definition of the mole.
\\$^3$ This is the value adopted internationally
for realizing representations of the volt using the Josephson effect.
\\$^4$ This is the value adopted internationally
for realizing representations of the ohm using the quantum Hall effect.
\end{table*}

\begin{table*}[!]
\caption{Values of some x-ray-related quantities
based on the 2010 CODATA adjustment of the values of the constants.}
\label{tab:xrayvalues}
\hbox to 7 in {
\begin{tabular}{llllll}
\toprule
& & & & Relative std. \\
~~~~~~Quantity & Symbol & ~~~~Numerical value & Unit & uncert. $u_{\rm r}$ \\
\colrule
Cu x unit: $\lambda({\rm CuK}{\rm \alpha}_{\rm 1}) / 1\,537.400 $ & ${\rm xu}
  ({\rm CuK}{\rm \alpha}_{\rm 1})$ & $ 1.002\,076\,97(28)\times 10^{-13}$ & m &
  $ 2.8\times 10^{-7}$
\vbox to 12 pt {} \\
Mo x unit: $\lambda({\rm MoK}{\rm \alpha}_{\rm 1}) / 707.831 $ & ${\rm xu}({\rm
  MoK}{\rm \alpha}_{\rm 1})$ & $ 1.002\,099\,52(53)\times 10^{-13}$ & m & $
  5.3\times 10^{-7}$ \\
{\aa}ngstrom star$: \lambda({\rm WK}{\rm \alpha}_{\rm 1}) / 0.209\,010\,0 $ &
  \AA$^{\ast}$ & $ 1.000\,014\,95(90)\times 10^{-10}$ & m & $ 9.0\times 10^{-7}$
  \\
lattice parameter$^1$ of Si \ (in vacuum, 22.5 $^\circ$C)~~ & $a$ & $
  543.102\,0504(89)\times 10^{-12}$ & m & $ 1.6\times 10^{-8}$ \vbox to 9 pt {}
  \\
\{220\} lattice spacing of Si $a/\sqrt{8}$ & $d_{\rm 220}$ & $
  192.015\,5714(32)\times 10^{-12}$ & m & $ 1.6\times 10^{-8}$ \\
\,\, (in vacuum, 22.5 $^\circ$C)&&&&\\
molar volume of Si \ $M({\rm Si})/\rho({\rm Si})=N_{\rm A}a^{3}\!/8$ & $V_{\rm
  m}$(Si) & $ 12.058\,833\,01(80)\times 10^{-6}$ & m$^{3}$ mol$^{-1}$ & $
  6.6\times 10^{-8}$ \\
\, \, (in vacuum, 22.5 $^\circ$C)&&&&\\
\botrule
\end{tabular}
}
$^1$ This is the lattice parameter (unit cell edge length)
of an ideal single crystal of naturally occurring
Si free of impurities and imperfections, and is deduced from
measurements on extremely pure and nearly perfect single crystals of Si
by correcting for the effects of impurities.
\end{table*}

\thispagestyle{empty}
\def\hsp{\hbox to 15 pt {}}
\begin{table*}[!]
\caption{The values in SI units of some non-SI units based on the 2010
CODATA adjustment of the values of the constants.}
\label{tab:units}
\begin{tabular}{l@{\hsp}l@{\hsp}l@{\hsp}l@{\hsp}l@{\hsp}l}
\toprule
& & & & Relative std. \\
\s{35}Quantity & \s{-3}Symbol & \s{17}Numerical value & \s{2}Unit & uncert.
  $u_{\rm r}$ \\
\colrule
\multicolumn {5} {c} { \vbox to 12 pt {} Non-SI units accepted for use with the
  SI} \\
electron volt: ($e/{\rm C}$) {\rm J} & eV & $ 1.602\,176\,565(35)\times 10^{-19}
  $ & J & $ 2.2\times 10^{-8}$ \\
(unified) atomic mass unit: ${1\over12}m(^{12}$C)~~ & u & $
  1.660\,538\,921(73)\times 10^{-27}$ & kg & $ 4.4\times 10^{-8}$ \\
\multicolumn {5} {c} {} \\
\multicolumn {5} {c} {Natural units (n.u.)} \\
n.u. of velocity & $c,c_0$ & 299\,792\,458 & m s$^{-1}$ & exact \\
n.u. of action: $h/2\rmpi$~~ & $\hbar$ & $ 1.054\,571\,726(47)\times 10^{-34}$ &
  J s & $ 4.4\times 10^{-8}$ \\
& & $ 6.582\,119\,28(15)\times 10^{-16}$ & eV s & $ 2.2\times 10^{-8}$ \\
& $\hbar c$ & $ 197.326\,9718(44)$ & MeV fm & $ 2.2\times 10^{-8}$ \\
n.u. of mass & $m_{\rm e}$ & $ 9.109\,382\,91(40)\times 10^{-31}$ & kg & $
  4.4\times 10^{-8}$ \\
n.u. of energy & $m_{\rm e}c^2$ & $ 8.187\,105\,06(36)\times 10^{-14}$ & J & $
  4.4\times 10^{-8}$ \\
& & $ 0.510\,998\,928(11)$ & MeV & $ 2.2\times 10^{-8}$ \\
n.u. of momentum & $m_{\rm e}c$ & $ 2.730\,924\,29(12)\times 10^{-22}$ & kg m
  s$^{-1}$ & $ 4.4\times 10^{-8}$ \\
& & $ 0.510\,998\,928(11)$ & MeV/$c$ & $ 2.2\times 10^{-8}$ \\
n.u. of length: $\hbar/m_{\rm e}c$& $\lbar_{\rm C}$ & $
  386.159\,268\,00(25)\times 10^{-15}$ & m & $ 6.5\times 10^{-10}$ \\
n.u. of time & $\hbar/m_{\rm e}c^2$ & $ 1.288\,088\,668\,33(83)\times 10^{-21}$
  & s & $ 6.5\times 10^{-10}$ \\
\multicolumn {5} {c} {} \\
\multicolumn {5} {c} {Atomic units (a.u.)} \\
a.u. of charge & $e$ & $ 1.602\,176\,565(35)\times 10^{-19}$ & C & $ 2.2\times
  10^{-8}$ \\
a.u. of mass & $m_{\rm e}$ & $ 9.109\,382\,91(40)\times 10^{-31}$ & kg & $
  4.4\times 10^{-8}$ \\
a.u. of action: $h/2\rmpi$& $\hbar$ & $ 1.054\,571\,726(47)\times 10^{-34}$ & J
  s & $ 4.4\times 10^{-8}$ \\
a.u. of length: Bohr radius (bohr)~~& & & & \\
\, $\alpha/4\rmpi R_\infty$& $a_0$ & $ 0.529\,177\,210\,92(17)\times 10^{-10}$ &
  m & $ 3.2\times 10^{-10}$ \\
a.u. of energy: Hartree energy (hartree)~~ & & & & \\
\, $e^2\!/4\rmpi\epsilon_0a_0=2R_\infty hc = \alpha^2m_{\rm e}c^2$ & $E_{\rm h}$
  & $ 4.359\,744\,34(19)\times 10^{-18}$ & J & $ 4.4\times 10^{-8}$ \\
a.u. of time & $\hbar/E_{\rm h}$ & $ 2.418\,884\,326\,502(12)\times 10^{-17}$ &
  s & $ 5.0\times 10^{-12}$ \\
a.u. of force & $E_{\rm h}/a_0$ & $ 8.238\,722\,78(36)\times 10^{-8}$ & N & $
  4.4\times 10^{-8}$ \\
a.u. of velocity: $\alpha c$ & $a_0E_{\rm h}/\hbar$ & $
  2.187\,691\,263\,79(71)\times 10^{6}$ & m s$^{-1}$ & $ 3.2\times 10^{-10}$ \\
a.u. of momentum & $\hbar/a_0$ & $ 1.992\,851\,740(88)\times 10^{-24}$ & kg m
  s$^{-1}$ & $ 4.4\times 10^{-8}$ \\
a.u. of current & $eE_{\rm h}/\hbar$ & $ 6.623\,617\,95(15)\times 10^{-3}$ & A &
  $ 2.2\times 10^{-8}$ \\
a.u. of charge density & $e/a_0^3$ & $ 1.081\,202\,338(24)\times 10^{12}$ & C
  m$^{-3}$ & $ 2.2\times 10^{-8}$ \\
a.u. of electric potential & $E_{\rm h}/e$ & $ 27.211\,385\,05(60)$ & V & $
  2.2\times 10^{-8}$ \\
a.u. of electric field & $E_{\rm h}/ea_0$ & $ 5.142\,206\,52(11)\times 10^{11}$
  & V m$^{-1}$ & $ 2.2\times 10^{-8}$ \\
a.u. of electric field gradient & $E_{\rm h}/ea_0^2$ & $
  9.717\,362\,00(21)\times 10^{21}$ & V m$^{-2}$ & $ 2.2\times 10^{-8}$ \\
a.u. of electric dipole moment & $ea_0$ & $ 8.478\,353\,26(19)\times 10^{-30}$ &
  C m & $ 2.2\times 10^{-8}$ \\
a.u. of electric quadrupole moment & $ea_0^2$ & $ 4.486\,551\,331(99)\times
  10^{-40}$ & C m$^2$ & $ 2.2\times 10^{-8}$ \\
a.u. of electric polarizability & $e^2a_0^2/E_{\rm h}$ & $
  1.648\,777\,2754(16)\times 10^{-41}$ & C$^2$ m$^2$ J$^{-1}$ & $ 9.7\times
  10^{-10}$ \\
a.u. of 1$^{\rm st}$ hyperpolarizability & $e^3a_0^3/E_{\rm h}^2$ & $
  3.206\,361\,449(71)\times 10^{-53}$ & C$^3$ m$^3$ J$^{-2}$ & $ 2.2\times
  10^{-8}$ \\
a.u. of 2$^{\rm nd}$ hyperpolarizability & $e^4a_0^4/E_{\rm h}^3$ & $
  6.235\,380\,54(28)\times 10^{-65}$ & C$^4$ m$^4$ J$^{-3}$ & $ 4.4\times
  10^{-8}$ \\
a.u. of magnetic flux density & $\hbar/ea_0^2$ & $ 2.350\,517\,464(52)\times
  10^{5}$ & T & $ 2.2\times 10^{-8}$ \\
a.u. of magnetic dipole moment: $2\mu_{\rm B}$ & $\hbar e/m_{\rm e}$ & $
  1.854\,801\,936(41)\times 10^{-23}$ & J T$^{-1}$ & $ 2.2\times 10^{-8}$ \\
a.u. of magnetizability & $e^2a_0^2/m_{\rm e}$ & $ 7.891\,036\,607(13)\times
  10^{-29}$ & J T$^{-2}$ & $ 1.6\times 10^{-9}$ \\
a.u. of permittivity: $10^7/c^2$ & $e^2/a_0E_{\rm h}$ & $
  1.112\,650\,056\ldots\times 10^{-10}$ & F m$^{-1}$ & exact \\
\botrule
\end{tabular}
\end{table*}

\begingroup
\squeezetable
\begin{table*}[!]
\def\hsp{\hbox to 16 pt {}}
\caption{The values of some energy equivalents derived from the
relations $E=mc^2 = hc/\lambda = h\nu = kT$, and based on the 2010
CODATA adjustment of the values of the constants; 1~eV~$=(e/{\rm C})$~J,
1~u $= m_{\rm u} = \textstyle{1\over12}m(^{12}{\rm C}) =
10^{-3}$~kg~mol$^{-1}\!/N_{\rm A}$, and $E_{\rm h} = 2R_{\rm \infty}hc =
\alpha^2m_{\rm e}c^2$ is the Hartree energy (hartree).}
\label{tab:enconv1}
\begin{tabular} {l@{\hsp}l@{\hsp}l@{\hsp}l@{\hsp}l}
\toprule
\multicolumn{5} {c} {Relevant unit} \\
\colrule
    &   \s{30}J &  \s{30}kg &  \s{30}m$^{-1}$  &   \s{30}Hz  \vbox to 12 pt {}
  \\
\colrule
        &                    &                    &               &
  \\
1~J    & $(1\ {\rm J})=$    &  (1 J)/$c^2=$     &       (1 J)/$hc=$   &
  (1 J)/$h=$              \\
 & 1 J   & $ 1.112\,650\,056\ldots\times 10^{-17}$ kg  & $
  5.034\,117\,01(22)\times 10^{24}$ m$^{-1}$ & $ 1.509\,190\,311(67)\times
  10^{33}$ Hz \\
        &        &             &           &  \\
1~kg    &        (1 kg)$c^2=$   &   $(1 \ {\rm kg})=$       &        (1
  kg)$c/h=$ &        (1 kg)$c^2/h=$          \\
 &  $ 8.987\,551\,787\ldots\times 10^{16}$ J & 1 kg   & $
  4.524\,438\,73(20)\times 10^{41}$ m$^{-1}$ & $ 1.356\,392\,608(60)\times
  10^{50}$ Hz \\
 &           &            &         &   \\
1~m$^{-1}$   &  (1 m$^{-1})hc=$   &       (1 m$^{-1})h/c=$    &  $(1$ m$^{-1})=$
  &   (1 m$^{-1})c=$  \\
 &  $ 1.986\,445\,684(88)\times 10^{-25}$ J  & $ 2.210\,218\,902(98)\times
  10^{-42}$ kg  & 1 m$^{-1}$  & $ 299\,792\,458$ Hz \\
 &           &            &         &   \\
1~Hz   &  (1 Hz)$h=$  &  (1 Hz)$h/c^{2}=$   &        (1 Hz)/$c=$  &  $(1$ Hz$)=$
  \\
 &  $ 6.626\,069\,57(29)\times 10^{-34}$ J  & $ 7.372\,496\,68(33)\times
  10^{-51}$ kg & $ 3.335\,640\,951\ldots\times 10^{-9}$ m$^{-1}$ & 1 Hz \\
 &           &            &         &   \\
1~K  &  (1 K)$k=$  &   (1 K)$k/c^{2}=$  &    (1 K)$k/hc=$ &  (1 K)$k/h=$   \\
 &  $ 1.380\,6488(13)\times 10^{-23}$ J  & $ 1.536\,1790(14)\times 10^{-40}$ kg
  & $ 69.503\,476(63)$ m$^{-1}$ & $ 2.083\,6618(19)\times 10^{10}$ Hz \\
         &            &            &           &        \\
1~eV    &  (1 eV) =  &  $(1~{\rm eV})/c^{2}=$  &   $(1~{\rm eV})/hc=$   &
  $(1~{\rm eV})/h=$   \\
 &  $ 1.602\,176\,565(35)\times 10^{-19}$ J  & $ 1.782\,661\,845(39)\times
  10^{-36}$ kg & $ 8.065\,544\,29(18)\times 10^{5}$ m$^{-1}$ & $
  2.417\,989\,348(53)\times 10^{14}$ Hz \\
         &          &             &         &     \\
1~u   &   $(1~{\rm u})c^{2}=$   &  (1 u) =   &  $(1~{\rm u})c/h=$  &  $(1~{\rm
  u})c^{2}/h=$    \\
 &  $ 1.492\,417\,954(66)\times 10^{-10}$ J  & $ 1.660\,538\,921(73)\times
  10^{-27}$ kg & $ 7.513\,006\,6042(53)\times 10^{14}$ m$^{-1}$ & $
  2.252\,342\,7168(16)\times 10^{23}$ Hz \\
         &          &             &        &       \\
1~$E_{\rm h}$   &  $(1~E_{\rm h})=$  & $(1~E_{\rm h})/c^2=$ &  $(1~E_{\rm h}
  )/hc=$  &  $(1~E_{\rm h})/h=$   \\
 &  $ 4.359\,744\,34(19)\times 10^{-18}$ J  & $ 4.850\,869\,79(21)\times
  10^{-35}$ kg & $ 2.194\,746\,313\,708(11)\times 10^{7}$ m$^{-1}$ & $
  6.579\,683\,920\,729(33)\times 10^{15}$ Hz \\
\botrule
\end{tabular}
\end{table*}
\endgroup

\begingroup
\squeezetable
\begin{table*}[!]
\def\hsp{\hbox to 16 pt {}}
\caption{The values of some energy equivalents derived from the
relations $E=mc^2 = hc/\lambda = h\nu = kT$, and based on the 2010
CODATA adjustment of the values of the constants; 1~eV~$=(e/{\rm C})$~J,
1~u $= m_{\rm u} = \textstyle{1\over12}m(^{12}{\rm C}) =
10^{-3}$~kg~mol$^{-1}\!/N_{\rm A}$, and $E_{\rm h} = 2R_{\rm \infty}hc =
\alpha^2m_{\rm e}c^2$ is the Hartree energy (hartree).}
\label{tab:enconv2}
\begin{tabular} {l@{\hsp}l@{\hsp}l@{\hsp}l@{\hsp}l}
\toprule
\multicolumn{5} {c} {Relevant unit} \\
\colrule
    &   \s{30}K &  \s{30}eV &  \s{30}u  &   \s{30}$E_{\rm h}$  \vbox to 12 pt {}
  \\
\colrule
        &                    &                    &               &
  \\
1~J     & (1 J)/$k=$   &  (1 J) =   &       (1 J)/$c^2$ = &   (1 J) =
  \\
 &  $ 7.242\,9716(66)\times 10^{22}$ K  & $ 6.241\,509\,34(14)\times 10^{18}$ eV
  & $ 6.700\,535\,85(30)\times 10^{9}$ u & $ 2.293\,712\,48(10)\times 10^{17}$
  $E_{\rm h}$ \\
        &        &             &           &  \\
1~kg    &        (1 kg)$c^2/k=$    &   (1 kg)$c^2$ =    &    (1 kg) =  &  (1
  kg)$c^2=$   \\
 &  $ 6.509\,6582(59)\times 10^{39}$ K  & $ 5.609\,588\,85(12)\times 10^{35}$ eV
  & $ 6.022\,141\,29(27)\times 10^{26}$ u  & $ 2.061\,485\,968(91)\times 10^{34}
  $ $E_{\rm h}$ \\
 &           &            &         &   \\
1~m$^{-1}$   &  (1 m$^{-1})hc/k=$  &       (1 m$^{-1})hc=$  &  (1 m$^{-1})h/c$ =
  &   (1 m$^{-1})hc=$ \\
 &  $ 1.438\,7770(13)\times 10^{-2}$ K  & $ 1.239\,841\,930(27)\times 10^{-6}$
  eV  & $ 1.331\,025\,051\,20(94)\times 10^{-15}$ u & $
  4.556\,335\,252\,755(23)\times 10^{-8}$ $E_{\rm h}$ \\
         &           &            &            &      \\
1~Hz   &  (1 Hz)$h/k=$  &  (1 Hz)$h=$  &   (1 Hz)$h/c^2$ = & (1 Hz)$h=$  \\
 &  $ 4.799\,2434(44)\times 10^{-11}$ K  & $ 4.135\,667\,516(91)\times 10^{-15}$
  eV  & $ 4.439\,821\,6689(31)\times 10^{-24}$ u & $
  1.519\,829\,846\,0045(76)\times 10^{-16}$ $E_{\rm h}$ \\
           &             &          &           &        \\
1~K  & $(1$ K$)=$  &   (1 K)$k=$ &    (1 K)$k/c^2=$ &  (1 K)$k=$   \\
 & 1 K   &   $ 8.617\,3324(78)\times 10^{-5}$ eV & $ 9.251\,0868(84)\times
  10^{-14}$ u & $ 3.166\,8114(29)\times 10^{-6}$ $E_{\rm h}$ \\
         &            &            &           &        \\
1~eV    &  (1 eV)/$k=$  &  $(1$ eV$)=$  &   $(1~{\rm eV})/c^2=$  &     $(1~{\rm
  eV})=$   \\
 &  $ 1.160\,4519(11)\times 10^{4}$ K  & 1 eV  & $ 1.073\,544\,150(24)\times
  10^{-9}$ u & $ 3.674\,932\,379(81)\times 10^{-2}$ $E_{\rm h}$ \\
         &          &             &         &     \\
1~u   &   $(1~{\rm u})c^{2}/k=$   &  $(1~{\rm u})c^2=$ &  $(1$ u$)=$  &
  $(1~{\rm u})c^2=$   \\
 &  $ 1.080\,954\,08(98)\times 10^{13}$ K  & $ 931.494\,061(21)\times 10^{6}$ eV
  & 1 u & $ 3.423\,177\,6845(24)\times 10^{7}$ $E_{\rm h}$ \\
         &          &             &        &       \\
1~$E_{\rm h}$   &  $(1~E_{\rm h})/k=$  & $(1~E_{\rm h})=$ &  $(1~E_{\rm h}
  )/c^2=$ & $(1~E_{\rm h})=$    \\
 &  $ 3.157\,7504(29)\times 10^{5}$ K  & $ 27.211\,385\,05(60)$ eV  & $
  2.921\,262\,3246(21)\times 10^{-8}$ u  & $1~E_{\rm h}$  \\
\botrule
\end{tabular}
\end{table*}
\endgroup

\clearpage

\section{Summary and Conclusion}
\label{sec:c}

The focus of this section is (i) comparison of the 2010 and 2006
recommended values of the constants and  identification of those new
results that have contributed most to the changes in the 2006 values;
(ii) presentation of several conclusions that can be drawn from the 2010
recommended values and the input data on which they are based; and (iii)
identification of new experimental and theoretical work that can advance
our knowledge of the values of the constants.

Topic (iii) is of special importance in light of the adoption by the
24th General Conference on Weights and Measures (CGPM) at its meeting in
Paris in October 2011 of Resolution 1 entitled ``On the possible future
revision of the International System of Units, the SI,'' available on
the BIPM Web site at
bipm.org/utils/common/pdf/24\_CGPM\_Resolutions.pdf.

In brief, this resolution notes the intention of the CIPM to propose,
possibly to the 25th CGPM in 2014, a revision of the SI.  The ``New
SI,'' as it is called to distinguish it from the current SI, will be the
system of units in which seven reference constants, including the Planck
constant $h$, elementary charge $e$, Boltzmann constant $k$, and
Avogadro constant $N_{\rm A}$, have exact assigned values.  Resolution 1
also looks to CODATA to provide the necessary values of these four
constants for the new definition.  Details of the proposed New SI may be
found in \citet{2011191} and the references cited therein; see also
\citet{2011232,2010054}.

\subsection{Comparison of 2010 and 2006 CODATA recommended
values}
\label{ssec:crv}

\def\pad{{\hbox to 10 pt {}}}
\def\hsp{{\hbox to 14 pt {}}}
\def\pa{{\hbox to 5 pt {}}}
\def\s#1{\hbox to #1 pt {}}
\begin{table}
\caption{Comparison of the 2010 and 2006 CODATA adjustments of the
values of the constants by the comparison of the corresponding
recommended values of a representative group of constants.  Here $D_{\rm
r}$ is the 2010 value minus the 2006 value divided by the standard
uncertainty $u$ of the 2006 value (i.e., $D_{\rm r}$ is the change in
the value of the constant from 2006 to 2010 relative to its 2006
standard uncertainty).}
\label{tab:10vs06}
\begin{tabular}{@{\hsp} c @{\hsp} l @{\hsp} c @{\hsp} r @{\hsp} }
\toprule
Quantity & 2010 rel. std.  & Ratio 2006 $u_{\rm r}$    & $D_{\rm r}$ \pad \\
         & uncert. $u_{\rm r}$ & to 2010 $u_{\rm r}$ &    \\
\colrule
\vbox to 10pt{}$\alpha$  &  \pa $ 3.2\times 10^{-10}$ & \pa $ 2.1$ & $ 6.5$\pad
  \\
$R_{\rm K}$  &  \pa $ 3.2\times 10^{-10}$ & \pa $ 2.1$ & $ -6.5$\pad  \\
$a_{\rm 0}$  &  \pa $ 3.2\times 10^{-10}$ & \pa $ 2.1$ & $ 6.5$\pad   \\
$\lambda_{\rm C}$  &  \pa $ 6.5\times 10^{-10}$ & \pa $ 2.1$ & $ 6.5$\pad  \\
$r_{\rm e}$  &  \pa $ 9.7\times 10^{-10}$ & \pa $ 2.1$ & $ 6.5$\pad \\
$\sigma_{\rm e}$  &  \pa $ 1.9\times 10^{-9}$ & \pa $ 2.1$ & $ 6.5$\pad   \\
$ h $  &  \pa $ 4.4\times 10^{-8}$ & \pa $ 1.1$ & $ 1.9$\pad  \\
$m_{\rm e}$  &  \pa $ 4.4\times 10^{-8}$ & \pa $ 1.1$ & $ 1.7$\pad \\
$m_{\rm h}$  &  \pa $ 4.4\times 10^{-8}$ & \pa $ 1.1$ & $ 1.7$\pad \\
$m_{\rmssalpha}$  &  \pa $ 4.4\times 10^{-8}$ & \pa $ 1.1$ & $ 1.7$\pad  \\
$N_{\rm A}$  &  \pa $ 4.4\times 10^{-8}$ & \pa $ 1.1$ & $ -1.7$\pad  \\
$E_{\rm h}$  &  \pa $ 4.4\times 10^{-8}$ & \pa $ 1.1$ & $ 1.9$\pad  \\
$c_1$  &  \pa $ 4.4\times 10^{-8}$ & \pa $ 1.1$ & $ 1.9$\pad \\
$e$  &  \pa $ 2.2\times 10^{-8}$ & \pa $ 1.1$ & $ 1.9$\pad \\
$K_{\rm J}$  &  \pa $ 2.2\times 10^{-8}$ & \pa $ 1.1$ & $ -1.8$\pad \\
$F$  &  \pa $ 2.2\times 10^{-8}$ & \pa $ 1.1$ & $ -1.4$\pad \\
$\gamma^{\,\prime}_{\rm p}$  &  \pa $ 2.5\times 10^{-8}$ & \pa $ 1.1$ & $
  -1.3$\pad \\
$\mu_{\rm B}$  &  \pa $ 2.2\times 10^{-8}$ & \pa $ 1.1$ & $ 2.3$\pad \\
$\mu_{\rm N}$  &  \pa $ 2.2\times 10^{-8}$ & \pa $ 1.1$ & $ 2.3$\pad \\
$\mu_{\rm e}$  &  \pa $ 2.2\times 10^{-8}$ & \pa $ 1.1$ & $ -2.3$\pad \\
$\mu_{\rm p}$  &  \pa $ 2.4\times 10^{-8}$ & \pa $ 1.1$ & $ 2.2$\pad \\
$R$  &  \pa $ 9.1\times 10^{-7}$ & \pa $ 1.9$ & $ -0.7$\pad  \\
$k$  &  \pa $ 9.1\times 10^{-7}$ & \pa $ 1.9$ & $ -0.7$\pad  \\
$V_{\rm m}$  &  \pa $ 9.1\times 10^{-7}$ & \pa $ 1.9$ & $ -0.7$\pad \\
$c_2$  &  \pa $ 9.1\times 10^{-7}$ & \pa $ 1.9$ & $ 0.7$\pad \\
$\sigma$  &  \pa $ 3.6\times 10^{-6}$ & \pa $ 1.9$ & $ -0.7$\pad \\
$ G $  &  \pa $ 1.2\times 10^{-4}$ & \pa $ 0.8$ & $ -0.7$\pad \\
$R_\infty$  &  \pa $ 5.0\times 10^{-12}$ & \pa $ 1.3$ & $ 0.2$\pad \\
$m_{\rm e}/m_{\rm p}$  &  \pa $ 4.1\times 10^{-10}$ & \pa $ 1.1$ & $ 0.0$\pad \\
$m_{\rm e}/m_{\rmssmu}$  &  \pa $ 2.5\times 10^{-8}$ & \pa $ 1.0$ & $ -0.4$\pad
  \\
$A_{\rm r}({\rm e})$  &  \pa $ 4.0\times 10^{-10}$ & \pa $ 1.1$ & $ 0.1$\pad \\
$A_{\rm r}({\rm p})$  &  \pa $ 8.9\times 10^{-11}$ & \pa $ 1.2$ & $ 0.4$\pad \\
$A_{\rm r}({\rm n})$  &  \pa $ 4.2\times 10^{-10}$ & \pa $ 1.0$ & $ 0.1$\pad \\
$A_{\rm r}({\rm d})$  &  \pa $ 3.8\times 10^{-11}$ & \pa $ 1.0$ & $ -0.2$\pad \\
$A_{\rm r}({\rm t})$  &  \pa $ 8.2\times 10^{-10}$ & \pa $ 1.0$ & $ 0.0$\pad \\
$A_{\rm r}({\rm h})$  &  \pa $ 8.3\times 10^{-10}$ & \pa $ 1.0$ & $ -0.2$\pad \\
$A_{\rm r}({\rmalpha})$  &  \pa $ 1.5\times 10^{-11}$ & \pa $ 1.0$ & $ 0.0$\pad
  \\
$d_{\rm 220}$  &  \pa $ 1.6\times 10^{-8}$ & \pa $ 1.6$ & $ -1.0$\pad \\
$g_{\rm e}$  &  \pa $ 2.6\times 10^{-13}$ & \pa $ 2.8$ & $ 0.5$\pad \\
$g_{\rmssmu}$  &  \pa $ 6.3\times 10^{-10}$ & \pa $ 1.0$ & $ -0.3$\pad \\
$\mu_{\rm p}/\mu_{\rm B}$  &  \pa $ 8.1\times 10^{-9}$ & \pa $ 1.0$ & $ 0.0$\pad
  \\
$\mu_{\rm p}/\mu_{\rm N}$  &  \pa $ 8.2\times 10^{-9}$ & \pa $ 1.0$ & $ 0.0$\pad
  \\
$\mu_{\rm n}/\mu_{\rm N}$  &  \pa $ 2.4\times 10^{-7}$ & \pa $ 1.0$ & $ 0.0$\pad
  \\
$\mu_{\rm d}/\mu_{\rm N}$  &  \pa $ 8.4\times 10^{-9}$ & \pa $ 1.0$ & $ 0.0$\pad
  \\
$\mu_{\rm e}/\mu_{\rm p}$  &  \pa $ 8.1\times 10^{-9}$ & \pa $ 1.0$ & $ 0.0$\pad
  \\
$\mu_{\rm n}/\mu_{\rm p}$  &  \pa $ 2.4\times 10^{-7}$ & \pa $ 1.0$ & $ 0.0$\pad
  \\
$\mu_{\rm d}/\mu_{\rm p}$  &  \pa $ 7.7\times 10^{-9}$ & \pa $ 1.0$ & $ 0.0$\pad
  \\
\botrule

\end{tabular}
\end{table}

Table~\ref{tab:10vs06} compares the 2010 and 2006 recommended values of
a representative group of constants.  The fact that the values of many
constants are obtained from expressions proportional to the
fine-structure constant $\alpha$, Planck constant $h$, or molar gas
constant $R$ raised to various powers leads to the regularities observed
in the numbers in columns 2 to 4. For example, the first six quantities
are obtained from expressions proportional to $\alpha^a$, where $|a| =
1,~2,~3$, or $6$.  The next 15 quantities, $h$ through the magnetic
moment of the proton $\mu_{\rm p}$, are calculated from expressions
containing the factor $h^a$, where $|a| = 1$ or $1/2$.  And the five
quantities $R$ through the Stefan Boltzmann constant $\sigma$ are
proportional to $R^a$, where $|a| =  1$ or $4$.

Further comments on some of the entries in Table~\ref{tab:10vs06} are as
follows.

(i) The large shift in the 2006 recommended value of $\alpha$ is mainly
due to the discovery and correction of an error in the numerically
calculated value of the eighth-order coefficient $A_1^{(8)}$ in the
theoretical expression for $a_{\rm e}$; see Sec.~\ref{sssec:ath}.  Its
reduction in uncertainty is due to two new results.  The first is the
2008 improved value of $a_{\rm e}$ obtained at Harvard University with a
relative standard uncertainty of $2.4\times10^{-10}$ compared to the
$7.0\times10^{-10}$ uncertainty of the earlier Harvard result used in
the 2006 adjustment.  The second result is  the 2011 improved LKB
atom-recoil value of $h/m(^{87}{\rm Rb})$ with an uncertainty of
$1.2\times10^{-9}$ compared to the $1.3\times10^{-8}$ uncertainty of the
earlier LKB result used in 2006.  The much reduced uncertainty of
$g_{\rm e}$ is also due to the improved value of $\alpha$.

(ii) The change in the 2006 recommended value of $h$ is due to the 2011
IAC result for $N_{\rm A}$ with a relative standard uncertainty of
$3.0\times10^{-8}$ obtained using $^{28}{\rm Si}$ enriched single
crystals.  It provides a value of $h$ with the same uncertainty, which
is smaller than the $3.6\times10^{-8}$ uncertainty of the value of $h$
from the 2007 NIST watt-balance measurement of $K_{\rm J}^2R_{\rm K}$;
the latter played the dominant role in determining the 2006 recommended
value.  The two  differ by about 18 parts in $10^8$, resulting in a
shift of the 2006 recommended value by nearly twice its uncertainty. In
the 2006 adjustment inconsistencies among some of the electrical and
silicon crystal data (all involving natural silicon) led the Task Group
to increase the uncertainties of these data by the multiplicative factor
1.5 to reduce the inconsistencies to an acceptable level.  In the 2010
adjustment, inconsistencies among the data that determine $h$ are
reduced to an acceptable level by using a multiplicative factor of 2.
Consequently the uncertainties of the 2006 and 2010 recommended values
of $h$ do not differ significantly.

(iii) The 2006 recommended value of the molar gas constant $R$ was
determined by the 1988 NIST speed-of-sound result with a relative
standard uncertainty of $1.8\times10^{-6}$, and to a much lesser extent
the 1979 NPL speed-of-sound result   with an uncertainty of
$8.4\times10^{-6}$ obtained with a rather different type of apparatus.
The six new data of potential interest related to $R$ that became
available during the 4 years between the 2006 and 2010 adjustments have
uncertainties ranging from $1.2\times10^{-6}$ to $12\times10^{-6}$ and
agree with each other as well as with the NIST and NPL values. Further,
the self-sensitivity coefficients of four of the six were sufficiently
large for them to be included in the 2010 final adjustment, and they are
responsible for the small shift in the 2006 recommended value and the
reduction of its uncertainty by nearly a factor of 2.

(iv) Other constants in Table~\ref{tab:10vs06} whose changes are worth
noting are the Rydberg constant $R_\infty$, proton relative atomic mass
$A_{\rm r}({\rm p})$, and \{220\} natural Si lattice spacing $d_{220}$.
The reduction in uncertainty of $R_\infty$ is due to improvements in the
theory of H and D energy levels and the 2010 LKB result for the $1{\rm
S_{1/2}}-3{\rm S_{1/2}}$ transition frequency in hydrogen with a
relative standard uncertainty of $4.4\times10^{-12}$.  For $A_{\rm
r}({\rm p})$, the reduction in uncertainty is due to the 2008 Stockholm
University (SMILETRAP) result for the ratio of the cyclotron frequency
of the excited hydrogen molecular ion to that of the deuteron, $f_{\rm
c}({\rm H_2^{+*}})/f_{\rm c}({\rm d})$, with a relative uncertainty of
$1.7\times10^{-10}$. The changes in $d_{220}$ arise from the omission of
the 1999 PTB result for $h/m_{\rm n}d_{220}({\scriptstyle {\rm W}04})$,
the 2004 NMIJ result for $d_{220}({\scriptstyle {\rm NR3}})$, the 2007
INRIM  results for $d_{220}({\scriptstyle {\rm W4.2a}})$, and
$d_{220}({\scriptstyle {\rm MO^*}})$, and the inclusion of the new 2008
INRIM result for $d_{220}({\scriptstyle {\rm MO^*}})$ as well as the new
2009 INRIM results for $d_{220}({\scriptstyle {\rm W}04})$ and
$d_{220}({\scriptstyle {\rm W4.2a}})$.

\subsection{Some implications of the 2010 CODATA recommended
values and adjustment for metrology and physics}
\label{ssec:imp?}

\emph{Conventional electric units}. The adoption of the conventional
values $K_{\rm J-90} = 483\,597.9~{\rm GHz/V}$ and $R_{\rm K-90} =
25\,812.807~{\Omega}$ for the Josephson and von Klitzing constants in
1990 can be viewed as establishing conventional, practical units of
voltage and resistance, $V_{90}$ and ${\it \Omega}_{90}$, given by
$V_{90} = (K_{\rm J-90}/K_{\rm J})$ V and ${\it \Omega}_{90} = (R_{\rm
K}/R_{\rm K-90})~{\rm \Omega}$.  Other conventional electric units
follow from $V_{90}$ and ${\it \Omega}_{90}$, for example, $A_{90} =
V_{90}/{\it \Omega}_{90}$, $C_{90} = A_{90}$~s, $W_{90} = A_{90}V_{90}$,
$F_{90} = C_{90}/V_{90}$, and $H_{90} = {\it \Omega}_{90}$~s, which are
the conventional, practical units of current, charge, power,
capacitance, and inductance, respectively \cite{2001027}.  For the
relations between $K_{\rm J}$ and $K_{\rm J-90}$, and $R_{\rm K}$ and
$R_{\rm K-90}$, the 2010 adjustment gives
\begin{eqnarray}
K_{\rm J} &=&
K_{\rm J-90} [1-6.3(2.2)\times10^{-8}] \, , \\
R_{\rm K} &=& R_{\rm K-90}
[1+1.718(32)\times10^{-8}] \, ,
\end{eqnarray}
which lead to
\begin{eqnarray}
V_{90} &=& [1 + 6.3(2.2) \times  10^{-8}]~{\rm V},
\label{eq:c901} \\ {\it \Omega}_{90} &=& [1 + 1.718(32) \times
10^{-8}]~{\rm \Omega}, \label{eq:c902} \\ A_{90} &=& [1 - 4.6(2.2)
\times  10^{-8}]~{\rm A}\, , \label{eq:c903} \\ C_{90} &=& [1 - 4.6(2.2)
\times  10^{-8}]~{\rm C}\, , \label{eq:c904} \\ W_{90} &=& [1 + 10.8(5.0)
\times  10^{-8}]~{\rm W}\, , \label{eq:c905} \\ F_{90} &=& [1 - 1.718(32)
\times  10^{-8}]~{\rm F}\, , \label{eq:c906} \\ H_{90} &=& [1 + 1.718(32)
\times  10^{-8}]~{\rm H}\, .  \label{eq:c907}
\end{eqnarray}
Equations~(\ref{eq:c901}) and (\ref{eq:c902}) show that $V_{90}$ exceeds
V and ${\it \Omega}_{90}$ exceeds $\Omega$ by $6.3(2.2) \times 10^{-8}$
and $1.718(32) \times 10^{-8}$, respectively.  This means that measured
voltages and resistances traceable to the Josephson effect and $K_{\rm
J-90}$ and the quantum Hall effect and $R_{\rm K-90}$, respectively, are
too small relative to the SI by these same fractional amounts.  However,
these differences are well within the $40 \times 10^{-8}$ uncertainty
assigned to $V_{90}/$V and the $10 \times 10^{-8}$ uncertainty assigned
to ${\it \Omega}_{90}/\Omega$ by the Consultative Committee for
Electricity and Magnetism (CCEM) of the CIPM \cite{2001223,1989052}.

\emph{Josephson and quantum Hall effects}.  Although there is extensive
theoretical and experimental evidence for the exactness of the
Josephson and quantum-Hall-effect relations $K_{\rm J}=2e/h$
and $R_{\rm K}=h/e^2$, and some of the input data available for the 2010
adjustment provide additional supportive evidence for these expressions,
some other data are not supportive.  This dichotomy reflects the rather
significant inconsistencies among a few key data, particularly the
highly accurate IAC enriched silicon XRCD result for  $N_{\rm A}$, and
the comparably accurate NIST watt-balance result for $K_{\rm J}^2R_{\rm
K}$, and will only be fully resolved when the inconsistencies are
reconciled.

\emph{The New SI}. Implementation of the New SI requires that the four
reference constants $h$, $e$, $k$, and $N_{\rm A}$ must  be known with
sufficiently small uncertainties to meet current and future measurement
needs.  However, of equal if not greater importance, the causes of any
inconsistencies among the data that provide their values must be
understood. Although the key data  that provide the 2010 recommended
value of $k$ would appear to be close to meeting both requirements, this
is not the case for $h$, $e$, and $N_{\rm A}$, which are in fact
interrelated. We have
\begin {eqnarray}
N_{\rm A}h &=& {cA_{\rm r}({\rm e})M_{\rm u}\alpha^2\over 2R_{\infty}}\, ,
\label{eq:nah} \\
e &=& \left({2\alpha h \over \mu_{0}c}\right)^{1/2}\, .
\label{eq:ealh}
\end{eqnarray}

Since the combined relative standard uncertainty of the 2010 recommended
values of the constants on the right-hand-side of Eq.~(\ref{eq:nah}) is
only $7.0\times 10^{-10}$, a measurement of $h$ with a given relative
uncertainty, even as small as  $5\times 10^{-9}$, determines $N_{\rm A}$
with essentially the same relative uncertainty. Further, since the
recommended value of $\alpha$ has a relative uncertainty of only
$3.2\times 10^{-10}$, based on Eq.~(\ref{eq:ealh}) the relative
uncertainty of $e$ will be half that of $h$ or $N_{\rm A}$. For these
reasons, the 2010 recommended values of $h$ and $N_{\rm A}$ have the
same $4.4\times 10^{-8}$ relative uncertainty, and the uncertainty of
the recommended value of $e$ is $2.2\times 10^{-8}$.  However, these
uncertainties are twice as large as they would have been if there were
no disagreement between the watt-balance values of $h$ and the enriched
silicon XRCD value of $N_{\rm A}$. This disagreement led to an increase
in the uncertainties of the relevant data by a factor of 2.  More
specifically, if the data had been consistent the uncertainties of the
recommended values of $h$ and $N_{\rm A}$ would be $2.2\times 10^{-8}$
and $1.1\times 10^{-8}$ for $e$. Because these should be sufficiently
small for the New SI to be implemented, the significance of the
disagreement and the importance of measurements of $h$ and $N_{\rm A}$
are apparent.

\emph{Proton radius}. The proton rms charge radius $r_{\rm p}$
determined from the  Lamb shift in muonic hydrogen disagrees
significantly with values determined from H and D transition frequencies
as well as from electron-proton scattering experiments.  Although the
uncertainty of the muonic hydrogen value is significantly smaller than
the uncertainties of these other values, its negative impact on the
internal consistency of the theoretically predicted and experimentally
measured frequencies, as well as on the value of the Rydberg constant,
was deemed so severe that the only recourse was to not include it in the
final least-squares adjustment on which the 2010 recommended values are
based.

\emph{Muon magnetic moment anomaly}.  Despite extensive new theoretical
work, the long-standing significant difference between the theoretically
predicted, standard-model value of  $a_{\rmssmu}$ and the experimentally
determined value remains unresolved.  Because the difference is from 3.3
to possibly 4.5 times the standard uncertainty of the difference,
depending on the way the all-important hadronic contribution to the
theoretical expression for $a_{\rmssmu}$ is evaluated, the theory was
not incorporated in the 2010 adjustment. The recommended values of
$a_{\rmssmu}$ and those of other constants that depend on it are,
therefore, based  on experiment.

\emph{Electron magnetic moment anomaly, fine-structure constant, and
QED}. The most accurate value of the fine-structure constant $\alpha$
currently available from a single experiment has a relative standard
uncertainty of $3.7\times 10^{-10}$; it is obtained by equating the QED
theoretical expression for the electron magnetic moment anomaly $a_{\rm
e}$ and the most accurate experimental value of $a_{\rm e}$, obtained
from measurements on a single electron in a Penning trap. This value of
$\alpha$ is in excellent agreement with a competitive experimental value
with an uncertainty of $6.6\times 10^{-10}$.  Because the latter is
obtained from the atom-recoil determination of the quotient
$h/m(^{87}{\rm Rb})$ using atom-interferometry and is only weakly
dependent on QED theory, the agreement provides one of the most
significant confirmations of quantum electrodynamics.

\emph{Newtonian constant of gravitation}.  The situation regarding
measurements of $G$ continues to be problematic and has become more so
in the past 4 years. Two new results with comparatively small
uncertainties have become available for the 2010 adjustment, leading to
an increase in the scatter among the now 11 values of $G$.  This has
resulted in a $20\,\%$ increase in the uncertainty of the 2010
recommended value compared to that of its 2006 predecessor. Clearly,
there is a continuing problem for the determination of this important,
but poorly-known, fundamental constant; the uncertainty of the 2010
recommended value is now 120 parts in $10^6$.

\subsection{Suggestions for future work}
\label{ssec:sfw}

For evaluation of the fundamental constants, it is desirable not only to
have multiple results with competitive uncertainties for a given
quantity, but also to have one or more results obtained by a different
method.  If the term ``redundant'' is used to describe such an ideal set
of data, there is usually only limited redundancy among the key data
available for any given CODATA adjustment.

With this in mind, based on the preceding discussion, our suggestions
are as follows.

(i) Resolution of the disagreement between the most accurate
watt-balance result for $K_{\rm J}^2R_{\rm K}$ and the XRCD result for
$N_{\rm A}$. Approaches to solving this problem might include new
measurements of $K_{\rm J}^2R_{\rm K}$ using watt balances of different
design (or their equivalent) with uncertainties at the 2 to 3 parts in
$10^8$ level, a thorough review by the researchers involved of their
existing measurements of this quantity, tests of the exactness of the
relations $K_{\rm J}=2e/h$ and $R_{\rm K}=h/e^2$, independent
measurements of the isotopic composition of the enriched silicon
crystals and their $d_{220}$ lattice spacing used in the determination
of $N_{\rm A}$ (these are the two principal quantities for which only
one measurement exists), and a thorough review by the researchers
involved of the many corrections required to obtain $N_{\rm A}$ from the
principal quantities measured.

(ii) Measurements of $k$ (and related quantities such as $k/h$) with
uncertainties at the 1 to 3 parts in $10^6$ level using the techniques
of dielectric gas thermometry, refractive index gas thermometry, noise
thermometry, and Doppler broadening, because these methods are so very
different from acoustic gas thermometry, which is the dominant method
used to date.

(iii) Resolution of the discrepancy between the muonic hydrogen inferred
value of $r_{\rm p}$ and the spectroscopic value from H and D transition
frequencies.  Work underway on frequency measurements in hydrogen as
well as the analysis of $\rmmu^-{\rm p}$ and $\rmmu^-{\rm d}$ data and
possible measurements in $\rmmu^-{\rm h}$ and $\rmmu^-\rmalpha$ should
provide additional useful information.  Independent evaluations of
electron scattering data to determine $r_{\rm p}$ are encouraged as well
as verification of the theory of H, D, and muonic hydrogen-like energy
levels.

(iv) Independent calculation of the eighth- and tenth-order coefficients
in the QED expression for $a_{\rm e}$, in order to increase confidence
in the value of $\alpha$ from $a_{\rm e}$.

(v) Resolution of the disagreement between the theoretical expression
for $a_{\rmssmu}$ and its experimental value.  This discrepancy along
with the discrepancy between theory and experiment in muonic hydrogen
are two important problems in muon-related physics.

(vi) Determinations of $G$ with an uncertainty of one part in $10^5$
using new and innovative approaches that might resolve the disagreements
among the measurements made within the past three decades.

\section{Acknowledgments}

We gratefully acknowledge the help of our many colleagues throughout the
world who provided the CODATA Task Group on Fundamental Constants with
results prior to formal publication and for promptly and patiently
answering our many questions about their work.  We wish to thank Barry
M. Wood, the current chair of the Task Group, as well as our fellow
Task-Group members, for their invaluable guidance and suggestions during
the course of the 2010 adjustment effort.

\clearpage
\section*{Nomenclature}
\label{sec:nom}
\begin{longtable}{lp{6.75cm}}
\endfirsthead
\endhead
\endlastfoot{}
\endfoot{}

AMDC & Atomic Mass Data Center,
Centre de Spectrom\'etrie Nucl\'eaire et de Spectrom\'etrie de Masse
(CSNSM), Orsay, France \\

$A_{\rm r}(X)$ & Relative atomic mass of $X$: $A_{\rm r}(X) =
m(X)/m_{\rm u}$ \\

$A_{90}$  & Conventional unit of electric current: \newline
$A_{90} = V_{90}/{\it \Omega}_{90}$  \\

\AA$^\ast$ & \AA ngstr\"om-star:
${\rm \lambda}({\rm WK}{\rmalpha}_{1}) = 0.209\,010\,0 \ {\rm \AA}^\ast$
\\

$a_{\rm e}$ & Electron magnetic moment anomaly: \newline $a_{\rm e} =
(|g_{\rm e}|-2$)/2 \\

$a_{\rmssmu}$ & Muon magnetic moment anomaly: \newline $a_{\rmssmu} =
(|g_{\rmssmu}| -2$)/2  \\

BIPM & International Bureau of
Weights and Measures, S\`evres, France \\

BNL & Brookhaven National Laboratory, Upton, New York, USA \\

CERN & European Organization for Nuclear Research, Geneva, Switzerland \\

CIPM &  International Committee for Weights and Measures \\

CODATA \ \ & Committee on Data for Science and
        Technology of the International Council
        for Science  \\

$CPT$ & Combined charge conjugation, parity inversion, and time reversal \\

$c$ & Speed~of~light~in~vacuum \\

d & Deuteron (nucleus of deuterium D, or $^2$H) \\

$d_{220}$ & $\{220\}$ lattice spacing of an ideal crystal of naturally occurring
silicon \\

$d_{220}({\scriptstyle X })$ & $\{220\}$ lattice spacing of crystal $X$ of
  naturally occurring silicon \\

$E_{\rm b}$ & Binding energy \\

e   & Symbol for either member of the electron-positron
pair; when necessary, e$^-$ or e$^+$ is used to indicate
the electron or positron  \\

$e$ & Elementary charge: absolute value of the charge of the electron \\

$F$ & Faraday constant: $F$ = $N_{\rm A}e$ \\

FSU & Florida State University, Tallahassee, Florida, USA \\

FSUJ & Friedrich-Schiller University, Jena, Germany \\

${\cal F}_{90}$ & ${\cal F}_{90} = (F/A_{90})$~A \\

$G$ & Newtonian constant of gravitation \\

$g$ & Local acceleration of free fall \\

$g_{\rm d}$ & Deuteron $g$-factor: $g_{\rm d} = \mu_{\rm d}/\mu_{\rm N}$ \\

$g_{\rm e}$ & Electron $g$-factor: $g_{\rm e} = 2\mu_{\rm e}/\mu_{\rm B}$ \\

$g_{\rm p}$ & Proton $g$-factor: $g_{\rm p} = 2\mu_{\rm p}/\mu_{\rm N}$ \\

$g^\prime_{\rm p}$ & Shielded proton $g$-factor:
$g_{\rm p}^\prime = 2\mu_{\rm p}^\prime/\mu_{\rm N}$ \\

$g_{\rm t}$ & Triton $g$-factor: $g_{\rm t} = 2\mu_{\rm t}/\mu_{\rm N}$ \\

$g_{X}(Y)$ & $g$-factor of particle $X$ in the ground (1S) state of
hydrogenic atom $Y$ \\

$g_{\rmssmu}$ & Muon $g$-factor: $g_{\rmssmu} = 2\mu_{\rmssmu}
  /(e\hbar/2m_{\rmssmu})$ \\

GSI & Gesellschaft f\"ur Schweironenforschung, Darmstadt, Germany \\

HD & HD molecule (bound state of hydrogen and deuterium atoms) \\

HT & HT molecule (bound state of hydrogen and tritium atoms) \\

h & Helion (nucleus of $^{3}$He) \\

$h$ & Planck constant; $\hbar = h/2\rmpi$ \\

HarvU & Harvard University, Cambridge, Massachusetts, USA \\

IAC & International Avogadro Coordination \\

ILL & Institut Max von Laue-Paul Langevin, Grenoble, France \\

INRIM  & Istituto Nazionale di Ricerca Metrologica, Torino, Italy \\

IRMM & Institute for Reference Materials and Measurements, Geel, Belgium \\

KRISS & Korea Research Institute of Standards and Science, Taedok Science Town,
  Republic of Korea \\

KR/VN & KRISS-VNIIM collaboration \\

$K_{\rm J}$ & Josephson constant: $K_{\rm J} = 2e/h$ \\

$K_{\rm J -90}$ & Conventional value of the Josephson constant
$K_{\rm J}$:
$K_{\rm J -90} = 483\,597.9$~GHz V$^{-1}$ \\

$k$ & Boltzmann constant: $k = R/N_{\rm A}$ \\

LAMPF & Clinton P. Anderson Meson Physics Facility at Los Alamos National
  Laboratory, Los Alamos, New Mexico, USA \\

LKB & Laboratoire Kastler-Brossel, Paris, France \\

LK/SY & LKB and SYRTE collaboration \\

LNE & Laboratoire national de m\'etrologie et d'essais, Trappes, France \\

METAS & Federal Office of Metrology, Bern-Wabern, Switzerland \\

MIT & Massachusetts Institute of Technology, Cambridge, Massachusetts, USA \\

MPQ & Max-Planck-Institut f\"ur Quantenoptik, Garching, Germany \\

$M(X)$ & Molar mass of $X$: $M(X) = A_{\rm r}(X) M_{\rm u}$ \\

Mu & Muonium (${\rmmu}^{+} {\rm e}^{-}$ atom) \\

$M_{\rm u}$ & Molar mass constant: $M_{\rm u} = 10^{-3}~{\rm kg~mol^{-1}}$ \\

$m_{\rm u}$ & Unified atomic mass constant: $m_{\rm u} = m(^{12}{\rm C})/12$ \\

$m_{X}$, $m(X)$ & Mass of $X$ (for the electron e, proton p, and
other elementary particles, the first symbol is used, i.e.,
$m_{\rm e}$, $m_{\rm p}$, etc.) \\

$N_{\rm A}$ & Avogadro constant \\

NIM & National Institute of Metrology, Beijing, China (People's Republic of) \\

NIST & National Institute of Standards and Technology, Gaithersburg,
Maryland and Boulder, Colorado, USA  \\

NMI & National Metrology Institute, Lindfield, Australia \\

NMIJ & National Metrology Institute of Japan, Tsukuba, Japan \\

NPL & National Physical Laboratory, Teddington, UK \\

n & Neutron \\

PTB & Physikalisch-Technische Bundesanstalt, Braunschweig and Berlin, Germany \\

p & Proton \\

$\overline {\rm p} \,^A$He$^+$ & Antiprotonic helium ($^A$He$^{+}$
+ $\overline {\rm p}$ atom, $A= 3 \mbox{ or } 4$)  \\

QED & Quantum electrodynamics \\

$p(\chi^2|\nu)$ & Probability that an observed value of chi-square for
$\nu$ degrees of freedom would exceed $\chi^2$ \\

$R$ & Molar gas constant \\

$\overline R$ & Ratio of muon anomaly difference frequency to
free proton NMR frequency \\

$R_{\rm B}$ & Birge ratio: $R_{\rm B} = (\chi^{2}/\nu) ^\frac {1}{2}$ \\

$r_{\rm d}$ & Bound-state rms charge radius of the deuteron \\

$R_{\rm K}$ & von Klitzing constant: $R_{\rm K} = h/e^{2}$ \\

$R_{\rm K-90}$ & Conventional value of the von Klitzing constant
$R_{\rm K}$:
$R_{\rm K-90} = 25\,812.807~{\rm \Omega}$ \\

$r_{\rm p}$ & Bound-state rms charge radius of the proton \\

$R_\infty$ & Rydberg constant: $R_\infty = m_{\rm e}c\alpha^{2}
/2h$ \\

$r(x_i,x_j)$ & Correlation coefficient of estimated values $x_i$ and $x_j$:
$r(x_i,x_j) = u(x_i,x_j)/[u(x_i) u(x_j)]$ \\

$S_{\rm c}$ & Self-sensitivity coefficient \\

SI & Syst\`eme international d'unit\'es (International System of Units) \\

StanfU & Stanford University, Stanford, California, USA \\

StockU & Stockholm University, Stockholm, Sweden \\

StPtrsb & St. Petersburg, Russian Federation \\

SYRTE & Syst\`emes de r\'ef\'erence Temps Espace, Paris, France \\

$T$ & Thermodynamic temperature \\

Type A & Uncertainty evaluation by the statistical analysis of series
of observations \\

Type B & Uncertainty evaluation by means other than the statistical analysis of
  series
of observations \\

$t_{90}$ & Celsius temperature on the International Temperature Scale of 1990
  (ITS-90) \\

t & Triton (nucleus of tritium T, or $^3$H) \\

USus & University of Sussex, Sussex, UK \\

UWash & University of Washington, Seattle, Washington, USA \\

u & Unified atomic mass unit (also called the dalton, Da): 1 u = $m_{\rm u}$ =
  $m(^{12}$C)/12 \\

$u(x_i)$ & Standard uncertainty (i.e., estimated standard deviation) of an
estimated value $x_i$ of a quantity $X_i$ (also simply $u$) \\

$u_{\rm r}(x_i)$ & Relative standard uncertainty of an
estimated value $x_i$ of a quantity $X_i$: \newline
$u_{\rm r}(x_i) = u(x_i)/|x_i|, \ x_i \ne 0$ (also simply $u_{\rm r}$) \\

$u(x_i,x_j)$ & Covariance of estimated values $x_i$ and $x_j$ \\

$u_{\rm r}(x_i,x_j)$ & Relative covariance of estimated values $x_i$ and $x_j$:
$u_{\rm r}(x_i,x_j) = u(x_i,x_j)/(x_i x_j)$ \\

$V_{\rm m}({\rm Si)}$ & Molar volume of naturally occurring silicon \\

VNIIM & D. I. Mendeleyev All-Russian Research Institute for Metrology, St.
  Petersburg, Russian
Federation  \\

$V_{90}$ & Conventional unit of voltage based on the Josephson effect and
$K_{\rm J -90}$: $V_{90} = (K_{\rm J -90}/K_{\rm J}$) V \\

$W_{90}$ & Conventional unit of power:
$W_{90} = V^{2}_{90}/{\it \Omega}_{90}$ \\

XROI & Combined x-ray and optical interferometer \\

xu(CuK${\rmalpha}_1$) & Cu x unit: ${\rm \lambda}$(CuK${\rmalpha}_1$) =
1\,537.400 xu(CuK${\rmalpha}_1$) \\

xu(MoK${\rmalpha}_1$) & Mo x unit: ${\rm \lambda}$(MoK${\rmalpha}_1$) =
707.831 xu(MoK${\rmalpha}_1$) \\

$x(X)$ & Amount-of-substance fraction of $X$ \\

YaleU & Yale University, New Haven, Connecticut, USA \\

$\alpha$ & Fine-structure constant: $\alpha = e^2/4\rmpi\epsilon_0\hbar c
\approx 1/137 $ \\

${\rmalpha}$ & Alpha particle (nucleus of $^{4}$He) \\

$\it\Gamma ^{\prime}_{X-{\rm 90}}$(lo) & $\it\Gamma ^{\prime}_{X-{\rm 90}}({\rm
  lo})
= (\gamma_{X}^\prime \, A_{\rm 90})$~A$^{-1}$, $X$ = p or h \\

$\it\Gamma ^{\prime}_{\rm p-90}$(hi) & $\it\Gamma ^{\prime}_{\rm p-90}({\rm hi})
= (\gamma_{\rm p}^\prime / A_{\rm 90})$~A \\

$\gamma_{\rm p}$ & Proton gyromagnetic ratio:~$\gamma_{\rm p} =
2\mu_{\rm p} / \hbar$ \\

$\gamma_{\rm p}^\prime$ & Shielded proton gyromagnetic ratio:
$\gamma_{\rm p}^\prime = 2\mu^\prime_{\rm p}/\hbar$ \\

$\gamma^{\prime}_{\rm h}$ & Shielded helion gyromagnetic ratio:
$\gamma ^{\prime}_{\rm h} = 2|\mu^{\prime}_{\rm h}|/ \hbar $ \\

$\Delta\nu_{\rm Mu}$ & Muonium ground-state hyperfine splitting \\

$\delta_{\rm e}$ & Additive correction to the theoretical expression
for the electron magnetic moment
anomaly $a_{\rm e}$ \\

$\delta_{\rm Mu}$ & Additive correction to the theoretical expression for the
  ground-state
hyperfine splitting of muonium ${\rm \Delta \nu}_{\rm Mu}$ \\

$\delta_{\overline {\rm p} \,{\rm He}}$ & Additive correction to the
theoretical expression for a particular transition frequency of antiprotonic
  helium \\

$\delta_{X} (n{\rm L}_{j})$ & Additive correction to the theoretical expression
  for an energy
level of either hydrogen H or deuterium D with quantum numbers $n$, L, and $j$
  \\

$\delta_{\rmssmu}$ & Additive correction to the theoretical expression for the
  muon magnetic moment anomaly $a_{\rmssmu}$ \\

$\epsilon_{\rm 0}$ & Electric constant: $\epsilon_{\rm 0} = 1/\mu_{\rm 0} c^2$
  \\

$\doteq$ & Symbol used to relate an input datum to its observational equation \\

${\rm \lambda}({X\,{\rm K}}{\rmalpha}_1)$ & Wavelength of K${\rmalpha}_1$ x-ray
  line of element $X$ \\

${\rm \lambda}_{\rm meas}$ & Measured wavelength of the 2.2 MeV capture
${\rmgamma}$-ray emitted in the
reaction ~n + p $\rightarrow$ d + ${\rmgamma}$ \\

${\rmmu}$ & Symbol for either member of the muon-antimuon pair; when necessary,
  ${\rmmu}^{-}$ or ${\rmmu}^{+}$
is used to indicate the negative muon or positive muon \\

$\mu_{\rm B}$ & Bohr magneton: $\mu_{\rm B} = e \hbar/2m_{\rm e}$ \\

$\mu_{\rm N}$ & Nuclear magneton: $\mu_{\rm N} = e\hbar/2m_{\rm p}$ \\

$\mu_{X}(Y)$ & Magnetic moment of particle $X$ in atom or molecule $Y$. \\

$\mu_{\rm 0}$ & Magnetic constant:
$\mu_{\rm 0} = 4{\rmpi}\times 10^{-7}$~N/A$^{2}$ \\

$\mu_{X}$,~$\mu^\prime_{X}$ &
Magnetic moment, or shielded magnetic moment, of particle $X$ \\

$\nu$ & Degrees of freedom of a particular adjustment \\

$\nu(f_{\rm p})$ & Difference between muonium hyperfine splitting Zeeman
  transition frequencies
$\nu_{34}$ and $\nu_{12}$ at a magnetic flux density $B$ corresponding to the
  free
proton NMR frequency $f_{\rm p}$ \\

$\sigma$ & Stefan-Boltzmann constant: $\sigma = 2{\rmpi}^{5}k^{4}/(15h^{3}c^{2}
  )$ \\

${\rmtau}$ & Symbol for either member of the tau-antitau pair;
when necessary, ${\rmtau}^{-}$ or
${\rmtau}^{+}$ is used to indicate the negative tau or positive tau \\

${\rm \chi}^{2}$ & The statistic ``chi square'' \\

${\it \Omega}_{90}$ & Conventional unit of resistance based on the quantum Hall
effect and $R_{\rm K -90}: {\it \Omega}_{90} = (R_{\rm K}/R_{\rm K -90})~{\rm
  \Omega}$

\end{longtable}

\end{document}